\documentclass[a4paper,12pt,oneside,onecolumn,openany,final]{memoir}

\setlrmarginsandblock{3cm}{2.5cm}{*}
\setulmarginsandblock{2.5cm}{3cm}{*}
\setsecnumdepth{subsection}
\chapterstyle{brotherton}
\newsubfloat{figure}
\checkandfixthelayout

\usepackage[UKenglish]{babel}
\usepackage[utf8]{inputenc}
\usepackage[T1]{fontenc}

\usepackage{graphicx} 
\usepackage{amsmath,amssymb}
\usepackage{slashed}
\usepackage{xcolor} 
\usepackage{feynmp}
\usepackage[maxauthors=3, usetitle, usedoi, linkdoi, super=false]{rsc}

\begin{document}

\begin{titlingpage}
\begin{center}
\includegraphics[height=4cm]{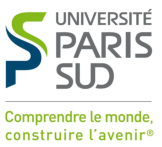}
\hfill
\includegraphics[height=4cm]{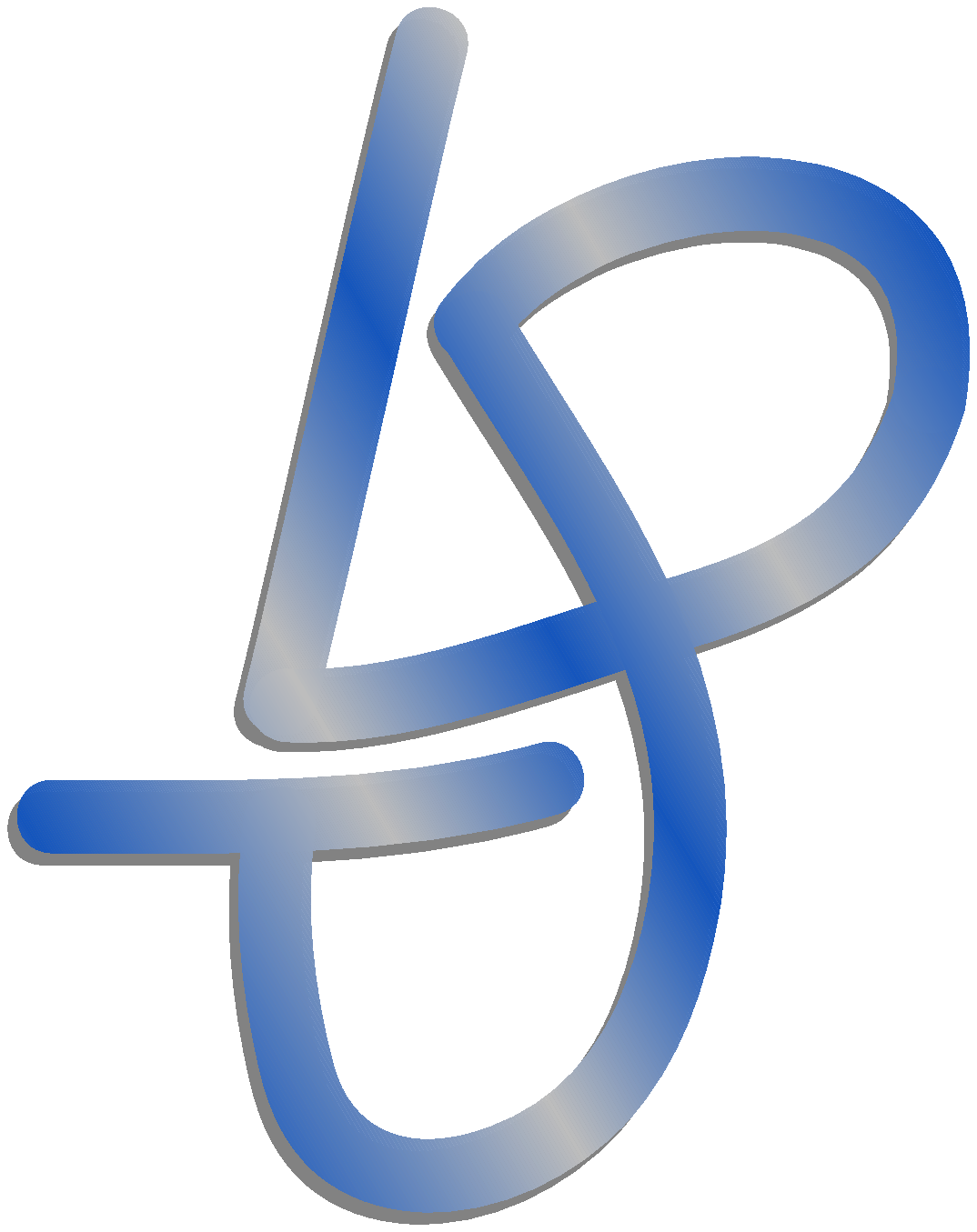}
\end{center}
\vspace{0.5cm}
\begin{flushleft}
\begin{minipage}[c]{\textwidth}
\centering
\LARGE \textbf{UNIVERSITÉ PARIS-SUD XI}\\
\large ÉCOLE DOCTORALE 517 ``PARTICULES, NOYAUX ET COSMOS''\\
Laboratoire de Physique Théorique - UMR 8627
\end{minipage}
\end{flushleft}
\vspace{0.5cm}
\begin{flushleft}
\begin{minipage}[c]{\textwidth}
\centering
\LARGE \textbf{THÈSE DE DOCTORAT}\\
\LARGE \textbf{ EN PHYSIQUE THÉORIQUE}\\
\vspace*{0.5cm}
\normalsize soutenue publiquement le 04 juillet 2013
\end{minipage}
\end{flushleft}
\vspace{0.3cm}
\begin{flushleft}
\begin{minipage}[c]{\textwidth}
\centering
\normalsize par
\end{minipage}
\end{flushleft}
\vspace{0.3cm}
\begin{flushleft}
\begin{minipage}[c]{\textwidth}
\centering
\LARGE \textbf{Cédric WEILAND}
\end{minipage}
\end{flushleft}
\hspace*{-1cm}
\begin{center}
\begin{tabular}{|c|}
\hline \vspace{-10pt}\\
\LARGE \textbf{Effects of fermionic singlet neutrinos}\\
\LARGE \textbf{on high- and low-energy observables}\\
\hline
\end{tabular}
\end{center}
\begin{flushleft}
\begin{tabular}{lll}
\normalsize
\textbf{Directrice de thèse:} & Asmâa ABADA & Professeur \vspace{0.2cm}\\
\underline{\textbf{Composition du jury}} &  & \vspace{0.2cm}\\
\textit{Président du jury:} & Achille Stocchi & Professeur\\
\textit{Rapporteurs:} & María José Herrero Solans & Professeur\\
 & Thomas Schwetz-Mangold & Chargé de recherche\\
\textit{Examinateurs:} & Martin Hirsch & Chargé de recherche\\
 & Stéphane Lavignac & Chargé de recherche
\end{tabular}
\end{flushleft}
\end{titlingpage}

\thispagestyle{plain}
\setcounter{page}{3} 
\vspace*{\fill}
\includegraphics[height=1cm]{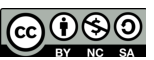}\\
Thèse publiée sous licence Creative Commons Attribution Pas d'Utilisation Commerciale Partage à l'Identique 3.0

\newpage
\thispagestyle{plain}
\begin{center}

\large EFFETS DES NEUTRINOS SINGULETS FERMIONIQUES SUR LES OBSERVABLES DE HAUTE ET BASSE ÉNERGIE\\
\vspace{0.5cm}
\normalsize \textbf{Résumé}

\end{center}

Dans cette thèse, nous étudions à la fois des observables de basse et de haute \hyphenation{é-ner-gie}énergie liées à la présence de neutrinos massifs. Les oscillations de neutrino ont apporté des preuves indiscutables en faveur de l'existence d'angles de mélange et de masses non-nuls. Néanmoins, la formulation originale du Modèle Standard ne permet pas d'expliquer ces observations, d'où la nécessité d'introduire de nouveaux modèles. Parmi de nombreuses possibilités, nous nous concentrons ici sur le seesaw inverse, un mécanisme générant des neutrinos massifs via l'ajout de fermions singulets de jauge au Modèle Standard. Ce modèle offre une alternative attractive aux réalisations habituelles du seesaw puisqu'il a des couplages de Yukawa potentiellement naturels ($\mathcal{O}(1)$) tout en conservant l'échelle de la nouvelle physique à des énergies accessibles au LHC. Parmi de nombreux effets, ce scénario peut générer de larges écarts à l'universalité leptonique. Nous avons étudié ces signatures et trouvé que les rapports $R_K$ et $R_\pi$ constituent de nouvelles contraintes pour le seesaw inverse. Nous nous sommes aussi intéressés à l'intégration de l'inverse seesaw dans différents modèles supersymétriques. Ceci conduit à une augmentation de la section efficace de  divers processus violant la saveur leptonique du fait de contributions plus importantes venant des diagrammes pingouins comportant un boson de Higgs ou $Z^0$. Finalement, nous avons aussi montré que les nouveaux canaux de désintégration ouverts par la présence de neutrinos stériles dans les modèles de seesaw inverse supersymétriques peuvent significativement relaxer les contraintes sur la masse et les couplages d'un boson de Higgs CP-impair.\\

\begin{flushleft}
\textbf{Mots Clés:} Modèle Standard, Neutrinos, Inverse Seesaw, Supersymétrie, Violation de la saveur leptonique, Violation de l'universalité leptonique\\
\vspace{0.5cm}
Thèse préparée au LABORATOIRE DE PHYSIQUE THÉORIQUE D'ORSAY, Bâtiment 210, Université Paris-Sud 11, 91405 Orsay Cedex
\end{flushleft}

\newpage
\thispagestyle{plain}
\begin{center}

\large EFFECTS OF FERMIONIC SINGLET NEUTRINOS ON HIGH- AND LOW-ENERGY OBSERVABLES\\
\vspace{0.5cm}
\normalsize \textbf{Abstract}

\end{center}

In this doctoral thesis, we study both low- and high-energy  observables related to massive neutrinos.   Neutrino oscillations have provided indisputable evidence in favour  of non-zero neutrino masses and mixings.   However, the original formulation of the Standard Model cannot  account for these observations, which calls for the introduction of new Physics. Among many possibilities, we focus here on  the inverse seesaw, a neutrino mass generation mechanism in which the Standard Model is extended with fermionic gauge  singlets. This model offers an attractive alternative to the usual seesaw  realisations since it can potentially have natural Yukawa couplings ($\mathcal{O}(1)$) while keeping the new  Physics scale at energies within reach of the LHC. Among the many possible effects, this scenario can lead to deviations  from lepton flavour universality. We have investigated these signatures and found that the ratios $R_K$  and $R_\pi$ provide new, additional constraints on the inverse seesaw. We have also considered the embedding of the inverse seesaw in supersymmetric models. This leads to increased rates for  various lepton flavour violating processes, due to enhanced contributions from penguin diagrams mediated by the Higgs and $Z^0$  bosons. Finally, we also found that the new invisible decay channels associated with the sterile neutrinos present in the supersymmetric inverse seesaw could significantly weaken the constraints on the mass and couplings of a light CP-odd Higgs boson.\\

\begin{flushleft}
\end{flushleft}
\vspace{0.5cm}
Thesis prepared at the LABORATOIRE DE PHYSIQUE THÉORIQUE D'ORSAY, Bâtiment 210, Université Paris-Sud 11, 91405 Orsay Cedex

\chapter*{Remerciements}

Je voudrais tout d'abord remercier Henk Hilhorst, le directeur du Laboratoire de Physique Théorique d'Orsay, pour son accueil chaleureux au sein du laboratoire mais aussi le personnel administratif qui aura grandement facilité mon séjour et les nombreuses démarches liées à la vie d'un jeune chercheur. Je garderai d'agréables souvenirs de mes trois années passées au LPT, que ce soit pour l'ambiance détendue mais studieuse ou la disponibilité des chercheurs du laboratoire. Je pense spécialement à Damir Becirevic qui aura pris le temps de discuter avec moi quand l'heure de la recherche d'un postdoc fut venue.

Je tiens tout particulièrement à remercier les membres de mon jury de thèse qui ont accepté de lire ma thèse et de se déplacer, pour certains depuis l'étranger. Merci à Achille Stocchi d'avoir assuré la présidence de ce jury, merci à María José Herrero Solans et à Thomas Schwetz-Mangold d'avoir accepté d'être rapporteurs et de relire mon manuscrit dans un délai aussi court, merci à Martin Hirsch et à Stéphane Lavignac pour leur présence et les discussions qui suivirent ma soutenance. Enfin, merci pour toutes les remarques, suggestions et corrections qui contribuèrent à améliorer la qualité de ce manuscrit.

Parmi les membres du jury, Asmâa Abada tient une place particulière. Elle a en effet accepté d'être ma directrice de thèse, veillant sur ma croissance scientifique, partageant son expérience et me délivrant de judicieux conseils. Elle a aussi supporté mes nombreuses questions, y répondant toujours de bon c{\oe}ur, ainsi que mon entêtement à vouloir explorer certains sujets. Pour tout cela, je lui suis reconnaissant et redevable. 

Je voudrais aussi remercier Debottam Das et Avelino Vicente avec qui j'ai collaboré durant ma thèse. Nous avons eu de nombreuses discussions enrichissantes et ils ont su me transmettre leurs connaissances. Un grand merci à Ana Teixeira en particulier. Elle m'a appris beaucoup en Physique mais aussi sur le monde de la recherche. Je n'aurais pu espérer de meilleure collaboratrice pour ma dernière année de thèse.

J'ai également une pensée pour tous les doctorants du laboratoire et les discussions toujours passionnantes que nous avons eues, qu'elles aient été scientifiques ou non. En particulier, je remercie ceux qui ont partagé mon bureau: Julien Baglio, Adrien Besse et Michele Lucente. Ils ont supporté mes tics et défauts, faisant toujours preuve d'un bon esprit rafraîchissant. Merci aussi à Guillaume Toucas qui m'a beaucoup appris sur la QCD et a toujours été partant pour une bonne bière.

Je souhaiterais aussi remercier Hitoshi Murayama, l'IPMU et la JSPS de m'avoir invité au Japon. Ce séjour fut enrichissant à tout point de vue. J'y découvris une nouvelle culture et un nouveau domaine de la physique théorique. J'eus l'occasion de partager une machiya avec des gens géniaux (Sophie Blondel, James Mott et Jonathan Davies) et de présenter mes travaux lors de Neutrino 2012. Enfin, j'y ai rencontré de nombreux autres jeunes (ou moins jeunes) chercheurs dont l'enthousiasme fut communicatif. J'ai une pensée particulière pour Richard Ruiz, compagnon de travail mais aussi camarade lors de nombreuses sorties tokyoïtes. J'ai bien conscience de ne pouvoir nommer tout ceux qui rendirent ce séjour inoubliable, qu'ils m'en pardonnent.

Une composante indéniable du succès de cette thèse a été le soutien de ma famille, qu'elle en soit remerciée, tout particulièrement ma mère. Enfin, ces trois années auraient été complètement différentes sans la présence en Île-de-France de Romain Vasseur, Mandy Muller, Alexandre Lazarescu et Wahb Ettoumi. Ils ont été les cobayes malgré eux de mes expérimentations culinaires et mixologiques lors de nombreuses soirées mais aussi les meilleurs compagnons de voyage possibles pour découvrir New York et l'Ouest américain. Nous avons partagé nos joies et nos peines durant notre doctorat, ils ont été mes amis durant ces six dernières années.

\chapter*{Acknowledgements}

I would like to begin by thanking Henk Hilhorst, the director of the Laboratoire de Physique Théorique d'Orsay, for his warm welcome to the laboratory but also the administrative staff who made my stay easier and helped with the numerous procedures that arise in the life of a young researcher. I will keep pleasant memories of the three years that I spent in the LPT, be it for the relaxed but studious atmosphere or the availability of the researchers. I think in particular of Damir Becirevic who spent a lot of time discussing with me when I was looking for a postdoc.

I want to deeply thank the members of my examination committee who agreed to read my thesis and had to travel, some of them from foreign countries. Thanks to Achille Stocchi who presided this committee, thanks to María José Herrero Solans and Thomas Schwetz-Mangold who agreed to report on my thesis on such a short time, thanks to Martin Hirsch and Stéphane Lavignac for coming and for the discussions we had after my defence. Last but not least, thanks for all the remarks, suggestions and corrections that helped improve this manuscript.

Among the member of my jury, Asmâa Abada has a special place. She accepted to be my PhD advisor, watching over my scientific growth, sharing her experience and giving me sensible pieces of advice. She also endured my numerous questions, always answering them wholeheartedly, and my stubbornness to explore some topics. For all this, I am grateful and deeply indebted to her.

I would also like to thank Debottam Das and Avelino Vicente, who I collaborated during my PhD. We had numerous discussion and they knew how to pass their knowledge. Big thanks to Ana Teixeira in particular. She taught me a lot about Physics and the world of academia. I could not have hoped for a better collaborator during the last year of my PhD.

I have a thought for all the PhD students of the laboratory and the always fascinating discussions that we had, be them scientific or not. I especially thank those who shared my office: Julien Baglio, Adrien Besse and Michele Lucente. They had to deal with my tic and faults, but they always did it with a refreshingly good spirit. I also thank Guillaume Toucas who taught me a lot about QCD and was always happy to grab a good beer.

I would also like to think Hitoshi Murayama, the IPMU and the JSPS for inviting me to Japan. This stay taught me so many things and was one of the greatest opportunity I had during my PhD. I have discovered a new culture and a new field of theoretical Physics. I had the possibility to share a machiya with amazing peoples (Sophie Blondel, James Mott and Jonathan Davies) and to present my work during Neutrino 2012. Finally, I met there many young (and not so young) researcher whose enthusiasm was contagious. I especially think of Richard Ruiz, a comrade at work but also a partner in crime when we went out to discover Tokyo. I am aware that I cannot list all the persons who made this stay impossible to forget, I ask for their forgiveness if their name does not appear here.

The support of my family, especially my mother, is an indisputable part of the success of this PhD and I want to thank them for it. To conclude, this past three years would have been much different if Romain Vasseur, Mandy Muller, Alexandre Lazarescu and Wahb Ettoumi had not lived in the Paris area. They were test subjects for my cooking and mixing experiments during innumerable parties but they were also the best travel companions when discovering New York and the American West. We shared our joys and pains during our PhD, they have been my friends for the past six years.

\newpage
\tableofcontents

\newpage
\listoffigures

\newpage
\listoftables

\newpage

\chapter*{List of Publications}
\addcontentsline{toc}{chapter}{List of Publications}

The following papers were written during the 3 years of my PhD.\\
Les papiers suivants ont été écrits durant les 3 années de ma thèse.\vspace{0.5cm}\\

\textbf{Published papers / Articles publiés :}\vspace{0.5cm}\\
\textbf{``A possible connection between neutrino mass generation and the lightness of a NMSSM pseudoscalar''},\\
Asmaa Abada, Gautam Bhattacharyya, Debottam Das and Cédric Weiland\\
\textbf{Phys. Lett. B700 (2011) 351-355}, arXiv:1011.5037[hep-ph]\vspace{0.2cm}\\
\textbf{``Enhanced Higgs Mediated Lepton Flavor Violating Processes in\\ the Supersymmetric Inverse Seesaw Model''},\\
Asmaa Abada, Debottam Das and Cédric Weiland\\
\textbf{JHEP03(2012)100}, arXiv:1111.5836[hep-ph]\vspace{0.2cm}\\
\textbf{``Enhancing lepton flavour violation in the supersymmetric inverse seesaw beyond the dipole contribution''},\\
Asmaa Abada, Debottam Das, Avelino Vicente, Cédric Weiland\\
\textbf{JHEP09(2012)015}, arXiv:1206.6497[hep-ph]\vspace{0.2cm}\\
\textbf{``Tree-level lepton universality violation in the presence of sterile neutrinos: impact for $R_K$ and $R_\pi$''},\\
A. Abada, D. Das, A.M. Teixeira, A. Vicente and C. Weiland\\
\textbf{JHEP02(2013)048}, arXiv:1211.3052[hep-ph]\vspace{0.5cm}\\

\textbf{Proceedings / Comptes-rendus :}\vspace{0.5cm}\\
\textbf{``Enhanced Higgs Mediated Lepton Flavor Violating Processes in the\\ Supersymmetric Inverse Seesaw Model''},\\
Cédric Weiland\\
\textbf{Proceedings of the International School of Physics ”Enrico Fermi”, Course CLXXXII}, arXiv:1205.6400[hep-ph]\vspace{0.2cm}\\
\textbf{``A natural connection between neutrino mass generation and the lightness of a next-to-minimal supersymmetric Standard Model pseudoscalar''},\\
Debottam Das, Asmaa Abada, Gautam Bhattacharyya, Cédric Weiland\\
\textbf{Pramana 79 (2012) 867-870}\vspace{0.2cm}\\
\textbf{``Enhanced lepton flavour violation in the supersymmetric inverse seesaw''},\\
C Weiland\\
\textbf{J. Phys.: Conf. Ser. 447 (2013) 012037}, arXiv:1302.7260[hep-ph]\vspace{0.2cm}\\
\textbf{``Lepton universality in kaon decays''},\\
C. Weiland\\
arXiv:1306.2894[hep-ph]\\

\newpage
\thispagestyle{plain}
\vspace*{7cm}
\begin{epigraphs}
 \qitem{Die Forschung war aufs vernehmlichste daran erinnert worden, dass das Fundament der Wissenschaft, wenn ihr Gebäude höher und höher geführt wird, gleichzeitig in die Tiefe sinken muss, wenn es sein Gewicht noch tragen soll. Denn der Boden, in dem dieses Gebäude ruht, ist ja nicht der Fels einer sicheren vor aller Wissenschaft stehenden Erkenntnis, sondern ist das fruchtbare Erdreich der Sprache, die aus Handeln und Erfahren sich bildet.}{\textit{Ordnung der Wirklichkeit}\\ \textsc{Werner Heisenberg}}
\end{epigraphs}

\chapter*{Introduction}

Despite being the most common matter fermions that we know, neutrinos remains mysterious particles. While they are approximately as numerous as the photons in the Universe (for every electron there exist more than ten billion neutrinos), they are so elusive that a light-year of lead would only stop half of the neutrinos emitted by the Sun. Their existence is crucial to supernovae, which produce heavy elements, but also to nuclear $\beta$-decays which are at the heart of the thermonuclear reactions of stellar evolution. Their density also has important consequences in the evolution of the early Universe.

Focusing on Particle Physics, neutrinos are unique particles since they only interact weakly. Their masses are a million times smaller than the electron mass and, being  electrically neutral, they have the unique possibility to be their own antiparticle. This has fuelled the interest in the experimental determination of their properties but also in theoretical models that could explain neutrino masses and mixings. The seesaw mechanisms are very attractive options since they can generate naturally small neutrino masses and could possibly address the problem of the baryonic asymmetry of the Universe. Among them, the inverse seesaw seems very appealing since its naturally low scale can lead to sizeable effects in a number of observables that can be tested at the current generation of experiments.

This dissertation is divided into three parts, the first one presenting the Standard Model of Particle Physics and the neutrino sector. In the first chapter, we focus on the Standard Model, providing a brief historical introduction before proceeding to describe the model in detail. In Chapter~\ref{ChapNuExp} after a historical account of experimental and theoretical progresses in the neutrino sector, we present the basic principle behind neutrino oscillations and the current experimental efforts to determine the parameters associated with the neutrino sector.

In the second part of this thesis, we discuss neutrino mass generation mechanisms and their impact on lepton universality tests. The Chapter~\ref{chap3} introduces the theory of massive fermions and the impossibility to generate neutrino masses in the Standard Model. This calls for new Physics: one possibility is the seesaw mechanism in its different realisations, which can be embedded in the Standard Model or in extended frameworks. Among the seesaw realisations, we have focused here on the inverse seesaw. In Chapter~\ref{chaptLFU}, we describe how some of the best predictions of the Standard Model, which have been measured with an uncertainty below one percent, can be modified by the inclusion of extra fermionic singlets.

Finally, in the last part, we consider extended frameworks. In particular, we focus on supersymmetric extensions of the Standard Model which addresses some of the issues of the Standard Model. In Chapter~\ref{chapSUSY}, we introduce supersymmetric models while, in Chapter~\ref{CPoddLFV}, we discuss how embedding the inverse seesaw in supersymmetric frameworks can lead to interesting experimental signatures like invisible CP-odd Higgs boson decays and lepton flavour violation.

The original results derived during this PhD thesis are collected in Chapters~\ref{chaptLFU} and~\ref{CPoddLFV}.

\let\oldafterpartskip\afterpartskip 
\renewcommand*{\afterpartskip}{\epigraph{All opinions are not equal. Some are a very great deal more robust, sophisticated and well supported in logic and argument than others.}{\textit{The Salmon of Doubt}\\ \textsc{Douglas Adams}}
\vfil\clearpage}
\part{{}The Standard Model and Neutrino Physics}

\chapter{Introducing the Standard Model}

The Standard Model (SM) of Particle Physics is one of the most successful and thoroughly tested physical theory. Building on Quantum Mechanics and Special Relativity, it gives a coherent picture of the microscopic world with its fundamental blocks and their interactions in the language of quantum field theory and gauge theory. Describing experimental results with a remarkable precision, it can also make definite predictions. Since it contains falsifiable statements, the SM thus verifies Karl Popper's criterion of demarcation between scientific and unscientific theories. 

Being convinced of its importance in modern Physics, we now provide a short historical introduction to the Standard Model. In a second section, we will brush a theoretical portrait of this forty years old model. 

\section{A tapestry ``made by many artisans''\label{SMhistory}}

As Sheldon Glashow, Abdus Salam and Steven Weinberg reminded us in their respective Nobel lectures in 1979, the Standard Model was not devised by one person. Its formulation was a collective effort, spanning decades and making use of many different contributions. This short historical account will focus on some of the theoretical ideas and experimental results which made the development of this ``integral work of art''~\cite{Glashow:1979pj} possible. It is of the utmost importance to keep in mind that the SM did not originate from a clear, definite and evident research program. It is the combination of theories that were devised  in parallel, exchanging concepts and building on the successes of each others.

\subsubsection{Quantum electrodynamics}

The first among those theories was Quantum Electrodynamics (QED). Its foundations were laid in 1927 when Paul Dirac introduced the annihilation and creation operators~\cite{Dirac:1927dy}. The following year,  trying to get rid of the negative energy solutions that appeared in the Klein-Gordon equation, a relativistic version of the Schrödinger equation, he formulated the relativistic wave equation for the electron~\cite{Dirac:1928hu} and subsequently used it to derive the magnetic moment of that particle~\cite{Dirac:1928ej}. However, when he computed the self-energy of the electron in 1931, Robert Oppenheimer found that this theory diverges at high-energies~\cite{PhysRev.35.461}. The existence of divergences in QED and in other field theories would prove crucial to the development of the Standard Model, since it would later drive the search for a theory of weak interactions beyond the one formulated by Enrico Fermi.

A first way to overcome low-energy divergences was proposed in 1937 by Felix Bloch and Arnold Nordsieck~\cite{PhysRev.52.54}. Later, in 1947, Hans Bethe introduced the procedure of renormalization in his calculation of the energy shift of the levels of a hydrogen atom~\cite{PhysRev.72.339}, which showed excellent agreement with the previous measurement by Willis Lamb and Robert Retherford~\cite{PhysRev.72.241}. The same year, Henry Foley and Polykarp Kusch measured the anomalous moment of the electron~\cite{PhysRev.73.412}, which would be calculated by Julian Schwinger one year later~\cite{Schwinger:1948iu} and found to agree within 3\%.

Building on a first paper by Shin'ichir\={o} Tomonaga~\cite{Tomonaga:1946zz} that introduced a covariant formulation of QED, Richard Feynman~\cite{PhysRev.74.1430, *PhysRev.76.769}, J.~S. Schwinger~\cite{PhysRev.74.1439, *PhysRev.75.651}, Takao Tati and S.~I. Tomonaga~\cite{PTP.2.101, *PTP.3.391} wrote down, in 1948, a fully covariant and renormalized version of QED. The following year, Freeman Dyson demonstrated the equivalence of Feynman's, Schwinger's and Tomonaga's theories~\cite{PhysRev.75.486} and linked covariant QED~\cite{PhysRev.75.1736} with the $S$~matrix formalism developed by Werner Heisenberg in 1943~\cite{heisenberg1943beobachtbaren}. With this, a complete description of QED had been devised and the most important subsequent developments would be the Ward identities~\cite{Ward:1950xp} and the introduction of the renormalization group by Ernst Stückelberg and André Petermann~\cite{stuckelberg1953normalisation}, which was later explored by Murray Gell-Mann and  Francis Low~\cite{PhysRev.95.1300}.

\subsubsection{Early theories of weak interactions}

While QED progressively provided a complete picture of particles interacting through electromagnetism, the elaboration of a theory describing the weak force was a very eventful journey. Many of the early preconceptions about Physics had to be cast away, new particles were predicted and, in the end, it was found that the weak force was only a low-energy component of the electroweak theory.

Assuming the existence of the neutrino proposed by Wolfgang Pauli four years earlier, E. Fermi wrote down in 1934 a four-fermion interaction~\cite{fermi1934versuch, *fermi1934tentativo} inspired  by the electromagnetic current,
\begin{equation}
 \mathcal{L}_{weak}=\frac{G_F}{\sqrt{2}}(\bar{\psi}_p \gamma^\mu \psi_n)(\bar{\psi}_e \gamma_\mu \psi_\nu)\,.
\end{equation}
This would be extended in 1936 by George Gamow and Edward Teller to include scalar, pseudo-scalar, vector, axial and tensor structures~\cite{PhysRev.49.895}. In 1949, Jayme Tiomno and John Wheeler~\cite{RevModPhys.21.153}, independently from Tsung-Dao Lee, Marshall Rosenbluth and Chen-Ning Yang~\cite{PhysRev.75.905}, proposed that Fermi weak interactions were universal, which is to say that the coupling constant is the same for different processes like $\beta$ decay or muon capture.

Until then it was thought that parity, the reversal of spatial axes, was a fundamental symmetry of nature, leaving the laws of Physics invariant. However, what was supposed to be two mesons decaying into final states with opposite parity, $\theta^+ \rightarrow \pi^+ \pi^0$ and $\tau^+ \rightarrow \pi^+ \pi^+ \pi^-$, seemed to be a single particle since they had the same decay width and the same mass as measured by Luis Alvarez and Sulamith Goldhaber~\cite{alvarez1955lifetime} and Robert Birge, James Peterson, Donald Stork and Marian Whitehead~\cite{PhysRev.100.430}. That situation was known as the $\theta-\tau$ puzzle. T.~D. Lee and C.~N. Yang suggested in 1956 that parity could be violated in weak interactions along with possible experimental tests~\cite{PhysRev.104.254}. This was confirmed by the measure of the angular distribution of electrons in $\beta$ decay of $^{60}\mathrm{Co}$ the following year by Chien-Shiung Wu \textit{et al.}~\cite{PhysRev.105.1413}.  Then, in 1958, the violation of parity was incorporated in the theory of weak interactions by R.~P.  Feynman and M. Gell-Mann~\cite{PhysRev.109.193}, Robert Marshak and George Sudarshan~\cite{PhysRev.109.1860.2} and Jun Sakurai~\cite{sakurai1958mass}. This was done by considering a universal $V-A$ current for the Fermi interaction. Thus the weak interactions would violate parity (and charge conjugation) but conserve their product, CP. But that symmetry was found, a few years later, to be violated too.

\subsubsection{Non-Abelian gauge theories and spontaneous symmetry breaking}

From a gauge theory point of view, QED is relatively simple since it is based on the abelian group $\mathrm{U}(1)$. However this can only describe interactions with one intermediate vector boson. In a seminal paper written in 1954, C.~N. Yang and Robert Mills extended the concept of gauge theories
to non-abelian groups~\cite{PhysRev.96.191}. They tried to construct a theory of the strong interactions based on the isotopic spin conservation group $\mathrm{SU}(2)$. Their model succeeded in having three intermediate vector bosons with electric charge $0$ and $\pm e$ when, at the time, pions were thought to be the mediators of the strong force. However, they predicted a massless vector boson while pions were known to have a non-zero mass of approximately $135\;\mathrm{GeV}$. A. Salam and John Ward introduced in 1961 the gauge principle as a method for building interacting field theories~\cite{Salam:1961en}. However, gauge theories  were not really well thought of at that time. Indeed, the vector bosons must remain massless in order to preserve gauge invariance, which contradicted the idea that the bosons mediating the forces should be massive since weak and strong interactions are short-ranged.

This apparent contradiction was solved three years later when Gerald Guralnik, Carl Hagen and Tom Kibble~\cite{PhysRevLett.13.585}, Robert Brout and François Englert~\cite{PhysRevLett.13.321} and Peter Higgs~\cite{Higgs1964132, *PhysRevLett.13.508} proposed a mechanism for spontaneous symmetry breaking that gives rise to massive vector bosons. Their work built on an earlier discovery by Yoichiro Nambu~\cite{PhysRevLett.4.380} and Jeffrey Goldstone~\cite{Goldstone1961} that a symmetry can be conserved at the current level while the vacuum is not invariant under the action of the corresponding generators, spontaneously breaking the symmetry. Moreover, if a continuous global symmetry is broken in that way, new massless scalar bosons appear in the theory. In 1967, T. Kibble extended this mechanism to non-Abelian gauge theories~\cite{Kibble:1967sv}. The same year, Ludvig Faddeev and Victor Popov devised a method for the calculation of Feynman diagrams in Yang-Mills theories~\cite{Faddeev:1967fc}. As for QED, the renormalizability of  those theories remained an open question for some time. Indeed, it was only in 1971 that Gerard 't Hooft demonstrated that massless and massive Yang-Mills theories with spontaneous symmetry breaking are renormalizable.

\subsubsection{The quark model and quantum chromodynamics}

While theorists  were developing the tools of gauge theories, experimentalists were discovering new particles at an increasing rate, especially since they were shifting from the study of cosmic ray showers to accelerator based experiments. In fact, the fifties saw an avalanche of new particles. Trying to make sense of all these new states, M. Gell-Mann~\cite{GellMann:1961ky} and Yuval Ne'eman~\cite{Ne'eman:1961cd} made use of the ``eightfold way'', the $\mathrm{SU}(3)$ flavour symmetry, to classify known hadrons. But this was clearly an approximate symmetry because the mesons grouped in the same octet had different masses. This independently led M. Gell-Mann~\cite{GellMann:1964nj} and George Zweig~\cite{Zweig:1981pd, *Zweig:1964jf} to suggest that hadrons were made of quarks and anti-quarks in 1964. This was confirmed four years later by deep inelastic scattering experiments at SLAC~\cite{PhysRevLett.23.930, *PhysRevLett.23.935}. The large angle at which electrons were sometimes deviated led James Bj{\o}rken and R.~P. Feynman to interpret neutron and proton as being made of point-like particles~\cite{Bjorken:1968dy, *PhysRevLett.23.1415}.

However, the quark model suffered from a problem: baryons are fermions but, for example, the $\Delta^{++}$ is made of three up quarks and has a positive parity, which is forbidden by the Pauli exclusion principle. To solve this issue, Oscar Greenberg~\cite{Greenberg:1964pe}, Y. Nambu and Moo-Young Han~\cite{PhysRev.139.B1006} introduced, in 1964, an extra quantum number, known today as colour, with a $\mathrm{SU}(3)$ symmetry. Y. Nambu and M.~Y. Han also assigned the interactions to an octet of vector bosons, the gluons. At the same time, J.~D. Bj{\o}rken and S.~L. Glashow~\cite{Bjorken:1964gz} proposed the existence of a fourth quark, named charm, to improve the Gell-Mann--Okubo mass formula (and obtain the correct mass of the $\rho$ meson), restore the lepton--quark symmetry and give a better description of weak interactions. Building on the quark--lepton symmetry, S.~L. Glashow, Jean Iliopoulos and Luciano Maiani predicted the existence of the charm quark through the suppression of the flavour changing neutral current by the GIM mechanism~\cite{PhysRevD.2.1285}. While this prediction was made in 1970, the charm quark existence was only confirmed four years later by the parallel discovery at SLAC and Brookhaven of the $J/\psi$~\cite{PhysRevLett.33.1406, *PhysRevLett.33.1404}.

The formulation of the electroweak theory and its successes restored due interest in Yang-Mills theories during the early seventies. Using the renormalization group method introduced by M. Gell-Mann and F.~E. Low~\cite{PhysRev.95.1300}, David Gross and Frank Wilczek~\cite{PhysRevLett.30.1343}, and independently David Politzer~\cite{PhysRevLett.30.1346}, showed that, in Yang-Mills theories with a small number of fermions, the $\beta$ function\footnote{The $\beta$ function describes the shift of the renormalized coupling due to a change in the renormalization scale. The renormalization group equation gives $\beta(\bar{g})=\frac{\mathrm{d}\bar{g}}{\mathrm{d}\log(Q/M)}$ with the initial condition for the running coupling $\bar{g}(M)=g$.} is negative, driving the effective coupling to zero. This remarkable property, named asymptotic freedom, is very important since it prevents the apparition of a Landau pole\footnote{A Landau pole corresponds to a diverge in the running of a renormalized coupling, with the coupling becoming infinite at a finite energy scale.}  in the theory. Subsequently, the Lagrangian of quantum chromodynamics (QCD) was written down and the non-observation of the massless gluon was linked to colour confinement in unbroken $\mathrm{SU}(3)$~\cite{PhysRevLett.31.494, *PhysRevD.8.3633, *Fritzsch1973365}.

\subsubsection{CP violation and the CKM matrix}

The first experiments studying strange hadrons found that strangeness changing weak decays with $\Delta s=1$ were strongly suppressed. In order to explain this and write down a universal hadronic current, Nicola Cabibbo introduced in 1963 a mixing between the down and the strange quarks~\cite{PhysRevLett.10.531},
\begin{equation}
\left( \begin{array}{c} d' \\ s' \end{array} \right) = \left( \begin{array}{c c} \cos \theta_C & \sin \theta_C \\ -\sin\theta_C & \cos\theta_C \end{array} \right) \left( \begin{array}{c} d \\ s\end{array} \right)\,,
\end{equation}
where $d'$, $s'$ are the interaction eigenstates while $d$ and $s$ are the mass eigenstates. The following year, an experiment made a surprising discovery. While it was known that parity was violated, it was thought at that time, as suggested by Lev Landau~\cite{Landau1957127}, that CP was a true symmetry of nature. However by studying the decay of $K^0_L$ in two pions, James Christenson, James Cronin, Val Fitch and René Turlay found the first evidence of CP violation. However, the four-quark picture that emerged from the GIM mechanism was not able to account for CP violation. This could only be accommodated if a third generation of quarks existed as shown by Makoto Kobayashi and Toshihide Maskawa in 1973~\cite{Kobayashi01021973}. The two generation mixing matrix parametrized by the Cabbibo angle became then a unitary mixing matrix parametrized by three mixing angles and a complex phase responsible for CP violation.

\subsubsection{Unifying electromagnetism with the weak interaction}

Even though the $V-A$ theory appeared to give an appropriate description of the weak interaction, it was plagued with theoretical inconsistencies. The Fermi interaction is a four-fermions interaction and, as such, its coupling constant $G_F$ has a dimension of $[m]^{-2}$. Therefore, the cross section 
for a process described by the Fermi theory grows as $\sigma \sim G_F^2 s$ with $s$ the invariant mass of the colliding particles, which is clearly divergent and ends up violating unitarity. In 1957, J. Schwinger~\cite{Schwinger1957407}, T.~D. Lee and C.~N. Yang~\cite{PhysRev.108.1611} introduced the idea of a possible intermediate vector boson for weak interactions, making the new coupling constant dimensionless. However, this only delays unitarity violation since the cross section for the intermediate boson scattering is still divergent. Besides, to account for the short range of the weak interaction, the mediators should be massive. But adding a mass term for a gauge boson to a Yang-Mills theory would explicitly break gauge invariance. As mentioned earlier, the Higgs mechanism provided a way out of this issue, immediately exploited by A. Salam and J.~C. Ward, giving rise to the electroweak Lagrangian and predicting the $W^\pm$ mass~\cite{Salam1964168}. Three years later, in 1967, S. Weinberg also formulated a Lagrangian for the electroweak theory and predicted the mass of the $W^\pm$ and $Z^0$ bosons~\cite{PhysRevLett.19.1264}, the latter having already been introduced by S.~L. Glashow in 1961~\cite{Glashow1961579}. When combined with QCD, the electroweak theory forms what is now called the Standard Model of particle physics.

\subsubsection{Recent developments}

The newly formulated Standard Model quickly received a strong experimental support from the observation of weak neutral current in the Gargamelle experiment at CERN in 1973~\cite{Hasert1973138, *Hasert1973121}, an observation that was confirmed the following year at Fermilab~\cite{PhysRevLett.32.800}. The first particle from the third generation, the $\tau$ lepton, was produced at SLAC~\cite{Perl:1975bf} in 1975. Two years later, the first meson containing a third generation quark, the $\Upsilon$, was discovered at Fermilab~\cite{Herb:1977ek}, while evidence for the existence of gluons appeared two years later in the process $e^+ e^- \rightarrow 3\; \mathrm{jets}$~\cite{Barber:1979yr, *Brandelik:1979bd, *Berger:1979cj, *Bartel:1979ut}. The electroweak theory received an important experimental confirmation in 1983 when the weak bosons $W^\pm$ and $Z^0$ were observed by the UA1 and UA2 collaborations at CERN~\cite{Arnison:1983rp, *Arnison:1983mk, *Banner:1983jy, *Bagnaia:1983zx}. The last missing quark, the top, was discovered in 1995 by the CDF and {D\O}  experiments~\cite{Abe:1995hr, *Abachi:1995iq}. Finally, last year saw the observation of a new boson by the ATLAS and CMS collaborations with properties compatible with the final piece of the Standard Model, the Higgs boson~\cite{Aad:2012tfa, *Chatrchyan:2012ufa}.

With all these developments, the Standard Model is now a full-fledged theory whose particles have all been observed. It has been thoroughly tested and only one of its prediction has been found to strongly disagree with experiments yet: the absence of neutrino mass. In the following section, we will describe the picture that has been forged over the years.

\section{The Standard Model in a nutshell}

After quickly explaining the idea behind gauge theories, we will introduce the Standard Model gauge group. Then, we will present its particle content and the concept of spontaneous symmetry breaking. Finally, we will explicitly write the Standard Model Lagrangian and describe how masses are generated, discussing the predictions coming from the electroweak symmetry breaking~\cite{Peskin:1995ev, *itzykson2012quantum, *ryder1996quantum, *ChengAndLi}.

\subsubsection{Gauge theories and the Standard Model gauge group}

Gauge theories have the very interesting  property that their dynamics is determined by an underlying symmetry. When the Lagrangian is invariant under local symmetry transformations associated with a Lie group,  the Hermitian matrices $t^a\,(a=1,...,N)$, which form a unitary representation of the $N$ generators of the Lie algebra, satisfy the commutation relations
\begin{equation}
 [t^a,t^b]=\imath f^{abc} t^c\,,
\end{equation}
where $f^{abc}$ are the structure constants. If all the latter are zero, then the group is said to be Abelian. Under a transformation described by the parameters $\theta^a(x)\,(a=1,...,N)$, a matter field transforms as
\begin{equation}
 \psi(x)\rightarrow\psi\prime(x)=e^{\imath \theta^a(x) t^a} \psi(x)\,.
\end{equation}
However, the kinetic term in the Lagrangian for the matter field cannot be invariant under this local transformation since the field derivative transforms differently.

This issue is solved by introducing a set of $N$ real spin-1 gauge fields $A_\mu^a\,(a=1,...,N)$ and replacing the ordinary derivative by the covariant derivative defined by
\begin{equation}
 D_\mu=\partial_\mu+\imath g t^a A_\mu^a\,,
\end{equation}
where $g$ is a real coupling constant and the vector field $A_\mu^a$ transforms as
\begin{equation}
 t^a A_\mu^a\rightarrow t^a A_\mu^a\prime= e^{\imath \theta^a(x) t^a} \left(t^a A_\mu^a - \frac{\imath}{g}\partial_\mu\right) e^{-\imath \theta^a(x) t^a}\,.
\end{equation}
To make this vector gauge field dynamical, a field strength should be added. So that the Lagrangian is invariant under local gauge transformations, the field strengh must have the form
\begin{equation}
 F_{\mu\nu}^a=\partial_\mu A^a_\nu - \partial_\nu A^a_\mu - g f^{abc} A_\mu^b A_\nu^c\,.
\end{equation}
It is worth noting that the invariance of the Lagrangian under local gauge transformation requires the introduction of gauge bosons, and determines their coupling to matter up to a universal scale given by the coupling constant. This invariance also forbids mass terms for the gauge fields since they would explicitly break the local symmetry. This is true as long as the symmetry remains unbroken. Conversely, explicit mass terms for matter fields are allowed.

Let us now focus on the Standard Model. As mentioned in the historical introduction, it has two sectors that do not mix. The first one, quantum chromodynamics, describes strong interactions via the unbroken gauge group $\mathrm{SU}(3)_c$, where $c$ stands for colour, the associated charge. The second one, the electroweak theory, describes the weak and electromagnetic interactions through the product $\mathrm{SU}(2)_L\times\mathrm{U}(1)_Y$. $L$ denotes the fact that the weak interaction maximally violates parity by acting only on left-handed fermions, while $Y$ is the hypercharge, related to the electrical charge $Q$ through the Gell-Mann--Nishijima relation
\begin{equation}
\label{GMN}
 Q=I_3+\frac{Y}{2}\,,
\end{equation}
where $I_3$ is the third component of weak isospin. As we will see below, the electroweak symmetry is broken at low energy to $\mathrm{U}(1)_{em}$, which allows the weak interaction bosons to be massive while the photon remains massless.

\subsubsection{The particle content}

Built on the gauge group $\mathrm{SU}(3)_c\times\mathrm{SU}(2)_L\times\mathrm{U}(1)_Y$, the Standard Model should include the gauge fields associated with this symmetry group. Let us first consider the unbroken $\mathrm{SU}(3)_c$. It has eight generators, which, in its 3-dimensional unitary representation, are given by the eight $3\times3$ Hermitian traceless Gell-Mann matrices $\lambda_a$. Those are associated with eight vector gauge boson, the gluons, which are massless since $\mathrm{SU}(3)_c$ remains unbroken. If we look at the electroweak theory, things are slightly more complicated: $\mathrm{SU}(2)_L$ has three generators, which correspond in the 2-dimensional unitary representation to the Pauli matrices
\begin{equation}
 \sigma_1=\left(\begin{array}{rr} 0 & 1 \\1 & 0\end{array}\right)\,,\;\sigma_2=\left(\begin{array}{rr} 0 & -\imath \\ \imath & 0 \end{array}\right)\,,\;
 \sigma_3=\left(\begin{array}{rr} 1 & 0 \\0 & -1\end{array}\right)\,.
\end{equation}
They add three gauge bosons while a fourth one comes from $\mathrm{U}(1)_Y$. However, at low energy $\mathrm{SU}(2)_L\times\mathrm{U}(1)_Y$ is spontaneously broken to $\mathrm{U}(1)_{em}$, which means that one gauge boson should remain massless. Moreover, $\mathrm{U}(1)_Y$ and $\mathrm{U}(1)_{em}$ correspond to different charges, even if they are related through the Gell-Mann--Nishijima relation. This indicates that the massless spin-1 field will be a mixture of spin-1 fields from $\mathrm{SU}(2)_L\times\mathrm{U}(1)_Y$. We will explicitly show this below when describing the Higgs mechanism. The only other boson in the Standard Model is a spin-0 particle, the Higgs boson, which is required to spontaneously break $\mathrm{SU}(2)_L\times\mathrm{U}(1)_Y$. The bosonic content of the Standard Model, in unitary gauge after electroweak symmetry breaking, is given in table~\ref{SMbosons}.
\begin{table}[tb]
  \begin{center}
    \begin{tabular}{|c|c|c|c|c|}
     \hline
	Field & Mass (GeV) & Spin & $\mathrm{U}(1)_{em}$ charge & $\mathrm{SU}(3)_c$ rep.\\
     \hline
     \hline
	$G$ & $0$ & $1$ & $0$ & $\mathbf{8}$\\
     \hline
	$W^\pm$ & $80.385 \pm 0.015$ & $1$ & $\pm1$ & $\mathbf{1}$\\
	$Z^0$ & $91.1876 \pm 0.0021$ & $1$ & $0$ & $\mathbf{1}$\\
     \hline
	$\gamma$ & $0$ & $1$ & $0$ & $\mathbf{1}$\\
     \hline
	$\phi^0$ & $\begin{array}{lr} ATLAS: & 125.5\pm0.7\\ CMS: & 125.8\pm0.6\end{array}$ & $0$ & $0$ & $\mathbf{1}$\\
     \hline
    \end{tabular}
    \caption[Bosonic content of the SM]{\label{SMbosons} The bosonic content of the Standard Model after electroweak symmetry breaking. The masses are extracted from the Review of Particle Physics~\cite{Beringer:1900zz}, with the exception of the so called ``Higgs'' mass~\cite{ATLAS-CONF-2013-014, *CMS-PAS-HIG-13-002}.}
 \end{center}
\end{table}

Fermions belong to irreducible unitary representations of the Lie groups. As a consequence, they can be classified according to the representations they belong to. For example,  leptons and quarks form singlets and triplets of $\mathrm{SU}(3)_c$, respectively, while antiquarks are assigned to the conjugate $\mathbf{3^*}$ representation. However, the number of generations is unconstrained whereas the number of fermions for each generation and the representation they belong to must ensure anomaly cancellation. Three generations are present in the Standard Model in order to account for the observed particles and CP violation. All particles have two chiralities, with the exception of neutrinos which have no right-handed component, making them massless by construction\footnote{However, this has been proven to be in disagreement with the experimental observation of neutrino oscillations. We will return to this issue in the next chapter.}. Left-handed fields are grouped in $\mathrm{SU}(2)$ doublets with the third component of weak isospin written $I_3$.  All the Standard Model fundamental fermions have a $1/2$ spin, an electrical charge given by the Gell-Mann--Nishijima relation (eq.~\ref{GMN}) and are listed in table~\ref{SMfermionsF}.
\begin{table}[t]
  \begin{center}
    \begin{tabular}{|c|c|c|c|c|}
     \hline
	Field & Mass (GeV) & $\mathrm{U}(1)_{em}$ charge & $I_3$ & $\mathrm{SU}(3)_c$ rep.\\
     \hline
     \hline
	$\nu_e$ & $<2\times 10^{-9}$ & $\phantom{-}0$ & $\phantom{-}1/2$ & $\mathbf{1}$\\
	$e$ & $(5.10998928 \pm 0.00000011)\times 10^{-4}$ & $-1$ & $-1/2$ & $\mathbf{1}$\\
     \hline
	$\nu_\mu$ & $<1.9\times 10^{-4}$ & $\phantom{-}0$ & $\phantom{-}1/2$ & $\mathbf{1}$\\
	$\mu$ & $(1.056583715 \pm 0.000000035)\times 10^{-1}$ & $-1$ & $-1/2$ & $\mathbf{1}$\\
     \hline
	$\nu_\tau$ & $<1.82\times10^{-2}$ & $\phantom{-}0$ & $\phantom{-}1/2$ & $\mathbf{1}$\\
	$\tau$ & $1.77682 \pm 0.00016$ & $-1$ & $-1/2$ & $\mathbf{1}$\\
     \hline
     \hline
	$u$ & $(2.27 \pm 0.14)\times 10^{-3}\,(\overline{\mathrm{MS}})$ & $\phantom{-}2/3$ & $\phantom{-}1/2$ & $\mathbf{3}$\\
	$d$ & $(4.78 \pm 0.09)\times 10^{-3}\,(\overline{\mathrm{MS}})$ & $-1/3$ & $-1/2$ & $\mathbf{3}$\\
     \hline
	$c$ & $1.275 \pm 0.004\,(\overline{\mathrm{MS}})$ & $\phantom{-}2/3$ & $\phantom{-}1/2$ & $\mathbf{3}$\\
	$s$ & $(9.43 \pm 0.12)\times 10^{-2}\,(\overline{\mathrm{MS}})$ & $-1/3$ & $-1/2$ & $\mathbf{3}$\\
     \hline
	$t$ & $173.5 \pm 1.0$ & $\phantom{-}2/3$ & $\phantom{-}1/2$ & $\mathbf{3}$\\
	$b$ & $4.18 \pm 0.03\,(\overline{\mathrm{MS}})$ & $-1/3$ & $-1/2$ & $\mathbf{3}$\\
     \hline
    \end{tabular}
    \caption[Fermionic content of the SM]{\label{SMfermions} The fermionic content of the Standard Model. All masses are extracted from the Review of Particle Physics~\cite{Beringer:1900zz}.}
 \end{center}
\end{table}

\subsubsection{The Standard Model Lagrangian}

With the description of the gauge group and the particle content of the SM given in the previous sections, we can now write down the corresponding Lagrangian. However, as we have mentioned for bosons, explicit mass terms violate gauge invariance. This is also true for fermionic Dirac mass terms. To respect Lorentz invariance, these terms should relate the two chiralities of a fermion through $m(\overline{\psi_R}\psi_L+\overline{\psi_L}\psi_R)$. However, left-handed and right-handed fields belong to different $\mathrm{SU}(2)_L$ representations and their product is not gauge invariant. A way out of this issue is to introduce a scalar field,  which belongs to a $\mathrm{SU}(2)_L$ doublet $\phi$, that will take a non-zero vacuum expectation value (vev). This is, in fact, the same scalar field that generates masses for the weak bosons through the Higgs mechanism, which will be described in more detail in the next section.

In the following, we will write the SM Lagrangian before electroweak symmetry breaking. The field strengths for the eight gluons $G^a_\mu\,(a=1,...,8)$, the three gauge bosons $W^a_\mu\,(a=1,2,3)$ associated with $\mathrm{SU}(2)_L$ and the $\mathrm{U}(1)_Y$ vector boson $B_\mu$ are given by
\begin{eqnarray}
 G^a_{\mu\nu}&=&\partial_\mu G^a_\nu - \partial_\nu G^a_\mu - g_s f^{abc} G_\mu^b G_\nu^c\,,\\
 W_{\mu\nu}^a&=&\partial_\mu W^a_\nu - \partial_\nu W^a_\mu - g \varepsilon^{abc} W_\mu^b W_\nu^c\,,\\
 B_{\mu\nu}&=&\partial_\mu B_\nu - \partial_\nu B_\mu\,,
\end{eqnarray}
where $g_s$ and $g$ are the $\mathrm{SU}(3)_c$ and $\mathrm{SU}(2)_L$ coupling constants, respectively, $f^{abc}=-\imath \mathrm{Tr}([\lambda_a,\lambda_b]\lambda_c)/4$ and $\varepsilon^{abc}$ is the three-dimensional Levi-Civita tensor. For completeness, we also define the $\mathrm{U}(1)_Y$ coupling constant as $g\prime$. With this, it is now possible to express the covariant derivative as
\begin{equation}
 D_\mu=\partial_\mu+\imath g_s \frac{\lambda^a}{2} G_\mu^a+\imath g \frac{\sigma^a}{2} W_\mu^a+\imath g\prime \frac{Y}{2} B_\mu\,.
\end{equation}
For example, a left-handed neutrino will (only) see the term associated with the strong coupling drop from the covariant derivative.

The SM Lagrangian can be decomposed as follows
\begin{equation}
 \mathcal{L}_{SM} = \mathcal{L}_{gauge} + \mathcal{L}_{matter} + \mathcal{L}_{Higgs} + \mathcal{L}_{Yukawa}\,,
\end{equation}
where, noting the $\mathrm{SU}(2)_L$ doublets $L=\binom{\nu_L}{e_L}$, $Q=\binom{u_L}{d_L}$ and $\phi=\binom{\phi^+}{\phi^0}$, the different contributions are
\begin{eqnarray}
 \mathcal{L}_{gauge} &=& -\frac{1}{4} G^a_{\mu\nu} G^{a\mu\nu} -\frac{1}{4} W^a_{\mu\nu} W^{a\mu\nu} -\frac{1}{4} B_{\mu\nu} B^{\mu\nu}\,,\\
 \mathcal{L}_{matter} &=& \imath \sum_{i=e,\mu,\tau} \overline{L_i} \slashed{D} L_i + \imath \sum_{i=1,2,3} \overline{Q_i} \slashed{D} Q_i +  \imath \sum_{i=e,\mu,\tau} \overline{\ell_{Ri}} \slashed{D} \ell_{Ri}\\ \nonumber
    & & + \imath \sum_{i=u,c,t} \overline{q_{Ri}} \slashed{D} q_{Ri} + \imath \sum_{i=d,s,b} \overline{q_{Ri}} \slashed{D} q_{Ri}\,,\\
 \mathcal{L}_{Higgs} &=& (D^\mu \phi)^\dagger (D_\mu \phi) - \mu^2 \phi^\dagger \phi - \lambda (\phi^\dagger \phi)^2\,,\label{HiggsSM}\\
 \mathcal{L}_{Yukawa} &=& -\sum_{i,j=e,\mu,\tau} \left( Y^{ij}_\ell \overline{L_i} \phi e_{Rj} + h.c. \right) -
\sum_{i=1,2,3,j=u,c,t} \left( Y^{ij}_u \overline{Q_i} \widetilde{\phi} q_{Rj} + h.c. \right)\\ \nonumber
    & & - \sum_{i=1,2,3,j=d,s,b} \left( Y^{ij}_d \overline{Q_i} \phi q_{Rj} + h.c. \right)\,,
\end{eqnarray}
with $\widetilde{\phi}=\imath \sigma_2 \phi^*$. But what happens after electroweak symmetry breaking when the Higgs scalar develops a non-zero vev ?

\subsubsection{Spontaneous symmetry breaking and the Higgs mechanism}

Let us first recall that even if a Lagrangian is invariant under a given global continuous symmetry, the corresponding theory might not respect this symmetry. Symmetries can be realized in two ways in Nature. The first one is the Wigner-Weyl mode where the vacuum is also invariant under the symmetry considered. This would, for example, imply that particles that are related through the symmetry are degenerate in mass. The second case corresponds to the Nambu-Goldstone mode where the ground state of the theory is not invariant. As a consequence, a massless scalar will appear in the theory for every generator of the broken symmetry.

\begin{figure}[tb]
\centering
 \includegraphics[width=0.5\textwidth]{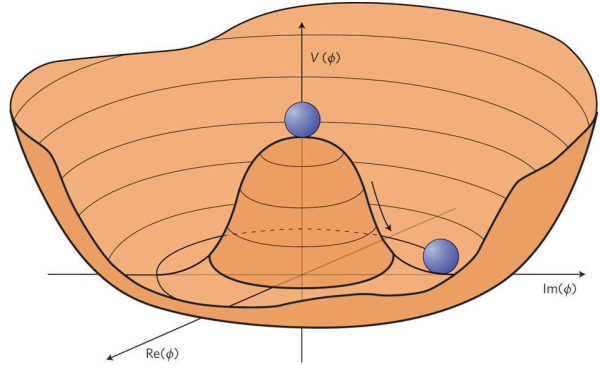}
\caption[Scalar potential of the SM]{The scalar potential of the Standard Model: spontaneous symmetry breaking occurs when the field leaves the unstable extremum at $\phi=0$ to reach a minimum in a randomly chosen direction, thus acquiring a non-zero vev. Reprinted by permission from Macmillan Publishers Ltd: Nature Physics~\cite{2011NatPh72A}, copyright 2011.}
\label{HiggsPotential}
\end{figure}

Applying this idea to local symmetries is the basis of the Higgs mechanism. When one considers the scalar potential of the SM, pictured in~\ref{HiggsPotential} 
\begin{equation}
 V(\phi)=\mu^2 \phi^\dagger \phi + \lambda (\phi^\dagger \phi)^2\,,
\end{equation}
the condition $\mu^2 < 0$ is required to have a vacuum state that breaks $\mathrm{SU}(2)_L$ invariance. Thus, the vev of the Higgs doublet is
\begin{equation}
 \langle \phi \rangle = \frac{1}{\sqrt{2}} \binom{0}{v}\,,
\end{equation}
where
\begin{equation}
v=\sqrt{\frac{-\mu^2}{\lambda}}\,.
\end{equation}

The gauge boson masses can be derived by considering the kinetic term of the Higgs doublet. The covariant derivative reads
\begin{equation}
  D_\mu=\partial_\mu+\imath g \frac{\sigma^a}{2} W_\mu^a+\imath g\prime \frac{1}{2} B_\mu\,,
\end{equation}
which at the Higgs vev leads to the mass terms
\begin{eqnarray}
 |D_\mu \phi|^2 & = & \frac{1}{2} \left| \begin{pmatrix} \partial_\mu + \frac{\imath}{2} (g W^{3}_\mu + g\prime B_\mu) & \frac{\imath}{2} g (W^{1}_\mu - \imath W^{2}_\mu) \\ \frac{\imath}{2} g (W^{1}_\mu + \imath W^{2}_\mu) & \partial_\mu - \frac{\imath}{2} (g W^{3}_\mu - g\prime B_\mu)  \end{pmatrix} \binom{0}{v} \right|^2\\ \nonumber
&=& \frac{1}{2} \frac{v^2}{4} \left[g^2 (W^{1}_\mu + \imath W^{2}_\mu)(W^{1}_\mu - \imath W^{2}_\mu) + (g W^{3}_\mu - g\prime B_\mu)^2\right]\,.
\label{WZmassterms}
\end{eqnarray}
If the physical fields are defined as
\begin{equation}
 W^\pm_\mu = \frac{1}{\sqrt{2}} (W^{1}_\mu \mp \imath W^{2}_\mu)\,,\;Z^0_\mu=\frac{g W^{3}_\mu - g\prime B_\mu}{\sqrt{g^2+g\prime^2}}\,,\;
A_\mu = \frac{g\prime W^{3}_\mu + g B_\mu}{\sqrt{g^2+g\prime^2}}\,,
\end{equation}
then one can immediately see from eq.~(\ref{WZmassterms}) that there are three massive vector bosons, the two $W^\pm$ with a mass of
\begin{equation}
 m_W=\frac{gv}{2}\,,
\end{equation}
and the $Z^0$ with a mass of
\begin{equation}
 m_Z=\frac{\sqrt{g^2+g\prime^2}}{2} v\,,
\end{equation}
and a massless photon in accordance to the fact that $\mathrm{U}(1)_{em}$ remains unbroken. It is also possible to define the weak mixing angle $\theta_w$ which relates $W^3_\mu$ and $B_\mu$ to $Z^0_\mu$ and $A_\mu$ through the relation
\begin{equation}
 \tan \theta_w=\frac{g\prime}{g}\,.
\end{equation}
From this and the fermion couplings to the weak gauge bosons, the electric charge of the positron is given by 
\begin{equation}
  e = g \sin \theta_w\,,
\end{equation}
and the Fermi constant reads
\begin{equation}
 \frac{G_F}{\sqrt{2}}=\frac{g^2}{8 m_W^2}\,,
\end{equation}
which, from the experimental measurement of the Fermi constant and the $W^\pm$ mass, can be translated to $v\simeq 246\;\mathrm{GeV}$.

It is worth noting that the Higgs scalar doublet being complex, it has four degrees of freedom. Three of them are Goldstone bosons that end up combining with the $W^\pm$ and $Z^0$ to give them their longitudinal component. This has a consequence at high-energy, which is the Goldstone boson equivalence theorem~\cite{Cornwall:1974km, *Vayonakis:1976vz}: at energies much higher than $m_W$, the leading-order amplitude of a process involving a longitudinally polarized on-shell vector boson is equal to the one where the vector boson has been replaced with the corresponding Goldstone boson.

Finally, it is straightforward to find that fermion masses arise from the Yukawa terms. However, Yukawa couplings are generally complex matrices that should be decomposed by biunitary transformations of $\psi_L$ and $\psi_R$. If the resulting diagonal matrix is written $y^{ij}_\alpha$ then the corresponding fermion masses are given by
\begin{equation}
 m_i=\frac{y^{ii}_\alpha v}{\sqrt{2}}\,,
\end{equation}
while the product of the unitary transformations of the left-handed up and down quarks results in the CKM matrix~\cite{Kobayashi01021973}. The leptonic sector having only one Yukawa coupling matrix in the SM, it is possible to define a basis, the flavour basis, in which the charged lepton mass matrix and the charged lepton interactions with the neutrinos are diagonal. However, this forbids the possibility of neutrino oscillations, contradicting experimental observations. This will be discussed in more details in the next chapter, along some possible ways out of this issue.

\chapter{The neutrino sector\label{ChapNuExp}}

Due to their unique properties, neutrino are fascinating particles: they are the only neutral elementary fermions in the Standard Model (SM), which opens the possibility for them to be of Majorana nature. They are only sensitive to weak interactions, rendering them very elusive. While all fermions have masses around or above the MeV, their masses lie at the eV scale, or below. Neutrinos can have a crucial role in a wide variety of phenomena, ranging from the formulation of $\beta$ decay to supernovae dynamics and Universe and stellar evolutions. All these features contribute to their uniqueness among all other SM fermions.

In this chapter, we will summarise some important discoveries in neutrino Physics before proceeding to a survey of different topics and current related experimental issues.

\section{A brief history of neutrinos\label{NuHistory}}

\begin{figure}[!htp]
\centering
 \includegraphics[width=1.06\textwidth]{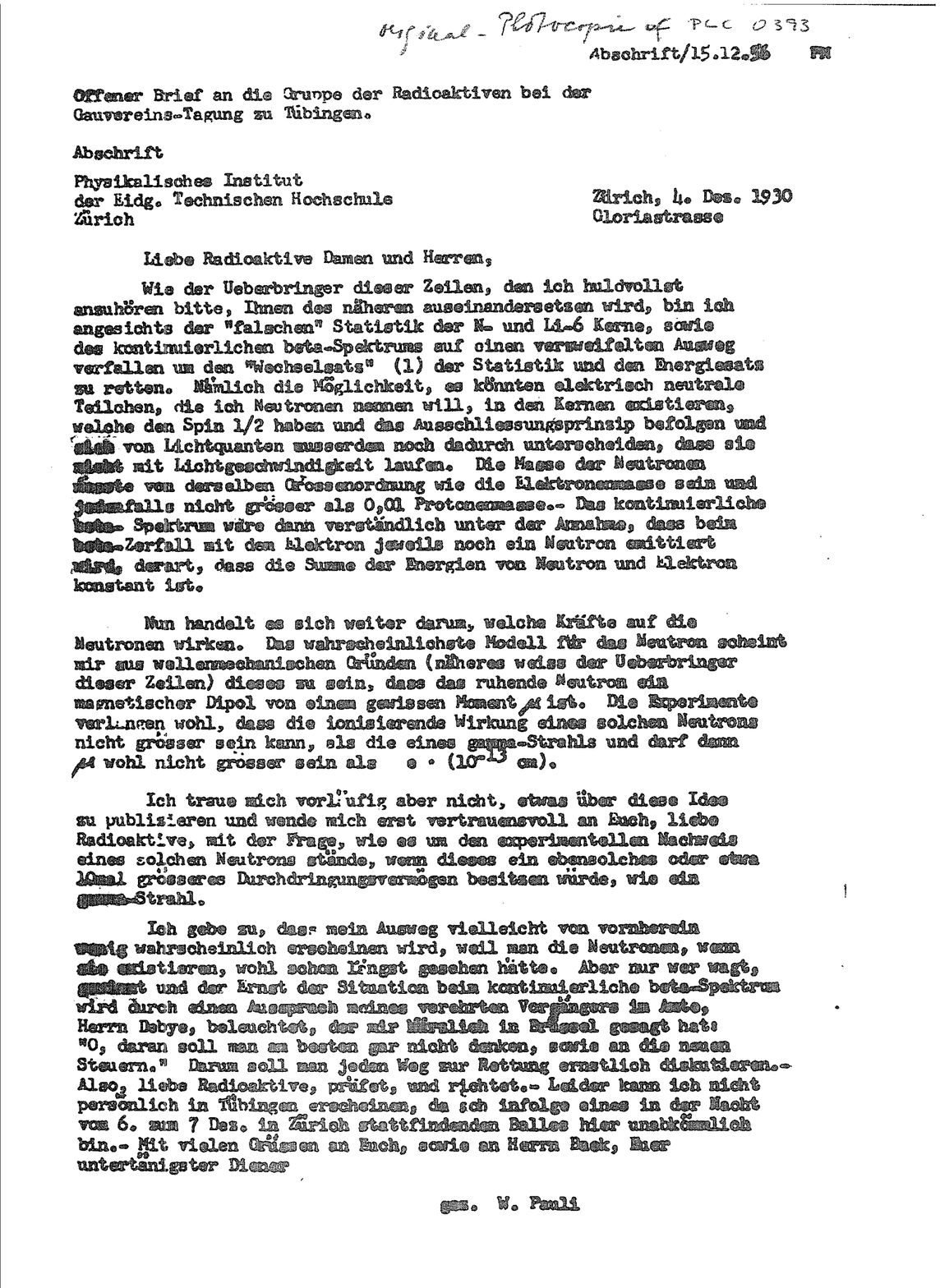}
\caption[Original Pauli's letter]{Pauli's open letter to the Tübingen congress. From the Pauli Letter Collection at CERN, reference  \texttt{meitner\_0393}, reproduced by permission of the Pauli Committee.}
\label{PauliLetter}
\end{figure}

The beginning of neutrino history can be traced back to the $\beta$ decay problem and to its solution. In 1914, James Chadwick observed for the first time that the $\beta$ spectrum was continuous~\cite{Chadwick:1914zz}, which was confirmed in 1927 by Charles Ellis and William Wooster~\cite{Ellis01031927, *Ellis01121927}. This proved to be a very serious issue at that time since the electron spectrum should be monochromatic if the $\beta$ decays corresponded to two body decays. Such a puzzle even led Niels Bohr to suggest that energy might not be conserved after all. The solution came, in 1930,  from W. Pauli who proposed, in a now famous letter (see fig.~\ref{PauliLetter}),
the existence of a third particle, the neutron, later to be renamed neutrino (``little'' neutron in Italian) by E. Fermi. But it was only three years later that W. Pauli  publicly presented and committed his idea to paper~\cite{pauli1933rapports}. In 1934, E. Fermi included the neutrino in its four-fermion description of the $\beta$ decay~\cite{fermi1934versuch, *fermi1934tentativo} while, during the same year, H.~A. Bethe and Rudolf Peierls came to the discouraging conclusion that, with a cross-section for the inverse $\beta$ decay below $10^{-43} \mathrm{cm}^{2}$, ``there is no practically possible way of observing the neutrino''~\cite{1934NatureBethe}. This period of intense theoretical development received another important contribution in 1937, when Ettore Majorana formulated a theory where the neutrino and its antiparticle are one and the same~\cite{Majorana1937symmetry}.

The next important period in neutrino history occurred during the fifties with the first observation of the electron anti-neutrino by Frederick Reines and Clyde Cowan. The former started addressing the problem of neutrino detection in 1951, coming to the conclusion that the best source of neutrinos would be an atom bomb, conclusion shared by E. Fermi~\cite{Reines19941}.
\begin{figure}[t]
\centering
\subbottom[First neutrino experiment of F. Reines and C.L. Cowan][Conceptual proposal of the first experiment.]{\includegraphics[width=0.49\textwidth]{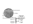}\label{BombExp}}
\hfill
\subbottom[Schematic of Savannah River detector][Schematic of Savannah River detector.]{\includegraphics[width=0.49\textwidth]{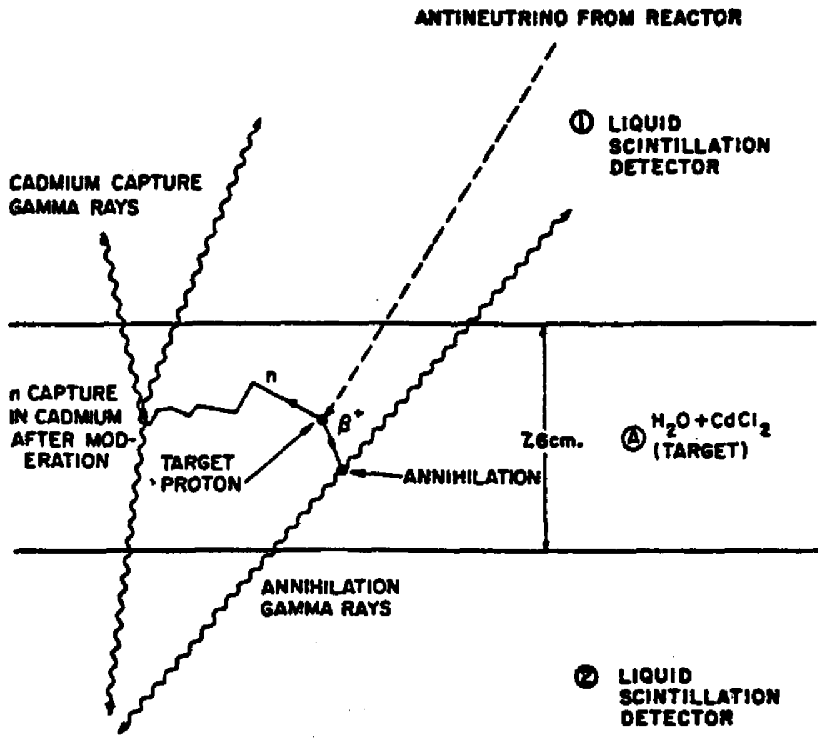}\label{SavannahRiver}}
\caption[F. Reines and C.~L. Cowan experiments]{The experiments designed to detect neutrinos by F. Reines and C.~L. Cowan. Reprinted from~\cite{Reines19941}, copyright 1994, with permission from Elsevier.} \label{NuExp}
\end{figure}
With C.~L. Cowan, F. Reines subsequently designed  the experiment presented in fig.~\ref{BombExp}, which involved an atom bomb and a ton-scale detector (using liquid scintillator) that should free fall in vacuum in order to escape the shock wave from the blast. Surprisingly enough, the experiment, which used inverse $\beta$ decay to detect neutrinos, was approved by the Los Alamos National Laboratory direction. However, F. Reines and C.~L. Cowan preferred to use a nuclear reactor as a source, finding that the background could be sufficiently reduced to make the experiment viable, provided that they searched for the positron annihilation signal and the delayed signal from the neutron capture by cadmium. A first experiment conducted at the Hanford nuclear reactor found evidence of the neutrino but was plagued by a high level of background due to cosmic rays~\cite{PhysRev.92.830}. The experiment was later moved to the Savannah River Plant which, being $12$~meters underground, offered a shielded location. This allowed the team to confirm the observation of the neutrino and to measure the inverse $\beta$ decay cross-section~\cite{Cowan20071956, *CowanNature}. A schematic description of the experiment is given in fig.~\ref{SavannahRiver}.

This observation was followed by two important theoretical developments in 1957. First, A. Salam~\cite{Salam1957parity}, L.~D. Landau~\cite{landau1957ob}, T.~D. Lee and C.~N. Yang~\cite{PhysRev.105.1671} introduced the two component theory of neutrino according to which the neutrino has to be either right or left-handed. It had been found that electrons emitted in weak decays were mostly left-handed~\cite{PhysRev.106.386}, implying that if weak currents were vectorial or axial, then the neutrino was necessarily left-handed. Conversely, if  weak currents were scalar or tensorial, then the neutrino had to be right-handed. Second, Bruno Pontecorvo realised the possibility of neutrino oscillations~\cite{pontecorvo1957mesonium, *pontecorvo1957inverse}, although he was considering neutrino--antineutrino oscillations at the time. The following year, Maurice Goldhaber, Lee Grodzins and Andrew Sunyar found the first evidence of a negative helicity for neutrinos~\cite{PhysRev.109.1015}, implying that weak currents have a $V-A$ structure.

In 1962, a group at the Brookhaven AGS accelerator observed the muon neutrino for the first time~\cite{Danby:1962nd}. Three years later, atmospheric neutrinos from cosmic rays were detected for the first time in mines in South Africa~\cite{Reines:1965qk} and India~\cite{Achar1965196}. Finally, the end of the decade saw the start of one of the longest ever running experiments, the Homestake experiment, based on a radiochemical detection method first proposed by B. Pontecorvo in 1946~\cite{pontecorvochalk} and L.~W. Alvarez in 1949~\cite{Alvarez:1949zz}.

During the 1970s, the first indication of neutrino oscillations appeared at the Homestake experiment~\cite{Cleveland:1998nv}. A deficit in the expected neutrino flux from the Sun emerged, pointing towards either an inconsistency in the Standard Solar Model or to neutrino oscillations. This was known as the solar neutrino problem. In 1989, the MARK-II collaboration obtained the first evidence of the existence of three active light neutrinos~\cite{Abrams:1989yk}, which would later be confirmed by experiments at the Large Electron--Positron Collider (LEP)~\cite{ALEPH:2005ab}. The definitive confirmation of neutrino oscillations would come in 1998 from the Super-Kamiokande collaboration~\cite{Fukuda:1998mi}: they observed a zenith dependant deficit of muon neutrinos, consistent with the two-flavour oscillation $\nu_\mu \rightarrow \nu_\tau$ hypothesis. Around the same time, the DONUT collaboration announced the discovery of the tau neutrino~\cite{Kodama:2000mp}. The oscillation hypothesis was further confirmed by the Sudbury Neutrino Observatory (SNO) collaboration~\cite{Ahmad:2001an, *Ahmad:2002jz}, while the KamLAND experiment, in 2005, observed with a very high significance a distortion in the neutrino energy spectrum in agreement with $\bar \nu_e$ oscillations~\cite{Araki:2004mb}.

In the following section, we will briefly describe the basic theoretical framework behind neutrino oscillations. We will also introduce and define notions that will be useful when summarising current issues in the last section.

\section{Theoretical interlude on neutrino oscillations}

The simplest explanation for neutrino oscillations is to consider that they are massive and non-aligned with the charged leptons, which corresponds to a mixing in charged currents
\begin{equation}
 J^\mu_W=\frac{1}{\sqrt{2}}\overline{\nu_i}U^*_{ji}\gamma^\mu P_L \ell_j\,,
\end{equation}
where $U$ is the Pontecorvo-Maki-Nakagawa-Sakata~(PMNS) matrix, from the names of B. Pontecorvo who proposed neutrino oscillations and Ziro Maki, Masami Nakagawa and Shoichi Sakata who introduced this mixing matrix~\cite{Maki:1962mu, *Pontecorvo:1967fh}. Here, $\nu_i$ and $\ell_j$ are mass eigenstates, which means that they are eigenstates of the Casimir operator $P^2$, $P_\mu$ being the 4-momentum operator. They are, equivalently, eigenvectors of their respective mass matrices, using the approach introduced in Chapter~\ref{chap3}. It is also possible to define flavour eigenstates as the neutrinos associated with the transition $\ell_\alpha^- \rightarrow \nu_\alpha$, in the basis where charged leptons are diagonal. The flavour eigenstates are related to the mass eigenstates through
\begin{equation}
 |\nu_\alpha \rangle = \sum_{i=1}^3 U^*_{\alpha i} |\nu_i \rangle\,,
\end{equation}
when neutrinos are described by plane waves and experiments are not sensitive to the neutrino mass differences in the production and detection processes. Nevertheless, neutrino production and detection are localized and neutrino are properly described by wave packets.

A neutrino that is produced or detected in association with a charged lepton is a flavour eigenstate, which corresponds to a superposition of mass eigenstates. However, if the mass eigenstates are non-degenerate, the associated plane waves will have different time evolutions via the Schrödinger equation. This results in a superposition of mass eigenstates that is different from the initial flavour eigenstates, generating neutrino oscillations. Following~\cite{giunti2007fundamentals}, the transition probability for ultrarelativistic neutrinos is
\begin{equation}
  P_{\nu_\alpha \rightarrow \nu_\beta}(L,E)=\sum_{k,j=1}^{3} U^*_{\alpha k} U_{\beta k} U_{\alpha j} U^*_{\beta j} \mathrm{exp}\left(-\imath \frac{\Delta m^2_{kj}L}{2E}\right)\,,\label{NuOscillation}
\end{equation}
where $L$ is the distance between the source and the detector, $E\simeq|\vec{p}|$ is the neutrino energy and the squared mass difference is
\begin{equation}
 \Delta m^2_{kj} = m^2_k - m^2_j\,.
\end{equation}
It is clear from eq.~(\ref{NuOscillation}) that the oscillation probability is non-zero only if mixing is present and the neutrino masses are different. 

The PMNS~matrix can be parametrized as
\begin{equation}
 U_{PMNS}= U_D \times \mathrm{diag}(1\,, e^{\imath \alpha_{21}/2}\,, e^{\imath \alpha_{31}/2})\,,
\end{equation}
where $\alpha_{21}$ and $\alpha_{31}$ are two physical Majorana phases that are absent if the neutrinos are Dirac fermions. The part of the PMNS matrix parametrized similarly to the CKM matrix is
\begin{equation}
U_D=\left(\begin{array}{c c c} c_{13}c_{12} & c_{13}s_{12} & s_{13}e^{-i\delta}\\ -c_{23}s_{12}-s_{23}s_{13}c_{12}e^{i\delta} & c_{23}c_{12}-s_{23}s_{13}s_{12}e^{i\delta} & s_{23}c_{13}\\ s_{23}s_{12}-c_{23}s_{13}c_{12}e^{i\delta} & -s_{23}c_{12}-c_{23}s_{13}s_{12}e^{i\delta} &
c_{23}c_{13}\end{array}\right)\,.
\label{PMNS}
\end{equation}
where $c_{ij}=\cos \theta_{ij}$, $s_{ij}=\sin \theta_{ij}$ and $\delta$ is the CP violating phase. It is worth noting that while the PMNS and CKM matrix are similar, they belong to two different mixing regimes, the CKM matrix being much more hierarchical than the PMNS matrix. Another interesting feature is that neutrino oscillation probabilities do not depend on Majorana phases. The dependence of the oscillation probability~(\ref{NuOscillation}) on the mixing matrix can be rewritten
\begin{equation}
 U^*_{D\alpha k}\, e^{-\imath \alpha_{k1}/2}\,  U_{D\beta k}\, e^{\imath \alpha_{k1}/2}\, U_{D\alpha j}\, e^{-\imath \alpha_{j1}/2}\, U^*_{D\beta j1}\, e^{-\imath \alpha_{j}/2} = U^*_{D\alpha k}   U_{D\beta k} U_{D\alpha j} U^*_{D\beta j}\,,
\end{equation}
with $\alpha_{11}=0$ and all the Majorana phases cancelling out. As a consequence, experiments based on oscillations cannot test the Majorana or Dirac nature the neutrino.

\begin{figure}[!t]
\centering
 \includegraphics[width=0.8\textwidth]{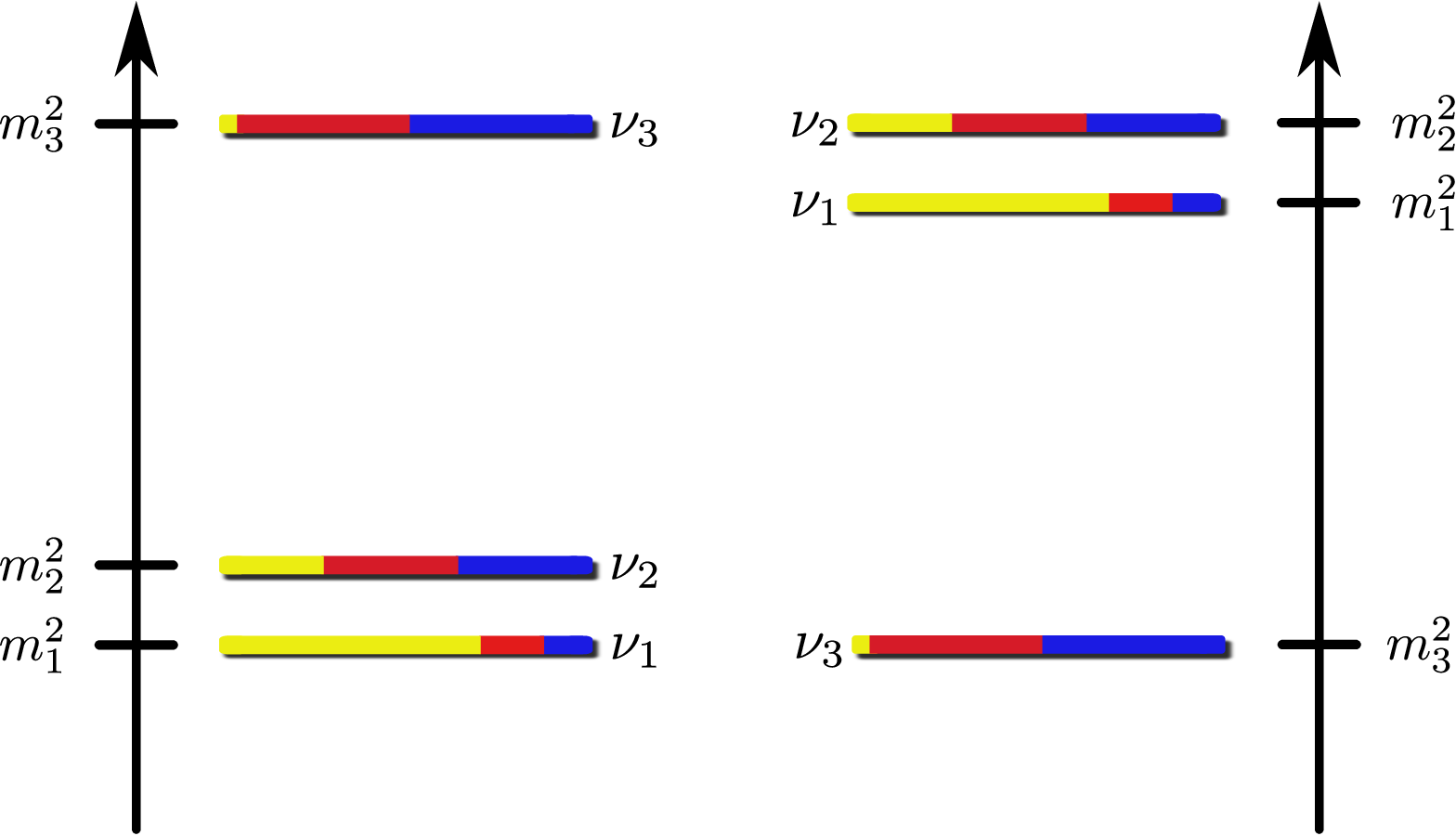}
\caption[Neutrino hierarchies]{Mass eigenstates composition and ordering in the normal (left, NH) and inverted (right, IH) hierarchy of the neutrino spectrum. For each eigenstate, the contribution from the electron neutrino is given in yellow, red for the muon neutrino and blue for the tau neutrino. Figure by kismalac on Wikimedia Commons.}
\label{NeutrinoHierarchy}
\end{figure}

Nowadays, most of the oscillation parameters have been experimentally measured, with the exception of a CP violating phase in the lepton mixing matrix. Three groups have published fits to the data from atmospheric, solar, reactor and accelerator experiments~\cite{Tortola:2012te, Fogli:2012ua, GonzalezGarcia:2012sz}. The global fit collaboration NuFIT established~\cite{GonzalezGarcia:2012sz} 
\begin{align}
 \sin^2\theta_{12}&=0.306^{0.012}_{-0.012}\,, 					 & \Delta m^2_{12}&=7.45^{+0.19}_{-0.16}\times10^{-5}\;\mathrm{eV}^2\,,\nonumber\\
 \sin^2\theta_{23}&=0.437^{+0.061}_{-0.031}\,, 					& \Delta m^2_{32}&=-2.410^{+0.062}_{-0.063}\times10^{-3}\;\mathrm{eV}^2\;(IH)\,,\nonumber\\
\sin^2\theta_{13}&=0.0231^{+0.0023}_{-0.0022}\,,				& \Delta m^2_{31}&=+2.421^{+0.022}_{-0.023}\times10^{-3}\;\mathrm{eV}^2\;(NH)\,,
\end{align}
where NH and IH refer to the normal and inverted hierarchies of the neutrino spectrum, respectively. Both hierarchies are depicted in figure~\ref{NeutrinoHierarchy}
and correspond to possible mass orderings allowed due to the absence of a decisive measurement of $\Delta m^2_{32}$ sign.

Recently, huge improvements have been made in the determination of the oscillation parameters leading to the global fit presented above and they will be described in the next section. Moreover, neutrinos coming from many different sources, astrophysical or terrestrial, have been observed and we will briefly survey the different dedicated experiments in the following section.

\section{The current experimental status\label{SectCurrent}}

The discovery of neutrinos fuelled numerous impressive experimental achievements, from the detection of neutrinos emitted by a supernova $168000$ light-years away to the measurement of a squared mass difference of less than $10^{-65}\;\mathrm{kg}^2$. We summarize in this section the current experimental situation starting from neutrino oscillation experiments, moving afterwards to mass measurements and ending with results from cosmology and astrophysics.

\subsubsection{Solar neutrinos and the MSW effect}

The foundation of the current solar model can be traced back to an article by H.~A. Bethe~\cite{Bethe:1939bt} where he described the major thermonuclear reactions occurring in ordinary stars. These are the CNO cycle and $pp$ chains which results in the conversion of four protons into a $^4\mathrm{He}$ nucleus, accompanied by the emission of two positrons, two electron neutrinos and photons. This led to the publication in 1964 of back-to-back articles by John Bahcall and Raymond Davis~\cite{Bahcall:1964gx, *Davis:1964hf} that triggered the search for solar neutrinos. The pioneering experiment took place at the Homestake Gold Mine in South Dakota,  continuously  operating between 1970 and 1994. While succeeding in detecting neutrinos emitted by the Sun, the Homestake experiment was at the origin of the solar neutrino problem. Indeed, the experiment measured a  neutrino capture rate of $2.56 \pm 0.16 \mathrm{(statistical)} \pm 0.16 \mathrm{(systematic)}\;\mathrm{SNU}$~\cite{Cleveland:1998nv}, where a Solar Neutrino Unit is defined as $1\;\mathrm{SNU}=10^{-36} \mathrm{events}.\mathrm{atom}^{-1}.\mathrm{s}^{-1}$. This roughly corresponded to a third of the capture rate predicted by the Standard Solar Model (SSM). However, as the SSM was being increasingly substantiated by helioseismic measurements, it became difficult to find an astrophysical explanation to the solar neutrino problem.

The solution to the solar neutrino problem came from neutrino oscillations and the Mikheev–Smirnov–Wolfenstein (MSW) effect. In 1978, Lincoln Wolfenstein introduced an effective potential due to coherent forward elastic scattering in matter~\cite{Wolfenstein:1977ue}. All neutrino flavours can interact through neutral currents with the medium, which modifies the total Hamiltonian in matter. However, since this potential is the same for all flavours, it only results in a common phase shift that does not affect the oscillation probability. On the contrary, only electron neutrinos can interact with electrons through charged currents, which introduces an effective potential that modifies the oscillation probabilities involving electron neutrinos. Later, in 1985, Stanislav Mikeev and Alexei Smirnov realised that for a specific electron density the transition becomes resonant with a maximal mixing angle~\cite{Mikheev:1986gs, *Mikheev:1986wj}.

\begin{figure}[!t]
\centering
 \includegraphics[angle=270,width=0.7\textwidth]{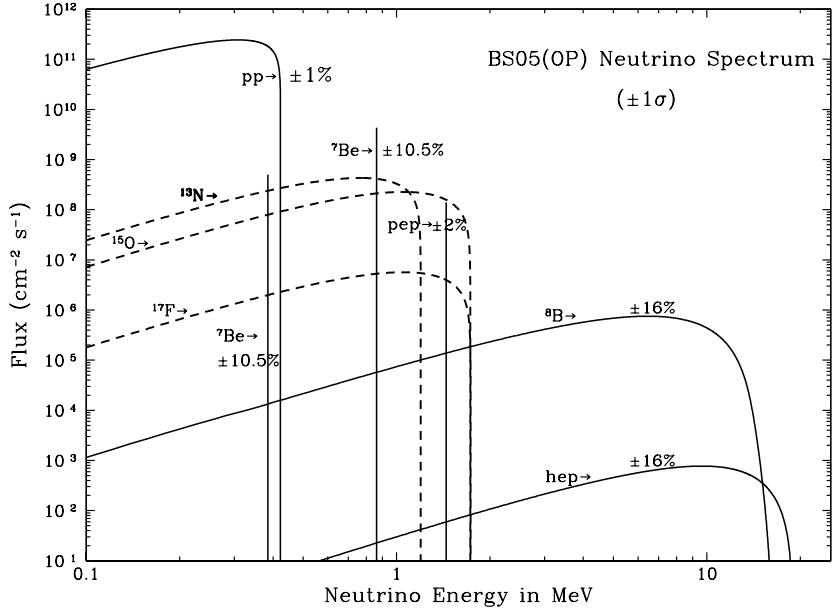}
\caption[Energy spectra of solar neutrinos]{Energy spectra of solar neutrinos according to the production mechanism for the BS05 solar model. The differential fluxes for continuous sources are in $\mathrm{cm}^{-2}.\mathrm{s}^{-1}.\mathrm{MeV}^{-1}$. Reproduced by permission of the American Astronomical Society from~\cite{Bahcall:2004pz}.}
\label{SolarNu}
\end{figure}

\begin{figure}[t]
\centering
 \includegraphics[width=0.5\textwidth]{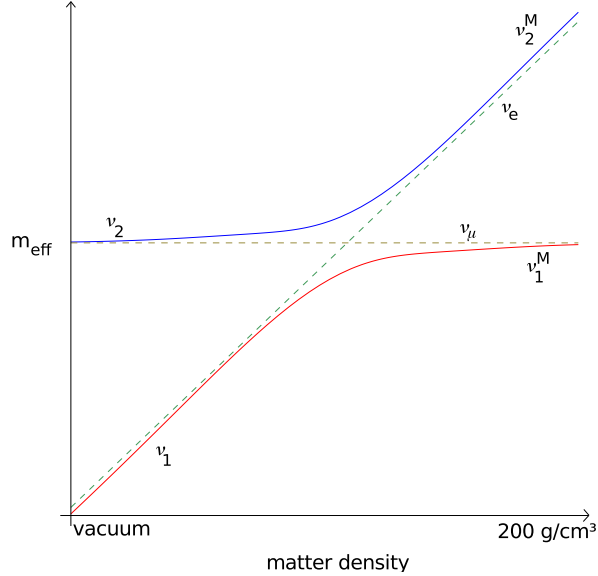}
\caption[MSW effect in the Sun]{Illustration of the level crossing in the Sun due to the MSW effect.}
\label{MSWsun}
\end{figure}

Solar neutrinos have different energies, depending on their production mechanism within the Sun as depicted in fig.~\ref{SolarNu}.
Since low-energy neutrinos ($pp$, $pep$, $^7\mathrm{Be}$, CNO) have an oscillation length shorter than the size of the solar core, the MSW effect is averaged and the oscillation probability is driven by the vacuum effects. However, this is not true for more energetic neutrinos ($^8\mathrm{B}$, $hep$). In fact, at the center of the Sun, the electron neutrino $\nu_e$ nearly coincides with the mass eigenstate $\nu_2^M$, see fig.~\ref{MSWsun}.
If the MSW resonance region is crossed adiabatically, there is no transition between different mass eigenstates and the neutrino emerges from the Sun as a mass eigenstate $\nu_2$ which has a large $\nu_\mu$ component. As a consequence, through neutrino oscillations and the MSW effect, electron neutrinos produced in the Sun oscillate into other flavours, which explains the observed deficit.

This hypothesis received a definite confirmation from the SNO experiment (Sudbury Neutrino Observatory) which measured the rate of three different reactions in heavy water
\begin{align}
 \nu_x + d &\rightarrow p + n + \nu_x\,, &\mathrm{(neutral\;current)}\\
 \nu_x + e^- &\rightarrow \nu_x + e^-\,, &\mathrm{(elastic\;scattering)}\\
 \nu_e + d &\rightarrow p + p + e^-\,. &\mathrm{(charged\;current)}
\end{align}
The detection is made through a combination of Cherenkov light emitted by the electron in the last two reactions, and scintillation from the neutron capture by heavy water for the first reaction. This allowed to confirm that the total $^8\mathrm{B}$ neutrino flux was in agreement with the SSM prediction and to determine the mixing angle $\theta_{12}$ as well as the squared mass difference $\Delta m^2_{12}$~\cite{Aharmim:2011vm}. The SNO experiment, located in the Creighton Mine in Sudbury (Ontario, Canada), stopped data taking in November 2006. It is currently being upgraded into SNO+, a kton scale liquid scintillator detector. There are currently four running experiments that make use of solar neutrinos: Super-Kamiokande (SK), Borexino, the Imaging Cosmic And Rare Underground Signals (ICARUS) experiment and the KAMioka Liquid scintillator Anti-Neutrino Detector (KamLAND). Super-Kamiokande is a $50\;\mathrm{kton}$ water Cherenkov detector built in the Kamioka mine in Japan. Neutrinos are detected via the Cherenkov light emitted by the final charged lepton in elastic scattering and inverse $\beta$ decay processes. It is worth noting that Super-Kamiokande is able to distinguish between different flavours through the topology of the Cherenkov ring. However, due to the relatively high threshold of this detection method, only $^8\mathrm{B}$ neutrinos can be detected. The results concerning solar neutrinos of the first three running phases can be found in~\cite{Hosaka:2005um, *Cravens:2008aa, *Abe:2010hy}. Borexino is located at the Laboratori Nazionali del Gran Sasso (LNGS) in Italy and detects neutrinos through their elastic scattering on electrons in a $278\;\mathrm{tons}$ organic liquid scintillator target. Due to its low threshold, it has been possible to detect $^8\mathrm{B}$, $^7\mathrm{Be}$ and $pep$ neutrinos and has been able to constrain the MSW effect in Earth by searching for a day-night asymmetry~\cite{Bellini:2008mr, *Bellini:2011rx, *Collaboration:2011nga, *Bellini:2011yj}. ICARUS is a $760\;\mathrm{tons}$ liquid argon time projection chamber (TPC), also located at the LNGS. Neutrinos interact via neutral currents, charged currents and quasi-elastic scattering. The charged particle emitted in the process is detected via the scintillation of the liquid argon and the ionization along the track. ICARUS solar program is detailed in~\cite{Arneodo:2000fa} and focuses on $^8\mathrm{B}$ neutrinos. Finally, KamLAND, as its name states, is situated at the Kamioka Observatory. $^8\mathrm{B}$ neutrinos are detected through their elastic scattering on the electrons of a $1\;\mathrm{kton}$ target made of organic liquid scintillator~\cite{Abe:2011em}.

\subsubsection{Atmospheric neutrinos}

\begin{figure}[!t]
\centering
 \includegraphics[width=\textwidth]{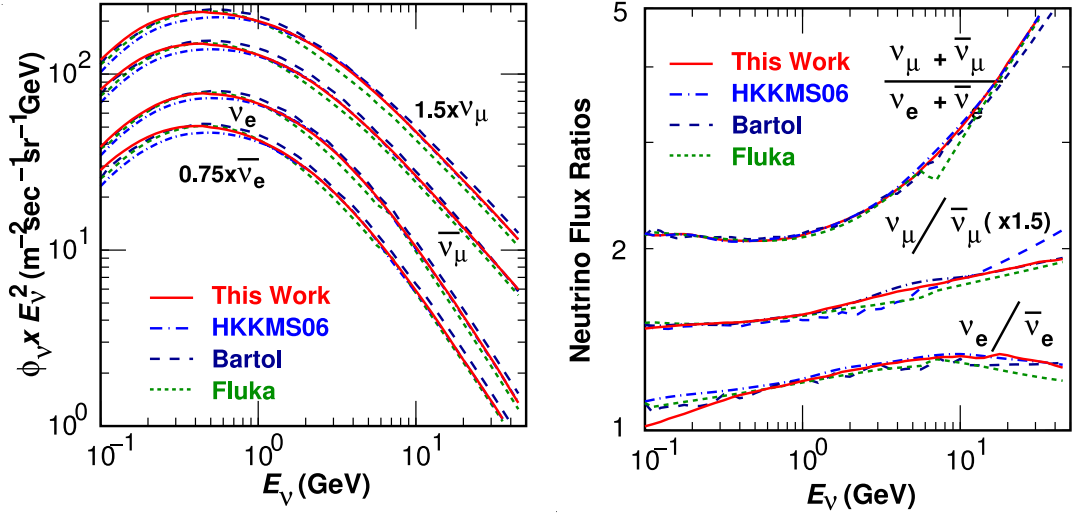}
\caption[Atmospheric neutrino fluxes]{All direction average of the atmospheric neutrino fluxes: (left) absolute values for each type of neutrino, (right) ratios defined in eq.~\ref{atmratios}. The figure is taken from~\cite{Honda:2011nf}; copyright 2011 by the American Physical Society.}
\label{AtmosphericNu}
\end{figure}

Atmospheric neutrinos are produced when cosmic rays reach the Earth atmosphere. Primary cosmic rays are mostly made of proton and alpha particles and generate secondary cosmic rays in the form of atmospheric showers by interacting with nuclei in the atmosphere. These secondary rays contain many hadrons that decay into muons, which subsequently decay, leading to neutrinos with an energy between $100\;\mathrm{MeV}$ and $100\;\mathrm{GeV}$. The corresponding fluxes can be seen in fig.~\ref{AtmosphericNu}.
For energies below $1\;\mathrm{GeV}$, most muons decay in the atmosphere and the neutrino production is driven by
\begin{align}
 \pi^+ (\pi^-) &\rightarrow \mu^+ (\mu^-) + \nu_\mu (\overline{\nu_\mu})\,,\nonumber\\
 \mu^+ (\mu^-) &\rightarrow e^+ (e^-) + \nu_e (\overline{\nu_e}) + \overline{\nu_\mu} (\nu_\mu )\,,
\end{align}
which leads to the following flux ratios
\begin{equation}
 \frac{\phi_{\nu_\mu}+\phi_{\overline{\nu_\mu}}}{\phi_{\nu_e}+\phi_{\overline{\nu_e}}}\simeq 2\,, \quad \frac{\phi_{\nu_\mu}}{\phi_{\overline{\nu_\mu}}}\simeq1 \,, \quad \frac{\phi_{\nu_e}}{\phi_{\overline{\nu_e}}}\simeq\frac{\phi_{\mu^+}}{\phi_{\mu^-}}\,.
\label{atmratios}
\end{equation}
At higher energies, the fraction of muons that have decayed decreases, which in turn increases the ratio $(\phi_{\nu_\mu}+\phi_{\overline{\nu_\mu}})/(\phi_{\nu_e}+\phi_{\overline{\nu_e}})$. Moreover, the length dependence of this ratio can be probed by looking at the zenith-angle dependence of the measurement.

As already mentioned in Section~\ref{NuHistory}, the first atmospheric neutrinos were detected in 1965~\cite{Reines:1965qk, Achar1965196} and a deviation from the expected value of the ratio $(\phi_{\nu_\mu}+\phi_{\overline{\nu_\mu}})/(\phi_{\nu_e}+\phi_{\overline{\nu_e}})$ prompted the SK collaboration to claim an evidence for neutrino oscillations in 1998~\cite{Fukuda:1998mi}. Currently, six experiments have an atmospheric neutrino program: SK, the Main Injector Neutrino Oscillation Search (MINOS), ICARUS, the Astronomy with a Neutrino Telescope and Abyss environmental RESearch (ANTARES) experiment, IceCube and the Baikal NT200 neutrino telescope. SK has had a very rich program involving atmospheric neutrinos, measuring the mixing angle $\theta_{23}$ and the squared mass difference $\Delta m^2$, searching for CP violation, non-standard neutrino interactions and tau neutrinos~\cite{Wendell2010md, *Abe:2011ph, *Mitsuka:2011ty, *Abe:2012jj}. MINOS is an accelerator-based experiment whose far detector is a magnetized steel-scintillator calorimeter. Muon neutrinos are detected through their charged current interactions and the magnetic field  is used to determine the charge of the produced muons, allowing to distinguish between neutrino and antineutrino events. MINOS can also detect electron neutrinos via the shower induced by charged and neutral interactions. The corresponding interaction rate can be used to determine the mixing angle and squared mass difference for atmospheric neutrinos~\cite{Adamson:2012gt}. ICARUS also has a program dedicated to atmospheric neutrino searches with the possibility to observe charged and neutral current processes involving all neutrino flavours~\cite{Arneodo:2001tx}. ANTARES, IceCube and Baikal are neutrino telescopes with similar operating principles: high-energy neutrinos interact with a medium generating charged particles that will emit Cherenkov light along their track. By measuring the light cone characteristics, like its opening or its fuzziness, these experiments can distinguish between different event types. Located in the  Mediterranean Sea, off the coast of Toulon, ANTARES can detect muons with energies higher than $20\;\mathrm{GeV}$ and this ability has been used to measure the atmospheric mixing angle~\cite{AdrianMartinez:2012ph}. As for IceCube, it makes use of the South Pole ice cap as a detection medium and covers now around $1\;\mathrm{km}^3$. It recently measured the electron neutrino flux in the energy range from $80\;\mathrm{GeV}$ to $6\;\mathrm{TeV}$ and observed atmospheric neutrino oscillations using a sample of low-energy muons between $20\;\mathrm{GeV}$ and $100\;\mathrm{GeV}$~\cite{Aartsen:2012uu, *Aartsen:2013jza}. Finally, Baikal is similar to ANTARES, except for its location since it is more than $1\;\mathrm{km}$ below the surface of lake Baikal in Russia~\cite{Aynutdinov:2009zzb}.

\subsubsection{Reactor neutrinos}

Nuclear reactors have been instrumental in probing neutrino properties. They are important sources of neutrinos since they yield roughly $10^{20} \overline{\nu_e}.\mathrm{s}^{-1}.\mathrm{GW_{th}}^{-1}$, produced in the $\beta$ decay of fission products. However, nuclear reactors have some drawbacks as neutrino sources. First, the neutrino flux is isotropic, which means that it decreases as $L^{-2}$, making it rapidly quite small and difficult to distinguish from natural backgrounds. Second, the emitted neutrinos have energies of a few MeV, with a rapidly decreasing flux when the energy increases. As a consequence, only electron antineutrino disappearance can be studied, since the heavier charged leptons cannot be produced at such energies ($m_\mu\simeq 106\;\mathrm{MeV}\,,\;m_\tau\simeq 1777\;\mathrm{MeV}$). Moreover, the detection through inverse $\beta$ decay has an energy threshold of roughly $m_n-m_p+m_e\simeq1.8\;\mathrm{MeV}$, reducing the flux of detectable neutrinos to a quarter of the emitted flux. Finally, the spectrum of emitted $\overline{\nu_e}$ is very difficult to compute because many isotopes contribute to the decay chains. A recent calculation~\cite{Mueller:2011nm} led to a new evaluation of the flux, approximately $3\%$ higher than the previous one, in disagreement with the measurements of short-baseline (SBL, $10\;\mathrm{m}<L<100\;\mathrm{m}$) reactor experiments. This issue is known as the reactor anomaly.

There are currently four running reactor experiments, all of them measuring long-baseline (LBL, $L\geq1\;\mathrm{km}$) oscillations: KamLAND, Double Chooz, the 
Reactor Experiment for Neutrino Oscillation (RENO) and Daya Bay. KamLAND is unique since it is the only one with a very long baseline, having a flux-weighted average distance of $180\;\mathrm{km}$ between the detector and the 53 nuclear power reactors in Japan. KamLAND was able to provide a precise measurement of the squared mass difference for the solar neutrino oscillations and has recently used the extensive shutdown of Japanese nuclear reactors to study the background of this measurement~\cite{Abe:2008aa, *Gando:2013nba}. The three other experiments all use a common approach with near detectors close to the nuclear reactors and far detectors roughly $1\;\mathrm{km}$ away. Moreover, the near detectors are scaled down versions of the far detectors, both filled with gadolinium-doped liquid scintillator, and the disappearance probability is measured between them. This allows the experiments to remove many systematic uncertainties, especially those associated with the flux of neutrinos produced in the power reactors. This was instrumental to measure the last mixing angle $\theta_{13}$~\cite{An:2012bu, *Abe:2012tg, *Abe:2013sxa, *Ahn:2012nd}.

\subsubsection{Accelerator neutrinos}

Accelerator neutrinos may be produced through three different methods: decays in flight of pions and kaons, muon decays at rest and beam dumps. Hadron decays produce a beam of muon neutrinos and antineutrinos, whose composition can be selected through focusing horns and which is in turn used to search for muon neutrino disappearance and electron and tau neutrino appearance. Muon decay at rest was used in $\overline{\nu_\mu}\rightarrow\overline{\nu_e}$ experiments, while beam dumps were used in historical experiments like DONUT, which discovered the tau neutrino~\cite{Kodama:2000mp}, or Gargamelle, which discovered weak neutral currents~\cite{Hasert1973138, *Hasert1973121}.

\begin{figure}[!t]
\centering
 \includegraphics[width=0.6\textwidth]{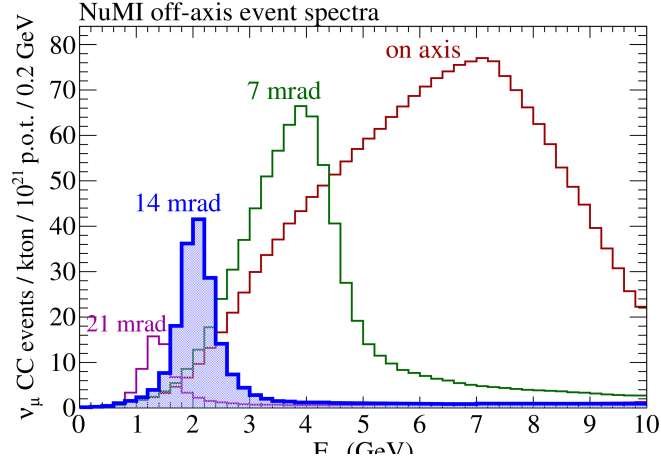}
\caption[Neutrino energy spectra from the NuMI beam]{Neutrino energy spectra in an off-axis detector for various positions. Reprinted from~\cite{Patterson:2012zs}, copyright 2013, with permission from Elsevier.}
\label{OffAxis}
\end{figure}

Present accelerator neutrino experiments can be separated in two classes. The first contains five experiments with long baselines: ICARUS, the Oscillation Project with Emulsion-tRacking Apparatus (OPERA), MINOS, the NuMI Off-axis $\nu_e$ Appearance (NO$\nu$A) experiment and the Tokai to Kamioka (T2K) experiment. All use beams from in flight decay of hadrons that were produced by the collision of a proton beam with a graphite target. Protons extracted from the SPS at $400\;\mathrm{GeV}$ generate the CERN Neutrinos to Gran Sasso (CNGS) beam, which is then sent to ICARUS and OPERA located $730\;\mathrm{km}$ away from CERN. The high energy of the beam is crucial since it allows the detection of tau neutrinos through charged current interactions. ICARUS has recently published a result~\cite{Antonello:2012pq} that severely constrains an earlier claim from the LSND experiment of an oscillation with a squared mass difference at the eV scale~\cite{Aguilar:2001ty}. OPERA has been optimized for the detection of $\tau$ leptons produced in the detector. Thus, it is has two Super-Modules composed of a target section followed by a magnetic muon spectrometer. The target is made of bricks, in which lead plates and nuclear emulsion films alternate, separated along the beam direction by scintillator strips. The collaboration has recently announced the observation of $\tau$ leptons~\cite{Agafonova:2013dja}  and has also restricted the parameter space compatible with the LSND signal by searching for $\nu_e$ appearance~\cite{Agafonova:2013xsk}. The Neutrinos at the Main Injector, or NuMI,  beam is produced at Fermilab from $120\;\mathrm{GeV}$ protons and then sent to MINOS and NO$\nu$A. As previously mentioned, the MINOS far detector is a magnetized steel-scintillator calorimeter, $735\;\mathrm{km}$ away from Fermilab. But MINOS also has a near detector, which is a smaller replica of the far one, providing essentially the same advantages that the two detector configuration conveys for reactor experiments. Its observation of $\nu_\mu$ and $\overline{\nu_\mu}$ disappearance provided a measurement of the atmospheric squared mass difference~\cite{Adamson:2007gu, *Adamson:2012rm} but MINOS has also searched for $\nu_e$ and $\overline{\nu_e}$ appearance and non-standard interactions~\cite{Adamson:2011ku, *Adamson:2013ue, *Adamson:2013ovz}. NO$\nu$A has a $14\;\mathrm{kton}$ highly segmented liquid scintillator far detector, with tracking capabilities, located $810\;\mathrm{km}$ away from the source and $14\;\mathrm{mrad}$ off the beam axis, and a similar $0.3\;\mathrm{kton}$ near detector. This configuration is very interesting since it gives a narrower neutrino spectrum in the far detector as can be seen in fig.~\ref{OffAxis}. Data taking should begin in 2013 in neutrino mode and the experiment will, in the end, search for $\nu_\mu$ and $\overline{\nu_\mu}$ disappearance and $\nu_e$ and $\overline{\nu_e}$ appearance~\cite{Patterson:2012zs}. The last LBL experiment is T2K, which uses a beam from J-PARC in Tokai where $50\;\mathrm{GeV}$ protons are extracted from the Main Ring. Its far detector is SK, $22\;\mathrm{mrad}$ off the beam axis with a $295\;\mathrm{km}$ baseline between Tokai and Kamioka. T2K uses 2 near detectors, one on-axis to precisely monitor the beam and another one off-axis to study neutrino interactions and measure the neutrino flux. Its physic program is similar to the one from MINOS or NO$\nu$A and the collaboration recently published results for $\nu_\mu$  disappearance and $\nu_e$  appearance~\cite{Abe:2012gx, *Abgrall:2013xua}.

The second class of accelerator-based experiments searches for short baseline oscillations, with the aim of confirming, or excluding, the LSND anomaly. Currently taking data, MiniBooNE is the first stage of the Booster Neutrino Experiment (BooNE) using only one detector $541\;\mathrm{m}$ from the target on which $8\;\mathrm{GeV}$ protons from the Fermilab Booster collide. The second stage of the experiment will be the operation of a second detector, only $200\;\mathrm{m}$ from the target. Neutrinos are detected through Cherenkov light and scintillation generated when they interact within $806\;\mathrm{tons}$ of mineral oil. MiniBooNE has seen evidence of $\nu_e$ and $\overline{\nu_e}$ appearance for which there is a small overlap in the oscillation parameter space with LSND~\cite{AguilarArevalo:2012va, *Aguilar-Arevalo:2013pmq}. The MiniBooNE collaboration also worked with the SciBar Booster Neutrino Experiment (SciBooNE) collaboration, searching for $\nu_\mu$ and $\overline{\nu_\mu}$ disappearance~\cite{Mahn:2011ea, *Cheng:2012yy}. The SciBooNE detector is located $100\;\mathrm{m}$ after the target and uses a scintillator tracker, an electromagnetic calorimeter and a muon detector to measure the neutrino beam flux for MiniBooNE.

A last accelerator experiment, that is actively taking data, is the Main INjector ExpeRiment for $\nu$-A (MINER$\nu$A). Detecting neutrinos from the NuMI beam, it aims at  improving the current measurement of neutrino detection cross-sections and at studying the neutrino--nucleus interactions~\cite{Fields:2013qxa, *Fiorentini:2013ezn}.

\subsubsection{Geo-neutrinos}

\begin{figure}[!t]
\centering
 \includegraphics[width=0.6\textwidth]{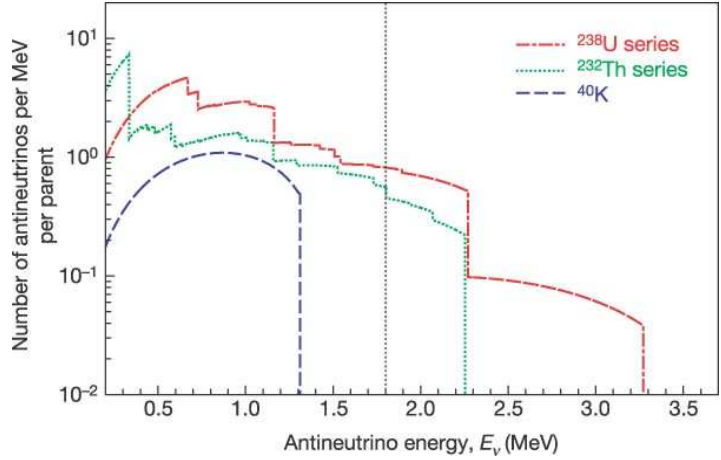}
\caption[Geo-neutrino energy distributions]{Expected geo-neutrino energy distributions. The vertical dotted line is the energy threshold for detection through inverse beta decay. Reprinted by permission from Macmillan Publishers Ltd: Nature~\cite{Araki:2005qa}, copyright 2005.}
\label{GeoNu}
\end{figure}

Geo-neutrinos are electron antineutrinos from natural radioactivity. They are mostly produced in $\beta$ decays of $^{40}\mathrm{K}$ and isotopes from the $^{238}\mathrm{U}$ and $^{232}\mathrm{Th}$ decay chains. As can be seen from their energy distribution  on fig.~\ref{GeoNu},
electron antineutrinos emitted by $^{40}\mathrm{K}$ are not energetic enough to be detected. As of today, geo-neutrinos have been detected in two experiments: KamLAND and Borexino~\cite{Araki:2005qa, Bellini:2013nah}. Studies of geo-neutrinos are interesting since they could help distinguish between models with different crust or mantle compositions and deduce the Earth radiogenic heat power.

\subsubsection{Direct mass measurements}

Neutrino oscillations imply that neutrinos are massive with non-zero mixing angles. However, oscillation experiments can only measure squared mass differences and thus, the question of the absolute mass scale of neutrinos remains open. Many experiments have been devoted to this search and the methods employed include neutrinoless double $\beta$ decay, which will be described in more detail below, monochromatic lines in the muon energy spectrum in pion decays~\cite{Assamagan:1995wb} and the measurement of the electron spectrum in $\beta$ decay. The latter technique was originally proposed by E. Fermi and Francis Perrin in 1933~\cite{fermi1934versuch, *fermi1934tentativo, perrin1933possibilite}. The $Q$-value of a $\beta$ decay is defined by
\begin{equation}
 Q = m_i-m_f-m_e\,,
\end{equation}
where $m_i$ and $m_f$ are respectively the mass of the initial and final nucleus and $m_e$ is the electron mass. Neglecting the recoil energy of the final nucleus with respect to its mass energy, the kinetic energy of the electron, defined as $T= E_e - m_e$, is maximal when the neutrino has a zero momentum in the rest frame of the decaying nucleus
\begin{equation}
 T_{\mathrm{max}}=Q - m_{\nu_e}\,.
\end{equation}
From the above formula, one can see that the neutrino mass will directly influence the end-point of the electron energy spectrum. The latter was derived assuming no neutrino mixing. Taking into account lepton mixing in charged current interactions parametrized by the PMNS matrix given in eq.~\ref{PMNS}, the differential decay rate is modified via the following substitution~\cite{Shrock:1980vy, McKellar:1980cn, *Kobzarev:1980nk}
\begin{equation}
 (Q-T)\sqrt{(Q-T)^2-m_{\nu_e}^2} \rightarrow (Q-T)\sum_{i=1}^3|U_{ei}|^2\sqrt{(Q-T)^2-m_i^2}\,,
\label{Kurie}
\end{equation}
 where $m_i\;(i=1,2,3)$ are the masses of neutrino eigenstates. This leads to a shift of the end-point value to $T_{\mathrm{max}}=Q - m_1$ in the normal hierarchy and $T_{\mathrm{max}}=Q - m_3$ in the inverted hierarchy. It also introduces kinks at $T=Q - m_i$, whose sizes are proportional to $|U_{ei}|^2$. Finally, if an experiment does not observe the effects of neutrino masses, it implies that its resolution in $T$ is much larger than the neutrino masses. Eq.~(\ref{Kurie}) can then be expanded in powers of $m_k/(Q-T)$, leading to 
\begin{equation}
 (Q-T)\sum_{i=1}^3|U_{ei}|^2\sqrt{(Q-T)^2-m_i^2}\simeq(Q-T)\sqrt{(Q-T)^2-m_\beta^2}\,,
\end{equation}
with an effective electron neutrino mass in $\beta$ decay given by
\begin{equation}
 m_\beta^2=\sum_{i=1}^3|U_{ei}|^2 m_i^2\,.
\end{equation}
This effective neutrino mass, because of its dependence on the mass ordering, can then be used to distinguish between the normal and inverted hierarchies of the neutrino spectrum if measured.

There are two main experimental techniques to measure the electron spectrum end-point. The first one was chosen by the two experiments that have given the best upper limit of the effective electron neutrino mass, the Mainz and Troitsk experiments~\cite{Kraus:2004zw, *Aseev:2011dq},
\begin{equation}
 m_\beta < 2.05\;\mathrm{eV}\quad\mathrm{at}\;95\%\;\mathrm{CL}\,.
\end{equation}
These experiments both used tritium as the $\beta$ decaying isotope, because its low $Q$-value ($18.6\;\mathrm{keV}$) allows the use of electromagnetic spectrometers with an energy resolution as low as  $3-4\;\mathrm{eV}$. The two collaborations have subsequently merged and work on the construction of the Karlsruhe Tritium Neutrino (KATRIN) experiment, the next generation of spectrometer experiment, which is designed to reach a sensitivity of $0.4\;\mathrm{eV}$ at $90\%$ CL within three years of running~\cite{Angrik:2005ep}. The other experimental approach makes use of cryogenic microcalorimeters. In this case, all the energy released by the $\beta$ decay is detected, except for the energy carried away by the neutrino, which reduces uncertainties by removing all atomic and molecular final state effects. The currently running MARE experiment built thermal detectors made of $^{187}\mathrm{Re}$, which has one of the lowest known transition energies $Q\simeq2.47\;\mathrm{keV}$. This is a strong experimental advantage since the number of decays near the spectrum end-point is proportional to $Q^{-3}$. As a consequence, the useful fraction of events is roughly 350 times higher in rhenium than in tritium. The aim of the experiment is to reach a sensitivity of $0.2\;\mathrm{eV}$ by deploying an array of 50000 detectors~\cite{collaboration2006mare, *Nucciotti:2010tx}.

\subsubsection{Neutrinoless double beta decay}

\begin{figure}[!t]
\centering
\subbottom[Two neutrinos double $\beta$ decay][$2\beta_{2\nu}^-$]{
\begin{fmffile}{2nu2beta}
\begin{fmfgraph*}(180,105)
\fmflabel{$d$}{d1}
\fmflabel{$d$}{d2}
\fmflabel{$u$}{u1}
\fmflabel{$e^-$}{e1}
\fmflabel{$\overline{\nu_e}$}{nu1}
\fmflabel{$\overline{\nu_e}$}{nu2}
\fmflabel{$e^-$}{e2}
\fmflabel{$u$}{u2}
\fmfforce{(0.0w,1.0h)}{d1}
\fmfforce{(0.0w,0.0h)}{d2}
\fmfforce{(1.0w,1.0h)}{u1}
\fmfforce{(1.0w,0.0h)}{u2}
\fmfforce{(1.0w,0.8h)}{e1}
\fmfforce{(1.0w,0.6h)}{nu1}
\fmfforce{(1.0w,0.4h)}{nu2}
\fmfforce{(1.0w,0.2h)}{e2}
\fmfforce{(0.6w,0.7h)}{We1}
\fmfforce{(0.6w,0.3h)}{We2}
\fmf{fermion, tension=0.6}{d1,W1}
\fmf{fermion, tension=0.3}{W1,u1}
\fmf{fermion, tension=0.6}{d2,W2}
\fmf{fermion, tension=0.3}{W2,u2}
\fmffreeze
\fmf{wiggly,label=$W$}{W1,We1}
\fmf{wiggly,label=$W$}{W2,We2}
\fmf{fermion}{We1,e1}
\fmf{fermion}{nu1,We1}
\fmf{fermion}{We2,e2}
\fmf{fermion}{nu2,We2}
\end{fmfgraph*}
\end{fmffile}
\label{2nu2beta}
}
\hfill
\subbottom[Neutrinoless double $\beta$ decay][$2\beta_{0\nu}^-$]{
\begin{fmffile}{0nu2beta}
\begin{fmfgraph*}(180,105)
\fmflabel{$d$}{d1}
\fmflabel{$d$}{d2}
\fmflabel{$u$}{u1}
\fmflabel{$e^-$}{e1}
\fmflabel{$e^-$}{e2}
\fmflabel{$u$}{u2}
\fmfforce{(0.0w,1.0h)}{d1}
\fmfforce{(0.0w,0.0h)}{d2}
\fmfforce{(1.0w,1.0h)}{u1}
\fmfforce{(1.0w,0.0h)}{u2}
\fmfforce{(1.0w,0.7h)}{e1}
\fmfforce{(1.0w,0.3h)}{e2}
\fmfforce{(0.6w,0.7h)}{We1}
\fmfforce{(0.6w,0.3h)}{We2}
\fmf{fermion, tension=0.6}{d1,W1}
\fmf{fermion, tension=0.3}{W1,u1}
\fmf{fermion, tension=0.6}{d2,W2}
\fmf{fermion, tension=0.3}{W2,u2}
\fmffreeze
\fmf{wiggly,label=$W$}{W1,We1}
\fmf{wiggly,label=$W$}{W2,We2}
\fmf{fermion}{We1,e1}
\fmf{fermion}{We2,e2}
\fmf{plain, label=$\nu_e$}{We1,We2}
\end{fmfgraph*}
\end{fmffile}
\label{Onu2beta}
}
\caption[Double $\beta$ decays]{Tree-level quark diagrams of double $\beta$ decays.} \label{2beta}
\end{figure}
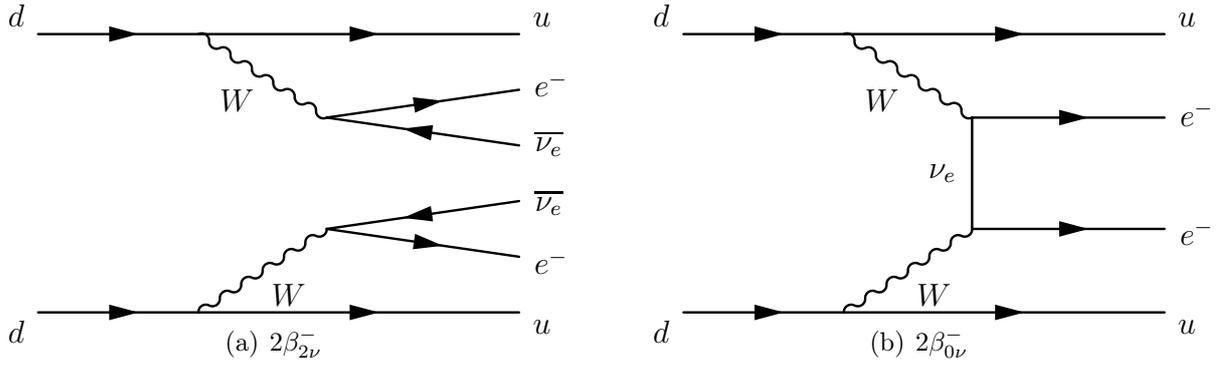

\begin{figure}[!t]
\centering
 \includegraphics[width=0.8\textwidth]{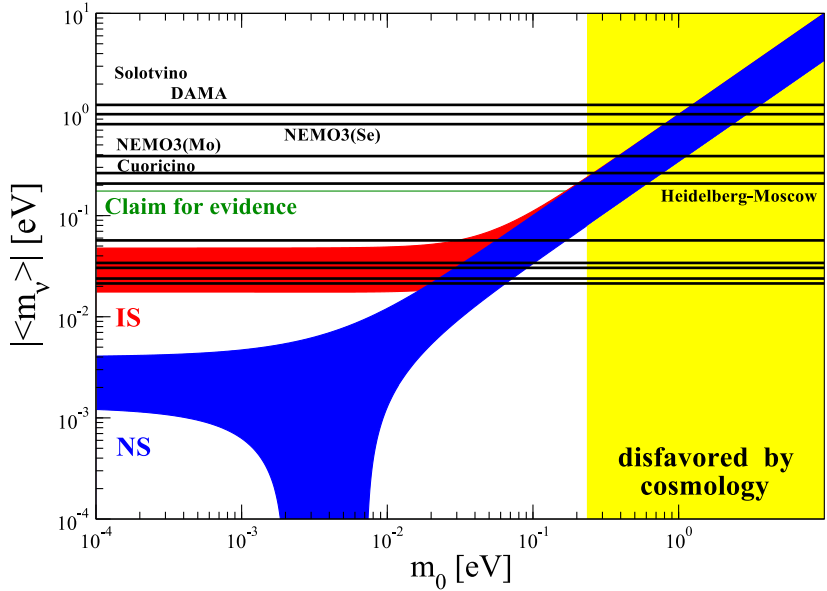}
\caption[Allowed effective Majorana mass range in $2\beta_{0\nu}$ decays]{Allowed effective Majorana mass range as a function of the lightest neutrino mass $m_0$. Here $|\!<\!m_\nu\!>\!|=m_{2\beta}$ and the allowed range is determined from the oscillation parameters. The current limits and expected sensitivities of some experiments are given by the horizontal lines. The figure is taken from~\cite{Vergados:2012xy}, copyright IOP Publishing. Reproduced by permission of IOP Publishing. All rights reserved.}
\label{0nu2betaNHIH}
\end{figure}

Neutrinoless double $\beta$ decays ($2\beta_{0\nu}$) have first been proposed by Wendell Furry in 1939~\cite{Furry:1939qr}. Since $2\beta_{0\nu}$ decays require a particle-antiparticle  and helicity matching, they are forbidden in the Standard Model. Indeed, such decays would violate lepton number conservation. Nevertheless, they can be present if the neutrinos are massive Majorana particles. These processes then occur at tree-level through the diagram in fig.~\ref{Onu2beta}.
Moreover, the helicity matching implies that the $2\beta_{0\nu}$ decay rate depends on the neutrino mass. Taking neutrino mixing into account, the decay amplitude is proportional to the effective Majorana mass~\cite{giunti2007fundamentals}
\begin{equation}
 m_{2\beta}=\sum_{i=1}^3 U_{ei}^2 m_i\,.
\end{equation}
The ability of $2\beta_{0\nu}$ processes to give information on the neutrino mass scale makes these experiments attractive but, more important, its observation would prove that neutrinos are Majorana particles. Indeed, as was demonstrated by Joseph Schecter, Jos\'e Valle and, later, Eiichi Takasugi~\cite{Schechter:1981bd, *Takasugi:1984xr}, the mechanism behind the $2\beta_{0\nu}$ decay can always generate a contribution to the Majorana mass term. This contribution could be cancelled at every order in perturbation theory. However, this is not natural since it would imply a fine-tuning of masses and mixings at every order, unless the Majorana mass term is forbidden by an additional global symmetry. Because the Lagrangian should be invariant under this global symmetry, the latter can only be a phase transformation. If it were a continuous symmetry, then it would correspond to $\mathrm{U}(1)_L$, which would be broken by the $2\beta_{0\nu}$ decay only for the two emitted electrons. But a broken $\mathrm{U}(1)_L$ cannot forbid a Majorana mass term. If it were discrete, then the invariance of weak charged-current interactions would imply
\begin{align}
\phi_W&=\phi_\nu-\phi_e\,,\\
\phi_W&=\phi_u-\phi_d\,,
\end{align}
while the $2\beta_{0\nu}$ is due to $2 d \rightarrow 2 u + 2 e^-$, which gives
\begin{equation}
 \phi_d=\phi_u+\phi_e\,.
\end{equation}
Combining these conditions together, one finds $\phi_\nu=0$ in contradiction with the assumption that the discrete symmetry forbids the Majorana mass term. As a consequence, an operator generating $2\beta_{0\nu}$ decays always gives rise to Majorana mass terms. A comparison between the theoretical predictions and the experimental limits can be found in fig.~\ref{0nu2betaNHIH}.
It is worth noting that in the case of the IH, the effective Majorana mass has a lower bound, which means that excluding $m_{2\beta} > 0.01\;\mathrm{eV}$ will rule out the possibility of an IH neutrino spectrum.

There is currently an intense experimental activity searching for $2\beta_{0\nu}$ decays. The experiments can be grouped according to the nuclei studied and the technology chosen. Two broad categories emerge when the experimental setup is considered, one that uses the source as a detector and another with separated sources and detector. The first category contains germanium detectors, bolometers, liquid scintillators, xenon TPC, while the second one is much less populated, with only one experiment that combines a tracker and a calorimeter. In germanium diodes, where the detector material also serves as the radioactive source, a decay is detected by collecting the emitted charges. The Heidelberg-Moscow Double Beta Decay Experiment, which reported a controversial positive signal~\cite{KlapdorKleingrothaus:2006ff, *KlapdorKleingrothaus:2000sn}, and the GERmanium Detector Array (GERDA), which is currently taking data and aims to probe values of $m_{2\beta}$ below $0.3\;\mathrm{eV}$~\cite{Gerda2013}, are both based on this technology. The MAJORANA experiment currently operates a demonstrator to test the feasibility of a ton-scale experiment based on germanium detectors~\cite{Phillips:2011db}. The main representative of the experiments using bolometers, where the decay is detected through the temperature variation of the detector, is the Cryogenic Underground Observatory for Rare Events (CUORE/CUORICINO). CUORICINO is a tower of  $\mathrm{Te0}_2$ crystals, the first stage of the CUORE experiment, which will have 19 towers similar to CUORICINO and a sensitivity below $0.1\;\mathrm{eV}$. CUORICINO data taking has been completed in 2008 and allowed the collaboration to put an upper bound $m_{2\beta} <0.3-0.7\;\mathrm{eV}$, depending on the nuclear matrix element evaluation considered~\cite{Andreotti:2010vj}. An improvement on this bolometric technique would be to use scintillating crystals, since the background could be strongly reduced by the coincidence requirement between the light and heat signal. A few experiments (LUCIFER~\cite{Beeman:2013vda}, AMORE~\cite{Bhang:2012gn}, MOON~\cite{Fushimi:2010zzb}) are currently characterising different crystals in order to evaluate their potential. Another search strategy is based on large detectors using liquid scintillator with dissolved source isotopes, as KamLAND-Zen or SNO+~\cite{Hartnell:2012qd}, or using the Xenon scintillator as the source, as in XMASS~\cite{Abe:2013tc} or DAMA-LXe~\cite{Bernabei:2002bn}. KamLAND-Zen dissolved $^{136}\mathrm{Xe}$ in its organic scintillator and already completed the analysis of the data from its first phase, resulting in a strong tension with the Heidelberg-Moscow claim since KamLAND-Zen has derived $m_{2\beta} <0.12-0.25\;\mathrm{eV}$~\cite{Gando:2012zm}. Xenon TPC are used by the Enriched Xenon Observatory 200 (EXO 200), which published a very competitive upper-limit last year: $m_{2\beta} <0.14-0.38\;\mathrm{eV}$~\cite{Auger:2012ar}. The last experiment that we will mention is the Neutrino Ettore Majorana Observatory (NEMO). Using a completely different technique, where the source is made of thin foils distributed within the detector, it enjoys a large background rejection since both the energy spectrum and the electron tracks can be measured. However, it is limited by the mass of the radioactive sources that it can contain. NEMO-3 finished taking data in 2011 and most of the complete dataset has been analysed, yielding the upper bound $m_{2\beta} <0.31-0.96\;\mathrm{eV}$~\cite{Barabash:2012gc}. The collaboration wishes to keep working on this technology since it is complementary to the source-detector approach and a demonstrator for SuperNEMO is currently being built.

\subsubsection{Astrophysical neutrinos}

\begin{figure}[!t]
\centering
 \includegraphics[width=0.8\textwidth]{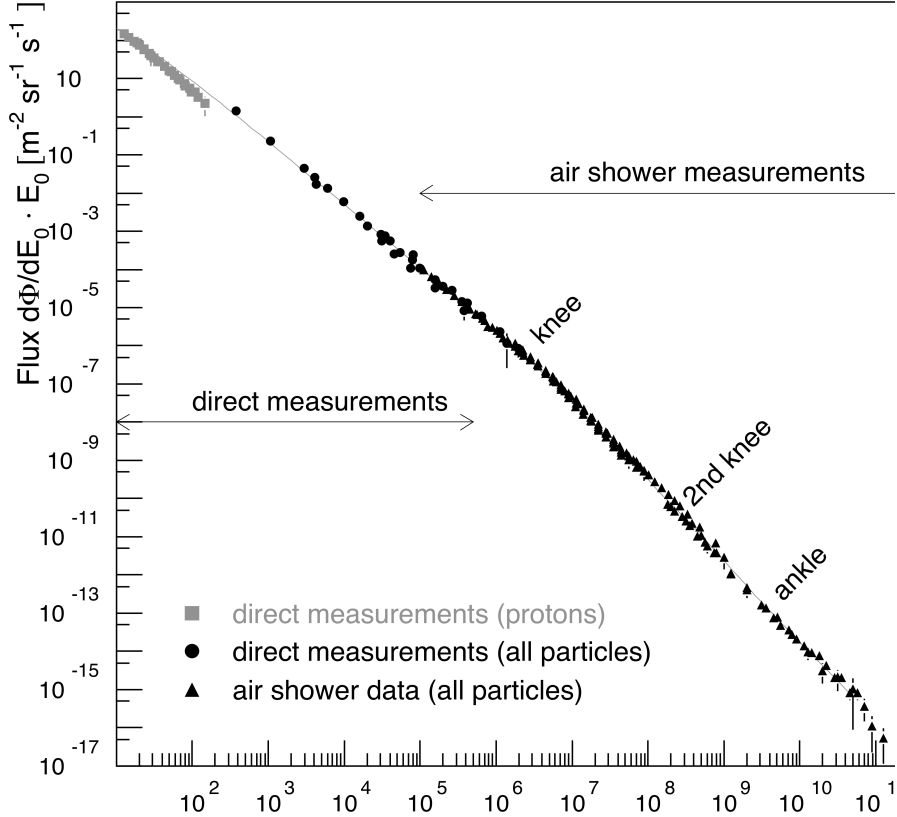}
\caption[Energy spectrum of cosmic rays]{Cosmic ray energy spectrum. Direct measurements refers to flux measurements with detectors outside the atmosphere. Reprinted from~\cite{Bluemer:2009zf}, copyright 2009, with permission from Elsevier.}
\label{CosmicRayFlux}
\end{figure}

As mentioned in the subsection discussing atmospheric neutrinos, many neutrinos are produced when cosmic rays interact with the atmosphere. As can be seen in fig.~\ref{CosmicRayFlux},
most of the cosmic rays are protons with energies below $100\;\mathrm{GeV}$ and these are the sources of the neutrinos studied by experiments searching for atmospheric oscillations. However, cosmic rays can be far more energetic and high-energy neutrinos are expected to be produced. Up to the knee, much of the cosmic rays are supposed to come from galactic sources like supernovae remnants or microquasars. At energies above the PeV, the contributions from extra-galactic sources, such as active galactic nuclei~(AGN) or gamma-ray bursts~(GRB), become important. However, at the highest energies, above $10^{10}\mathrm{GeV}$, the spectrum exhibits a sharp cut-off, the GZK cut-off, that was predicted by Kenneth Greisen, George Zatsepin and Vadim Kuzmin in 1966~\cite{Greisen:1966jv, *Zatsepin:1966jv}. This is due to the interaction of cosmic rays with photons of the Cosmic Microwave Background (CMB). It has recently been confirmed by the atmospheric shower observatory Pierre Auger~\cite{Kampert:2012vh}. The Baikal~\cite{Aynutdinov:2005dq}, ANTARES~\cite{AdrianMartinez:2012rp, *AdrianMartinez:2012tf, *ANTARES:2012uba, *Adrian-Martinez:2013gel, *Adrian-Martinez:2013sga} and IceCube~\cite{Abbasi:2011ke, *Abbasi:2012cu, *Abbasi:2012zw, *IceCube:2012sj} telescope experiments have recently put upper limits on the neutrino flux from GRB, binary systems or point-like sources within our galaxy. Last year, IceCube also reported the observation of two PeV neutrino events~\cite{Aartsen:2013dba}. Another type of experiments uses radio emissions due to the Askaryan effect~\cite{askaryan1962cerenkov} when ultra-high-energy (UHE) neutrinos interact in a dense dielectric medium and emit a cone of coherent radiation, analogous to Cherenkov light. The interaction can be searched for in the South Pole ice as done by the Antarctic Impulsive Transient Antenna~(ANITA)~\cite{Gorham:2010kv}, the Radio Ice Cherenkov Experiment~(RICE)~\cite{Kravchenko:2011im} and the Askaryan Radio Array~\cite{Allison:2011wk}. Other experiments like the NuMoon project~\cite{Buitink:2013gk} or  the Goldstone Lunar Ultra-high energy neutrino Experiment~(GLUE)~\cite{Saltzberg:2000bk} use lunar regolith as the interaction medium. UHE neutrinos can also be searched through the horizontal air-shower they produce in the atmosphere close to the detector or when they interact with dense matter in its vicinity. Using this detection method, the Pierre Auger observatory has not yet reported any candidate event~\cite{Abreu:2011zze}.

Another important source of astrophysical neutrinos is the explosion of supernovae. Supernovae come either from thermonuclear explosions or from the gravitational collapse of the core. The latter scenario is interesting for neutrino Physics since, in these supernovae, $99\%$ of the emitted energy is carried by neutrinos~\cite{Haxton:2012bk}. Moreover, they constitute unique many body systems dominated by neutrinos, where the MSW effect arises from neutrino-neutrino interactions. Supernovae also are important laboratories for nucleosynthesis, in which neutrinos can have major effects. In 1987, the neutrino burst from the explosion of SN1987A was detected by four experiments: Baksan, IMB, Kamiokande and LSD, using different technologies~\cite{Alekseev:1987ej, *Bionta:1987qt, *Hirata:1987hu, *Dadykin:1987ek}. The absence of detection of any other supernova in our galaxy (since SN1987A) puts an upper limit on the rate of core-collapse supernovae in the Milky Way. The detection of the neutrino burst from SN1987A has also been used to constrain neutrino masses~\cite{Loredo:2001rx}, neutrino mixings with heavier neutrinos, the electron antineutrino lifetime~\cite{Schramm:1987ra, *Hirata:1988ad}, the electron neutrino magnetic moment~\cite{Goldman:1987fg, *Lattimer:1988mf, *Barbieri:1988nh} and the neutrino charge~\cite{Barbiellini:1987zz}. Extra-galactic supernovae can also generate a relic neutrino background that has been searched for by SK~\cite{Malek:2002ns}. MiniBooNE, SNO, LVD, Borexino, SK and IceCube even joined together to create an early warning system and serve as trigger for other experiments~\cite{1367-2630-6-1-114}.

Finally, a constraint on the neutrino magnetic momentum can be derived from stellar evolution. When a star with a mass comparable with the Sun leaves the main sequence to become a red giant, its core contracts and, when its temperature and density are high enough, it starts consuming helium. However, if the cooling is enhanced by the neutrino magnetic moment, the core will grow for a longer period, which will change the ratio of red giants to horizontal branch stars. As a consequence, the neutrino magnetic moment has to be smaller than $\mu_{ij}\lesssim10^{-12}\mu_B$, $\mu_B$ being the Bohr magneton~\cite{Raffelt:1990pj}.

\subsubsection{Neutrinos in cosmology}

Since they can carry much of the energy and entropy and can modify the composition of the cosmological fluid through weak interactions, neutrinos are major players in the early Universe. Before decoupling around $T\sim1\;\mathrm{MeV}$, neutrinos participate in Big Bang Nucleosynthesis through reactions that convert neutrons into protons and \textit{vice versa}. The production of $^4\mathrm{He}$ is very sensitive to the neutron-to-proton ratio which depends on the freeze-out temperature. Increasing the number of neutrino species will contribute to the number of relativistic quantum degrees of freedom, parametrized by an effective number of neutrinos $N_{eff}$, thus increasing the decoupling temperature and ultimately the amount of $^4\mathrm{He}$. Taking also into account the abundance of deuterium yields the limit $N_{eff}<4.08$~\cite{Mangano:2011ar}.

After their decoupling, neutrinos will free stream and remain relativistic much longer than the other fermions. This will leave imprints in the CMB that can be used to determine the effective number of light relativistic degrees of freedom and put an upper bound on neutrino masses. The recently released data from Planck gives $N_{eff}=3.36^{+0.68}_{-0.64}$ and $\sum m_\nu < 0.98\;\mathrm{eV}$ at $95\%$ CL~\cite{Ade:2013lta}. Increasing the number of neutrino species will increase the expansion rate before decoupling, enhancing the first acoustic peak in the angular power spectrum and shifting all the peaks towards higher multipole moments. Increasing the neutrino mass will increase the amount of hot dark matter, modifying the angular spectrum through an early integrated Sachs-Wolfe effect\footnote{The Sachs-Wolfe effect corresponds to a gravitational redshift of the CMB photon. In the case of the integrated Sachs-Wolfe effect, the CMB photons travel towards us through area of different gravitational potential and the time evolution of these gravitational potentials causes the redshift. The adjective early insists here on the integrated Sachs-Wolfe effect occurring during the radiation domination era.} for example. Being a form of hot dark matter, massive neutrinos will also reduce the clustering at small scales, modifying the matter power spectrum. This can be verified through the measurement of distant quasar spectra, especially the effect of intermediate hydrogen clouds known as the Lyman-$\alpha$ forest~\cite{Abazajian:2011dt}. Combining Planck data with the baryon acoustic oscillations and the Hubble parameter measurement gives $\sum m_\nu < 0.23\;\mathrm{eV}$ and $N_{eff}=3.52^{+0.48}_{-0.45}$~\cite{Ade:2013lta} under the assumption of a flat Universe.\\
 
\begin{table}[t]
  \begin{center}
    \begin{tabular}{|c|c|c|c|}
     \hline
	Type & L (in km) & E (in MeV) & $\Delta m^2$ (in $\mathrm{eV}^2$) \\
     \hline
     \hline
	Atmospheric & $10^4$ & $10^3$ & $10^{-4\phantom{0}}$ \\
	LBL accelerator & $10^3$ & $10^3$ & $10^{-3\phantom{0}}$ \\
        SBL accelerator & $1$ & $10^3$ & 1 \\
	LBL reactor & $1$ & $1$ & $10^{-3\phantom{0}}$ \\
	SBL reactor & $10^{-2}$ & $1$ & $10^{-1\phantom{0}}$ \\
        Solar & $10^8$ & $1$ & $10^{-11}$ \\
        KamLAND & $10^2$ & $1$ & $10^{-5\phantom{0}}$ \\
     \hline
    \end{tabular}
    \caption[Oscillation experiments and their sensitivities]{\label{OscillationExp} Order of magnitude estimate of the squared mass difference sensitivities for different type of oscillation experiments.}
 \end{center}
\end{table}

\noindent To conclude this chapter, we can say that the last two decades have been very rich in discoveries, confirming the existence of three active massive neutrinos with masses at the eV scale or below, which are a mixture of the flavour eigenstates as depicted in fig.~\ref{NeutrinoHierarchy}. We summarised the different types of neutrino oscillation experiments in table~\ref{OscillationExp} together with their key parameters. Nevertheless, some questions remain open when it comes to the absolute mass of neutrinos, their hierarchy, the $\theta_{23}$ quadrant, the CP violating phases of the leptonic mixing matrix, the short-baseline anomaly observed by LSND and MiniBooNE and the reactor anomaly which seem to point towards the existence of an extra sterile neutrino with a mass around the eV. Other more theoretical issues are still to be clarified as, for instance, the mass generation mechanism or the possible explanation of the baryonic asymmetry of the Universe through leptogenesis.

\renewcommand*{\afterpartskip}{\epigraph{There is a theory which states that if ever anyone discovers exactly what the Universe is for and why it is here, it will instantly disappear and be replaced by something even more bizarre and inexplicable.\\
There is another theory which states that this has already happened.}{\textit{The Restaurant at the End of the Universe}\\ \textsc{Douglas Adams}}
\vfil\clearpage}
\part{{}Fermionic singlets and massive neutrinos}

\chapter{Neutrino mass generation mechanisms\label{chap3}}

As discussed in the previous chapters, a tremendous amount of data has strongly substantiated the hypothesis of oscillating massive neutrinos. In this chapter, we discuss why massive neutrinos necessarily require an extension of the Standard Model. Then, we describe how neutrino masses can be generated, (mostly) following the  conventions of~\cite{giunti2007fundamentals}.

\section{Massive neutrinos and the Standard Model\label{massiveNu}}

The Standard Model has been built along the principle of minimality, which means with a minimal field content and the simplest gauge structure in agreement with observations. As a consequence, the gauge group is $\mathrm{SU}(3)_c\times\mathrm{SU}(2)_L\times\mathrm{U}(1)_Y$ and all the fermions have right-handed (RH) and left-handed (LH) components, except for the three neutrinos. What put them aside ? Neutrinos only interact through the weak interaction, which maximally violates parity by coupling only to the LH component of a fermion. Moreover, when the Standard Model was built, there was no need to incorporate massive neutrinos, since no mass effect had been seen, either in direct kinematic searches from meson decays or from chirality flips, with neutrinos always producing LH leptons and antineutrinos RH antileptons. As a consequence, the neutrino was described in the Standard Model by the two-component theory~\cite{Salam1957parity, landau1957ob, PhysRev.105.1671}, being purely left-handed.

In the two component theory, the neutrino is described by a (two-component) complex spinor, known as a Weyl spinor. These spinors form an irreducible representation of the Clifford algebra $Cl_{(1,3)}(\mathbb{C})$\footnote{ The Clifford algebra $Cl_{(1,3)}(\mathbb{C})$ is the algebra of the spin representation of the Lorentz group $\mathrm{SO}(1,3)$.}.  However, it is more practical to work with bispinors since the fundamental representation of $Cl_{(1,3)}(\mathbb{C})$ is 4-dimensional. Then, the most generic bispinor is the combination of two Weyl spinors 
\begin{equation}
 \psi=\binom{\chi}{\eta}\,.
\end{equation}
In the chiral representation, the Dirac matrices are given by
\begin{equation}
\gamma^0=\left(\begin{array}{c c} 0 & -\mathbf{1}\\ -\mathbf{1} & 0 \end{array}\right)\,,\quad
\vec{\gamma}=\left(\begin{array}{c c} 0 & \vec{\sigma}\\ -\vec{\sigma} & 0 \end{array}\right)\,,\quad
\gamma^5=\left(\begin{array}{c c} \mathbf{1} & 0\\ 0 & -\mathbf{1} \end{array}\right)\,,\label{ChiralRep}
\end{equation}
and the RH and LH projections are
\begin{equation}
 \psi_R=P_R \psi= \frac{\mathbf{1}+\gamma^5}{2} \binom{\chi}{\eta}=\binom{\chi}{0}\,,\quad
 \psi_L=P_L \psi= \frac{\mathbf{1}-\gamma^5}{2} \binom{\chi}{\eta}=\binom{0}{\eta}\,.
\end{equation}
The (free) Dirac equation 
\begin{equation}
 (\imath \slashed{\partial} - m) \psi=0\,,
\end{equation}
applied to this bispinor is
\begin{equation}
 \left(\begin{array}{c c} -m & \imath (-\partial_0 + \vec{\sigma}.\vec{\gamma})\\ \imath (-\partial_0 - \vec{\sigma}.\vec{\gamma}) & -m \end{array}\right)\binom{\chi}{\eta}=0\,,
\end{equation}
which can be rewritten using the chiral components
\begin{align}
 \imath\slashed{\partial}\psi_L&=m \psi_R\,,\label{DiracPsiL}\\
 \imath\slashed{\partial}\psi_R&=m \psi_L\,.\label{DiracPsiR}
\end{align}
It is worth noting that in the massless limit, the right-hand side of eq.~(\ref{DiracPsiL}) and~(\ref{DiracPsiR}) vanishes. The equations thus decouple and correspond to the description of two different particles. However, when the fermion is massive, this bispinor is named a Dirac spinor which, using chiral components, is written as
\begin{equation}
 \psi=\psi_L+\psi_R\,,
\end{equation}
for which the corresponding mass term in the Lagrangian reads
\begin{equation}
 m(\overline{\psi_L}\psi_R+\overline{\psi_R}\psi_L)\,.
\end{equation}
Being built from two different Weyl spinors, the Dirac spinor has 4 degrees of freedom.

It is interesting to consider how the Dirac spinor behaves under charge conjugation. In the chiral representation defined by eq.~\ref{ChiralRep}, the charge conjugation matrix is given by
\begin{equation}
 \mathcal{C}= \imath \gamma^2 \gamma^0\,,
\end{equation}
and the charge conjugated field reads
\begin{equation}
 \psi^C=\xi \mathcal{C} \overline{\psi}^T\,,\label{ChargeConjugation}
\end{equation}
where $\xi$ is an arbitrary phase factor that can be reabsorbed by rephasing the field $\psi$. If we explicitly rewrite eq.~(\ref{ChargeConjugation}) using the chiral components of the Dirac spinor, we find
\begin{align}
 \psi_L^C&=\imath \gamma^2 \psi_L^*\,,\label{ConjRight}\\
 \psi_R^C&=\imath \gamma^2 \psi_R^*\,.
\end{align}
Interestingly, this allows to conclude that $\psi_L^C$ is right-handed. Would it be possible to use this property to define a bispinor from a single Weyl spinor? The answer to this question can be found by considering the Dirac equation for an electromagnetically interacting field $\psi$:
\begin{align}
 &(\imath\slashed{\partial}-q\slashed{A}-m)\psi=0\,,\label{DiracCharged}\\
 &(\imath\slashed{\partial}+q\slashed{A}-m)\psi^C=0\,.\label{DiracCharged2}
\end{align}
It is clear that for charged particles, eqs.~(\ref{DiracCharged}) and~(\ref{DiracCharged2}) are different and indeed describe two different particles. However, neutrinos are singlets of the unbroken gauge group $\mathrm{U}(1)_{em}$ after electroweak symmetry breaking (EWSB). Thus eqs.~(\ref{DiracCharged}) and~(\ref{DiracCharged2}) are identical and one finds that these fields could obey the Majorana condition
\begin{equation}
 \psi=\psi^C\,.
\end{equation}
Using Weyl spinors, this condition can be rewritten
\begin{equation}
 \binom{\xi}{\eta}=\imath \left(\begin{array}{c c} 0 & \sigma^2 \\ -\sigma^2  & 0 \end{array}\right) \binom{\xi^*}{\eta^*}=\imath \binom{\sigma^2 \eta^*}{-\sigma^2 \xi^*}\,,
\end{equation}
and the bispinor can be expressed with only one Weyl spinor $\eta$:
\begin{equation}
 \psi=\binom{\imath \sigma^2 \eta^*}{\eta}\,.
\end{equation}
This type of bispinor is called Majorana spinor and can also be defined using only the LH chiral component as
\begin{equation}
 \psi=\psi_L + \mathcal{C} \overline{\psi_L}^T\,,
\end{equation}
with a mass term in the Lagrangian
\begin{equation}
 \frac{1}{2} m (\overline{\psi_L^C}\psi_L + \overline{\psi_L} \psi_L^C)\,.\label{MajoranaMass}
\end{equation}
As a consequence, a Majorana spinor only has two degrees of freedom. Moreover, since in the massless limit both Dirac and Majorana spinors verify the same decoupled set of equations and only the LH component has gauge interactions, the only way to experimentally probe the nature (Dirac or Majorana) of neutrinos is to identify observables that have a dependence on the neutrino mass in their transition amplitude. Indeed, effects that come from kinematical factors are the same for Dirac and Majorana neutrinos, which explains why oscillations cannot probe the nature of the neutrino.

So far, we have seen that neutrinos are the only particles that can be described either by a Majorana or a Dirac spinor, let us turn to the Standard Model and  see how neutrinos could acquire a mass. Notice that the mass term in eq.~(\ref{MajoranaMass}) violates lepton number conservation by two units. However, lepton number conservation is an accidental symmetry of the SM arising from the gauge group, the particle field content and the renormalizability of the theory. In order to provide a Majorana mass term (\ref{MajoranaMass}) in the SM, one must violate lepton number. In a bottom-up approach, one can try to extend the SM by effective operators of dimension greater than four that parametrize the effects of the new Physics defined at a scale $\Lambda$
\begin{equation}
 \mathcal{L}_{eff}=\mathcal{L}_\mathrm{SM}+\delta \mathcal{L}^{d=5}+\delta \mathcal{L}^{d=6}+...\label{effectiveLagrangian}
\end{equation}
These operators respect the gauge symmetry of the Standard Model but combine SM fields in a non-renormalizable way, violating accidental symmetries like lepton number conservation. Neutrinos belongs to an $\mathrm{SU}(2)$ doublet with a non-zero hypercharge, the leptonic doublet $L$. In order to have a mass term in the Lagrangian which is invariant under $\mathrm{SU}(2)$, a naive idea would be to contract $L$ with itself but this gives a term $\bar \nu_L \nu_L=\bar \nu P_R P_L \nu=0$. One may also consider a term $\bar L L^C$, which has a hypercharge of $+2$ and should therefore be contracted with another field to form a gauge invariant. However, there is no field in the SM with such a hypercharge. The lowest dimensional operator able to generate neutrino masses was pointed out in 1979 by S. Weinberg~\cite{Weinberg:1979sa},
\begin{equation}
 \mathcal{O}_5=\frac{1}{2} f_{ij} \frac{(\overline{L_i^C}\widetilde{\phi}^*)(\widetilde{\phi}^\dagger L_j)}{\Lambda}\,,\label{WeinbergOperator}
\end{equation}
where $f_{ij}$ is a dimensionless coefficient for the operator, which is a complex number, $i,j$ being generation indices $i,j=1,...,3$. It is the only dimension 5 operator that can generate neutrino masses and it violates lepton number conservation by two units. After EWSB, the Majorana neutrino mass term is given by
\begin{equation}
 \frac{1}{2} f_{ij} \frac{v^2}{2\Lambda} (\overline{\nu^C_i} \nu_j + h.c.)\,.
\end{equation}
It is worth emphasing that every new Physics model that generates Majorana mass terms for the neutrinos gives rise to the Weinberg operator, eq.~(\ref{WeinbergOperator}), when the degrees of freedom above the electroweak scale are integrated out.

Since the Weinberg operator is common to all scenarios with Majorana neutrinos, higher-dimensional operators should be considered in order to disentangle the different mass generation mechanisms. A first dimension 6 operator would be the four-fermion interaction $(\bar L\gamma_\mu L) (\bar L \gamma^\mu L)$ that could lead to lepton flavour violating signals like \mbox{$\tau \rightarrow \mu \mu \mu$} for example. Another dimension 6 operator is $(\bar L \widetilde \phi)\slashed \partial (\widetilde \phi^\dagger L)$ that induces non-standard interactions for the neutrinos~\cite{Antusch:2008tz}. But what are the possible mass generation mechanisms that would lead to these operators ?

\section{Mass generation and the seesaw mechanisms}

As we have seen in the previous section, it is necessary to extend the SM in order to generate neutrino masses.  This can be done in many different ways, making the neutrino a Majorana or a Dirac fermion, extending the gauge sector or the field content, etc. But every mechanism should generate a small finite mass to explain the data. For example, one may want to consider the simplest SM extension that consists in the addition of RH neutrinos providing a Dirac mass term to the Lagrangian. But in that case, the Yukawa coupling of the neutrino will have to be $\mathcal{O}(10^{-11})$ in order to have neutrinos with masses around the eV scale. This is not really satisfactory since it adds six orders of magnitude to the mass hierarchy between the SM fermions, which already lacks a clear explanation. Besides, the RH neutrinos being gauge singlets, no symmetry forbids the existence of a Majorana mass term. This idea will actually prove useful later, when we will discuss the type I seesaw mechanism.

Generating a small neutrino mass using natural Yukawa couplings requires a mechanism to suppress the resulting neutrino mass. The mechanism in which this suppression occurs can arise from different types of new Physics. The absence of a RH neutrino can forbid the Dirac mass at tree-level, hence the neutrino mass has to be generated radiatively and will be suppressed by loop factors. This is the neutrino mass generation mechanism of the Zee-Babu model~\cite{Zee:1985id, *Babu:1988ki}, for example. The Dirac mass term could also be forbidden by a discrete symmetry like $Z_2$~\cite{Krauss:2002px}, which can also turn the right-handed neutrino into a dark matter candidate. As a whole, these models and their variants are know as radiative seesaws.

Another possibility is to forbid mass terms with a symmetry whose spontaneous breaking or small violation naturally generates small neutrino masses. This idea appears for instance in R-parity violating supersymmetric models~\cite{Aulakh:1982yn, *Hall:1983id, *Ross:1984yg}. It also occurs in models with a spontaneously broken extension of the SM gauge group~\cite{doi:10.1142/9789812562203}, like $\mathrm{SU}(3)_c\times\mathrm{SU}(2)_L\times\mathrm{U}(1)_Y\times\mathrm{U}(1)_{\mathrm{B-L}}$. 

In a larger framework involving extra-dimensions, the smallness of the neutrino mass can be related to the small overlap between the wave functions of the RH neutrinos and the lepton doublets~\cite{Grossman:1999ra, *ArkaniHamed:1998vp}. The hierarchy of the neutrino spectrum can be predicted in some scenarios, like~\cite{Moreau:2005kz}.

Finally the neutrino mass can be made small because the Dirac mass term is suppressed by a large mass scale like in the seesaw mechanisms or through the presence of multiple singlets as in the inverse seesaw. In the SM, neutrinos are part of an $\mathrm{SU(2)}$ doublet and the product of two doublets belongs either to the singlet or triplet representation of $\mathrm{SU(2)}$. Requiring the mass new terms to be renormalizable, the SM can be minimally extended by adding new fermions, belonging to singlets or triplets, or scalar triplets. The possibility of a scalar singlet is precluded by the condition of electric neutrality of the vacuum state. Indeed, the product $\bar L L^C$ has a hypercharge of $+2$, which would lead to an electrically charged vev for the scalar singlet. The three different new fields can be used to distinguish three types of seesaw mechanisms, which we will present below.

\subsubsection{Type I seesaw}

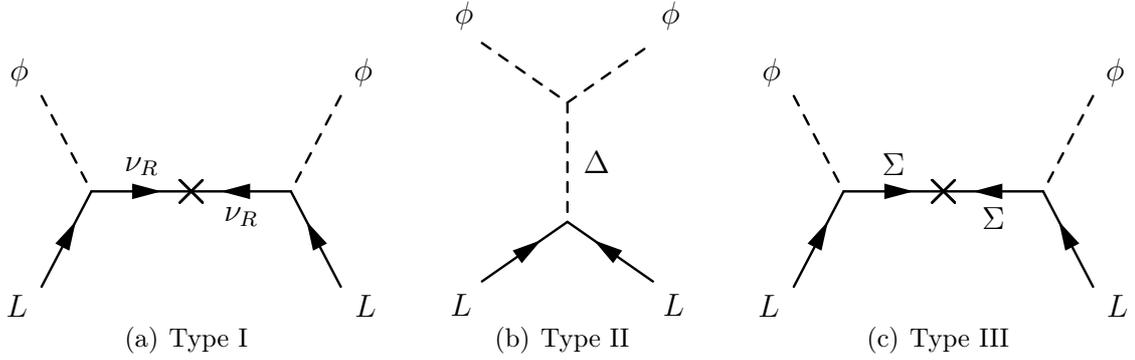
\begin{figure}[!t]
\centering
\subbottom[Type I seesaw][Type I]{
\begin{fmffile}{type1}
\begin{fmfgraph*}(140,80)
\fmflabel{$\phi$}{h1}
\fmflabel{$L$}{l1}
\fmflabel{$\phi$}{h2}
\fmflabel{$L$}{l2}
\fmfforce{0.1w,0.1h}{l1}
\fmfforce{0.1w,1.0h}{h1}
\fmfforce{0.9w,0.1h}{l2}
\fmfforce{0.9w,1.0h}{h2}
\fmfv{decor.shape=cross, decor.size=6thick}{middle}
\fmf{fermion}{l1,h1l1}
\fmf{fermion}{l2,h2l2}
\fmf{fermion,label=$\nu_R$, label.side=left}{h1l1,middle}
\fmf{fermion,label=$\nu_R$}{h2l2,middle}
\fmf{dashes}{h1,h1l1}
\fmf{dashes}{h2,h2l2}
\end{fmfgraph*}
\end{fmffile}
\label{type1}
}
\hfill
\subbottom[Type II seesaw][Type II]{
\begin{fmffile}{type2}
\begin{fmfgraph*}(80,100)
\fmflabel{$\phi$}{h1}
\fmflabel{$L$}{l1}
\fmflabel{$\phi$}{h2}
\fmflabel{$L$}{l2}
\fmfforce{0.1w,0.1h}{l1}
\fmfforce{0.1w,1.0h}{h1}
\fmfforce{0.9w,0.1h}{l2}
\fmfforce{0.9w,1.0h}{h2}
\fmf{fermion}{l1,l1l2}
\fmf{fermion}{l2,l1l2}
\fmf{dashes,label=$\Delta$}{h1h2,l1l2}
\fmf{dashes}{h1,h1h2,h2}
\end{fmfgraph*}
\end{fmffile}
\label{type2}
}
\hfill
\subbottom[Type III seesaw][Type III]{
\begin{fmffile}{type3}
\begin{fmfgraph*}(140,80)
\fmflabel{$\phi$}{h1}
\fmflabel{$L$}{l1}
\fmflabel{$\phi$}{h2}
\fmflabel{$L$}{l2}
\fmfforce{0.1w,0.1h}{l1}
\fmfforce{0.1w,1.0h}{h1}
\fmfforce{0.9w,0.1h}{l2}
\fmfforce{0.9w,1.0h}{h2}
\fmfv{decor.shape=cross, decor.size=6thick}{middle}
\fmf{fermion}{l1,h1l1}
\fmf{fermion}{l2,h2l2}
\fmf{fermion,label=$\Sigma$, label.side=left}{h1l1,middle}
\fmf{fermion,label=$\Sigma$}{h2l2,middle}
\fmf{dashes}{h1,h1l1}
\fmf{dashes}{h2,h2l2}
\end{fmfgraph*}
\end{fmffile}
\label{type3}
}
\caption[Simplest seesaw mechanisms]{Feynman diagrams of the 3 seesaw mechanisms.} \label{seesaw}
\end{figure}

In the original formulation of the SM, neutrinos only have one chirality. A minimal extension of the SM would be to incorporate RH components for the neutrinos. With this addition, a Dirac mass term can be constructed using the Higgs mechanism as for any other SM fermion. However, the RH neutrinos are gauge singlets under the SM gauge group: they have no strong or weak interactions and since they are electrically neutral, they have a zero hypercharge according to the Gell-Mann--Nishijima formula. Besides, no symmetry in the SM gauge group prevents the existence of a Majorana mass term for the RH neutrinos. Thus, with three RH neutrinos\footnote{In order to explain the oscillation data, the minimal number of RH neutrinos that has to be added to the SM is 2.} and before EWSB, the terms in the Lagrangian that will generate neutrino masses are
\begin{equation}
 \mathcal{L}_{\mathrm{type\;I}}= -\sum_{i,j} \left( Y^{ij}_\nu \overline{L_i} \widetilde{\phi} \nu_{Rj} + \frac{1}{2} M_{R}^{ij} \overline{\nu_{Ri}^C} \nu_{Rj} + h.c. \right)\,,
\end{equation}
where $i,j$ are generation indices that run between $1$ and $3$,  $Y_\nu$ is the neutrino Yukawa complex matrix and $M_R$ is a complex symmetric matrix. These are the mass terms in the Lagrangian of the type I seesaw~\cite{Minkowski:1977sc, *GellMann:1980vs, *yanagida1979horizontal, *Mohapatra:1979ia} which give after EWSB
\begin{equation}
 \mathcal{L}_{\mathrm{type\;I}}= -\sum_{i,j} \left( m_D^{ij} \overline{\nu_{Li}} \nu_{Rj} + \frac{1}{2} M_{R}^{ij} \overline{\nu_{Ri}^C} \nu_{Rj} + h.c. \right)\,,
\end{equation}
where the Dirac mass matrix is
\begin{equation}
 m_D=\frac{Y_\nu v}{\sqrt{2}}\,.
\end{equation}
Defining the vector column of LH fields $N_L=(\nu_{L1},...,\nu_{L3}\,,\;\nu_{R1}^C,...,\nu_{R3}^C)^T$, the mass terms can be rewritten as
\begin{equation}
\mathcal{L}_{\mathrm{type\;I}}= -\frac{1}{2} \overline{N_L} M_{\mathrm{type\;I}} N_L^C + h.c. = -\frac{1}{2} \overline{N_L} \left(\begin{array}{c c} 0 & m_D\\ m_D^T & M_R\end{array}\right) N_L^C + h.c.\,.
\end{equation}
In the limit where the values of $M_R$ are much larger than the values of $m_D$, the so called ``seesaw limit'', the neutrino mass matrix $M_{\mathrm{type\;I}}$ can be block-diagonalized~\cite{Kanaya:1980cw, *Schechter:1981cv} to give
\begin{equation}
 M_{\mathrm{light}}\simeq -m_D M_R^{-1} m_D^T\,,\quad\quad M_{\mathrm{heavy}}\simeq M_R\,.
\end{equation}
The masses of the light neutrinos, which are mostly composed of the LH field since the active-sterile mixing\footnote{The active-sterile mixing parametrizes the amount of the left-handed component in heavy mass eigenstates.} given by $m_D M_R^{-1}$ is suppressed, arise through the diagram of fig.~\ref{type1}.

Moreover, the Majorana mass term is not protected by any symmetry and can naturally take values around the Grand Unification scale, making the light neutrino mass small enough, even with $\mathcal{O}(1)$ Yukawa couplings. In that case, light neutrino masses at the $eV$ scale are generated by taking $M_R\sim10^{15}\;\mathrm{GeV}$ and $Y_\nu\sim\mathcal{O}(1)$. A seesaw scale around the TeV can also be considered, where the Yukawa couplings are of the order $Y_\nu\sim10^{-6}$. In the framework of the type I seesaw, the dimension five operator that generate neutrino masses and the dimension six operators (eq.~(\ref{effectiveLagrangian})) leading to low-energy signatures are correlated. As a consequence, the necessary suppression of the Weinberg operator to comply with the smallness of active neutrino masses also suppresses higher-order operators that give rise to non-standard interactions or charged lepton flavour violating processes. Moreover, the smallness of the Yukawa couplings when the seesaw scale is around $1\;\mathrm{TeV}$ reduces the production cross-section of heavy neutrinos at the LHC, making their production and indirect detection much more difficult.

\subsubsection{Type II seesaw}

The type II seesaw mechanism~\cite{Magg:1980ut, *Schechter:1980gr, *Wetterich:1981bx, *Lazarides:1980nt, *Mohapatra:1980yp}  is different from the type I in that it directly leads to a Majorana mass term from the leptonic doublet. It naturally arises in  GUT models like $\mathrm{SO}(10)$~\cite{doi:10.1142/9789812562203} for instance\footnote{In $\mathrm{SO}(10)$ extensions, RH neutrinos are already part of the $\{16\}$ spinorial representation. Hence type II actually means type I+II in this framework.}. The SM is extended by a scalar $\mathrm{SU}(2)$ triplet $\Delta$ with hypercharge $Y=2$
\begin{equation}
 \Delta=\left(\begin{array}{c c} \Delta^+/\sqrt{2} & \Delta^{++} \\ \Delta^0 & - \Delta^+/\sqrt{2} \end{array}\right)\,,
\end{equation}
which allows to write the following relevant terms of the Lagrangian 
\begin{equation}
 \mathcal{L}_{\mathrm{type\;II}}=-\sum_{i,j} \left( f_{ij} \overline{L_i^C} i \sigma_2 \Delta L_j +h.c. \right) - (\mu \widetilde{\phi}^T \Delta^* \imath \sigma^2 \phi +h.c.) - M_\Delta^2 \mathrm{Tr} (\Delta^\dagger \Delta) \,.
\end{equation}
where $f_{if}$ is a complex symmetric matrix playing the role of a Yukawa coupling, $\mu$ is a real dimensionful parameter characterising the violation of lepton number conservation and $M_\Delta$ is the mass scale of the new triplet. In the limit where $M_\Delta \gg v$, the neutral component of the triplet $\Delta^0$ acquires a vev
\begin{equation}
 v_\Delta\simeq\frac{\mu v^2}{M_\Delta^2}\,,
\end{equation}
and the neutrino mass matrix is given by
\begin{equation}
 M_{\mathrm{type\;II}}\simeq - 2 f v_\Delta\,.
\end{equation}
The neutrino mass is generated through the diagram~\ref{type2}. This scenario is quite different from the type I seesaw since the smallness of the neutrino mass is directly related to the smallness of the $\mu$ parameter. Because the size of this parameter controls the size of a symmetry breaking, its smallness is natural in the sense of 't Hooft~\cite{'tHooft:1979bh} since putting it to zero would increase the symmetry of the model. Due to the presence of the $\mu$ parameter in the light neutrino masses, the type II seesaw can lie at any scale with Yukawa couplings that are possibly of order $Y_\nu\sim1$. The introduction of new scalars can affect the Higgs sector, via the mixing of neutral scalars for example. Moreover, since the new scalars are charged under the SM gauge group, they also contribute to electroweak observables. A particular attention should be paid to the contribution to electroweak precision observables arising from the scalar sector, ensuring compatibility with current observations and experimental measurements.

\subsubsection{Type III seesaw}

Finally, a third type of seesaw can be constructed. It is similar to the type I seesaw, under the exception that the singlet neutrinos are replaced with fermionic $\mathrm{SU}(2)$ triplets of zero hypercharge
\begin{equation}
 \Sigma=\left(\begin{array}{c c} \Sigma^0/\sqrt{2} & \Sigma^{+} \\ \Sigma^- & - \Sigma^0/\sqrt{2} \end{array}\right)\,.
\end{equation}
The relevant part of the type III seesaw Lagrangian is~\cite{Foot:1988aq}
\begin{equation}
 \mathcal{L}_{\mathrm{type\;III}}= -\sum_{i,j} \left( \sqrt{2} Y^{ij}_\Sigma \overline{L_i} \Sigma \widetilde{\phi} + \frac{1}{2} M_{\Sigma}^{ij} \mathrm{Tr}(\overline{\Sigma^C_i} \Sigma_{j}) + h.c. \right)\,,
\end{equation}
and, as in the type I seesaw, the light neutrino mass, in the limit where the Majorana mass term dominates, $M_{\Sigma} \gg m_D$, reads
\begin{equation}
 M_{\mathrm{light}}\simeq -m_D M_\Sigma^{-1} m_D^T\,.
\end{equation}
As far as light neutrinos are concerned, the type III seesaw is similar to the type I, see fig~\ref{type3}. However, since the fermionic triplets also have gauge couplings, they have a richer phenomenology than singlet neutrinos. In particular, they allow for tree-level flavour changing neutral currents (FCNC) which are strongly constrained by experimental data.

As we have seen, there are many different mechanisms to generate neutrino masses. The seesaw mechanisms are very attractive models since they naturally suppress the neutrino mass term and explain the smallness of the neutrino mass within the same framework, without the need of any extra symmetry. In all of them, having a large new Physics scale makes the neutrino very light. However, because of the extra scale introduced by $\mu$, the type II seesaw also offers the possibility to link the smallness of the neutrino mass to the smallness of the dimensionful coupling that violates lepton number conservation. This property also appears in multi-singlet mechanisms like the inverse seesaw.

\section{The inverse seesaw}

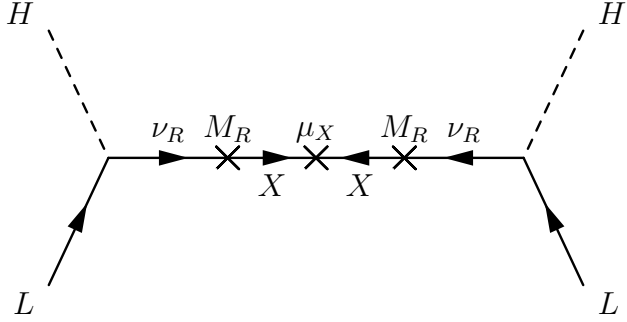
\begin{figure}
\begin{center}
\begin{fmffile}{ISS}
\begin{fmfgraph*}(200,120)
\fmflabel{$H$}{h1}
\fmflabel{$L$}{l1}
\fmflabel{$H$}{h2}
\fmflabel{$L$}{l2}
\fmfbottom{l1,l2}
\fmftop{h1,t1,t2,h2}
\fmfv{decor.shape=cross, decor.size=6thick, label=$\mu_X$, label.angle=90}{middle}
\fmfv{decor.shape=cross, decor.size=6thick, label=$M_R$, label.angle=90}{h1h3}
\fmfv{decor.shape=cross, decor.size=6thick, label=$M_R$, label.angle=90}{h2h4}
\fmf{fermion}{l1,h1l1}
\fmf{fermion}{l2,h2l2}
\fmf{fermion,label=$\nu_R$, label.side=left}{h1l1,h1h3}
\fmf{fermion,label=$\nu_R$}{h2l2,h2h4}
\fmf{fermion,tension=1.35,label=$X$}{h1h3,middle}
\fmf{fermion,tension=1.35,label=$X$, label.side=left}{h2h4,middle}
\fmf{dashes}{h1,h1l1}
\fmf{dashes}{h2,h2l2}
\fmffreeze
\end{fmfgraph*}
\end{fmffile}\\
\caption[Inverse seesaw mechanism]{Diagram generating the light neutrino mass in the inverse seesaw.}
\label{ISS}
\end{center}
\end{figure}

Multi-singlet neutrino mass mechanisms extend the Standard Model by adding \hyphenation{fer-mi-on-ic}fermionic singlets with equal lepton number. Then, depending on the mass terms considered, one can have either the inverse~\cite{Mohapatra:1986aw, *Mohapatra:1986bd, *Bernabeu1987303} or the linear seesaw~\cite{Akhmedov:1995ip, *Akhmedov:1995vm}. From now, we will focus here on the inverse seesaw, described by the diagram~\ref{ISS},
since the study of its phenomenology was the main purpose of my thesis. While it was historically introduced in supersymmetric $\mathrm{E}(6)$ models inspired by superstring theories or more recently in supersymmetric $\mathrm{SO}(10)$ Grand Unified Theory~(GUT)~\cite{Dev:2009aw}, the low-energy effective inverse seesaw Lagrangian before EWSB can be written
\begin{equation}
 \mathcal{L}_\mathrm{ISS} = - \sum_{i,j} \left( Y^{ij}_\nu \overline{\nu_{Ri}} \widetilde{\phi}^\dagger L_{j} + M_R^{ij} \overline{\nu_{Ri}} X_j + \frac{1}{2} \mu_{R}^{ij} \overline{\nu_{Ri}} \nu_{Rj}^C + \frac{1}{2} \mu_{X}^{ij} \overline{X_{i}^C} X_{j} + h.c. \right)\,,
\label{ISSlagrangian}
\end{equation}
where $Y_\nu$ is the neutrino Yukawa coupling, $M_R$ is a complex mass matrix that generates a lepton number conserving mass term and the two Majorana complex symmetric mass matrices $\mu_R$ and $\mu_X$  violate lepton number conservation by two units and have small elements. The smallness of the entries in the $\mu_R$ and $\mu_X$  matrices is natural since putting them to zero would restore the conservation of lepton number, increasing the symmetries of the model~\cite{'tHooft:1979bh}. Here, we have considered a generic framework with three generations of singlet pairs, $\nu_R$ ($L=+1$) and $X$ ($L=+1$), but the minimal model that fits oscillation data only requires two pairs of singlets~\cite{Malinsky:2009df}. After EWSB, in the basis of $(\nu_L\,,\;\nu_R^C\,,\;X)$, the $9\times 9$ neutrino mass matrix is given by
\begin{equation}
 M_{\mathrm{ISS}}=\left(\begin{array}{c c c} 0 & m_D^T & 0 \\ m_D & \mu_R & M_R \\ 0 & M_R^T & \mu_X \end{array}\right)\,.\label{ISSmatrix}
\end{equation}
To simplify the discussion, one can consider a model with only one generation: in the limit $\mu_R\,,\mu_X \ll m_D, M_R$, the mass eigenvalues are then given by
\begin{align}
 m_1 &= \frac{m_{D}^2}{m_{D}^2+M_{R}^2} \mu_X\,,\label{Mlight}\\
 m_{2,3} &= \pm \sqrt{M_{R}^2+m_{D}^2} + \frac{M_{R}^2 \mu_X}{2 (m_{D}^2+M_{R}^2)} + \frac{\mu_R}{2}\,.\label{Mheavy}
\end{align}
From eq.~(\ref{Mlight}, \ref{Mheavy}), it is easy to see that the smallness of the neutrino mass arises from the smallness of $\mu_X$ and, as in the type II seesaw, this smallness is natural. It is the same for $\mu_R$, which only enters the neutrino masses as a subdominant correction. As a consequence, and since none of the observables we studied, like lepton flavour violating decays or ratios of leptonic kaon decays, depends on $\mu_R$, we will neglect it for all practical purposes, like when computing contributions to the studied observables. Another observation is that the heavy neutrino eigenstates are nearly degenerate in mass, implying that they will behave as the two components of a pseudo-Dirac fermion.

Under the more constraining assumption $\mu_R\,,\mu_X \ll m_D \ll M_R$, the mass matrix $M_{\mathrm{ISS}}$ can be block diagonalized to give the light neutrino mass matrix~\cite{GonzalezGarcia:1988rw}
\begin{equation}
 M_{\mathrm{light}}\simeq m_D^T {M_R^T}^{-1} \mu_X M_R^{-1} m_D\,,
\label{LightMatrix}
\end{equation}
with the active-sterile mixing given by $m_D M_R^{-1}$. Since the smallness of light neutrino masses is linked to $\mu_X$ in the inverse seesaw, it is possible  to have at the same time $\mathcal{O}(1)$ Yukawa couplings and sterile neutrinos with masses at the TeV scale. This will have important phenomenological consequences like an enhanced active-sterile mixing or the possibility to directly produce the heavy neutrinos at the LHC.

The difference between the type I and inverse seesaw for indirect searches of new Physics can be better understood in the framework of effective theories. For example, in these, charged lepton flavour violating (cLFV) observables originate from a dimension 6 operator proportional to $Y_\nu^\dagger Y_\nu/M_R^2$ in both mechanisms. In the type I seesaw, the mass of the light neutrinos comes from the dimension 5 operator, which is proportional to $Y_\nu^T Y_\nu/M_R$, directly linking the size of the two operators in the absence of CP violation. In the inverse seesaw, the dimension 5 operator is further suppressed by a factor of $\mu_X/M_R$, which effectively decouples the two operators, allowing large cLFV effects while keeping the light neutrino masses small.

\section{Parametrization of the leptonic mixing}

Since the $n \times n$  neutrino mass matrix $M_\nu$ is complex and symmetric (because of the Majorana nature of the neutrino), it can be decomposed using the Takagi factorisation
\begin{equation}
 V_\nu^T M_\nu V_\nu = \mathrm{diag}(m_1\,,...\,,m_n)\,,
\end{equation}
where $V_\nu$ is a unitary matrix and $D$ is the diagonal matrix whose elements are the real positive square roots of the eigenvalues of $M_\nu^\dagger M_\nu$. So, if we consider for example an inverse seesaw mechanism with 3~pairs of singlet neutrinos, then $n=9$ and $V_\nu$ is a $9\times9$ unitary matrix, relating the weak eigenstates to the the mass eigenstates through
\begin{equation}
  \left(\begin{array}{c} \nu_L \\ \nu_R^C \\ X \end{array} \right) = V_\nu \left(\begin{array}{c} \nu_1 \\ \vdots \\ \nu_9 \end{array} \right)\,,
\end{equation}
while the charged lepton mass matrix can be factorised through the singular value decomposition $V_L^\dagger M_\ell V_R = \mathrm{diag}(m_e\,,m_\mu\,,m_\tau)$
\begin{equation}
 (e_L)_{\mathrm{weak}}= V_L \left(\begin{array}{c} e_L \\ \mu_L \\ \tau_L \end{array} \right)_{\mathrm{mass}}\,,\quad\quad\quad (e_R)_{\mathrm{weak}}= V_R \left(\begin{array}{c} e_R \\ \mu_R \\ \tau_R \end{array} \right)_{\mathrm{mass}}\,,
\end{equation}
where $V_R$ and $V_L$ are $3\times 3$ unitary matrices.

In general, neutrinos and charged leptons are not diagonal in the same basis. This misalignment introduces a mixing between different lepton flavours, which is the source of neutrino oscillations and other charged lepton flavour violating processes. At energies much lower than the $W^\pm$ mass, the charged-current Lagrangian reads
\begin{equation}
\Delta\mathcal{L}_W=\frac{g^2}{m_W^2} J_W^{\mu-} J_{W\mu}^{+}\,,
\end{equation}
with the charged current given by
\begin{align}
 J_{W_{ij}}^{\mu+} &= \frac{1}{\sqrt{2}} \overline{\nu_{i}} (V_\nu^\dagger V_L)_{ij} \gamma^\mu P_L \ell_j \nonumber \\
   &= \frac{1}{\sqrt{2}} \overline{\nu_{i}} (U^\dagger)_{ij} \gamma^\mu P_L \ell_j\,.
\label{Udef}
\end{align}
The mixing matrix $U$ is explicitly written as
\begin{equation}
 U_{ji}=\sum_{k=1}^3 V_{L_{kj}}^* V_{\nu_{ki}}\,,
\end{equation}
where $j$ runs between 1 and 3 for the charged leptons and $i$ between 1 and 9 for the neutrinos. Then $U$ is a rectangular matrix, which prevents it from being unitary. Indeed, we have
\begin{align}
 U U^\dagger = \mathbf{1}\,,\\
 U^\dagger U \neq \mathbf{1}\,,
\end{align}
because
\begin{align}
 (U^\dagger U)_{\alpha \alpha} &= \sum_{i=1}^{3} (U^\dagger)_{\alpha i} U_{i \alpha} \nonumber \\
      &= \sum_{i,j,k=1}^{3} V_{\nu_{j\alpha}}^* V_{L_{ji}} V_{L_{ki}}^* V_{\nu_{k\alpha}} \nonumber \\
      &= \sum_{j=1}^{3} V_{\nu_{j\alpha}}^* V_{\nu_{j\alpha}} \neq \mathbf{1}\,.
\end{align}
However, in the limit where extra neutrinos are very heavy and decouple, the leptonic mixing matrix is $3\times3$, unitary and correspond to the PMNS matrix.

The Casas-Ibarra parametrization~\cite{Casas:2001sr} is a method of reconstructing generic neutrino Yukawa textures compatible with the oscillation data. It was originally introduced in the type I seesaw~\cite{Casas:2001sr}, but it can be directly extended to the inverse seesaw. In the latter, the light neutrino mass matrix is given by eq.~(\ref{LightMatrix})
\begin{equation}
 M_{\mathrm{light}} \simeq Y_\nu^T {M_R^T}^{-1} \mu_X M_R^{-1} Y_\nu \frac{v^2}{2}\,,
\end{equation}
which is factorised using the PMNS matrix
\begin{equation}
 U_{PMNS}^T M_{\mathrm{light}} U_{PMNS} = \mathrm{diag}(m_1\,, m_2\,, m_3)\,.
\end{equation}
If we define the matrix $M=M_R \mu_X^{-1} M_R^T$, then the light mass matrix  can be recast in a form similar to the type I seesaw
\begin{equation}
 M_{\mathrm{light}}\simeq Y_\nu^T M^{-1} Y_\nu \frac{v^2}{2}\,.
\end{equation}
If the matrix $M$ is decomposed with a unitary matrix $V$ according to 
\begin{equation}
 M=V^\dagger \mathrm{diag}(M_1\,, M_2\,, M_3) V^*\,,
\end{equation}
the Casas-Ibarra parametrization can be directly  applied to the inverse seesaw mechanism:
\begin{equation}
 Y_\nu=\frac{\sqrt{2}}{v} V^\dagger \mathrm{diag}(\sqrt{M_1}\,,\sqrt{M_2}\,,\sqrt{M_3})\; R\; \mathrm{diag}(\sqrt{m_1}\,,\sqrt{m_2}\,,\sqrt{m_3}) U^\dagger_{PMNS}\,,
\label{CasasIbarraISS}
\end{equation}
where $R$ is a complex orthogonal matrix, which has 6 independent parameters. With the 6 independent parameters in the PMNS matrix and the 6 masses in the diagonal matrices, the right-hand side of eq.~(\ref{CasasIbarraISS}) has 18 independent parameters, which corresponds to the number of independent parameters in the Yukawa coupling. The $V$ matrix does not contribute to the above counting since it is completely determined by the requirement that it should decompose $M$ into a diagonal matrix. Then, assuming $R=\mathbf{1}$ is equivalent to hypothesis that $Y_\nu$ and $M$ are simultaneously diagonal, while assuming that the neutrino mass matrix is real results in $R \in \mathrm{O}_3(\mathbb{R})$.  The adapted Casas-Ibarra parametrization will prove extremely useful when studying the phenomenology of the inverse seesaw since it allows to scan the parameter space while making sure that every point comply with  the oscillation data, thus saving a lot of computing time.

In this chapter, we have presented the two types of spinors that can be used to describe fermions, motivated by the fact that, neutrinos being neutral particles, they could potentially be Majorana fermions. Since the experimental data presented in Chapter~\ref{ChapNuExp} requires neutrinos to have non-zero masses, we described mass generation mechanisms, focusing on seesaw mechanisms. Neutrino oscillations also implies that neutrino mix and we discussed lepton mixing and the Casas-Ibarra parametrization in the last section. However, there are numerous mass generation mechanisms which lead to very different experimental signatures at low-energies. In the following chapter, we will discuss how mechanisms that introduce new fermionic singlets lead to deviations from lepton universality.

\chapter{Impact of fermionic singlets on lepton universality tests\label{chaptLFU}}

A fundamental consequence of the Standard Model gauge structure is the flavour universality of the coupling constants. Since the gauge interactions do not distinguish between different generations, leptons from different families have identical gauge couplings. Any deviation from the expected SM estimates in lepton universality tests will point towards new Physics. Lepton Flavour Universality (LFU) can be violated due to a new Lorentz structure in the fermion interactions arising from a new field content, like a charged Higgs scalar, or due to corrections to the SM interactions. The first possibility has been intensively investigated in scenarios beyond the SM with extended Higgs sectors for leptonic pseudoscalar meson decays (e.g. two Higgs doublet models~\cite{Hou:1992sy}, supersymmetric extensions of the SM~\cite{Masiero:2005wr, *Girrbach:2012km, Fonseca:2012kr}). However, in these scenarios, the new tree-level corrections are lepton universal and loop-level corrections have to be considered. As seen in Chapter~\ref{chap3}, extensions of the SM with fermionic singlets induce corrections to the $W\ell\nu$ vertex leading to LFU violation in charged currents.

\section{Lepton universality tests}

Lepton universality has been extensively discussed in~\cite{davier1998lepton}, covering both the theoretical and experimental aspects of the subject. Since the charged leptons couple to the photon and to the weak gauge bosons, lepton universality tests can be categorised according to the gauge boson involved: photon, $W^\pm$ or $Z^0$. The universality of electric charge has been thoroughly tested for the muon. Measuring the energy difference between 1s and 2s levels in muonium, a bound system made from an electron and an antimuon, a charge ratio anomaly $1+q_{\mu^+}/q_{e^-}$ of $(1.1\pm 2.1)\times 10^{-9}$ was extracted~\cite{Meyer:1999cx}. Here, we are interested in the impact of sterile neutrinos. Since neutrinos only couple to weak gauge bosons, let us focus on tree-level universality tests involving $W^\pm$ and $Z^0$ bosons.

\subsubsection{Leptonic $Z^0$ couplings}

Lepton universality can be directly tested in $Z^0$ decays through ratios of decay widths into different leptons. Combined measurements from experiments at the electron-positron colliders LEP and SLC (Stanford Linear Collider) have given~\cite{ALEPH:2005ab}
\begin{align}
 \frac{\mathcal{B}(Z^0 \rightarrow \mu^+ \mu^-)}{\mathcal{B}(Z^0 \rightarrow e^+ e^-)}&=1.0009\pm0.0028\,,\\
 \frac{\mathcal{B}(Z^0 \rightarrow \tau^+ \tau^-)}{\mathcal{B}(Z^0 \rightarrow e^+ e^-)}&=1.0019\pm0.0032\,,
\end{align}
which are in agreement with the hypothesis of lepton universality at the $1\sigma$ level.

Other processes could be considered like leptonic meson and quarkonium decays, charged lepton production cross-sections at electron-positron colliders or hadron colliders via the Drell-Yan process. Unfortunately, all these observables can also be mediated by a photon, making the extraction of constraints on non-universal $Z^0$ couplings very challenging. On the opposite, $W^\pm$-mediated processes offer clear signatures and have been widely used to test lepton universality.

\subsubsection{Leptonic $W^\pm$ couplings}

The most straightforward observables one can think of are the leptonic decay widths of $W^\pm$ bosons and their ratios. They have been measured at LEP-II by the ALEPH, L3, DELPHI and OPAL collaborations~\cite{Heister:2004wr, *Achard:2004zw, *Abdallah:2003zm, *Abbiendi:2007rs}. The combined measurements of the branching ratios give~\cite{Alcaraz:2006mx}
\begin{align}
 \frac{\mathcal{B}(W \rightarrow \mu \bar{\nu}_{\mu})}{\mathcal{B}(W \rightarrow e \bar{\nu}_e)}&=0.994\pm0.020\,,\\
 \frac{\mathcal{B}(W \rightarrow \tau \bar{\nu}_{\tau})}{\mathcal{B}(W \rightarrow e \bar{\nu}_e)}&=1.074\pm0.029\,,\\
 \frac{\mathcal{B}(W \rightarrow \tau \bar{\nu}_{\tau})}{\mathcal{B}(W \rightarrow \mu \bar{\nu}_{\mu})}&=1.080\pm0.028\,,
\end{align}
showing a slight deviation from universality in the third generation. Assuming partial universality, this can be better quantified through the ratio
\begin{equation}
R_{\tau\ell}^W= \frac{2\mathcal{B}(W\to \tau \bar\nu_\tau)} 
{\mathcal{B}(W\to \mu \bar\nu_\mu)+\mathcal{B}(W\to e \bar\nu_e)}=1.077\pm0.026, 
\end{equation}
which corresponds to a $2.8\sigma$ deviation from the Standard Model prediction~\cite{Kniehl:2000rb}
\begin{equation}
 R_{\tau\ell}^W|_{SM}=0.999\,.
\end{equation}

Equally interesting are leptonic meson decays. For these, measuring ratios of decay widths is often more interesting since it reduces the experimental uncertainties, the ratios being independent of the flux of the decaying mesons~\cite{Lazzeroni:2012cx}. Moreover, considering these ratios also improves the precision of theoretical predictions since the hadronic uncertainties cancel to a good approximation. For pions and kaons, the only kinematically accessible charged leptons are the electron and muon. Denoting the decaying pseudoscalar meson by $P$, with $P=\pi\,,K$, the appropriate ratios are
\begin{equation}
 R_P =\frac{\Gamma (P^+ \to e^+ \nu)}{\Gamma (P^+ \to \mu^+\nu)}\,,
\end{equation}
where the most recent experimental measurements are
\begin{align}
 R_K^\text{exp}&=(2.488 \pm 0.010)\times 10^{-5}\,,\label{RKexp}\\
 R_\pi^\text{exp} &=(1.230 \pm 0.004)\times 10^{-4}\,,\label{Rpiexp}
\end{align}
from, respectively, the NA62 collaboration~\cite{Lazzeroni:2012cx} and the PDG evaluation~\cite{Beringer:1900zz} based on the measurements~\cite{Czapek:1993kc, *Britton:1992pg}. The SM prediction is given by~\cite{Cirigliano:2007xi}
\begin{equation}
 R_P^\mathrm{SM}=\left(\frac{m_e}{m_\mu}\right)^2\left(\frac{m_P^2-m_e^2}{m_P^2-m_\mu^2}\right)^2(1+\delta_\mathrm{QED})\,,
\end{equation}
where $\delta_\mathrm{QED}$ are QED corrections coming, for example, from long-range interactions and internal bremsstrahlung.  The experimental results eqs.~(\ref{RKexp}, \ref{Rpiexp}) can be compared with the SM predictions~\cite{Cirigliano:2007xi, Finkemeier:1995gi}
\begin{align}
 R_K^\text{SM}&=(2.477 \pm 0.001)\times 10^{-5}\,,\\
 R_\pi^\text{SM}&=(1.2354 \pm 0.0002)\times 10^{-4}\,.
\end{align}
Introducing a parameter $\Delta_{r_P}$ characterising the deviation from the SM prediction as
\begin{equation}
 R_P^\text{exp} \, = \,R_P^\text{SM} \, (1+\Delta r_P) \quad
\text{or equivalently}
\quad
\Delta r_P \, = \, \frac{R_P^\text{exp}}{R_P^\text{SM}} - 1\,,
\end{equation}
the experimental results can be rewritten as
\begin{equation}
 \Delta r_K \, = \, (4 \pm 4 )\, \times\, 10^{-3}\,, 
\quad \quad
 \Delta r_\pi \, = \, (-4 \pm 3 )\, \times\, 10^{-3}\,. 
\end{equation}
From this, it is clear that the SM predictions agree at the $1\sigma$ level with the observations. However, the smallness of the theoretical errors with respect to the uncertainty in experimental measurements has motivated new experiments that are expected to reduce the experimental uncertainty on $R_\pi$ by a factor five or more~\cite{Pocanic:2012gt,*Malbrunot:2012zz}. Beyond the Standard Model scenarios have been considered too. For instance, new contributions to $\Delta r_K$ that arise in supersymmetric models have shown to be below $10^{-3}$~\cite{Masiero:2005wr, *Girrbach:2012km, Fonseca:2012kr}.

Unfortunately, while pion and kaon decays seem very attractive due to the precise SM predictions and the small experimental uncertainties, they cannot test the universality of the $\tau$ coupling. A first solution is to consider heavier mesons. This is for example the case of the charged $B$ or $D_s$ mesons. The corresponding decay rates and those into other leptons are given in table~\ref{BandDrates},
\begin{table}[t]
  \begin{center}
    \begin{tabular}{|c|c|c|c|}
     \hline
	Decay channel & Branching ratio & SM prediction & Pull \\
     \hline
     \hline
	$D^+\rightarrow \mu^+ \nu$ & $(3.82\pm0.34)\times10^{-4}$~\cite{Eisenstein:2008aa} & $(4.18^{+0.13}_{-0.20})\times10^{-4}$~\cite{Charles:2011va} & $0.6\sigma$~\cite{Charles:2011va} \\
	$D_s^+\rightarrow \mu^+ \nu$ & $(5.9 \pm 0.33)\times10^{-3}$~\cite{Beringer:1900zz} & $(5.39^{+0.21}_{-0.22})\times10^{-3}$~\cite{Charles:2011va} & $1.3\sigma$~\cite{Charles:2011va} \\
	$D_s^+\rightarrow \tau^+ \nu$ & $(5.43 \pm 0.31)\times10^{-2}$~\cite{Beringer:1900zz} & $(5.44^{+0.05}_{-0.17})\times10^{-2}$~\cite{Charles:2011va} & $0.03\sigma$ \\
	$B^+\rightarrow \tau^+ \nu$ & $(1.15\pm0.23)\times10^{-4}$~\cite{Celis:2012dk} & $(0.757^{+0.098}_{-0.061})\times10^{-4}$~\cite{Charles:2011va} & $1.6\sigma$ \\
     \hline
    \end{tabular}
    \caption[Leptonic heavy meson decays]{\label{BandDrates} Measured branching ratios of heavy meson decays, corresponding SM predictions and pulls between the experimental and theoretical values.}
 \end{center}
\end{table}
with the corresponding SM predictions. It is worth noting that some measurements deviate from the theoretical values, even if the pulls are mild, below $2\sigma$. Similarly to $R_P$, one can construct another observable, $R_{D_s}$, defined as
\begin{equation}
 R_{D_s}=\frac{\Gamma (D_s^+ \to \tau^+ \nu)}{\Gamma (D_s^+ \to \mu^+\nu)}\,.
\label{RDs}
\end{equation}
Other ratios could also be considered, but are less attractive since they reintroduce the dependence on decay constants and CKM elements.

Another possibility to test the $\tau$ coupling universality is to consider $\tau$ decays into a pseudoscalar meson, namely a pion or a kaon, and a neutrino. For instance, the SM branching fractions of $\tau^- \rightarrow K^- \nu$ is given by
\begin{equation}
 \Gamma(\tau^- \rightarrow K^- \nu)=\frac{G_F^2 F_K^2}{8\pi m_\tau} |V_{us}|^2 (m_\tau^2-m_K^2)^2\,.
\end{equation}
The experimental measurements of the corresponding branching ratios are~\cite{Beringer:1900zz} 
\begin{align}
\mathcal{B}(\tau^-\rightarrow\pi^- \nu)&=0.1083\pm0.0006\,,\\
\mathcal{B}(\tau^-\rightarrow K^- \nu)&=(7.0\pm0.1)\times10^{-3}\,,
\end{align}
and ratios similar to $R_P$ can be built
\begin{equation}
 R_{P,\mu}^\tau=\frac{\mathcal{B}(\tau^-\rightarrow P^- \nu)}{\mathcal{B}(P^+ \to \mu^+\nu)}\,,\quad\mathrm{and}\quad  R_{P,e}^\tau=\frac{\mathcal{B}(\tau^-\rightarrow P^- \nu)}{\mathcal{B}(P^+ \to e^+\nu)}\,,
\label{RPltau}
\end{equation}
where the hadronic uncertainties cancel to a good approximation, as in $R_P$.

Another observable is the ratio between leptonic 3-body $\tau$ decays. It has the advantage of being purely leptonic and, as such, it is subject to much less theoretical uncertainties than meson decays. Unfortunately, two neutrinos are present in the final state, making the experimental reconstruction more challenging. The CLEO~\cite{Anastassov:1996tc} and BaBar~\cite{Aubert:2009qj} experiments measured this ratio to a precision better than that of the individual decay widths. The PDG average stands at~\cite{Beringer:1900zz}
\begin{equation}
 R_\tau=\frac{\mathcal{B}(\tau^-\rightarrow \mu^- \bar \nu_\mu \nu_\tau)}{\mathcal{B}(\tau^- \rightarrow e^- \bar \nu_e \nu_\tau)}=0.979 \pm 0.004\,.
\end{equation}

Finally, lepton universality can also be tested in ratios of semileptonic meson decays. However, hadronic matrix elements will not cancel out, making it harder to disentangle non-universality effects from hadron Physics. Nevertheless, we will consider the ratio recently measured by BaBar~\cite{Lees:2012xj} 
\begin{equation}
 R(D)=\frac{\mathcal{B}(\overline{B}\rightarrow D \tau^- \bar{\nu}_\tau )}{\mathcal{B}(\overline{B}\rightarrow D \ell^- \bar{\nu}_\ell)}=0.440\pm0.072\,,
\end{equation}
where $\overline{B}=B^-,\overline{B^0}$, $D=D^0,D^+$ and $\ell=e,\mu$. This should be compared with the SM prediction based on lattice estimation of the hadronic matrix element~\cite{Becirevic:2012jf}
\begin{equation}
 R(D)_{SM}=0.31\pm0.02\,,
\end{equation}
which deviates by $1.7\sigma$ from the experimental results. We will not consider other observables, like $\mathcal{B}(\overline{B}\rightarrow D^* \tau^- \bar{\nu}_\tau )$ or $\Gamma (K\rightarrow \pi \ell^- \bar{\nu}_\ell)$ since they either do not exhibit a large deviation from the SM prediction or their form factors have large theoretical uncertainties.

In this study, we minimally extend the SM with fermionic singlets, which modifies the leptonic charged currents coupling to the $W^\pm$ boson. It is worth mentionning that other studies using these observables have been conducted in supersymmetric models~\cite{Masiero:2005wr, *Girrbach:2012km, Fonseca:2012kr} or in the presence of a fourth generation~\cite{Lacker:2010zz}.

\section{Constraints on sterile neutrinos}

In this chapter we are interested in the effect of fermionic singlets coupled to leptons via Yukawa terms like in the type I or in the inverse seesaw mechanisms (see Chapter~\ref{chap3}). In the presence of additional states, the diagonalization of the neutrino mass matrix yields three light active neutrinos and other heavier states, mostly composed of sterile neutrinos, with couplings to the weak bosons suppressed by small elements of the leptonic mixing matrix. We stress, here, that sterile neutrinos can be much heavier than $1\;\mathrm{eV}$ and do not necessarily provide a solution to the neutrino anomalies discussed in Chapter~\ref{ChapNuExp}. There are strong experimental and observational bounds on the mass regimes and on the size of the active-sterile mixings that must be satisfied in addition to neutrino masses and mixings discussed in Chapter~\ref{ChapNuExp}.

First, there are robust laboratory bounds from direct sterile neutrinos searches~\cite{Atre:2009rg, Kusenko:2009up}. Below $100\;\mathrm{eV}$, sterile neutrinos can have an impact on neutrino oscillations~\cite{Smirnov:2006bu}. However, constraints in this mass range are not relevant for this study, since only neutrinos with a mass around the electron mass or above have a noticeable effect on the observables considered here. In the eV to MeV mass range, sterile neutrinos mixing with the electron neutrino will appear as kinks in the $\beta$ electron spectrum~\cite{Shrock:1980vy}, in a similar way to the other active neutrinos. In the MeV to GeV mass range, sterile neutrinos can be searched via monochromatic lines in the charged lepton spectra of two-body meson decays~\cite{Shrock:1980vy, Shrock:1980ct}. Another possibility is to try to detect visible decay products from sterile neutrinos~\cite{Kusenko:2004qc}, as the Belle collaboration recently did for sterile neutrinos produced in $B$ meson decays~\cite{Liventsev:2013zz}. Negative searches for the above signals can be
translated into bounds for $m_{\nu_s} - \theta_{i \alpha}$ combinations, where $\theta_{i \alpha}$ parametrizes the active-sterile mixing. Finally, the $Z^0$ invisible decay width could be modified~\cite{Dittmar:1989yg}, which adds further constraints. 

The non-unitarity of the leptonic mixing matrix is also subject to constraints: the rates for leptonic and hadronic processes with final state neutrinos usually depend on $\sum_i |U_{ji}|^2$, where the sum extends over all neutrino states kinematically accessible \mbox{($i=1,\dots,N_\text{max}$)}, and thus constrain the departure from unitarity $\sum_i |U_{ji}|^2 = 1$. Noting $\widetilde U_\text{PMNS}$ the $3\times3$ block of $U$ that corresponds to the mixing between the charged leptons and the active neutrinos, the deviation of $\widetilde U_\text{PMNS}$ from unitarity can be parametrized by \begin{equation}\label{eq:U:eta:PMNS}
\widetilde U_\text{PMNS} = (\mathbf{1} - \eta) U_\text{PMNS}\,.
\end{equation}
where $U_\text{PMNS}$ is the unitary mixing matrix that arises when only three massive neutrinos are present. Bounds on the non-unitarity parameter $\eta$ were derived using Non-Standard Interactions~\cite{Antusch:2008tz}. However, these are  not relevant when all the sterile neutrinos are lighter than the decaying particle since then all the neutrino mass eigenstates are accessible and unitarity is restored.

Unless the active-sterile mixings are negligible, the modified $W \ell \nu$ vertex may also contribute to lepton flavour violating processes\footnote{LFV is typically dipole dominated when the sterile neutrinos are light ($m_{\nu_s} \lesssim 300$ GeV), so that $\mu \to e \gamma$  is the most constraining LFV observable. For heavier sterile neutrinos,  other (model-dependent) contributions beyond the dipole might be more relevant~\cite{Abada:2012cq}.}, with potentially large rates.  The radiative $\mu \to e \gamma$ decay, searched for by the MEG experiment~\cite{Adam:2013mnn}, is the most stringent one\footnote{Recently, it has been also noticed that in the framework of a low-scale type I seesaw, the expected future sensitivity of $\mu- e$ conversion experiments can also play a relevant role in detecting or constraining sterile neutrino scenarios in the 2 GeV - 1000 TeV mass range~\cite{Alonso:2012ji}.}. The rate induced by sterile neutrinos must satisfy $\mathcal{B}(\mu \rightarrow e \gamma) \leq 5.7 \times 10^{-13}$.  Any change in the $W \ell \nu$ vertex will also affect other leptonic meson decays, in particular $B \to \ell \nu$. The following bounds are particularly relevant~\cite{Beringer:1900zz}:
\begin{align}
 \mathcal{B}(B \rightarrow e \nu) &< 9.8 \times 10^{-7}\,,\\
 \mathcal{B}(B \rightarrow \mu \nu) &< 10^{-6}\,,\\
 \mathcal{B}(B \rightarrow \tau \nu) &= (1.65 \pm 0.34) \times 10^{-4}\,.
\end{align}

Important constraints can also be derived from LHC Higgs searches~\cite{BhupalDev:2012zg} and electroweak precision data~\cite{delAguila:2008pw, Atre:2009rg}. LHC data on Higgs decays already provides some important bounds when the sterile states are slightly below $125$ GeV (due to the potential Higgs boson decays to left- and right-handed neutrinos). The active-sterile mixings can also introduce small deviations in the electroweak fits, which allows to constrain them. An effective approach was applied in~\cite{delAguila:2008pw}, assuming  very heavy sterile neutrinos, and thus these bounds will only be applied when all sterile neutrinos are heavier than the decaying particle.

If the neutrinos are Majorana fermions, constraints also arise from specific observables. Neutrinoless double beta decays will be affected through modifications of the effective mass $m_{2\beta}$ but also via new exchanges of heavy sterile neutrinos~\cite{Smirnov:2006bu}. New observables can also be searched for at colliders, for example same-sign dilepton and same-sign dimeson signals~\cite{Atre:2009rg}. Their non-observation puts limits in the $m_{\nu_s} - \theta_{i \alpha}$ plane.

Under the assumption of a standard cosmology, the most constraining bounds on sterile neutrinos stem from a wide variety of cosmological observations~\cite{Smirnov:2006bu, Kusenko:2009up}. Using Large Scale Structure (LSS), Lyman-$\alpha$, CMB, Big Bang nucleosynthesis and supernovae (SN1987A) data, one can also set relevant bounds on sterile neutrinos via effects analogous to those described in Section~\ref{SectCurrent} for active neutrinos. Active-sterile mixing also induces radiative decays $\nu_i \to \nu_j \gamma$, well constrained by cosmic X-ray searches.
\begin{figure}[!t]
 \centering
 \includegraphics[width=0.8\textwidth]{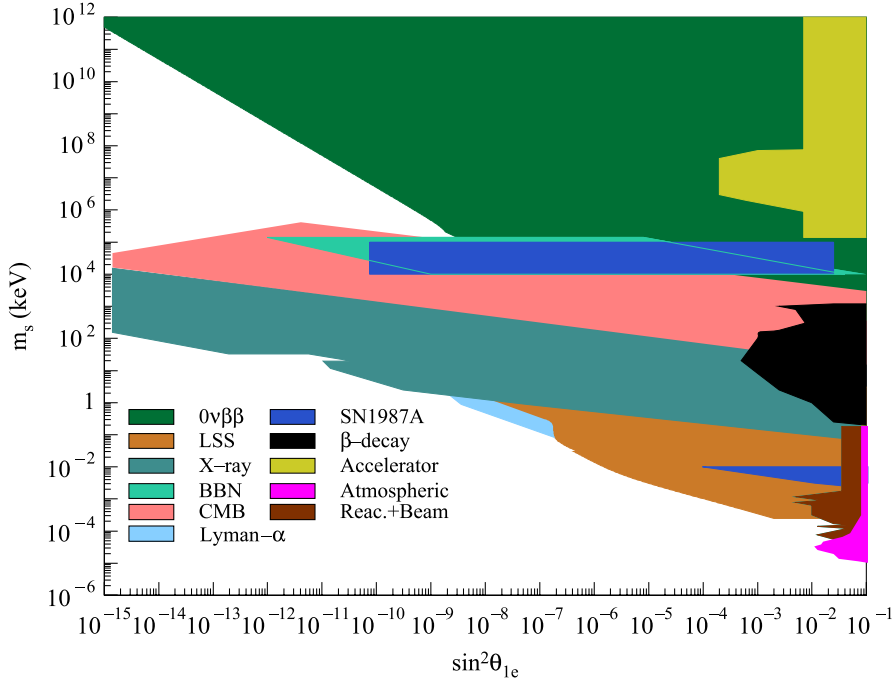}
 \caption[Limits on electron-sterile mixing]{Limits on the size of the electron component of a sterile neutrino, with respect to the sterile neutrino mass. Reprinted from~\cite{Kusenko:2009up}, copyright 2009, with permission from Elsevier.}
\label{SterileMixingE}
\end{figure}
Limits on electron-sterile mixing are given in figure~\ref{SterileMixingE} and similar plots for muon-sterile and tau-sterile mixings can be found in~\cite{Kusenko:2009up}. However, all the above cosmological bounds can be evaded if a non-standard cosmology is considered. In fact, the above cosmological constraints disappear in scenarios with a low reheating temperature~\cite{Gelmini:2008fq}. In our numerical analysis we will allow for the violation of the cosmological bounds, explicitly distinguishing results that are in conflict with them.

\section{$R_K$ in the inverse seesaw model\label{SectRk}}

We focus here on leptonic decays of light pseudoscalar mesons because of the precise SM predictions and the relatively small experimental uncertainties. We will first compute $\Delta r_P$ in a model-independent approach, allowing additional fermionic states to contribute. Then, we will consider the case of the inverse seesaw model in order to numerically illustrate the impact of sterile neutrinos on $\Delta r_P$~\cite{Abada:2012mc}.

\subsubsection{$\Delta r_P$ in the presence of sterile neutrinos}

\begin{figure}[!t]
\centering
\begin{fmffile}{LeptonicP}
\begin{fmfgraph*}(200,120)
\fmflabel{$\ell^+$}{l}
\fmflabel{$\nu$}{nu}
\fmfleft{K}
\fmfforce{(0.2w,0.7h)}{u}
\fmfforce{(0.2w,0.3h)}{s}
\fmfforce{(1.0w,0.8h)}{l}
\fmfforce{(1.0w,0.2h)}{nu}
\fmfpoly{filled=30, label=$K^+$, smooth, pull=?}{K,u,s} 
\fmf{fermion, label=$u$, label.side=left}{u,usW}
\fmf{fermion, label=$\bar s$,label.side=left}{usW,s}
\fmf{boson,tension=1.2,label=$W^+$}{usW,lnuW}
\fmf{fermion}{l,lnuW}
\fmf{fermion}{lnuW,nu}
\end{fmfgraph*}
\end{fmffile}
\label{LeptonicP}
\caption[Leptonic meson decay]{Feynman diagram of a pseudoscalar meson decaying into a charged lepton and a neutrino.}
\end{figure}
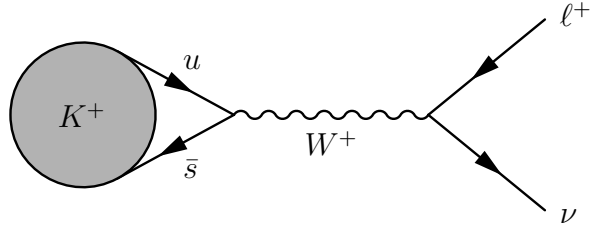

Let us consider the SM extended by $N_s$ additional sterile states and conduct a general calculation of the leptonic meson decay widths, where the meson is a pseudoscalar. The decay $K^+ \rightarrow \ell^+ \nu$ is described at the quark level by the diagram of fig.~\ref{LeptonicP}.
Since the decaying meson mass is much smaller than the $W^\pm$ mass, the intermediate boson propagator can be approximated by $\imath g_{\mu\nu} /m_W^2$ and the matrix element is
\begin{equation}
 \imath \mathcal{M}_{ij} = \langle \ell^+_j (q_1)\,\nu_i (q_2)|\imath \frac{G_F}{\sqrt{2}} U^*_{ji}\bar \nu_i \gamma_\mu (1-\gamma_5)\ell_j \bar s \gamma^\mu (1-\gamma_5) V^*_{us} u |K^+(p) \rangle\,,
\label{step1}
\end{equation}
with $V$ the CKM matrix, $G_F$ the Fermi coupling constant and no sum implied over the indices of the outgoing leptons $i,j$. Notice that now one has to consider all the final state neutrinos  $i=1,...,N_\mathrm{max}^{(\ell)}$, where $N_\text{max}^{(l)}$ denotes the heaviest neutrino mass eigenstate which is kinematically allowed. Considering a purely tree-level process, we neglect long-distance corrections and insert the vacuum state
\begin{equation}
 \imath \mathcal{M}_{ij} =\imath \frac{G_F}{\sqrt{2}} U^*_{ji} V^*_{us} \langle \ell^+_j (q_1)\,\nu_i (q_2)|\bar \nu_i \gamma_\mu (1-\gamma_5)\ell_j |0\rangle \langle 0| \bar s \gamma^\mu (1-\gamma_5)  u |K^+(p) \rangle\,.
\end{equation}
However, the isospin vector current is conserved and thus $\langle 0| \bar s \gamma^\mu u |K^+(p) \rangle=0$. Defining the meson decay constant by $\langle 0| \bar s \gamma^\mu \gamma_5  u |K^+(p) \rangle = \imath \sqrt{2} F_K p^\mu$ and contracting the current with the asymptotic states, the matrix element reads
\begin{equation}
 \imath \mathcal{M}_{ij} = - G_F F_K U^*_{ji} V^*_{us} \bar u_{\nu_i}(q_2) \slashed p (1-\gamma_5) v_{\ell_j}(q_1)\,,
\end{equation}
where $u$ and $v$ are the positive and negative frequency  (particle and antiparticle) solutions of the Dirac equation, respectively. This gives the spin averaged squared matrix element
\begin{equation}
 |\overline{\mathcal{M}_{ij}}|^2 = 4 G_F^2 F_K^2 |U_{ji}|^2 |V_{us}|^2 \left[m_K^2 (m_{\nu_i}^2+m_{\ell_j}^2)-(m_{\nu_i}^2-m_{\ell_j}^2)^2\right]\,,
\label{FinalStep}
\end{equation}
where $m_K$ is the mass of the charged kaon.

The decay rate is given by~\cite{Peskin:1995ev}
\begin{equation}
 \Gamma = \int \frac{1}{2m_K} \left(\frac{\mathrm{d}^3 q_1}{(2\pi)^3} \frac{1}{2 E_{\ell_j}} \frac{\mathrm{d}^3 q_2}{(2\pi)^3} \frac{1}{2 E_{\nu_i}} \right) |\overline{\mathcal{M}_{ij}}|^2 (2 \pi)^4 \delta^{(4)}(p-q_1-q_2)\,.
\end{equation}
Since the spin averaged squared matrix element is independent from the momenta of the leptons, it can be taken out of the integral leaving only the sum over the two-body Lorentz-invariant phase space, which is
\begin{equation}
 \int \mathrm{dLips}_2 = \frac{1}{16 \pi m_K^3} \left[(m_K^2 - m_{\ell_j}^2 - m_{\nu_i}^2)^2-4 m_{\ell_j}^2 m_{\nu_i}^2\right]^{1/2}\,.
\label{Lips2}
\end{equation}
This gives the decay rate into the mass eigenstates $\nu_i, \ell_j$
\begin{align}
 \Gamma =& \frac{G_F^2 F_K^2}{4 \pi m_K^3} |U_{ji}|^2 |V_{us}|^2 \left[m_K^2 (m_{\nu_i}^2+m_{\ell_j}^2)-(m_{\nu_i}^2-m_{\ell_j}^2)^2\right]\nonumber\\
 & \quad \quad \quad \quad \quad \quad \quad \quad \quad \quad\times\left[(m_K^2 - m_{\ell_j}^2 - m_{\nu_i}^2)^2-4 m_{\ell_j}^2 m_{\nu_i}^2\right]^{1/2}\,,
\end{align}
which can be easily applied to other pseudoscalar mesons by changing the meson mass $m_P$ and the corresponding decay constant.

The expression for $R_P$ is finally given by
\begin{equation}\label{eq:RPresult}
R_P = \frac{\sum_i F^{i1} G^{i1}}{\sum_k F^{k2} G^{k2}}\,, \quad \text{with}
\end{equation}
\begin{align}
 F^{ij} &= |U_{ji}|^2 \quad \text{and}\nonumber \\  
 G^{ij} &= \left[m_P^2 (m_{\nu_i}^2+m_{l_j}^2) - (m_{\nu_i}^2-m_{l_j}^2)^2 \right] \left[ (m_P^2 - m_{l_j}^2 -
  m_{\nu_i}^2)^2 - 4 m_{l_j}^2 m_{\nu_i}^2 \right]^{1/2}\,.
  \label{eq:FG}
\end{align}
The result of eq.~(\ref{eq:RPresult}) has a straightforward interpretation: $F^{ij}$ represents the impact of new interactions (absent in the SM), whereas $G^{ij}$ encodes the mass-dependent factors. Notice however that all states (charged and neutral fermions) do not necessarily contribute to $R_P$: this can be seen from inspection of $G^{ij}$, see eq.~(\ref{eq:FG}), which must be a positive definite quantity. 

The SM result can easily be recovered from eq.~(\ref{eq:RPresult}), in the limit $m_{\nu_i} = 0$ and $U_{ji} = \delta_{ji}$, 
\begin{equation} \label{eq:RMSM}
R_P^{SM} = \frac{m_e^2}{m_\mu^2}
\frac{(m_P^2-m_e^2)^2}{(m_P^2-m_\mu^2)^2} \,, 
\end{equation}
to which small electromagnetic corrections (accounting for internal bremsstrahlung and structure-dependent effects) should be added~\cite{Cirigliano:2007xi}.  

The general expression for $\Delta r_P$ now reads
\begin{equation}\label{eq:deltaRPresult}
\Delta r_P \,= \,\frac{m_\mu^2 (m_P^2 - m_\mu^2)^2}{m_e^2 (m_P^2 - m_e^2)^2}\,
\frac{\operatornamewithlimits{\sum}_{m=1}^{N_\text{max}^{(e)}} 
F^{m1}\, G^{m1}}
{\operatornamewithlimits{\sum}_{n=1}^{N_\text{max}^{(\mu)}} 
F^{n2}\, G^{n2}} -1 \,.
\end{equation}
Thus, depending on the masses of the new states (and their hierarchy) and most importantly, on their mixings to the active neutrinos, $\Delta r_P$ can considerably deviate from zero. In order to illustrate this, we consider two regimes: in the first (A), all sterile neutrinos are \textit{lighter} than the decaying meson, but heavier than the active neutrino states, i.e. $m_\nu^\text{active} \ll m_{\nu_{s}} \lesssim m_P$; in the second (B), all $\nu_{s}$ are \textit{ heavier} than $m_P$. Notice that in case (A), all the mass eigenstates can be kinematically available and one should sum over all $3+N_s$ states; furthermore there is an enhancement to $\Delta r_P$ arising from phase space factors, see eq.~(\ref{eq:FG}).

We further emphasise that scenarios (A) and (B) are in general experimentally indistinguishable concerning lepton flavour universality, the only exception corresponding to a very particular regime where the sterile neutrinos are very close in mass to the decaying pseudoscalar meson. In such a situation, the resulting charged lepton would either be less energetic and not pass the experimental kinematical cuts~\cite{Shrock:1980vy, Shrock:1980ct} or have a clearly reduced momentum.

\subsubsection{Numerical evaluation of $\Delta r_K$ in the inverse seesaw model}

We numerically evaluate the contributions to $R_K$ in the framework of the inverse seesaw model and address the two scenarios discussed before. The adapted Casas-Ibarra parametrization for $Y_\nu$, in eq.~(\ref{CasasIbarraISS}), ensures that neutrino oscillation data are satisfied (the best-fit values of the global analysis of~\cite{Tortola:2012te} are used and the CP violating phases of $U_\text{PMNS}$ is set to zero). For illustrative purposes, the $R$ matrix angles are taken to be real (thus no contributions to lepton electric dipole moments are expected) and randomly varied in the range ${\theta}_{i} \in [0,2 \pi]$. Although we do not discuss it here, we have verified that similar contributions to $\Delta r_K$ are found when considering the more general complex $R$ matrix case.

\begin{figure}[t]
\centering
\subbottom{\includegraphics[angle=270,width=0.48\textwidth]{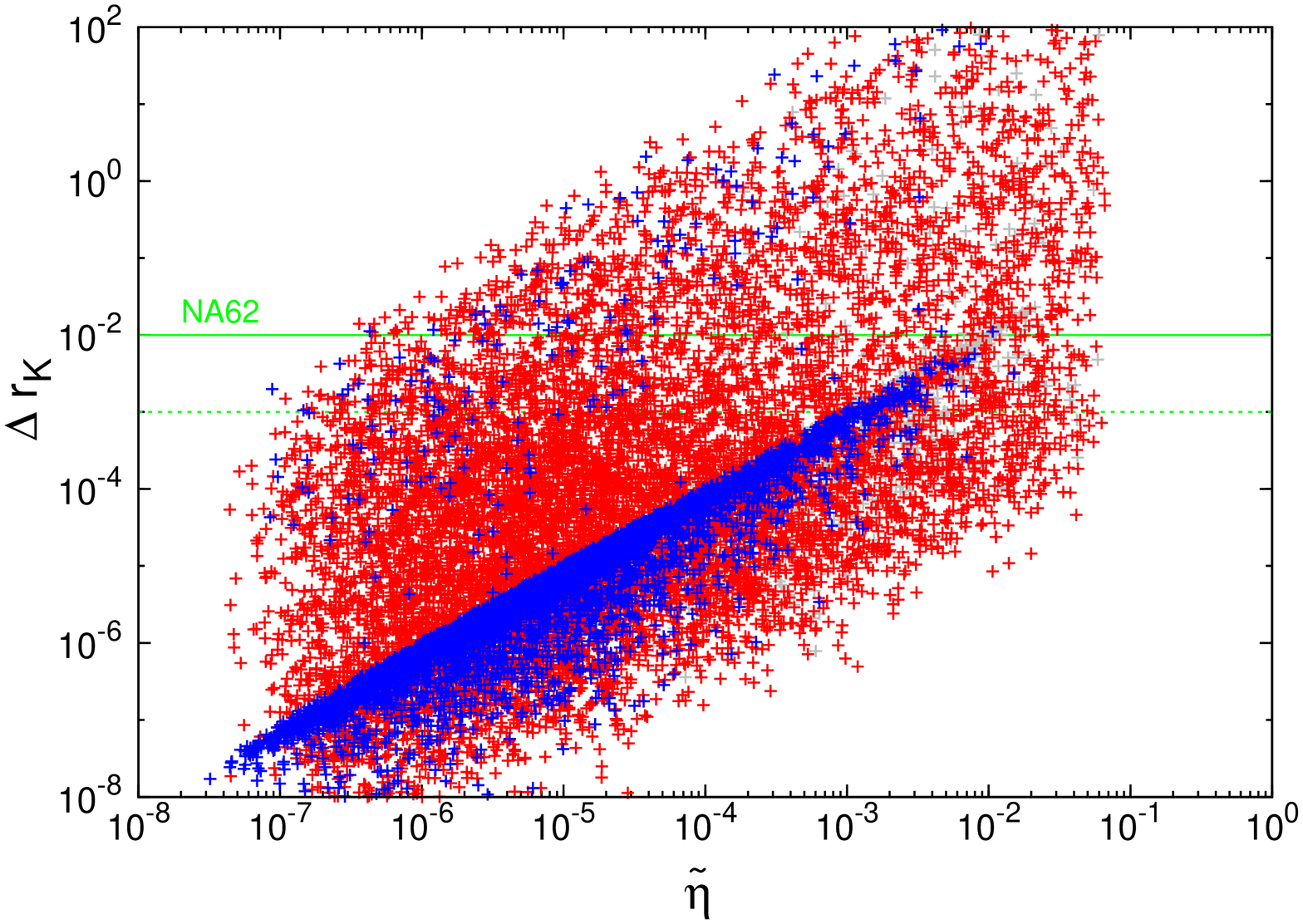}\label{scenarioA}}
\hfill
\subbottom{\includegraphics[angle=270,width=0.48\textwidth]{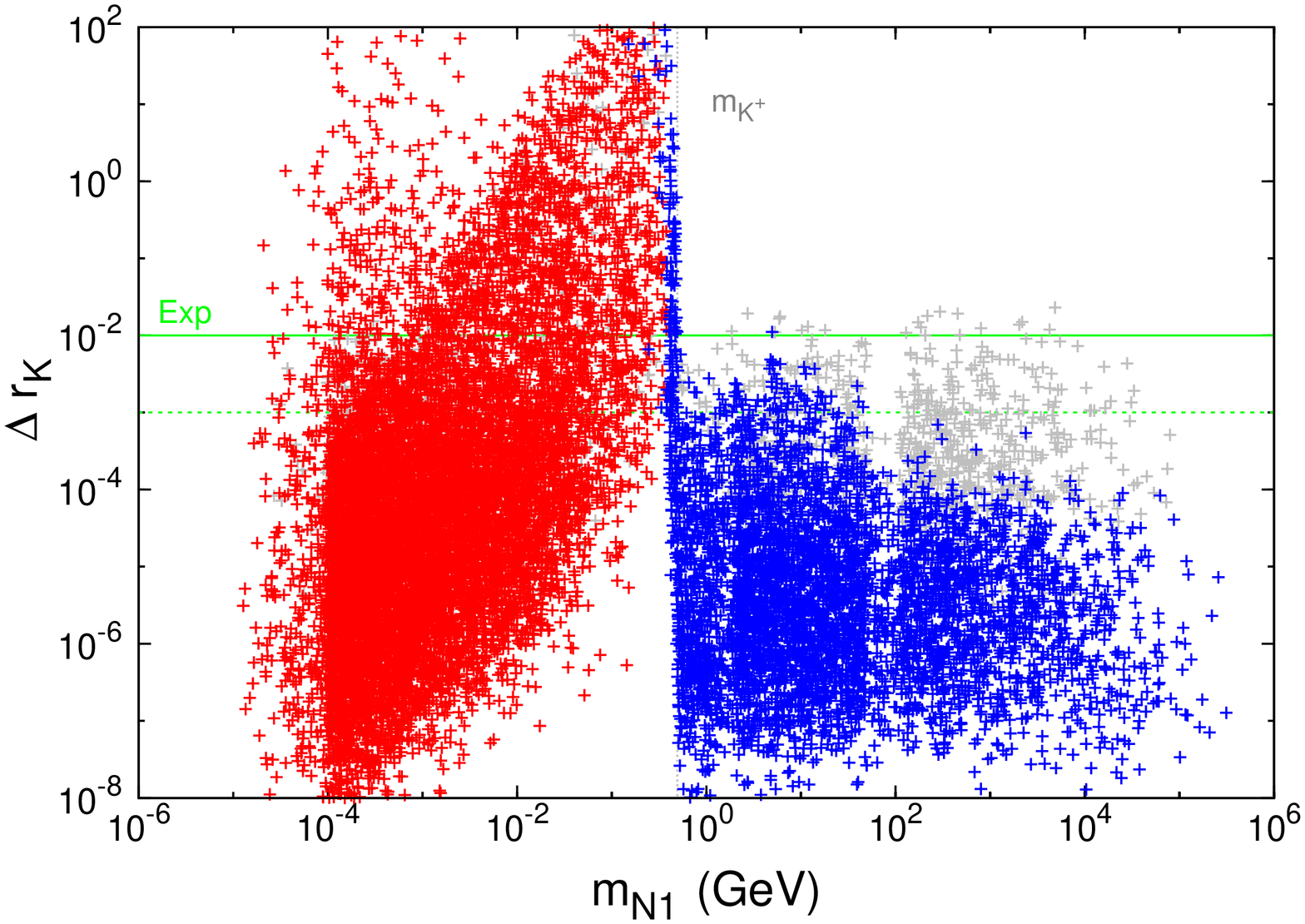}\label{scenarioB}}
\caption[$\Delta r_K$ in the inverse seesaw]{$\Delta r_K$ in the SM extended by the inverse seesaw as a function of $\tilde \eta = 1 - |\text{Det}(\tilde U_\text{PMNS})|$ (left) and as a function of $m_{N1}$ the mass of the lightest sterile neutrino (right). The upper (lower) green line denotes the current experimental limit (expected  sensitivity).  Grey points are excluded by the MEG limit on $\mathrm{Br}(\mu \rightarrow e \gamma)$, red points are excluded by cosmological observations while blue points are in agreement with all experimental limits and cosmological observations.} 
\label{RkResults}
\end{figure}

In figure~\ref{RkResults}, we collect our results for $\Delta r_K$ as a function of $\tilde \eta$, which parametrizes the departure from unitarity of the light active neutrino mixing sub-matrix $\tilde U_\text{PMNS}$,
\begin{equation}
 \tilde \eta = 1 - |\text{Det}(\tilde U_\text{PMNS})|\,,\label{tildeeta}
\end{equation}
and $m_{N1}$, the mass of the lightest sterile neutrino.
For the case of scenario (A), one can have very large contributions to $R_K$, which can even reach values $\Delta r_K \sim \mathcal{O}(1)$ (in some specific cases we find $\Delta r_K$ as large as $\sim 100$) as can be seen from fig.~\ref{scenarioB}.  The hierarchy of the sterile neutrino spectrum in case (A) is such that one can indeed have a significant amount of LFU violation, while still avoiding non-unitarity bounds.  Although this scenario would in principle allow to produce sterile neutrinos in light meson decays, the smallness of the associated $Y_\nu$ ($\lesssim\mathcal{O}(10^{-4})$), together with the loop function suppression, precludes the observation of LFV processes, even those with very good  associated experimental sensitivity, as is the case of $\mu \rightarrow e \gamma$.  The strong constraints from CMB and X-rays would exclude scenario (A); in order to render it viable, one would require a non-standard cosmology, see for instance~\cite{Gelmini:2008fq}

Despite the fact that in case (B) the hierarchy of the sterile states is such that non-unitarity bounds become very stringent (since the sterile neutrinos are not kinematically viable meson decay final states),  sizeable LFU violation is also possible, with deviations from the SM predictions as large as $\Delta r_K \sim \mathcal{O}(10^{-2})$. Contrary to case (A), whose results could also arise in other frameworks with light sterile neutrinos, the large deviations in (B) typically occur when all the singlet states are considerably heavier than the decaying meson, and reflect specific features of the ISS. As can be inferred from eq.~(\ref{LightMatrix}), in the inverse seesaw framework, one has $m_\nu \sim (Y_\nu \,v / M_R)^2 \, \mu_X$; hence, for ``low'' (when compared to, for instance,  the type I seesaw scale) $M_R$, light neutrino data can still be accommodated with large Yukawa couplings, $Y_\nu \sim \text{few} \times 10^{-1}$. As a consequence, large active-sterile mixings can occur, thus leading to an enhancement of $R_K$.  Even if  in this case one cannot produce sterile states in meson decays, the large $Y_\nu$ open the possibility of having larger contributions to LFV observables so that, for example, BR($\mu \to e \gamma)$ can be within MEG reach~\cite{Adam:2013mnn}.

Although we do not explicitly display it here, the prospects for $\Delta r_\pi$ are similar: in the same framework, one could have $\Delta r_\pi \sim \mathcal{O}(\Delta r_K)$, and thus $\Delta r_\pi \sim \mathcal{O}(1)$. Depending on the singlet spectrum, these observables can also be strongly correlated: if all the sterile states are heavier than the kaon, one finds $\Delta r_\pi \approx \Delta r_K$. The latter possibilities are a feature of the inverse seesaw mechanism (not possible in the unconstrained MSSM, for example) and are expected to be present in other low-scale seesaw models that allow for large active-sterile mixing angles.

The impact of this mechanism is not restricted to light meson decays. Moreover, there are currently some hints of lepton flavour universality violation in $D$ and $B$ meson decays. This has motivated the study of other observables, which we present in the following section.

\section{Other lepton universality tests}

\begin{figure}[!t]
\centering
\begin{fmffile}{SemileptonicTau}
\begin{fmfgraph*}(200,120)
\fmflabel{$\tau^-$}{tau}
\fmflabel{$\nu$}{nu}
\fmflabel{$\bar u$}{u}
\fmflabel{$s$}{s}
\fmfforce{(0.0w,0.3h)}{tau}
\fmfforce{(0.6w,0.0h)}{nu}
\fmfforce{(1.0w,1.0h)}{u}
\fmfforce{(1.0w,0.6h)}{s}
\fmfforce{(0.3w,0.3h)}{Wtaunu}
\fmfforce{(0.6w,0.8h)}{Wus}
\fmf{fermion}{tau,Wtaunu}
\fmf{fermion}{Wtaunu,nu}
\fmf{fermion}{u,Wus}
\fmf{fermion}{Wus,s}
\fmf{boson, label=$W^-$}{Wtaunu,Wus}
\end{fmfgraph*}
\end{fmffile}
\caption[Semileptonic $\tau$ decay]{Tree-level quark diagram for the semileptonic $\tau$ decay $\tau^- \rightarrow K^- \nu$.} \label{SemileptonicTau}
\end{figure}
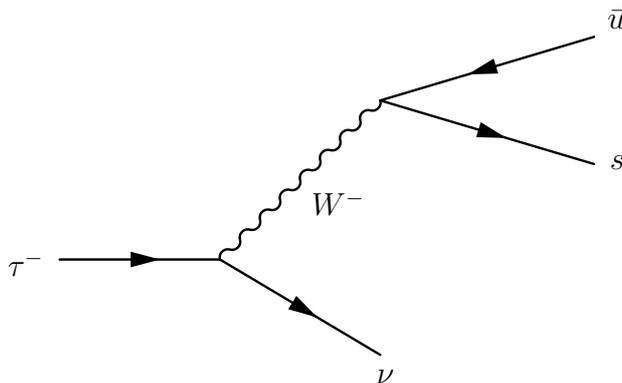

As we have seen, sterile neutrinos can lead to large deviations from the SM predictions in $R_K$ and $R_\pi$, and even saturate the experimental bounds. This fuels the interest in considering other lepton universality tests and searching for deviations with respect to the SM predictions. The expressions derived for $R_P$ can be directly generalized to $R_{D_s}$ (see eq.~(\ref{RDs})), which allows to probe universality violation in the $\tau$ sector. Another possibility to do this is to consider the ratios $R_{P,\ell}^{\tau}$ defined in eq.~(\ref{RPltau}). One needs then to evaluate the $\tau^- \rightarrow P^- \nu$ decay width in the presence of additional fermionic singlets. For the case of a charged kaon $K^-$ in the final state, the Feynman diagram is shown in fig.~\ref{SemileptonicTau} 
and the corresponding matrix element reads
\begin{equation}
 \imath \mathcal{M} = \langle \nu_i (q_2)\,K^-(q_1)|\imath \frac{G_F}{\sqrt{2}} U^*_{ji}\bar \nu_i \gamma_\mu (1-\gamma_5)\ell_j \bar s \gamma^\mu (1-\gamma_5) V^*_{us} u |\tau^-(p) \rangle\,.
\end{equation}
Proceeding similarly to eqs.~(\ref{step1}-\ref{FinalStep}), we obtain the spin-averaged squared matrix element
\begin{equation}
 |\overline{\mathcal{M}}|^2 = 2 G_F^2 F_K^2 |U_{\tau i}|^2 |V_{us}|^2 \left[(m_\tau^2-m_{\nu_i}^2)^2-m_K^2 (m_\tau^2+m_{\nu_i}^2)\right]\,,
\end{equation}
which is independent of any particle momentum. Thus, we can directly multiply it by the integrated two-body Lorentz-invariant phase space of eq.~(\ref{Lips2}) to find the following decay width
\begin{align}
 \Gamma=&\frac{G_F^2 F_K^2}{8 \pi m_\tau^3} |U_{\tau i}|^2 |V_{us}|^2 \left[(m_\tau^2-m_{\nu_i}^2)^2-m_K^2 (m_\tau^2+m_{\nu_i}^2)\right]\nonumber\\
  & \quad \quad \quad \quad \quad \quad \quad \quad \quad \quad\times\left[(m_\tau^2 - m_K^2 - m_{\nu_i}^2)^2-4 m_K^2 m_{\nu_i}^2\right]^{1/2}\,.
\end{align}
The expression for $R_{P,\ell}^\tau$ is finally given by
\begin{equation}
R_{P,\ell_j}^\tau = \frac{m_P^3}{2 m_\tau^3} \frac{\operatornamewithlimits{\sum}_{i=1}^{N_\text{max}^{(P)}} F^{i\tau} \tilde G^{i\tau}}{\operatornamewithlimits{\sum}_{k=1}^{N_\text{max}^{(\ell_j)}} F^{kj} G^{kj}}\,, \quad \text{with}
\end{equation}
\begin{equation}
 \tilde G^{i\tau} = \left[(m_\tau^2-m_{\nu_i}^2)^2-m_P^2 (m_\tau^2+m_{\nu_i}^2)\right] \left[(m_\tau^2 - m_P^2 - m_{\nu_i}^2)^2-4 m_P^2 m_{\nu_i}^2\right]^{1/2}\,,
\end{equation}
and $F^{kj}$ and $G^{kj}$ are given by eq.~(\ref{eq:FG}). Although more involved than the formula for $R_K$, the above expression is still free from hadronic uncertainties, so that at first order, any deviation from the SM prediction is due to the presence of the sterile neutrinos.

\begin{figure}[t]
\centering
\begin{fmffile}{Tau3l}
\begin{fmfgraph*}(200,120)
\fmflabel{$\ell_i^-$}{li}
\fmflabel{$\nu_\alpha$}{nua}
\fmflabel{$\ell_j^-$}{lj}
\fmflabel{$\nu_\beta$}{nub}
\fmfforce{(0.0w,0.3h)}{li}
\fmfforce{(0.6w,0.0h)}{nua}
\fmfforce{(1.0w,1.0h)}{nub}
\fmfforce{(1.0w,0.6h)}{lj}
\fmfforce{(0.3w,0.3h)}{Wtaunu}
\fmfforce{(0.6w,0.8h)}{Wus}
\fmf{fermion}{li,Wtaunu}
\fmf{plain}{Wtaunu,nua}
\fmf{plain}{nub,Wus}
\fmf{fermion}{Wus,lj}
\fmf{boson, label=$W^-$}{Wtaunu,Wus}
\end{fmfgraph*}
\end{fmffile}
\caption[Three-body lepton decay]{One of the two tree-level quark diagrams for the leptonic three-body muon and tau decays. The other one is its symmetric under the exchange of $\nu_\alpha$ and $\nu_\beta$.} \label{Tau3l}
\end{figure}
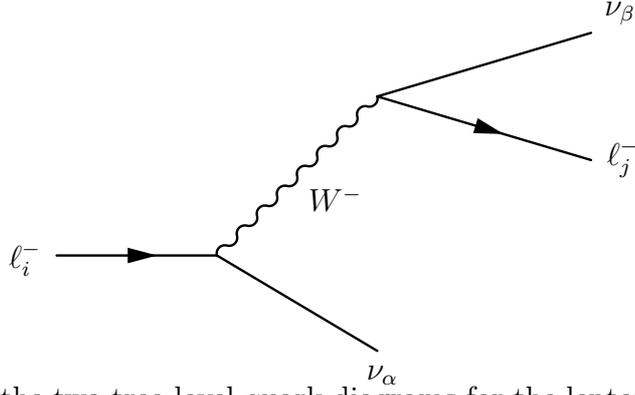

Lepton universality can also be tested by studying ratios of  leptonic three-body muon and tau decays. One should then compute the charged lepton decay widths in the channel $\ell_i^- \rightarrow \ell_j^- \nu \nu$. Since, until now, only one neutrino was present in the final state and escaped the detector without interacting, its Dirac or Majorana nature was of little concern. But here we have two neutrinos in the final state, and the statistics will be different according to the neutrino nature. Since we will illustrate our result in the inverse seesaw model, we present here the calculation for Majorana neutrinos. This decay is described by the Feynman diagram in fig.~\ref{Tau3l} 
and its symmetric under the exchange of $\nu_\alpha$ and $\nu_\beta$. The matrix element for the diagram in fig.~\ref{Tau3l} is
\begin{equation}
 \imath \mathcal{M}_1 = -\frac{\imath g^2}{8 m_W^2} U^*_{i\alpha} U_{j\beta} \bar u_{\nu_\alpha} \gamma^\mu (1-\gamma_5) u_{\ell_i} \bar u_{\ell_j} \gamma_\mu (1-\gamma_5) v_{\nu_\beta}\,,
\label{M_1}
\end{equation}
while the matrix element of the symmetric diagram reads
\begin{equation}
 \imath \mathcal{M}_2 = \frac{\imath g^2}{8 m_W^2} U^*_{i\beta} U_{j\alpha} \bar u_{\nu_\beta} \gamma^\mu (1-\gamma_5) u_{\ell_i} \bar u_{\ell_j} \gamma_\mu (1-\gamma_5) v_{\nu_\alpha}\,,
\end{equation}
where the factor $-1$ between $\mathcal{M}_1$ and $\mathcal{M}_2$ is due to the exchange of external fermions. The complete matrix element is the sum of the two contributions
\begin{equation}
 \imath \mathcal{M}_\mathrm{tot}=\imath \mathcal{M}_1 + \imath \mathcal{M}_2\,,
\end{equation}
and the spin-averaged squared matrix element is thus given by
\begin{equation}
 |\overline{\mathcal{M}_\mathrm{tot}}|^2=|\overline{\mathcal{M}_1}|^2 + |\overline{\mathcal{M}_2}|^2 + 2 \Re(\overline{\mathcal{M}_1 \mathcal{M}_2^*})\,.
\end{equation}
From eq.~(\ref{M_1}), the first contribution gives
\begin{align}
 |\overline{\mathcal{M}_1}|^2=&\frac{G_F^2}{4} |U_{i\alpha}|^2 |U_{j\beta}|^2 \mathrm{Tr}\left[(\slashed p_\alpha + m_{\nu_\alpha}) \gamma^\mu (1-\gamma_5) (\slashed p_i + m_{\ell_i}) \gamma^\nu (1-\gamma_5)\right]\nonumber\\
 &\times \mathrm{Tr}\left[(\slashed p_j + m_{\ell_j}) \gamma_\mu (1-\gamma_5) (\slashed p_\beta - m_{\nu_\beta}) \gamma_\nu (1-\gamma_5)\right]\,,
\end{align}
which reduces to
\begin{equation}
 |\overline{\mathcal{M}_1}|^2= 64 G_F^2 |U_{i\alpha}|^2 |U_{j\beta}|^2 [p_\alpha \cdot p_j][p_i \cdot p_\beta]\,.
\end{equation}
Likewise, one has for the symmetric diagram
\begin{equation}
 |\overline{\mathcal{M}_2}|^2= 64 G_F^2 |U_{i\beta}|^2 |U_{j\alpha}|^2 [p_\beta \cdot p_j][p_i \cdot p_\alpha]\,,
\end{equation}
while the interference term is given by
\begin{align}
 \mathcal{M}_1 \mathcal{M}_2^*= &\frac{G_F^2}{2} U^*_{i\alpha} U_{j\beta} U_{i\beta} U^*_{j\alpha} \bar u_{\nu_\alpha} \gamma^\mu (1-\gamma_5) u_{\ell_i} \bar u_{\ell_j} \gamma_\mu (1-\gamma_5) v_{\nu_\beta}\nonumber\\
 &\times \bar v_{\nu_\alpha} \gamma_\nu (1-\gamma_5) u_{\ell_j} \bar u_{\ell_i} \gamma^\nu (1-\gamma_5) u_{\nu_\beta}\,.
\end{align}
Usual spinor contractions cannot be straightforwardly used because of the mismatch between neutrino spinors. Dirac and Majorana spinors both verify~\cite{giunti2007fundamentals}
\begin{equation}
 u=C \bar v^T\,,\quad\mathrm{and}\quad v=C \bar u^T\,,
\end{equation}
which implies
\begin{equation}
 \bar u= - v^T C^\dagger \,,\quad\mathrm{and}\quad \bar v= - u^T C^\dagger\,.
\end{equation}
Thus the charged current can be recast as
\begin{equation}
 \bar u_{\ell_j} \gamma_\mu (1-\gamma_5) v_{\nu_\beta} = v_{\ell_j}^T \gamma_\mu^T (1-\gamma_5) \bar u_{\nu_\beta}^T = \bar u_{\nu_\beta} (1-\gamma_5) \gamma_\mu v_{\ell_j}\,,
\end{equation}
where the last equality is due to the fact that the current is a Hermitian covariant bilinear. Using this transformation of the current, we have
\begin{align}
 \mathcal{M}_1 \mathcal{M}_2^*= &\frac{G_F^2}{2} U^*_{i\alpha} U_{j\beta} U_{i\beta} U^*_{j\alpha} \bar u_{\nu_\alpha} \gamma^\mu (1-\gamma_5) u_{\ell_i} \bar u_{\nu_\beta} (1-\gamma_5) \gamma_\mu v_{\ell_j}\nonumber\\
 &\times \bar v_{\ell_j} (1-\gamma_5) \gamma_\nu u_{\nu_\alpha} \bar u_{\ell_i} \gamma^\nu (1-\gamma_5) u_{\nu_\beta}\,,
\end{align}
which, after averaging on the spins and simplifying Dirac matrices, gives
\begin{equation}
 \overline{\mathcal{M}_1 \mathcal{M}_2^*}= 32 G_F^2 U^*_{i\alpha} U_{j\beta} U_{i\beta} U^*_{j\alpha} m_{\nu_\alpha} m_{\nu_\beta} [p_i\cdot p_j]\,.
\end{equation}
Introducing the Dalitz variables~\cite{Dalitz:1954cq} defined as
\begin{align}
 s_{j\alpha}&=m_{j\alpha}^2=(p_j+p_\alpha)^2\,,\\
 s_{j\beta}&=m_{j\beta}^2=(p_j+p_\beta)^2\,,\\
 s_{\alpha\beta}&=m_{\alpha\beta}^2=(p_\alpha+p_\beta)^2\,,
\end{align}
the differential decay width reads
\begin{align}
 \mathrm{d}\Gamma=&\frac{G_F^2}{(2\pi)^3 m_{\ell_i}^3} (2-\delta_{\alpha\beta})\mathrm{d}s_{j\alpha}\mathrm{d}s_{j\beta}\left[\frac{1}{4} |U_{i\alpha}|^2 |U_{j\beta}|^2 (s_{j\alpha}-m_{\ell_j}^2-m_{\nu_\alpha}^2) (m_{\ell_i}^2+m_{\nu_\beta}^2-s_{j\alpha})\right.\nonumber\\
 &+\left.\frac{1}{4} |U_{i\beta}|^2 |U_{j\alpha}|^2 (s_{j\beta}-m_{\ell_j}^2-m_{\nu_\beta}^2) (m_{\ell_i}^2+m_{\nu_\alpha}^2-s_{j\beta})\right.\nonumber\\
 &+\left.\frac{1}{2} \Re(U^*_{i\alpha} U_{j\beta} U_{i\beta} U^*_{j\alpha}) m_{\nu_\alpha} m_{\nu_\beta} (s_{j\alpha} + s_{j\beta} - m_{\nu_\alpha}^2 - m_{\nu_\beta}^2) \right]\,,
\end{align}
where the $\delta_{\alpha\beta}$ appears because neutrinos are Majorana fermions. When the two mass eigenstates are identical, the sterile neutrinos cannot be distinguished and integrating over the whole phase space would lead to a double counting of the final states. This can be accounted for by a factor $(1-\delta_{\alpha\beta}/2)$. Then we have to evaluate the integration domain boundaries. If we integrate first over $s_{j\beta}$, they are
\begin{align}
 \mathrm{max}(s_{j\alpha})=&(m_{\ell_i}^2-m_{\nu_\beta}^2)^2\,,\\
 \mathrm{min}(s_{j\alpha})=&(m_{\ell_j}^2+m_{\nu_\alpha}^2)^2\,,\\
 \mathrm{max}(s_{j\beta})-\mathrm{min}(s_{j\beta})=&\frac{1}{s_{j\alpha}} \lambda^{1/2}(s^{1/2}_{j\alpha}, m_{\ell_j}, m_{\nu_\alpha}) \lambda^{1/2}(s^{1/2}_{j\alpha}, m_{\ell_i}, m_{\nu_\beta})\,,
\end{align}
where 
\begin{equation}
  \lambda(a,b,c) \, = \, (a^2 - b^2 -c^2)^2 -4\,b^2\,c^2\,.
\end{equation}
If we integrate $s_{j\alpha}$ first, they are
\begin{align}
 \mathrm{max}(s_{j\beta})=&(m_{\ell_i}^2-m_{\nu_\alpha}^2)^2\,,\\
 \mathrm{min}(s_{j\beta})=&(m_{\ell_j}^2+m_{\nu_\beta}^2)^2\,,\\
 \mathrm{max}(s_{j\alpha})-\mathrm{min}(s_{j\alpha})=&\frac{1}{s_{j\beta}}\lambda^{1/2}(s^{1/2}_{j\beta}, m_{\ell_j}, m_{\nu_\beta}) \lambda^{1/2}(s^{1/2}_{j\beta}, m_{\ell_i}, m_{\nu_\alpha})\,.
\end{align}
Rewriting the differential matrix element as explicitly symmetric under the exchange $\nu_\alpha \leftrightarrow \nu_\beta$ and summing over all the kinematically accessible neutrinos, the decay width of $\ell_i \rightarrow \ell_j \nu \nu$ reads
\begin{align}
 \Gamma &= \sum_{\alpha=1}^{\mathrm{N_{max}}(\ell_j)} \sum_{\beta=1}^{\alpha} \Gamma_{\alpha\beta}\,,\\
 \mathrm{with} & \nonumber\\
 \Gamma_{\alpha\beta} &= \frac{G_F^2 (2-\delta_{\alpha\beta})}{m_{\ell_i}^3 (2\pi)^3} \int_{(m_{\ell_j}+m_{\nu_\alpha})^2}^{(m_{\ell_i}-m_{\nu_\beta})^2} \mathrm{d}s_{j\alpha} \left[\frac{1}{4} |U_{i\alpha}|^2 |U_{j\beta}|^2 (s_{j\alpha} -m_{\ell_j}^2-m_{\nu_\alpha}^2) (m_{\ell_i}^2+m_{\nu_\beta}^2-s_{j\alpha}) \right.\nonumber\\
& \left. \quad \quad \quad \quad \quad + \frac{1}{2} \Re(U_{i\alpha}^* U_{j\beta} U_{i\beta} U_{j\alpha}^*) m_{\nu_\alpha} m_{\nu_\beta} \left(s_{j\alpha}-\frac{m_{\nu_\alpha}^2+m_{\nu_\beta}^2}{2}\right) \right]\nonumber\\
& \quad \quad \quad \quad \quad \times \frac{1}{s_{j\alpha}} \lambda^{1/2}(s^{1/2}_{j\alpha}, m_{\ell_j}, m_{\nu_\alpha}) \lambda^{1/2}(s^{1/2}_{j\alpha}, m_{\ell_i}, m_{\nu_\beta})\nonumber\\
& + \alpha \leftrightarrow \beta\,.\label{tau3l}
\end{align}
For comparison, in the case of Dirac neutrinos, the decay width would be given by
\begin{align}
 \Gamma &= \sum_{\alpha=1}^{N_{max}(\ell_j)} \sum_{\beta=1}^{N_{max}(\ell_j)} \Gamma_{\alpha\beta}\,,\\
 \mathrm{with} & \nonumber\\
 \Gamma_{\alpha\beta} &= \frac{G_F^2 |U_{i\alpha}|^2 |U_{j\beta}|^2}{2 m_{\ell_i}^3 (2\pi)^3} \int_{(m_{\ell_j}+m_{\nu_\alpha})^2}^{(m_{\ell_i}-m_{\nu_\beta})^2} \mathrm{d}s_{j\alpha} (s_{j\alpha} -m_{\ell_j}^2-m_{\nu_\alpha}^2) (m_{\ell_i}^2+m_{\nu_\beta}^2-s_{j\alpha}) \nonumber\\
& \quad \quad \quad \quad \times \frac{1}{s_{j\alpha}} \lambda^{1/2}(s^{1/2}_{j\alpha}, m_{\ell_j}, m_{\nu_\alpha}) \lambda^{1/2}(s^{1/2}_{j\alpha}, m_{\ell_i}, m_{\nu_\beta})\,.
\end{align}
Notice from eq.~(\ref{tau3l}) that the interference term is specific to Majorana neutrinos and proportional to their masses. This is to be compared with the SM prediction
\begin{equation}
 \Gamma=\frac{G_F^2 m_{\ell_i}^5}{192 \pi^3}\,.
\end{equation}
We explicitly verified that in the limit of massless particles in the final state the decay widths agree with the SM prediction.

Another observable that could lead to interesting results is the ratio of semileptonic pseudoscalar meson decays, for example
\begin{equation}
\frac{\Gamma(P\rightarrow P^\prime e \nu)}{\Gamma(P\rightarrow P^\prime \mu \nu)} \, .
\end{equation}
The total width of a semileptonic decay can be decomposed as 
\begin{equation}
\Gamma_\text{tot}\, = \, 
\Gamma_{c_1} \, + \,\Gamma_{c_2} \, + \,\Gamma_{c_3} \, + \,\Gamma_{c_4}\,,
\end{equation}
where each partial width is associated with the form factors 
$F^+ (q^2)$, $F^0 (q^2)$ (and combinations thereof), where $q$ denotes the momentum transfer, as follows
\begin{align}
\Gamma_{c_1, c_2}  \rightsquigarrow |F^+ (q^2)|^2\, ;
\quad 
\Gamma_{c_3}  \rightsquigarrow |F^0 (q^2)|^2\, ;
\quad 
\Gamma_{c_4} \rightsquigarrow 2 \Re(F^0 F^{+ *})\,.
\end{align}
The above widths can be written as
\begin{align}
& \Gamma_{c_1} = 
\frac{G_F^2}{192 \pi^3}
\frac{|V_{ij}|^2 |U_{\alpha \beta}|^2}{M^3}
\int_{\left(M_1+M_2\right)^2}^{\left(M-M_3\right)^2} dq^2 |F^+ (q^2)|^2 \lambda^{{3/2}}(q^2,M^2,M_3^2)\nonumber\\
& \hskip 5cm \times \lambda^{{3/2}}(q^2,M_1^2,M_2^2) \frac{1}{q^6}\,,
\nonumber \\
& \Gamma_{c_2}= 
\frac{G_F^2}{128 \pi^3}
\frac{|V_{ij}|^2 |U_{\alpha \beta}|^2}{M^3}
\int_{\left(M_1+M_2\right)^2}^{\left(M-M_3\right)^2} dq^2
|F^+ (q^2)|^2
\lambda^{{3/2}}(q^2,M^2,M_3^2) \lambda^{{1/2}}(q^2,M_1^2,M_2^2) \nonumber\\
&\hskip 5cm\times \frac{1}{q^6} \left[
q^2(M_1^2 +M_2^2) - (M_1^2 - M_2^2)^2
\right ],
\nonumber \\
& \Gamma_{c_3} =  
\frac{G_F^2}{128 \pi^3}
\frac{|V_{ij}|^2 |U_{\alpha \beta}|^2}{M^3}
\int_{\left(M_1+M_2\right)^2}^{\left(M-M_3\right)^2}dq^2
|F^0 (q^2)|^2 
\left (
\frac{\Delta M^2}{q^2}
\right )^2 \lambda^{{1/2}}(q^2,M^2,M_3^2) \nonumber\\
&\hskip 5cm\times \lambda^{{1/2}}(q^2,M_1^2,M_2^2) \frac{1}{q^2} \left[
q^2(M_1^2 +M_2^2) - (M_1^2 - M_2^2)^2
\right ],
\nonumber \\
& \Gamma_{c_4} = 0\,,
\end{align}
where $\Delta M^2  \, = \, M^2 -M_3^2$, with $M$ the mass of the decaying meson, $M_{1,2}$ the final state charged
and neutral leptons, and $M_3$ the final state meson. 

Finally, the $W^\pm$ decay width could also be modified by the presence of sterile neutrinos. It can be written as
\begin{equation}
 \Gamma(W \to \ell_i \nu) = \sum_j \Gamma_{VFF}(m_W,m_{\ell_i},m_{\nu_j},a_L^{ij},0)
\end{equation}
where the function $\Gamma_{VFF} = \Gamma_{VFF}(m_V,m_{F_1},m_{F_2},c_L,c_R)$ is
\begin{align}
 &\Gamma_{VFF} = \frac{\lambda^{1/2}(m_V, m_{F_1}, m_{F_2})}{48 \pi m_V^3} \\
 &\times \left[\left(|c_L|^2+|c_R|^2\right) \left(2 m_V^2-\frac{\left(m_{F_1}^2-m_{F_2}^2\right)^2}{m_V^2}-m_{F_1}^2-m_{F_2}^2\right)+12 m_{F_1}
   m_{F_2} \Re\left(c_L c_R^*\right)\right] \, . \nonumber
\end{align}
The coupling $a_L$ is given by $a_L^{ij} = 2^{3/4} m_W \sqrt{G_F} U_{ij}$.

As seen in leptonic kaon decays, the existence of sterile neutrinos can potentially lead to a significant violation of lepton flavour universality at tree-level in light meson decays. Provided that the active-sterile mixings are sufficiently large, the modified  leptonic interactions can generate large contributions to lepton universality tests, with measurable deviations from the standard model expectations. As an illustrative example, we have evaluated the contributions to $R_K$ in the inverse seesaw extension of the SM, for distinct hierarchies of the sterile states. In particular, we have studied the impact of non-unitarity in a low mass regime for the additional singlets, an inverse seesaw  mass regime  considerably lower than what had previously been addressed~\cite{Malinsky:2009gw, Antusch:2008tz}. Our analysis reveals that very large deviations from the SM predictions can be found with $\Delta r_K \sim \mathcal{O}(1)$ or even larger, well within reach of the NA62 experiment at CERN. This is in clear contrast with other models of new physics (for example unconstrained SUSY models, where one typically has $\Delta r_K \lesssim \mathcal{O}(10^{-3})$~\cite{Fonseca:2012kr}, and in models with four generations~\cite{Lacker:2010zz}).  We further notice that these large deviations are a generic and non fine-tuned feature of this model. It is worth emphasising that, in view of the potentially large new contributions to these observables, such an analysis of LFU violation in light meson decays actually allows to set bounds on the amount of unitarity violation (parametrized by $\tilde \eta$, eq.~(\ref{tildeeta})). The lepton universality tests mentioned above deserve a careful study, which is currently under way~\cite{inpreparation}. It will also be very interesting to search for correlations between the different observables that are specific to the presence of additional fermionic singlets.

The inverse seesaw model is a very attractive extension of the SM since it generates neutrino masses close to the electroweak scale. This allows the current generation of experiments to constrain this model, either from high-energy searches, like direct production of the heavy neutrinos or its potential impact on the Higgs sector, or from low-energy experiments at the high-intensity frontier, whose Physics programs include lepton universality tests or searches for charged lepton flavour violation. However, there exists other models that address problems of the SM not solved by the inverse seesaw, like the dark matter issue or the hierarchy problem. In the next chapters, we will consider supersymmetric models addressing the latter issues and discuss the implementation of the inverse seesaw in such a framework.

\renewcommand*{\afterpartskip}{\epigraph{Curiously enough, the only thing that went through the mind of the bowl of petunias as it fell was Oh no, not again. Many people have speculated that if we knew exactly why the bowl of petunias had thought that we would know a lot more about the nature of the Universe than we do now.}{\textit{The Hitchhiker's Guide to the Galaxy}\\ \textsc{Douglas Adams}}
\vfil\clearpage}
\part{{}Supersymmetric inverse seesaw models}

\chapter{Supersymmetric extensions of the Standard Model\label{chapSUSY}}

In the previous chapters we have discussed how the generation of neutrino masses calls for the introduction of new fields (such as new fermionic singlets) and we have also addressed a number of specific signatures of these extensions. However, other caveats of the SM have motivated the study of larger frameworks. A number of New Physics models have been considered during the last decades. Supersymmetric extensions of the SM are among them. In this chapter, we will introduce the concept of supersymmetry (SUSY) and present two supersymmetric models. But first, we will discuss some of the motivations that led to the introduction of supersymmetry and why it has attracted so much attention over the years.

\section{The appeal of supersymmetry}

The concept of supersymmetry has many attractive aspects. First, it gives a unified description of bosons and fermions, relating them through the generator of SUSY transformations. Thus, using the superfield formalism that will be introduced in the next section, they appear as different components of a supermultiplet and any supersymmetric Lagrangian has to be invariant under the exchange of the bosonic and fermionic components of the same supermultiplet. 

The second appealing aspect of supersymmetry comes from the fact that it extends the Poincaré algebra in a very specific way. In 1967, Sidney Coleman and Jeffrey Mandula derived a no-go theorem about extended spacetime symmetries~\cite{Coleman:1967ad}. Following~\cite{Wess1992supersymmetry}, it states that:

\textit{If $(1)$ the $S$~matrix is based on a local, relativistic quantum field theory in four-dimensional spacetime, $(2)$ there are only a finite number of different particles associated with one-particle states of a given mass, $(3)$ and there is an energy gap between the vacuum and the one particle  states, then the most general Lie algebra of symmetries of the $S$~matrix contains the Poincaré algebra and a finite number of Lorentz scalar operators that must belong to the Lie algebra of a compact Lie subgroup.}

Rephrasing the above statement, the most general Lie algebra of the $S$ matrix symmetries is the direct product of the Poincaré algebra with the internal symmetries, which should be described by a compact Lie algebra. In particular, this implies that the internal symmetry generators commute with the Poincaré generators. However, the Coleman-Mandula theorem only applies to Lie algebras, considering only bosonic operators that define an algebra via commutation relations. Supersymmetry evades the Coleman-Mandula theorem since it introduces a fermionic generator and anticommutation relations. Moreover, in 1975, Rudolf Haag, Jan Lopuszanski and Martin Sohnius proved that the SUSY algebra (see eq~(\ref{SUSYalgebraStart}-\ref{SUSYalgebraEnd})) is, in fact,  the most general extension of the Poincaré algebra assuming a single spinorial generator $Q$~\cite{Haag:1974qh}.

Third, supersymmetric theories have an improved ultra-violet (UV) behaviour. This is known as the non-renormalization theorem, which states that in SUSY theories couplings and masses are only renormalized through the wave function renormalization of the fields~\cite{Iliopoulos:1974zv}. Indeed, all loop corrections can be expressed as non-dynamical fields~\cite{Grisaru:1979wc}, known as D-terms, leading to kinetic terms that generate wave function renormalization. Moreover, this renormalization is at most logarithmically divergent in the UV cut-off. As a consequence, SUSY theories are free of quadratic divergences and thus provide a solution to the hierarchy problem that plagues the Standard Model. The latter issue comes from the fact that, in the SM, the Higgs mass is not protected by any symmetry and ends up being quadratically divergent in the cut-off. Assuming that the SM is valid up to the Planck scale $\Lambda_P\sim10^{18}\;\mathrm{GeV}$\footnote{Above the Planck scale, gravitational interactions become relevant and a theory of quantum gravity is required.}, the counter terms to the Higgs squared mass would have to be fine-tuned to $10^{-32}$ to cancel out the new radiative corrections to the Higgs mass, which is highly unnatural. Since the quadratic divergences are absent in SUSY theories, the counter terms do not need such a fine adjustment, making SUSY models more natural. Moreover, if the Higgs mass were to be driven to a very high scale, why would the electroweak scale be close to $100\;\mathrm{GeV}$ ? This question is the essence of the hierarchy problem.

\begin{figure}[t]
\centering
\subbottom[SM running gauge couplings][SM]{\includegraphics[width=0.49\textwidth]{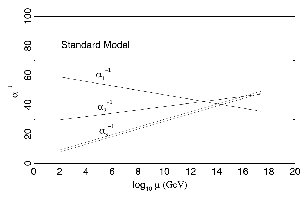}\label{runningSM}}
\hfill
\subbottom[MSSM running gauge couplings][MSSM]{\includegraphics[width=0.49\textwidth]{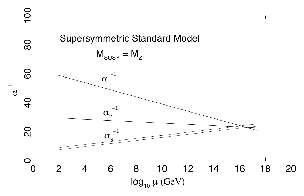}\label{runningMSSM}}
\caption[Running gauge couplings in the SM and MSSM]{Running gauge couplings in a supersymmetric and non-supersymmetric model as a function of the renormalization scale. Updated figures obtained upon private communications with the authors. The original figures appeared in~\cite{Langacker:1991an}.} \label{running}
\end{figure}

Fourth, supersymmetry improves the convergence of the running gauge couplings at high energy, as demonstrated in~\cite{Langacker:1991an}. In fact, there is an approximate convergence in the SM as shown in fig.~\ref{runningSM},
which would hint towards the idea of a Grand Unified Theory\footnote{At higher scales, there would be an unbroken phase of a larger gauge group where the couplings unify in a common value.}. Unfortunately, since the couplings do not meet in one point, GUT theories with direct breaking to the SM gauge group are excluded. However, the superpartners that are introduced in SUSY theories also contribute to the gauge coupling $\beta$ functions, improving the convergence of the running couplings. Actually, in the Minimal Supersymmetric extension of the Standard Model (MSSM), they unify within experimental error for $M_{GUT}\sim 10^{16}\;\mathrm{GeV}$ as seen in fig.~\ref{runningMSSM}.

Fifth, local supersymmetry offers a natural connection to gravity. Indeed, when supersymmetry is gauged, it is necessary to introduce a spin~$3/2$ particle whose superpartner is a spin~$2$ particle~\cite{baer2006weak, Wess1992supersymmetry}. The latter couples to the energy-momentum tensor as in general relativity, allowing the identification of the spin~$2$ field as the graviton~\cite{Deser:1976eh, *Freedman:1976xh}, whose superpartner is the gravitino. Unfortunately, as in general relativity, the new coupling is dimensionful, with a mass dimension of~$-1$, introducing non-renormalizable interactions in the theory.

Sixth, supersymmetry offers a framework to understand electroweak symmetry breaking. In the SM, the condition leading to EWSB, $\mu<0$, has to be enforced by hand. In many SUSY models, radiative corrections to the Higgs mass associated with the up-quark sector induce its running from a positive value at the ultraviolet scale, down to negative values in the infrared thus triggering electroweak symmetry breaking. This mechanism is named radiative electroweak symmetry breaking (REWSB) and it can take place over a wide range of parameters~\cite{Ibanez:1982fr, *Inoue:1982pi, *Inoue01071983, *Ibanez:1982ee, *Ellis:1983bp, *AlvarezGaume:1983gj}.

Finally, supersymmetric models with conserved R-parity\footnote{R-parity will be discussed in Section~\ref{SecMSSM}.}, a global $\mathbb{Z}_2$ symmetry under which SM fields are even while their superpartners are odd, may contain a candidate for dark matter~\cite{baer2006weak}. In R-parity conserving models, the Lightest Supersymmetric Particle (LSP), being the lightest particle with a conserved quantum number, is stable. Thus, if it is electrically neutral and does not interact strongly, it constitutes a natural dark matter candidate. Astrophysical and cosmological observations, like the galaxy rotation curves or matter distribution in the bullet cluster, point towards the existence of a non-baryonic, non-luminous type of matter. The most recent measurement of the CMB anisotropies by the Planck satellite yields a dark matter density $\Omega_{DM}\simeq0.26$, which is roughly five times the density of baryonic matter~\cite{Ade:2013lta}.

It is worth mentioning that SUSY extensions of the SM are not the only frameworks that have been considered. For example, many theories have been proposed to address the hierarchy problem. Among them are extra-dimensions, large~\cite{ArkaniHamed:1998rs, *ArkaniHamed:1998nn, Appelquist:2000nn} or warped~\cite{Randall:1999ee, *Randall:1999vf}, which lower the Planck scale, effectively reducing the fine-tuning, and even compositeness~\cite{PhysRevD.13.974, *PhysRevD.19.1277, *PhysRevD.20.2619}, where the Higgs field is a condensate from a new strongly interacting sector.

\section{A brief introduction to supersymmetry}

The idea of a symmetry relating fermions and bosons was first introduced by Hironari Miyazawa in 1966 as a symmetry between baryons and mesons~\cite{Miyazawa01121966}. It was later reinterpreted as a symmetry of spacetime and fundamental fields in 1971 by Yu Golfand and Evgeny Likhtman~\cite{Golfand:1971iw}, who introduced it as an extension of the Poincaré group algebra. Three years later, Julius Wess and Bruno Zumino wrote down the first four-dimensional supersymmetric quantum field theory with supersymmetry realised linearly~\cite{Wess:1974tw}. But it was only in the 1980s that the Minimal Supersymmetric Standard Model (MSSM), the simplest viable SUSY extension of the SM, was formulated and studied from a phenomenological point of view. In this section, we will use the conventions of~\cite{baer2006weak}.

\subsubsection{The supersymmetry algebra}

Denoting the generators of the Lorentz boost $M_{0i}=-K_i$, those of the rotations \mbox{$M_{ij}=\epsilon_{ijk} J_k$} and the ones of spacetime translations $P_\mu$, the Poincaré algebra is defined by the following commutation relations:
\begin{align}
 [P_\mu\,,P_\nu] &= 0\,,\label{SUSYalgebraStart}\\
 [M_{\mu\nu}\,,P_\lambda] &= \imath (g_{\nu\lambda}P_\mu - g_{\mu\lambda} P_\nu)\,,\\
 [M_{\mu\nu}\,,M_{\rho\sigma}]&=\imath (g_{\nu\rho} M_{\mu\sigma} +g_{\mu\sigma}M_{\nu\rho}-g_{\mu\rho}M_{\nu\sigma}-g_{\nu\sigma}M_{\mu\rho})\,, 
\end{align}
where $g_{\mu\nu}$ is the Minkowski metric with the signature $(+,-,-,-)$. The Poincaré algebra can then be enlarged by adding a new generator $Q$, which is a four-component Majorana spinor, and the resulting super-Poincaré algebra is defined by the above commutation relations extended by
\begin{align}
 \{Q_a\,,\bar{Q}_b\}&=2(\gamma^\mu)_{ab} P_\mu\,,\\
 [Q_a\,,P_\mu]&=0\,,\label{QP}\\
 [Q_a\,,M_{\mu\nu}]&=\frac{1}{2} (\sigma_{\mu\nu} )_{ab} Q_b\,,\label{SUSYalgebraEnd}
\end{align}
with $a\,,b$ spinorial indices and $\sigma^{\mu\nu}=\frac{\imath}{2} [\gamma^\mu\,,\gamma^\nu]$.

Two comments are in order. First, as mentioned in Section~\ref{massiveNu}, an irreducible representation of the Lorentz group, and by extension of the super-Poincaré group, is formed by Weyl spinors of a definite chirality. Thus, supersymmetric theories have to be built from either Weyl or Majorana spinors, which have as many independent component as the Weyl spinors. The usual convention is to choose all the fields in the left-handed representation of the super-Poincaré algebra. The right-handed component of the SM fermions can then be included through their charge conjugate given by eq.~(\ref{ConjRight}). Second, eq.~(\ref{QP}) implies that $[Q_a\,,P^2]=0$. Since the eigenvalue of the Casimir operator $P^2$ applied to a particle is the squared mass, particles related by SUSY transformations (i.e. particle and superparticle) have the same mass if supersymmetry remains unbroken.

As we have seen above, the Poincaré algebra can be enlarged by a spinorial generator, which can change a boson into a fermion and \textit{vice versa}. A convenient way to describe this is to group superpartners into one multiplet, known as a supermultiplet, making use of the superfield formalism.

\subsubsection{The superfield formalism}

Supersymmetric theories and transformations can be formulated in a very simple and elegant way using the superfield formalism. The first step is to enlarge the four-dimensional spacetime with a four-component Grassmannian coordinate $\theta$, which behaves like a Majorana spinor whose components are Grassmann (anticommuting) numbers
\begin{align}
 \{\theta_a\,,\theta_b\}=0\,,\label{GrassmanAnti}\\
 \{\theta_a\,,\psi_a\}=0\,,\\
 \bar{\theta}=\theta^T C\,.
\end{align}
It is easy to see from eq.~(\ref{GrassmanAnti}) that any product of $\theta$ that contains five or more $\theta$ vanishes. This enlarged spacetime is called a superspace and illustrates the deep connection that supersymmetric transformations have with more usual spacetime symmetries\footnote{The usual transformations of the Poincaré algebra are translations, rotations and Lorentz boost and act on the coordinates of the 4-dimensional spacetime. Through the superfield formalism, SUSY transformations correspond to the action on the extra Grassmannian coordinate as defined by eq.~(\ref{SUSYQ}).}. Fields can now be defined on this superspace and be expanded as a finite series of Grasmann variables. Following~\cite{baer2006weak}, a scalar superfield $\widehat{\Psi}(x,\theta)$ can be decomposed as
\begin{align}
 \widehat{\Psi}=&\mathcal{S}-\imath\sqrt{2}\bar{\theta}\gamma_5\psi -\frac{\imath}{2}(\bar{\theta}\gamma_5\theta)\mathcal{M} +\frac{1}{2}(\bar{\theta} \theta) \mathcal{N} + \frac{1}{2} (\bar \theta \gamma_5 \gamma_\mu \theta)V^\mu\nonumber\\
            &+\imath (\bar \theta \gamma_5 \theta)\left[\bar \theta \left(\lambda + \frac{i}{\sqrt{2}} \slashed{\partial} \psi\right) \right] -\frac{1}{4} (\bar{\theta}\gamma_5\theta)^2 \left[ \mathcal{D} - \frac{1}{2} \Box \mathcal{S} \right]\,,\label{SuperScalar}
\end{align}
where $\mathcal{S}$, $\mathcal{M}$, $\mathcal{N}$ and $\mathcal{D}$ are scalar fields, $\psi$ and $\lambda$ are spinor fields and $V^\mu$ is a vector field. From eq.~(\ref{SuperScalar}), one can see that multiplying by the Grassmann coordinate or deriving with respect to it corresponds to transitions between superpartners. A SUSY transformation with the Majorana spinor parameter $\alpha$ is then described by the generator $Q$ acting on the superfield $\widehat \Psi$ as:
\begin{equation}
 [\bar \alpha Q\,, \widehat \Psi]=\imath \left(\bar \alpha \frac{\partial}{\partial \bar \theta} + \imath \bar \alpha \slashed \partial \theta \right) \widehat \Psi\,.\label{SUSYQ}
\end{equation}
Thus, all the component fields of the scalar superfield transform into each other. But is there a reduced set of fields that would transform into each other and form an irreducible representation of the supersymmetry algebra?

The scalar superfield in eq.~(\ref{SuperScalar}) is reducible in the same way as a Dirac spinor by choosing $\lambda=0$, $\mathcal{D}=0$ and $V_\mu=\partial_\mu \xi$, with $\xi$ a scalar field. Then, the LH chiral superfield corresponds to the choice $\psi_R=0$, $V_\mu=\imath \partial_\mu \mathcal{S}$ and $\mathcal{N}=\imath \mathcal{M} = \imath \mathcal{F}$, which gives
\begin{align}
 \widehat{\mathcal{S}}_L=\mathcal{S}+\imath\sqrt{2}\bar{\theta}\psi_L +\imath(\bar{\theta}\theta_L)\mathcal{F}+ \frac{\imath}{2} (\bar \theta \gamma_5 \gamma_\mu \theta)\partial^\mu\mathcal{S}
            -\frac{1}{\sqrt{2}}(\bar \theta \gamma_5 \theta)\bar \theta  \slashed{\partial} \psi_L+\frac{1}{8} (\bar{\theta}\gamma_5\theta)^2 \Box \mathcal{S} \,,
\end{align}
with $\theta_L= \frac{1-\gamma_5}{2} \theta$. In the same way, the RH chiral superfield reads
\begin{align}
 \widehat{\mathcal{S}}_R=\mathcal{S}-\imath\sqrt{2}\bar{\theta}\psi_R -\imath(\bar{\theta}\theta_R)\mathcal{F}- \frac{\imath}{2} (\bar \theta \gamma_5 \gamma_\mu \theta)\partial^\mu\mathcal{S}
            -\frac{1}{\sqrt{2}}(\bar \theta \gamma_5 \theta)\bar \theta  \slashed{\partial} \psi_L+\frac{1}{8} (\bar{\theta}\gamma_5\theta)^2 \Box \mathcal{S} \,,
\end{align}
when the conditions $\psi_L=0$, $V_\mu=-\imath \partial_\mu \mathcal{S}$ and $\mathcal{N}=-\imath \mathcal{M} = -\imath \mathcal{F}$ are applied. Each of these chiral superfields forms an irreducible representation of the SUSY algebra and can be used to describe superfields that correspond to SM fermions and scalars. They can be rewritten in a simpler form
\begin{align}
 \widehat{\mathcal{S}}_L(\widehat x)=&\mathcal{S}(\widehat x) +\imath\sqrt{2}\bar{\theta}\psi_L(\widehat x) +\imath(\bar{\theta}\theta_L)\mathcal{F}(\widehat x)\,,\label{SuperScalarL}\\
 \widehat{\mathcal{S}}_R(\widehat x)=&\mathcal{S}(\widehat x^\dagger)-\imath\sqrt{2}\bar{\theta}\psi_R(\widehat x^\dagger) -\imath(\bar{\theta}\theta_R)\mathcal{F}(\widehat x^\dagger)\,, \label{SuperScalarR}
\end{align}
using the new variable $\widehat x_\mu=x_\mu + \frac{\imath}{2} \bar \theta \gamma_5 \gamma_\mu \theta$. Here, $\mathcal{F}$ is a non-dynamical field, named an auxiliary field, which is needed to ensure SUSY invariance.

Due to the second term in the right-hand side of eq.~(\ref{SUSYQ}), $\widehat \Psi$ and $\partial \widehat \Psi /\partial \bar \theta$ behave differently under a SUSY transformation. Supersymmetric covariant derivatives must be introduced in order to make the construction of a SUSY invariant Lagrangian easier. The condition that components of $D\widehat\Psi$ transform in the same way as those of $\widehat\psi$ imposes
\begin{equation}
 D=\frac{\partial}{\partial\overline\theta} - \imath \slashed \partial \theta\,,
\end{equation}
with the corresponding derivative defined as $\bar D= D^T C$,  which implies
\begin{equation}
 \bar D = - \frac{\partial}{\partial \theta} + \imath \bar \theta \slashed \partial\,.
\end{equation}
Left and right covariant derivatives can be defined too, through $D_L=P_L D$ and $D_R=P_R D$, giving
\begin{align}
 D_L=\frac{\partial}{\partial\widehat\theta_R} - \imath \slashed \partial \theta_R\,,\\
 D_R=\frac{\partial}{\partial\widehat\theta_L} - \imath \slashed \partial \theta_L\,.
\end{align}
These derivatives will be useful when expressing the superfield that contains the field strength.

But what is the representation of gauge fields? Starting with the scalar superfield from eq.~(\ref{SuperScalar}), the reality condition
\begin{equation}
 \widehat \Psi^\dagger = \widehat \Psi
\end{equation}
is applied in order to ensure that bosonic components are real and the fermionic components are  Majorana fields, making the field strength real. The next step is to apply the Wess-Zumino gauge condition, which sets $\mathcal{S}$, $\mathcal{M}$, $\mathcal{N}$ and $\psi$ to zero. Thus, the vector superfield reads
\begin{equation}
  \widehat{\Psi}_A=\frac{1}{2} (\bar \theta \gamma_5 \gamma_\mu \theta)V^\mu_A +\imath (\bar \theta \gamma_5 \theta) \bar \theta \lambda_A -\frac{1}{4} (\bar{\theta}\gamma_5\theta)^2 \mathcal{D}_A\,,\label{SuperVector}
\end{equation}
where the subscript $A$ runs over the gauge group generators $t_A\;(A=1\,,...\,,N)$ and $\mathcal{D}$ is an auxiliary field. From the vector superfield it is possible to define a curl superfield $\widehat W_A$\footnote{The curl superfield corresponds to the supersymmetric generalization of the field strength.}, which, in the Wess-Zumino gauge, reads 
\begin{equation}
\widehat W_A(\widehat x)=\lambda_{L_A}(\widehat x)+\frac{1}{2} \gamma^\mu \gamma^\nu F_{\mu\nu_A}(\widehat x)\theta_L-\imath\bar\theta\theta_L(\slashed D\lambda_R)_A-\imath\mathcal{D}_A(\widehat x)\theta_L\,,
\end{equation}
with $\slashed D_{AB}=\slashed \partial \delta_{AB} + \imath g (t_C^\dagger)_{AB} \slashed V_C$.

Equipped with the definition and representation of chiral superfields that can be used to describe chiral fermions and their superpartners, with vector superfields for the gauge mediators and even a curl superfield that contains a field strength, we can now move to the construction of a supersymmetric Lagrangian.

\subsubsection{The supersymmetric Lagrangian}

The SUSY transformation of the components of the superfields defined above leads to interesting results: the first one is that the $\mathcal{D}$ component of the scalar superfield transforms as a total derivative under SUSY transformation. Since all superfields, be them chiral or vector, are derived from this generic scalar superfield, the $D$-term of any product of superfields and their hermitian conjugates will only transform as a total derivative, making it a good candidate to write a SUSY Lagrangian. With the notation adopted, the $D$-term is given by $2\mathcal{D}$. From eq.~(\ref{SuperScalarL}), one verifies that LH chiral superfields only depend on $\widehat x$ and $\theta_L$. Thus the product of LH chiral superfields is a LH chiral superfield, whose $F$-term transforms as a total derivative, making it another candidate for a SUSY Lagrangian. Again, with the convention used, the $F$-term is defined by $-\mathcal{F}$.

These results will be taken into account when defining two potentials that will serve as building blocks to the supersymmetric Lagrangian. The first one is the Kähler potential, which, in a gauge theory, is defined as
\begin{equation}
 K=\widehat{\mathcal{S}}_L^\dagger e^{-2g t_A \widehat{\Psi}_A} \widehat{\mathcal{S}}_L\,,
\end{equation}
while the second is the superpotential noted $\widehat f$, a globally gauge-invariant product of no more than three LH chiral superfields to ensure renormalizability. Since the Hermitian conjugate of a LH chiral superfield has the form of a RH chiral superfield, the superpotential cannot contain Hermitian conjugates. This leads to its holomorphy, a property that suffices to establish the non-renormalization theorem~\cite{Seiberg:1993vc}. 

Starting from the Kähler potential, the superpotential, the vector and curl superfields, the supersymmetric Lagrangian can be defined as the sum of different contributions
\begin{equation}
 \mathcal{L}=\mathcal{L}_{GK}+\mathcal{L}_{\mathrm{gauge}}+\mathcal{L}_F+\mathcal{L}_{FI}\,,
\end{equation}
with $\mathcal{L}_{GK}$ the gauge kinetic terms, $\mathcal{L}_\mathrm{gauge}$ the kinetic and gauge interaction terms for fermions, $\mathcal{L}_F$ the superpotential contribution and $\mathcal{L}_{FI}$ the Fayet-Iliopoulos $D$-term which corresponds to the $D$-term from the vector superfield of an Abelian gauge group. These contributions are given by
\begin{align}
 \mathcal{L}_F&=\widehat f |_{-\bar \theta \theta_L} +h.c.\,,\\
 \mathcal{L}_{FI}&=\eta_a \mathcal{D}_a\,,\\
 \mathcal{L}_{GK}&=\frac{1}{2}\overline{\widehat{W}^C_A} \widehat{W}_A |_{-\bar \theta \theta_L}\,,\\
 \mathcal{L}_{\mathrm{gauge}}&=-2 K|_{(\bar \theta \theta)^2}\,,
\end{align}
where the choice of $-\bar \theta \theta_L$ instead of $\bar \theta \theta_L$ is purely conventional, $a$ runs over the different $\mathcal{U}(1)$ factor of the gauge group and $\eta_a$ are dimensionful coupling constants ($[\eta_a]=2$). The derivation $\partial / \partial \theta$ has been denoted by $|_\theta$.

Finally, the auxiliary fields $\mathcal{F}$ and $\mathcal{D}$ can be eliminated via their purely algebraic Euler-Lagrange equations
\begin{align}
 \mathcal{F}_i&=-\imath \left(\frac{\partial \widehat f}{\partial \widehat{\mathcal{S}}_i}\right)^\dagger_{\widehat{\mathcal{S}} = \mathcal{S}}\,,\\
 \mathcal{D}_A&=g\sum_i \mathcal{S}_i^\dagger t_A \mathcal{S}_i + \eta_A\,,
\end{align}
leading to the general master Lagrangian for supersymmetric theories.

As mentioned previously, the super-Poincaré algebra implies that $[Q_a\,,P^2]=0$, meaning that all components of a supermultiplet have the same mass if SUSY is unbroken. This means that the selectron, the scalar partner of the electron should have a mass of $0.51\;\mathrm{MeV}$~\cite{Beringer:1900zz}. However no scalar particle of this mass has been observed, the lower limit on the selectron  mass from negative searches being $107\;\mathrm{GeV}$ at a $95\%$ confidence level (CL)~\cite{Beringer:1900zz}. Thus, supersymmetry cannot be an exact symmetry of Nature and has to be broken.

\subsubsection{Supersymmetry breaking}

Since they  have proven very successful in particle physics as well as in condensed matter, spontaneous symmetry breaking mechanisms are very attractive. Using the elementary fields in supermultiplets, a spontaneous breaking of supersymmetry can be achieved if the F-term or the D-term develops a vev. The former has been proposed by Lochlainn O'Raifeartaigh in 1975~\cite{O'Raifeartaigh:1975pr}, while the latter was pointed out in 1974 by Pierre Fayet and Jean Iliopoulos~\cite{Fayet:1974jb}. However, building a phenomenologically viable model where this spontaneous symmetry breaking is realised explicitly has proven extremely difficult. As a consequence, most of the supersymmetric models studied nowadays consider that this breaking occurs in an ``hidden'' sector and is subsequently transmitted to our ``visible'' sector by a mediator, be it gravity (e.g. in minimal supergravity or mSUGRA~\cite{Chamseddine:1982jx, *Ohta:1982wn, *Barbieri:1982eh, *Hall:1983iz}), gauge interactions (in gauge mediated supersymmetry breaking or GMSB~\cite{Dine:1981gu, *Giudice:1998bp, *AlvarezGaume:1981wy, *Nappi:1982hm, *Dine:1994vc, *Dine:1993yw}) or an anomaly (in anomaly mediated supersymmetry breaking or AMSB~\cite{Randall:1998uk, *Giudice:1998xp, *Chacko:1999am}).

Since the exact dynamics behind supersymmetry breaking is unknown, its effects are parametrized by  new terms in the Lagrangian that explicitly break SUSY. However, this explicit breaking should be carefully introduced in order to retain some relations between tree-level couplings, not spoiling the improved ultra-violet behaviour of supersymmetric theories. This criterion is used to distinguish between hard and soft SUSY breaking operators. More precisely, SUSY breaking operators that do not introduce quadratical divergences are called soft while the others are known as hard. It is important to introduce only soft SUSY breaking operators in order to preserve the appeal of SUSY as a solution to the hierarchy problem by keeping couplings identical for particles belonging to the same supermultiplet. Marcus  Grisaru and Luciano Girardello studied and classified these operators in 1982~\cite{Girardello:1981wz}. As a result, the soft SUSY breaking Lagrangian generically contains the following operators
\begin{align}
 \mathcal{L}_{\mathrm{soft}}=&(\sum_i C_i \mathcal{S}_i + \sum_{i\,,j} B_{ij} \mu_{ij} \mathcal{S}_i \mathcal{S}_j + \sum_{i\,,j\,,k} A_{ijk} f_{ijk} \mathcal{S}_i \mathcal{S}_j \mathcal{S}_k + h.c.)\nonumber\\
     &- \sum_{i\,,j} \mathcal{S}_i^\dagger m^2_{ij} \mathcal{S}_j - \frac{1}{2} \sum_{A} M_A \bar{\lambda}_A \lambda_A -\frac{\imath}{2} \sum_A  M_A^\prime \bar{\lambda}_A \gamma_5 \lambda_A\,,
\end{align}
where the terms $\mu_{ij} \widehat{S}_i \widehat{S}_j$ and $f_{ijk} \widehat{S}_i \widehat{S}_j \widehat{S}_k$ appear in the superpotential. The various masses and coupling constants are generically complex and $M_A^\prime \bar{\lambda}_A \gamma_5 \lambda_A$ is CP-odd, introducing new sources of CP violation unless those are explicitly forbidden. The linear term $C_i \mathcal{S}_i$ only appears for singlet chiral superfields. The presence of singlet scalars introduces further restrictions that come from the possible quadratic divergences in tadpole graphs. Finally, in a theory free of gauge singlets, additional operators could be included like $\mathcal{S}_i \mathcal{S}_j \mathcal{S}_k^*$ or the mixing mass term between gauginos and fermions from chiral superfields in the adjoint representation of the gauge group.

After having briefly introduced the supersymmetric extension of the Poincaré algebra that relates bosons and fermions and described the method used to build supersymmetric models, let us move to specific SUSY models in the next sections.

\section{The Minimal Supersymmetric Standard Model\label{SecMSSM}}

The Minimal Supersymmetric Standard Model (MSSM) is the most minimal, viable SUSY extension of the Standard Model. It is based on the same gauge group $\mathrm{SU}(3)_c\times\mathrm{SU}(2)_L\times\mathrm{U}(1)_Y$ as the SM. Its field content, given in tables~\ref{MSSMgauge}
\begin{table}[t]
  \begin{center}
    \begin{tabular}{|c|c|c|c|}
     \hline
	Gauge group & Vector superfield & Spinor & Vector \\
     \hline
     \hline
	$U(1)_Y$ & $\widehat{B}$ & $\vphantom{\widetilde{b^{B^B}}}\widetilde{b}$ & $B_\mu$ \\
	$SU(2)_L$ & $\widehat{W}^i$ & $\widetilde{w}^i$ & $W^i_\mu$ \\
	$SU(3)_c$ & $\widehat{G}^\alpha$ & $\widetilde{g}^\alpha$ & $G^\alpha_\mu$ \\
     \hline
    \end{tabular}
    \caption[Vector superfields of the MSSM]{\label{MSSMgauge} Vector superfields of the MSSM. Tilded components are odd under R-parity and collectively known as gauginos.}
 \end{center}
\end{table}
and~\ref{MSSMmatter}
\begin{table}[t]
  \begin{center}
    \begin{tabular}{|c|c|c|c|c|c|}
     \hline
	Chiral superfield & Scalar & Spinor & $SU(3)_c$ & $SU(2)_L$ & $U(1)_Y$ \\
     \hline
     \hline
	$\widehat{Q}_i=\binom{\vphantom{\widetilde{b^{B}}}\widehat{u}_{L_i}}{\widehat{d}_{L_i}}$ & $\widetilde{Q}_i$ & $Q_i$ & $\mathbf{3}$ & $\mathbf{2}$ & $\phantom{-}\frac{1}{3}$ \\
	$\widehat{U}_i$ & $\widetilde{U}_i$ & $U_i$ & $\mathbf{3}^*$ & $\mathbf{1}$ & $-\frac{4}{3}$ \\
	$\widehat{D}_i$ & $\widetilde{D}_i$ & $D_i$ & $\mathbf{3}^*$ & $\mathbf{1}$ & $\phantom{-}\frac{2}{3}$ \\
     \hline
	$\widehat{L}_i=\binom{\vphantom{\widetilde{b^{B}}}\widehat{\nu}_{L_i}}{\widehat{\ell}_{L_i}}$ & $\widetilde{L}_i$ & $L_i$ & $\mathbf{1}$ & $\mathbf{2}$ & $-1$ \\
	$\widehat{E}_i$ & $\widetilde{E}_i$ & $E_i$ & $\mathbf{1}$ & $\mathbf{1}$ & $\phantom{-}2$ \\
     \hline
	$\widehat{H}_u=\binom{\vphantom{\widetilde{b^{B}}}\widehat{h}_{u}^+}{\widehat{h}_u^0}$ & $H_u$ & $\widetilde{H}_u$ & $\mathbf{1}$ & $\mathbf{2}$ & $\phantom{-}1$ \\
	$\widehat{H}_d=\binom{\widehat{h}_{d}^0}{\widehat{h}_d^-}$ & $H_d$ & $\widetilde{H}_d$ & $\mathbf{1}$ & $\mathbf{2}$ & $-1$ \\
     \hline
    \end{tabular}
    \caption[Chiral superfields of the MSSM]{\label{MSSMmatter} Chiral superfields of the MSSM and their gauge transformation properties, where $i$ is the generation index, running from $1$ to $3$. Tilded components are odd under R-parity and are known as squarks, sleptons and higgsinos, being the superpartners of quarks, leptons and Higgs bosons, respectively.}
 \end{center}
\end{table}
corresponds to the SM particles and their superpartners, the only exception being the Higgs sector, which contains two $\mathrm{SU}(2)$ doublets. The superpotential being a function of LH chiral superfields, RH fermions are introduced through their LH charge conjugates. For example, if the Dirac spinor of the electron is given by $e=e_L+e_R$, then the RH electron component is introduced via the Majorana spinor $E_1=(e_R)^C + e_R$. Thus, the selectron is given by $\widetilde{E}_1=\widetilde{e}_R^*$.

The Higgs sector is slightly more complicated in the MSSM for two reasons. First, being holomorphic, the superpotential does not contain conjugate supermultiplets. As a consequence, the use of $\widetilde{\phi}=\imath \sigma_2 \phi^*$ as in the SM is impossible and the $\mathbf{2}$  and $\mathbf{2}^*$ representations are no longer equivalent in SUSY. Second, the supersymmetrisation of the SM would add a fermionic partner to the SM Higgs doublet, a higgsino, with hypercharge $Y=1$.  The presence of this new charged fermion will spoil the cancellation of gauge anomalies that occurs in the SM, especially the Adler-Bell-Jackiw anomaly~\cite{Adler:1969gk, *Bell:1969ts} coming from triangular fermionic loops with an axial current. A solution to these issues lies in the introduction of a second Higgs chiral superfield with opposite hypercharge, which couples to the down-type superfields (this corresponds to a two Higgs doublet model (2HDM) of type II). The opposite hypercharge ensures anomaly cancellation while the Yukawa terms arise from the coupling to two different Higgs doublets, one that couples to the up-type fermions while the other one couples to down-type fermions~\cite{Fayet:1974pd, *Fayet:1976cr}.

In the early 1980's when the MSSM was formulated, a continuous $\mathrm{U}(1)$ symmetry was also introduced in SUSY models. After SUSY breaking, it led to a $\mathbb{Z}_2$ parity known as R-parity and defined by
\begin{equation}
 R=(-1)^{3(B-L)+2s}\,,
\end{equation}
where B and L are the baryonic and leptonic number, respectively, and $s$ is the spin. From a model building perspective, it could be the remnant of a broken continuous symmetry $\mathrm{U(1)_{R}}$, which would be present in unbroken SUSY models with conserved baryon and lepton numbers~\cite{Barbier:2004ez}. Moreover, from a phenomenological point of view, it has the desirable effect of forbidding B or L violating renormalizable operators in many SUSY models, while the non-renormalization theorem ensures that they do not reappear radiatively. This, in turn, ensures the proton stability, making sure that the model complies with the stringent experimental bounds, which give a proton mean life greater than $2.1\times10^{29}$~years~\cite{Beringer:1900zz} at $90\%$~CL.

Since the MSSM was built as an R-parity conserving model~\cite{Barbier:2004ez}, it is defined by the following superpotential
\begin{align}
\widehat{f}_\mathrm{MSSM}= \varepsilon_{ab} \left[
Y^{ij}_d \widehat{D}_i \widehat{Q}_j^b  \widehat{H}_d^a
              +Y^{ij}_{u}  \widehat{U}_i \widehat{Q}_j^a \widehat{H}_u^b 
              + Y^{ij}_e \widehat{E}_i \widehat{L}_j^b  \widehat{H}_d^a- \mu \widehat{H}_d^a \widehat{H}_u^b \right]\,,
\label{eq:SuperPotMSSM}
\end{align}
where $a\,,b$ are $\mathrm{SU}(2)$ indices, $i\,,j$ runs over the three generations and the colour indices have been omitted. Here and in the rest of this chapter, we use the convention that repeated indices are summed over. Since all $\mathrm{SU}(2)$ doublets belong to the $\mathbf{2}$ representation, the antisymmetric tensor $\varepsilon_{ab}$ is necessary to combine doublets in an invariant way. Keeping couplings identical for particles belonging to the same supermultiplet,  supersymmetry is broken via the soft SUSY breaking part of the MSSM Lagrangian, which reads
\begin{align}
 -\mathcal{L}_\mathrm{MSSM}^\mathrm{soft} =& m_{H_u}^2 H_u^\dagger H_u+m_{H_d}^2 H_d^\dagger H_d+(m_L^2)_{ij}\widetilde{L}_i^\dagger\widetilde{L}_j +(m_{E}^2)_{ij}\widetilde{E}_{i}^{\dagger}\widetilde{E}_{j} \nonumber \\
&+(m_Q^2)_{ij}\widetilde{Q}_i^\dagger\widetilde{Q}_j+(m_{U}^2)_{ij}\widetilde{U}_{i}^{\dagger}\widetilde{U}_{j}+(m_{D}^2)_{ij}\widetilde{D}_{i}^{\dagger}\widetilde{D}_{j} \nonumber \\
&+B \mu \varepsilon_{ab} H_d^a H_u^b + h.c.\nonumber \\
&-\varepsilon_{ab}\left[(A_uY_u)_{ij}\widetilde{U}_i \widetilde{Q}_j^a H_u^b + (A_dY_d)_{ij}\widetilde{D}_i \widetilde{Q}_j^b  H_d^a+ (A_eY_e)_{ij}\widetilde{E}_i \widetilde{L}_j^b  H_d^a+h.c.\right]\nonumber \\
&+M_1\overline{\tilde{b}}\tilde{b}+M_2\overline{\widetilde{w}}_i\widetilde{w}_i+M_3\overline{\widetilde{g}}_\alpha\widetilde{g}_\alpha + h.c.\,,\label{softMSSM}
\end{align}
where the scalar mass squared matrices are $3\times 3$ Hermitian, $M_i$ are complex gaugino mass parameters, the trilinear couplings $A$ are $3\times 3$ complex matrices and $m_{H_u}^2$, $m_{H_u}^2$ and $B$ are real parameters.  

After EWSB, the neutral components of the two Higgs doublets develop a vev. Noting $v_u$ and $v_d$ the vev of $H_u$ and $H_d$, they verify $v=\sqrt{v_u^2 + v_d^2}$, where $v$ is the SM Higgs vev. Then the vev $v_u$ and $v_d$ are related by
\begin{equation}
 \tan \beta = \frac{v_u}{v_d}\,.\label{tanbeta}
\end{equation}
Let us now discuss the low-energy spectrum of the MSSM, beginning with the Higgs sector where mixing occurs between different gauge eigenstates. Under the assumption of no CP violation in the Higgs sector, this results in two CP-even and one CP-odd neutral Higgs bosons and two charged scalars. The mixing of the neutral higgsino components with the neutral winos and binos gives rise to four neutralinos, which are Majorana fermions. The charged winos mix with the charged components of higgsinos leading to four charginos. This is summarised in table~\ref{mixingMSSM}.
\begin{table}[t]
  \begin{center}
    \begin{tabular}{|c|c|c|c|}
     \hline
	Sector & Spin & Gauge eigenstates & Mass eigenstates \\
     \hline
     \hline
	Higgs & 0 & $\vphantom{\widetilde{b^{B}}}{h}_{u}^+\,,{h}_{u}^0\,,{h}_{d}^-\,,{h}_{d}^0$ & $h\,,H\,,A\,,H^\pm$ \\
     \hline
	Neutralinos & 1/2 & $\vphantom{\widetilde{b^{B^B}}}\widetilde{W}^3\,,\widetilde{B}\,,\widetilde{h}_{u}^0\,,\widetilde{h}_{d}^0$ & $\widetilde{\chi}^0_1\,,\widetilde{\chi}^0_2\,,\widetilde{\chi}^0_3\,,\widetilde{\chi}^0_4$ \\
     \hline
	Charginos & 1/2 & $\vphantom{\widetilde{b^{B^B}}}\widetilde{W}^1\,,\widetilde{W}^2\,,\widetilde{h}_{u}^+\,,\widetilde{h}_{d}^-$ & $\widetilde{\chi}^\pm_1\,,\widetilde{\chi}^\pm_2$ \\
     \hline
    \end{tabular}
    \caption[Higgs bosons, charginos and neutralinos in the MSSM]{\label{mixingMSSM}Higgs bosons, charginos and neutralinos in the MSSM after EWSB. $h$ and $H$ are CP-even, while $A$ is CP-odd.}
 \end{center}
\end{table}
Since $\mathrm{SU}(3)$ remains unbroken, the gluino octet cannot mix with other fermions. It is a mass eigenstate with a tree-level mass $|M_3|$. We mentioned in the first section that R-parity conservation results in the presence of a dark matter candidate in the spectrum. In the MSSM, this candidate is very often the lightest neutralino $\widetilde{\chi}^0_1$. However, some regions of the parameter space will lead to a charged LSP, which excludes them since the DM candidate can only interact weakly.

In the SM, flavour changing neutral currents (FCNC) are forbidden at tree-level and suppressed at loop-level due to the GIM mechanism, in agreement with the experimental measurements of the $K_L$--$K_S$ mass difference or the decay rates of processes like $B_s \rightarrow \mu^+ \mu^-$ or $\mu \rightarrow e \gamma$. However, in the MSSM, the squark and slepton mass matrices are in general complex, with potentially large off-diagonal entries. Besides, the decompositions that give a diagonal quark or lepton mass matrix do not simultaneously make the squark or slepton mass matrix diagonal. This can results in large FCNC, which leads to the SUSY flavour problem. This issue can be solved in three different ways. First, the squarks and sleptons with the same quantum number can be degenerate, having the same mass. In this case, the loop function is the same for every flavour and the sum over every flavour make the off-diagonal terms vanish due to the unitarity of the mixing matrices. Second, the squark and quark (and similarly the slepton and lepton) mass matrices can be aligned, which means that they are decomposed by the same unitary transformation~\cite{Nir:1993mx}. Third, the squarks and sleptons can be very heavy, decoupling from the low-scale phenomenology, which suppresses their contributions. But the soft SUSY breaking parameters are, in general, complex, introducing new sources of CP violation. Nevertheless, those are experimentally very constrained, by precise measurement in the kaon sector for example, which leads to the SUSY CP problem. A common solution to both the SUSY flavour and CP problems is to assume that the squarks and sleptons masses are universal and real, while the trilinear couplings are proportional to the corresponding Yukawa matrices. A general analysis of these issues is given in~\cite{Gabbiani:1996hi}, while recent developments can be found in~\cite{Endo:2013bba, *Mahmoudi:2012un, Buchmueller:2012hv, Arbey:2011aa}.

This assumption of real and universal squarks and sleptons masses, with trilinear couplings proportional to the corresponding Yukawa matrices is naturally satisfied by gravity mediated SUSY breaking mechanism. Moreover, specific SUSY breaking mechanisms, like mSUGRA, dramatically reduces the number of free parameters\footnote{The MSSM has 124 paramaters, most of them coming from the soft SUSY breaking terms. Choosing a specific SUSY breaking mechanism will strongly reduce the number of free parameters in the soft SUSY breaking terms, thus allowing to conduct phenomenological analyses}.
In the following phenomenological studies, we will consider two of these frameworks. The first one is the constrained MSSM or CMSSM, which is inspired by a minimal gravity mediated SUSY breaking scenario. It has only five universal parameters at the GUT scale:
\begin{equation}
 m_0\,,m_{1/2}\,,A_0\,,\tan\beta\,,\mathrm{sign}(\mu)\,,
\end{equation}
where
\begin{align}
 &m_0^2=m_{H_u}^2=m_{H_d}^2\,,\label{CMSSM1}\\
 &m_0^2 \mathbf{1}= m_Q^2 = m_U^2= m_D^2= m_L^2 = m_E^2\,,\label{CMSSM2}\\
 &m_{1/2}=M_1=M_2=M_3\label{CMSSM3}\\
 &A_0 \mathbf{1}=A_u=A_d=A_e\,.\label{CMSSM4}
\end{align}
The second framework we considered is the non-universal Higgs masses (NUHM) scenario were the constraints on the soft SUSY breaking Higgs masses are relaxed according to
\begin{equation}
 m_0^2\neq m_{H_u}^2\neq m_{H_d}^2\,,
\end{equation}
leading to a scenario with six parameters
\begin{equation}
 m_0\,,m_{1/2}\,,m_{A}\,,A_0\,,\tan\beta\,,\mu\,.
\end{equation}
Starting with these conditions at the GUT scale, it is then necessary to run the various parameters down to the scale at which the phenomenological study is conducted via the renormalization group equations.

However, the MSSM is also constrained by many experimental measurement and observations. First, the recent discovery of a Higgs-like boson and the measurement of its mass, production rates and branching fractions puts stringent constraints on the parameter space~\cite{Djouadi:2013vqa, *Kadastik:2011aa, *Draper:2011aa, *Heinemeyer:2011aa, *Carena:2011aa, Ellis:2012aa, Arbey:2011ab}. The searches for heavier or charged Higgs bosons can also put additional limits~\cite{Arbey:2013jla}. The extra degrees of freedom can also lead to dangerous modifications of the electroweak precision tests~\cite{Heinemeyer:2004gx, *Zheng:2013bja}. Obviously, the non-observation of sparticles in direct searches also strongly constrains the available parameter space, putting lower bounds on their masses and upper bounds on their production rates~\cite{Moriond2012}. We mentioned above that, in the MSSM, the LSP is a candidate for dark matter if it is electrically neutral. The recent measurement of the dark matter relic density by the Planck collaboration~\cite{Ade:2013lta} limits the DM production in the early Universe, which in turn restricts the MSSM parameter space, especially when limits from direct searches are included~\cite{Arbey:2011aa, Buchmueller:2012hv, Han:2013gba, Ellis:2012aa}.

The MSSM is only the most minimal SUSY extension of the SM and suffers from some issues in itself. We will describe them and see how a specific extension of the MSSM addresses them.

\section{The Next-to-Minimal Supersymmetric Standard Model}

Although very appealing, the MSSM still have some caveats, among them the so-called $\mu$ problem~\cite{Kim:1983dt}, which comes from the fact that the dimensionful parameter $\mu$ that appears in the MSSM superpotential~(\ref{eq:SuperPotMSSM}) is not protected by any symmetry. As such, its natural values would be zero or close to the GUT or Planck scale, where new Physics would appear in the theory. However, EWSB and phenomenological constraints, like the lower bound on the chargino mass, require that $\mu$ should be of the order of the soft SUSY breaking terms, in the interval $100\;\mathrm{GeV}\leq\mu\leq M_\mathrm{SUSY}$~\cite{Ellwanger:2009dp}. The lower bound comes from negative searches of charginos, which give $m_{\widetilde{\chi}^\pm_1} \geq 94\mathrm{GeV}$ at $95\%$ CL~\cite{Beringer:1900zz}. Since~\cite{baer2006weak}
\begin{equation}
 m_{\widetilde{\chi}^\pm_1}^2+m_{\widetilde{\chi}^\pm_2}^2=\mu^2+M^2_2+2m_W^2\,,
\end{equation}
the lower bound on the lightest chargino mass can be directly translated into a lower bound on $\mu$, for arbitrary values of the other parameters. Besides, one of the EWSB symmetry breaking condition in the MSSM is~\cite{baer2006weak}
\begin{equation}
 \mu^2=\frac{m_{H_d}^2-m_{H_u}^2 \tan^2\beta}{\tan^2\beta-1} -\frac{m_Z^2}{2}\,,
\end{equation}
which implies that $\mu$ is smaller than the SUSY breaking scale $M_\mathrm{SUSY}$. Moreover, at tree-level in the MSSM, the Higgs boson mass is smaller than the mass of the $Z^0$ boson~\cite{baer2006weak}. This can only be reconciled with the mass measured by the CMS and ATLAS collaborations~\cite{ATLAS-CONF-2013-014, *CMS-PAS-HIG-13-002} at the price of large radiative corrections. But this reintroduces some fine-tuning in the MSSM, leading to the little fine tuning problem~\cite{Chankowski:1997zh, *Barbieri:1998uv, *Kane:1998im, *Giusti:1998gz}. 

A simple and elegant solution to the $\mu$ problem lies in the introduction of a singlet chiral superfield $\widehat{S}$, whose scalar component $S$ takes a vev induced by SUSY breaking and generates an effective $\mu$ term. This leads to the  Next-to-Minimal Supersymmetric Standard Model (NMSSM), which has been reviewed in~\cite{Ellwanger:2009dp} and was shown to reduce the amount of fine-tuning in the Higgs sector~\cite{Ellwanger:2011mu, *Kang:2012sy, *Ross:2012nr, *King:2012tr, *Barbieri:2013hxa, Cao:2012fz}. We will consider here a simple NMSSM formulation with the scale invariant superpotential
\begin{align}
\widehat{f}_\mathrm{NMSSM}= \varepsilon_{ab} \left[
Y^{ij}_d \widehat{D}_i \widehat{Q}_j^b  \widehat{H}_d^a
              +Y^{ij}_{u}  \widehat{U}_i \widehat{Q}_j^a \widehat{H}_u^b 
              + Y^{ij}_e \widehat{E}_i \widehat{L}_j^b  \widehat{H}_d^a- \lambda \widehat{S} \widehat{H}_d^a \widehat{H}_u^b \right] -\frac{\kappa}{3}\widehat{S}^3\,,
\label{eq:SuperPotNMSSM}
\end{align}
where $\lambda$ and $\kappa$ are dimensionless couplings. This corresponds to a model where the Lagrangian has a $\mathbb{Z}_3$ symmetry. The soft SUSY breaking Lagrangian are thus given by
\begin{align}
 -\mathcal{L}_\mathrm{NMSSM}^\mathrm{soft} =& m_{H_u}^2 H_u^\dagger H_u+m_{H_d}^2 H_d^\dagger H_d+ m_S^2 S^\dagger S +(m_L^2)_{ij}\widetilde{L}_i^\dagger\widetilde{L}_j +(m_{E}^2)_{ij}\widetilde{E}_{i}^{\dagger}\widetilde{E}_{j} \nonumber \\
&+(m_Q^2)_{ij}\widetilde{Q}_i^\dagger\widetilde{Q}_j+(m_{U}^2)_{ij}\widetilde{U}_{i}^{\dagger}\widetilde{U}_{j}+(m_{D}^2)_{ij}\widetilde{D}_{i}^{\dagger}\widetilde{D}_{j} \nonumber \\
&+\left(\lambda A_\lambda \varepsilon_{ab} H_d^a H_u^b S +\frac{\kappa}{3} A_\kappa S^3 + h.c.\right)\nonumber \\
&-\varepsilon_{ab}\left[(A_uY^u)_{ij}\widetilde{U}_i \widetilde{Q}_j^a H_u^b + (A_dY^d)_{ij}\widetilde{D}_i \widetilde{Q}_j^b  H_d^a+ (A_eY^e)_{ij}\widetilde{E}_i \widetilde{L}_j^b  H_d^a+h.c.\right]\nonumber \\
&+M_1\overline{\tilde{b}}\tilde{b}+M_2\overline{\widetilde{w}}_i\widetilde{w}_i+M_3\overline{\widetilde{g}}_\alpha\widetilde{g}_\alpha + h.c.\,,\label{softNMSSM}
\end{align}
with $A_\lambda$ and $A_\kappa$ the new trilinear couplings associated with the introduction of the singlet superfield $\widehat{S}$. When the singlet scalar develops a vev $s$, it will generate an effective $\mu$ term
\begin{equation}
 \mu_{\mathrm{eff}}=\lambda s\,,
\end{equation}
which will naturally be of the order of the SUSY breaking scale, since this is the scale of the vev $s$\footnote{We recall that for $A_\kappa^2 \geq 9 m_S^2$, the absolute minimum of $s$ is $s=\frac{1}{4\kappa} (-A_\kappa + \sqrt{A_\kappa^2-8m_S^2})$~\cite{Ellwanger:2009dp}.}. Thus it will provide a solution to the $\mu$ problem.

Another advantage of the scale invariant superpotential of eq.~(\ref{eq:SuperPotNMSSM}) is that it avoids the tadpole problem because no term linear in $\widehat{S}$ is present. Indeed, a singlet superfield can usually couple to the heavy fields that are present at the GUT or Planck scale, generating very large radiative corrections to terms linear in $\widehat{S}$ or $S$ in the superpotential or soft SUSY breaking Lagrangian, respectively. Nevertheless, if the heavy sector is invariant with respect to a discrete symmetry under which $\widehat{S}$ transforms, then these large radiative corrections are absent as long as the symmetry remains unbroken. Otherwise, they can reappear but at the scale of the discrete symmetry breaking. Thus having a conserved $\mathbb{Z}_3$, as we consider, prevents the tadpole problem. However, the introduction of a discrete symmetry can generate another problem: when a field charged under the symmetry takes a vev, domains with different vacua can arise creating domain walls, large anisotropies in the cosmic microwave background and spoiling nucleosynthesis~\cite{Ham:1996mi}. Finding a common solution to the tadpole and domain problem can be very difficult but solutions have been suggested making use of additional symmetries that would radiatively generate strongly suppressed $\mathbb{Z}_3$-symmetry breaking terms~\cite{Abel:1996cr, *Panagiotakopoulos:1998yw, *Panagiotakopoulos:1999ah, *Dedes:2000jp, *Hall:2012mx}.

Since the NMSSM contains an extra chiral superfield, its particle spectrum is an extended version of the MSSM one. The superfield being a singlet, it will only add neutral fields leading to 3 CP-even Higgs bosons $H_{1\,,2\,,3}$, 2 CP-odd  Higgs bosons $A_{1\,,2}$, and five neutralinos $\widetilde{\chi}^0_{1\,,...\,,5}$. However, the phenomenology is quite different from the MSSM since the dark matter candidate can be mostly singlino or have new annihilation channels via the extra Higgs bosons~\cite{Belanger:2005kh, *Gunion:2005rw, *Cerdeno:2007sn, *Kraml:2008zr, *Vasquez:2012hn}. The latter will also modify the behaviour of the Higgs sector, reducing the couplings of the SM-like Higgs boson or making the latter be $H_2$~\cite{Ellwanger:2011aa, *Belanger:2012tt, Cao:2012fz}. Moreover, the addition of extra degrees of freedom in the Higgs sector reduces the fine-tuning of the model. Finally, it is also possible to have a light CP-odd Higgs boson, which would then impact flavour Physics, especially $b$ Physics if $\tan \beta$ is quite high ($10$ and more).\\

In this chapter, we have exposed the basic ideas behind supersymmetry and why it is an attractive framework to extend the Standard Model. We have then introduced two of these extensions, namely the MSSM and in the NMSSM. Unfortunately, as the Standard Model, they contain massless neutrinos. We have presented the advantages of the inverse seesaw in chapter~\ref{chap3} and will discuss in the next chapter how it can be embedded in the MSSM or the NMSSM, leading to  models that address at the same time the issue of neutrino mass, the hierarchy problem and give a dark matter candidate.

\chapter{Phenomenology of supersymmetric inverse seesaw models\label{CPoddLFV}}

The simplest supersymmetrization of the SM cannot account for massive neutrinos. However, strong experimental evidence (see Chapter~\ref{ChapNuExp}) motivate the introduction of neutrino mass generation mechanism in supersymmetric models. The inverse seesaw is an attractive mechanism since its scale is naturally low, around the TeV. Hence, all the new Physics, coming from supersymmetry and the inverse seesaw mechanism, could lie at the TeV scale, leading to a very appealing phenomenology in neutrinoless double beta decay~\cite{Awasthi:2013we}, the Higgs sector~\cite{Abada:2010ym, Gogoladze:2012jp, *Bandyopadhyay:2012px, BhupalDev:2012zg, Banerjee:2013fga}, lepton flavour violating processes~\cite{Deppisch:2004fa, Hirsch:2009ra, Abada:2011hm, Abada:2012cq}, at the LHC or at a future linear collider~\cite{Mondal:2012jv, *Das:2012ze}, when the right-handed sneutrino is a dark matter candidate~\cite{Cerdeno:2011qv, *An:2011uq, *BhupalDev:2012ru, *DeRomeri:2012qd, Banerjee:2013fga} or in a leptogenesis scenario based on SUSY-GUT in $\mathrm{SO}(10)$~\cite{Blanchet:2010kw} for instance. In this chapter, we will discuss different phenomenological consequences of embedding the inverse seesaw in supersymmetric extensions of the SM. Among them, we will present the effect of the embedding the inverse seesaw in the MSSM on lepton flavour violating observables~\cite{Abada:2011hm, Abada:2012cq}. We will also address the impact of the inverse seesaw on the NMSSM with an emphasis on invisible Higgs decays (this analysis~\cite{Abada:2010ym} was conducted prior to the 2012 LHC results). 

\section{Charged lepton flavour violating observables}

Neutrino oscillations have provided indisputable evidence for flavour violation in the neutral
lepton sector. In the absence of any fundamental principle that prevents charged lepton flavour 
violation, one expects that extensions of the Standard Model accommodating neutrino masses and
mixings should also allow for lepton flavour violation in the charged lepton sector. 
Indeed, the additional new flavour dynamics and new field content present in many extensions of the 
SM may provide contributions to charged LFV  processes such as radiative (e.g. $\mu\to e\gamma$) 
and three-body lepton decays (for instance $\tau\to \mu\mu\mu$). 
These decays generally arise from higher order processes,
and so their branching ratios are expected to be small, making them difficult to observe. Since in the SM these signals are strongly suppressed,  any cLFV 
signal would provide clear evidence for new physics: mixing in the lepton sector and probably the presence of new particles, 
possibly shedding light  on  the origin of neutrino mass generation. 

There are a large number of
facilities~\cite{Bertl:2006up,Adam:2013mnn,Hayasaka:2010np,Hayasaka:2007vc,*PhysRevLett.104.021802,*PhysRevD.81.111101},
dedicated to the search of processes such as rare radiative decays,
3-body decays and muon-electron conversion in nuclei. Likewise,
rare leptonic and semi-leptonic meson decays also offer a rich testing
ground to experimentally probe cLFV. These low-energy searches are complementary to the LHC which, in
addition to directly searching for new physics states, also allows to
study numerous signals of cLFV at high-energy. One can have sizeable widths for
processes like $\chi_2^0 \to \chi_1^0 \ell_i^\pm \ell_j^\mp$,
flavoured slepton mass splittings (especially between the first and
second generation of left-handed sleptons) and finally the appearance
of new edges in same-flavour dilepton mass distributions~\cite{Abada:2010kj}.  Assuming a
unique source of LFV, namely neutrino mass generation, the interplay of low-
and high-energy LFV observables can strengthen or disfavour
the underlying model of new physics.  Illustrative examples of the
potential of this interplay can be found for instance
in~\cite{Abada:2010kj,Esteves:2010si,Abada:2011mg,Abada:2012re}.
However, there are other avenues that can be explored in this quest to
disentangle the underlying mechanism of neutrino mass generation, at
the origin of lepton flavour violation: this approach is based upon
exploring the correlation (or lack thereof) between different low-energy cLFV observables. The distinctive features of
the underlying model will be manifest in the nature and specific
hierarchy of the different contributions. For instance, in SUSY models
where $\gamma$-penguins provide the dominant contribution to radiative
and 3-body cLFV decays, one expects a strict correlation between
$\mathcal{B}(\mu \to e \gamma)$ and CR($\mu - e$, N). This is the case of
constrained Minimal Supersymmetry Standard Model (CMSSM) based
scenarios where additional lepton flavour violating sources have been
introduced. Deviations from strict universality (as is the case of
non-universal Higgs masses, NUHM), where for example Higgs-mediated
penguins might play a significant role in $\mu-e$ conversion, break
this strict correlation~\cite{Arganda:2007jw}.

The experiments looking for leptonic radiative decays are quite different depending on the lepton in the initial state. If it is a muon, the only decay is $\mu \rightarrow e \gamma$ which is studied by dedicated experiments such as MEG~\cite{Adam:2013mnn}. This collaboration has plans for an upgrade that would improve the sensitivity to $\mathcal{O}(10^{-14})$~\cite{Baldini:2013ke}. Radiative $\tau$ decays are studied at $B$ factories, which are also  $\tau$ factories, since the production cross-sections are very close at the $\Upsilon(4s)$ resonance. The current upper limits on $\mathcal{B}(\tau \to \mu \gamma)$ and $\mathcal{B}(\tau \to e \gamma)$ are given by the BaBar experiment, together with expected sensitivities at the future generation of $B$ factories, e.g. Belle II~\cite{Akeroyd:2004mj}.

For the same reasons 3-body decays of the $\tau$ lepton are also usually searched for at $B$ factories. The current upper limits come from the Belle experiment~\cite{Hayasaka:2010np,collaboration:2010ipa} because of its larger data sample compared to BaBar. Since these observables are currently not limited by the background, significant improvements are expected at Belle II~\cite{Akeroyd:2004mj}. The decay $\mu \to 3e$  has been investigated by the SINDRUM experiment~\cite{Bellgardt:1987du} and the future experiment Mu3e at PSI could reach a sensitivity of $10^{-15}$ (after upgrades $10^{-16}$)~\cite{Blondel:2013ia}.

Neutrinoless conversion in muonic atoms has also been studied for different nuclei  by the SINDRUM II collaboration~\cite{Dohmen:1993mp,Bertl:2006up} which has set the current upper limits. In the future, the sensitivity is expected to be greatly improved by different projects. Mu2e~\cite{Glenzinski:2010zz,Carey:2008zz} is a future experiment at Fermilab with expected sensitivities of respectively $10^{-17}$ (phase I) and $10^{-18}$ (phase II with Project X). On the other hand, the first experiment that could be built at J-PARC is DeeMe~\cite{Aoki:2010zz} with an expected sensitivity of $2\times 10^{-14}$ in 2015. Then COMET~\cite{Cui:2009zz} and 
PRISM/PRIME~\cite{mori1996experimental} would come with sensitivities of $10^{-15}$ (COMET Phase I, 2017), $10^{-17}$ (COMET phase II, 2021) and $10^{-18}$ (PRISM/PRIME) for a titanium nucleus. Current upper limits and future sensitivities for various cLFV observables are listed in table~\ref{cLFVexp}.

\begin{table}[t]
\begin{center}
 \begin{tabular}{|c|c|c|}
  \hline
    LFV Process & Present Bound & Future Sensitivity \\
  \hline
    $\tau \rightarrow \mu \mu \mu$ & $2.1\times10^{-8}$~\cite{Hayasaka:2010np} & $8.2 \times 10^{-10}$~\cite{O'Leary:2010af} \\
    $\tau^- \rightarrow e^- \mu^+ \mu^-$ &  $2.7\times10^{-8}$~\cite{Hayasaka:2010np} & $\sim 10^{-10}$~\cite{O'Leary:2010af} \\
    $\tau \rightarrow e e e$ & $2.7\times10^{-8}$~\cite{Hayasaka:2010np} &  $2.3 \times 10^{-10}$~\cite{O'Leary:2010af} \\
    $\mu \rightarrow e e e$ &  $1.0 \times 10^{-12}$~\cite{Bellgardt:1987du} &  $\sim10^{-16}$~\cite{Blondel:2013ia} \\
    $\tau \rightarrow \mu \eta$ & $2.3\times 10^{-8}$~\cite{collaboration:2010ipa} & $\sim 10^{-10}$~\cite{O'Leary:2010af}  \\
    $\tau \rightarrow \mu \eta^\prime$ & $3.8\times 10^{-8}$~\cite{collaboration:2010ipa} & $\sim 10^{-10}$~\cite{O'Leary:2010af} \\
    $\tau \rightarrow \mu \pi^{0}$ & $2.2\times 10^{-8}$~\cite{collaboration:2010ipa} & $\sim 10^{-10}$~\cite{O'Leary:2010af}  \\
    $B^{0}_{d} \rightarrow \mu \tau$ & $2.2\times 10^{-5}$~\cite{Aubert:2008cu} &  \\
    $B^{0}_{d} \rightarrow e \mu$ & $6.4\times 10^{-8}$~\cite{Aaltonen:2009vr} & $1.6\times 10^{-8}$~\cite{Bonivento:1028132}  \\
    $B^{0}_{s} \rightarrow e \mu$ & $2.0\times 10^{-7}$~\cite{Aaltonen:2009vr} & $6.5\times 10^{-8}$~\cite{Bonivento:1028132}  \\
    $\mu^-, \mathrm{Ti} \rightarrow e^-, \mathrm{Ti}$ &  $4.3\times 10^{-12}$~\cite{Dohmen:1993mp} & $\sim10^{-18}$~\cite{mori1996experimental} \\
    $\mu^-, \mathrm{Au} \rightarrow e^-, \mathrm{Au}$ & $7\times 10^{-13}$~\cite{Bertl:2006up} & \\
  \hline
 \end{tabular}
\end{center}
  \caption[Experimental limits on cLFV branching ratios]{Current experimental bounds and future sensitivities for some Higgs and $Z^0$-mediated cLFV observables.}\label{cLFVexp}
\end{table}

\section{Embedding the inverse seesaw in the MSSM}

While minimal extensions of the Standard Model can easily accommodate lepton flavour violation in the neutral lepton
sector (i.e. neutrino oscillations), the contributions of these models
to cLFV observables are in general extremely small\footnote{However, in specific scenarios, these contributions could be strongly enhanced~\cite{Ilakovac:1994kj, Cely:2012bz, Alonso:2012ji}.}. On the other hand, when such models - for example,
the seesaw in its different realisations - are embedded within a
larger framework, one can expect large contributions to cLFV
observables, well within experimental reach.  This is the case of
supersymmetric versions of the seesaw mechanism, which, in addition to
solving many theoretical and phenomenological issues, such
as the hierarchy problem, gauge coupling unification and dark matter,
can also account for neutrino masses and mixings.

However, one of the most upsetting caveats of these scenarios is that they prove to be extremely hard to test, and thus can
neither be confirmed nor excluded. This is due to the fact that, in order
to have Yukawa couplings sufficiently large to account
for measurable cLFV branching ratios, the typical scale of the extra
particles is in general very high, potentially close to the gauge
coupling unification scale, thus suppressing new contributions to cLFV processes.

This can be avoided if one simultaneously succeeds in having TeV-scale mediators, while preserving the possibility of large Yukawa couplings. 
From an effective theory point of view, this is equivalent to the decoupling of the coefficients associated with the dimension-five operator at the origin of neutrino masses and dimension-six operators, like the the four-fermion operators. In other words, the smallness of the light neutrino masses will be independant  from the amount of flavour violation. For instance, this is possible in the type II seesaw~\cite{Magg:1980ut, *Schechter:1980gr, *Wetterich:1981bx, *Lazarides:1980nt, *Mohapatra:1980yp}, as well as in the inverse seesaw~\cite{Mohapatra:1986aw, *Mohapatra:1986bd, *Bernabeu1987303} and their SUSY realisations~\cite{Rossi:2002zb, *Joaquim:2006uz, *Joaquim:2006mn, *Hirsch:2008gh, Deppisch:2004fa, Bazzocchi:2009kc, Dev:2009aw}.

\subsubsection{The model}

The inverse seesaw can be embedded in the Minimal Supersymmetric
extension of the SM by the addition of two extra gauge singlet
superfields, with opposite lepton numbers. When compared to other SUSY seesaw realisations, cLFV observables are 
enhanced in this framework, 
and such an enhancement can be attributed to 
large neutrino Yukawa couplings  ($Y_\nu \sim O(1)$), compatible with a seesaw 
scale $M_\mathrm{seesaw}$ close to the  
electroweak one, thus within LHC reach (see Chapter~\ref{chap3}). In what follows, we considered three pairs of singlet superfields, $\widehat{\nu}^C_i$ and
$\widehat{X}_i$ ($i=1,2,3$)\footnote{We use the notation:
  $\widetilde{\nu}^C=\widetilde{\nu}_R^* $.}  with lepton numbers
assigned to be $-1$ and $+1$, respectively, which are added to the
superfield content of the model~\cite{Abada:2011hm}. We nevertheless recall that neutrino data can be
successfully accommodated with only one generation of
$\widehat{\nu}^C$ and $\widehat{X}$~\cite{Hirsch:2009ra}. The SUSY inverse seesaw model is defined by the
following superpotential
\begin{equation}
\widehat f= \widehat f_\mathrm{MSSM} + \varepsilon_{ab} Y^{ij}_\nu \widehat{\nu}^C_i \widehat{L}^a_j \widehat{H}_u^b+M_{R_{ij}}\widehat{\nu}^C_i\widehat{X}_j+
\frac{1}{2}\mu_{X_{ij}}\widehat{X}_i\widehat{X}_j\,,
\label{eq:SuperPotMSSMISS}
\end{equation}
where $\widehat f_\mathrm{MSSM}$ is the MSSM superpotential in eq.~(\ref{eq:SuperPotMSSM}) and $i,j = 1,2,3$ are generation indices. The ``Dirac''-type right-handed neutrino mass term
$M_{R_{ij}}$ conserves lepton number, while the ``Majorana'' mass term
$\mu_{X_{ij}}$ violates it by two units. Since $M_{R_{ij}}$
conserves lepton number, in the limit $\mu_{X_{ij}}\rightarrow 0$,
lepton number conservation can be restored, making the smallness of $\mu_X$ natural in the sense of 't Hooft~\cite{'tHooft:1979bh}.  

The soft SUSY breaking Lagrangian is given by
\begin{align}
-\mathcal{L}^\mathrm{soft}&=-\mathcal{L}_\mathrm{MSSM}^\mathrm{soft} 
         +   \widetilde\nu^{C}_i m^2_{\widetilde \nu^C_{ij}}\widetilde\nu^{C*}_j
         + \widetilde X^{\dagger}_i m^2_{X_{ij}} \widetilde X_j
     + (A_{\nu}^{ij} Y_\nu^{ij} \varepsilon_{ab}
                 \widetilde\nu^C_i \widetilde L^a_j H_u^b +
                B_{M_R}^{ij} M_{R_{ij}}\widetilde\nu^C_i \widetilde X_j \nonumber\\
      &+\frac{1}{2}B_{\mu_X}^{ij} \mu_{X_{ij}} \widetilde X_i \widetilde X_j
      +\mathrm{h.c.}),
\label{eq:softSUSY}
\end{align} where ${\mathcal L}_\mathrm{MSSM}^\mathrm{ soft}$ collects the soft SUSY
breaking terms of the MSSM.   $B_{M_R}^{ij}$ and $B_{\mu_X}^{ij}$
are the new parameters involving the scalar partners of the sterile
neutrino states. Notice that while the former conserves lepton number,
the latter gives rise to a lepton number violating $\Delta L=2$ term.
Assuming a flavour-blind mechanism for SUSY breaking, we consider
universal boundary conditions for the soft SUSY breaking parameters at
some very high energy scale\footnote{This corresponds to the CMSSM scenario. In our subsequent analysis, we will relax some of these universality conditions, considering non-universal soft breaking terms for the Higgs sector, a scenario known as the NUHM.} (e.g. the gauge coupling unification scale
$\sim 10^{16}$ GeV), 
\begin{equation} m_\phi = m_0\,, M_\text{gaugino}= M_{1/2}\,,
A_{i}= A_0 \, \mathbb{I}\,,B_{\mu_X}=B_{M_R}= B_0 \, \mathbb{I}\,.  
\end{equation}
Without loss of generality, we choose a basis where $M_R$ is diagonal
at the SUSY scale, i.e.,
\begin{equation}
    M_{R} = \mathrm{diag}\;M_{R_{ii}}\,.
\end{equation}
In addition, in the numerical evaluations, we also assume
$\mu_{X}$ to be diagonal, a simplifying assumption motivated by
the fact that cLFV observables depend only indirectly on
$\mu_{X}$.

Finally, it is worth noting  that the effective right-handed sneutrino mass term is given by
\begin{equation}
M^2_{\widetilde \nu^C} \simeq m^2_{\widetilde \nu^C} + M_{R}^2 + { |Y_\nu|^2 v_u^2}\,.
\end{equation} 
Assuming $M_R \sim {\mathcal{O}}$(TeV), the effective mass term will 
not be very large, in clear contrast
to what occurs in the standard (type I) SUSY seesaw~\cite{Arganda:2005ji}. 
In our analysis, we are particularly interested in the role 
of such a light sneutrino (i.e.  
$M^2_{\widetilde \nu^C} \sim M^2_\text{SUSY}$) in the enhancement of the Higgs and $Z^0$
mediated contributions to 
lepton flavour violating observables.

\subsubsection{Higgs-mediated contributions}

For any seesaw realisation, the
neutrino Yukawa couplings could leave their imprints in the  SUSY soft-breaking slepton mass matrices, and 
consequently induce flavour violation at low energies due to the renormalisation group
evolution of the SUSY soft-breaking parameters. Even under the assumption of 
universal soft breaking terms at the GUT scale, radiative effects 
proportional to $Y_\nu$ induce flavour violation in the slepton mass 
matrices, which in turn gives rise to slepton mediated cLFV 
observables~\cite{Borzumati:1986qx, Hisano:1995cp, Hisano:1998fj}. 
As an example with CMSSM boundary conditions (eqs.~(\ref{CMSSM1}-\ref{CMSSM4})), in the leading logarithmic approximation, the RGE 
corrections to the left-handed slepton soft-breaking masses are given by
\begin{eqnarray}
(\Delta m_{\widetilde{L}}^2)_{ij}&\simeq&
-\frac{1}{8\pi^2}(3m_0^2+A_0^2) 
(Y_\nu^\dagger L Y_\nu)_{ij} \,, ~~ L=\ln\frac{M_{GUT}}{M_{R}} \,
\nonumber \\ 
&=&\xi (Y^\dagger_\nu Y_\nu)_{ij}\,,
\label{slepmixing}
\end{eqnarray}  
where, for simplicity, a degenerate right-handed neutrino spectrum, $M_{R_i}=M_{R}$ is assumed.

Compared to the standard (type I) SUSY seesaw, where 
$M_{R}\sim10^{14}$ GeV, the inverse seesaw is characterised by a right
handed neutrino mass scale $M_{R}\sim\mathcal{O}(\text{TeV})$ and this 
in turn leads to an enhancement of the factor $\xi$, see eq.~(\ref{slepmixing}), and hence will induce sizeable effects in all low-energy cLFV observables. Furthermore, having right-handed sneutrinos 
whose mass is of the same order of  the other sfermions, 
i.e.  $M^2_{\widetilde \nu^C} \sim M^2_\text{SUSY}$, 
the $\widetilde \nu^C$-mediated processes are no longer suppressed, and 
might even significantly contribute to the low-energy flavour 
violating observables.
Here, we focus on the impact of  such a light $\widetilde \nu^C$ 
in the Higgs mediated processes which are expected to be  important 
in the large $\tan \beta$ regime.

Although at tree level Higgs-mediated neutral currents are flavour conserving, 
non-holomorphic Yukawa interactions
of the type $\bar D_RQ_LH_u^*$ can be induced at the one-loop level, as first 
noticed in~\cite{Hall:1993gn}. Similarly, in the lepton sector, the origin of the Higgs-mediated flavour violating couplings
can be traced to a non-holomorphic Yukawa term of the form 
$\bar E_RLH_u^*$~\cite{Babu:2002et}. Other than the 
corrections to the 
lepton masses, these new couplings  give rise to additional 
contributions to several cLFV processes mediated by Higgs exchange. 
In particular $B_s \rightarrow \mu\tau$, $B_s \rightarrow e\tau$ (the so-called 
double penguin processes) were considered in~\cite{Dedes:2002rh}, while $\tau \rightarrow \mu \eta$ was 
studied in~\cite{Sher:2002ew}. 
A detailed analysis of the several $\mu-\tau$ lepton
flavour violating processes, namely $\tau \rightarrow \mu X$ 
($X = \gamma,e^+e^-,\mu^+\mu^-,\rho,\pi,\eta,\eta^\prime$) can be found in~\cite{Brignole:2004ah, Arganda:2005ji}.

Even though the flavour violating processes in the quark and lepton sectors 
have a similar diagrammatic origin, the 
source of flavour violation is different in each case. 
In the quark sector, trilinear soft SUSY breaking couplings 
involving up-type squarks
provide the dominant source of flavour violation~\cite{Babu:1999hn}, 
while in the lepton case, LFV stems
from the radiatively induced non-diagonal terms in the slepton masses 
(see eq.~(\ref{slepmixing}))~\cite{Babu:2002et}. 

In the standard SUSY seesaw (type I), the 
term ${\widetilde \nu^C_{i}}H_u{\widetilde L_{Lj}}$ is usually neglected, as it is suppressed by the
very heavy right handed sneutrino masses, ${M_{\widetilde \nu^C_{i}}} \sim 10^{14}$GeV.
However, in scenarios such as the inverse SUSY seesaw, where ${M_{\widetilde \nu^C_{i}}}\sim \mathcal{O}$(TeV), 
this term 
may provide the dominant contributions to Higgs-mediated lepton 
flavour violation. 

\begin{figure}[t]
\centering
\subtop{\includegraphics[width=0.49\textwidth]{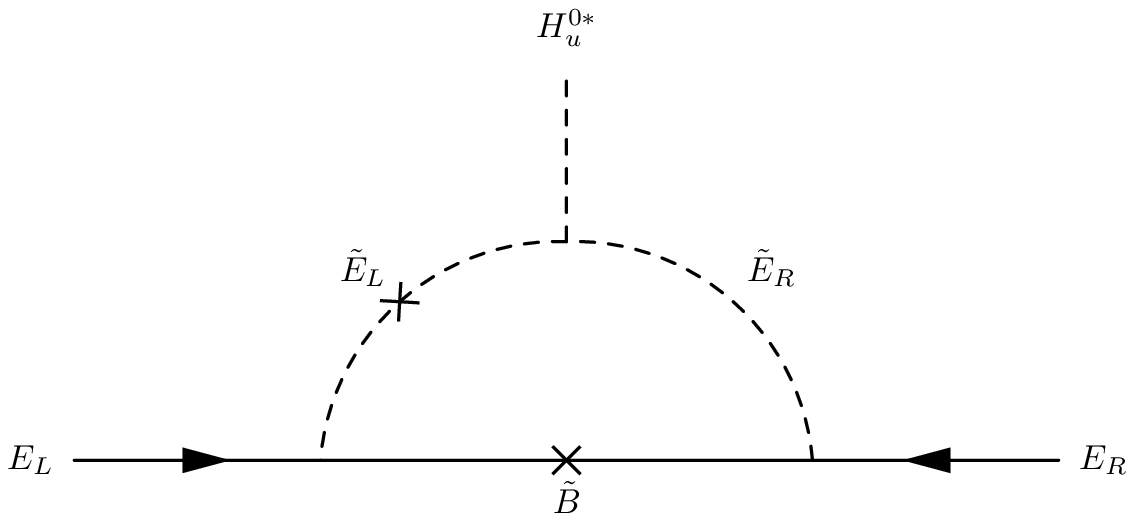}}
\hfill
\subtop{\includegraphics[width=0.49\textwidth]{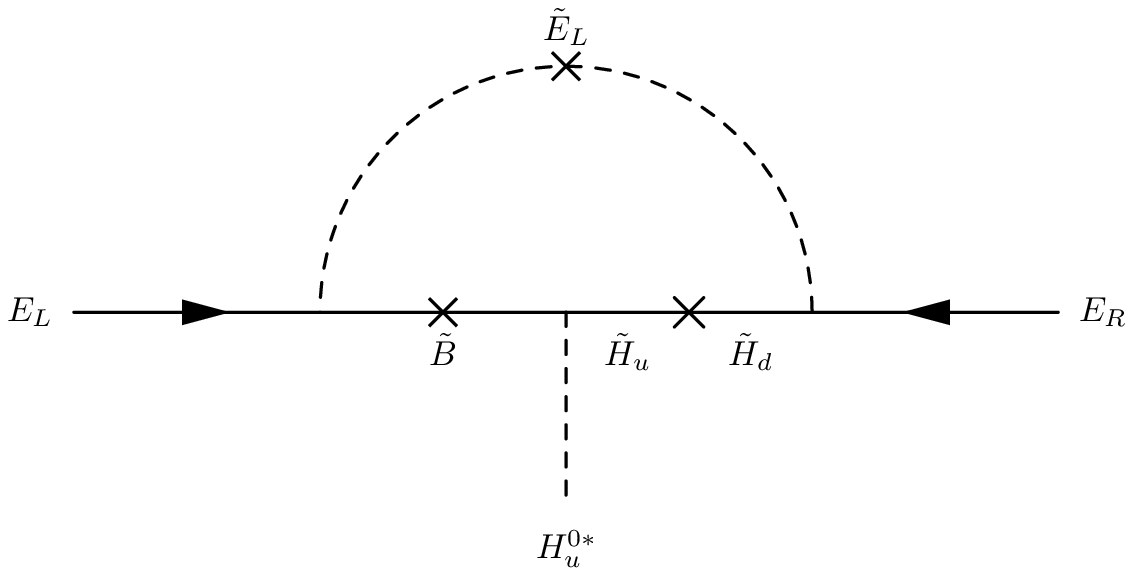}}
\subbottom{\includegraphics[width=0.49\textwidth]{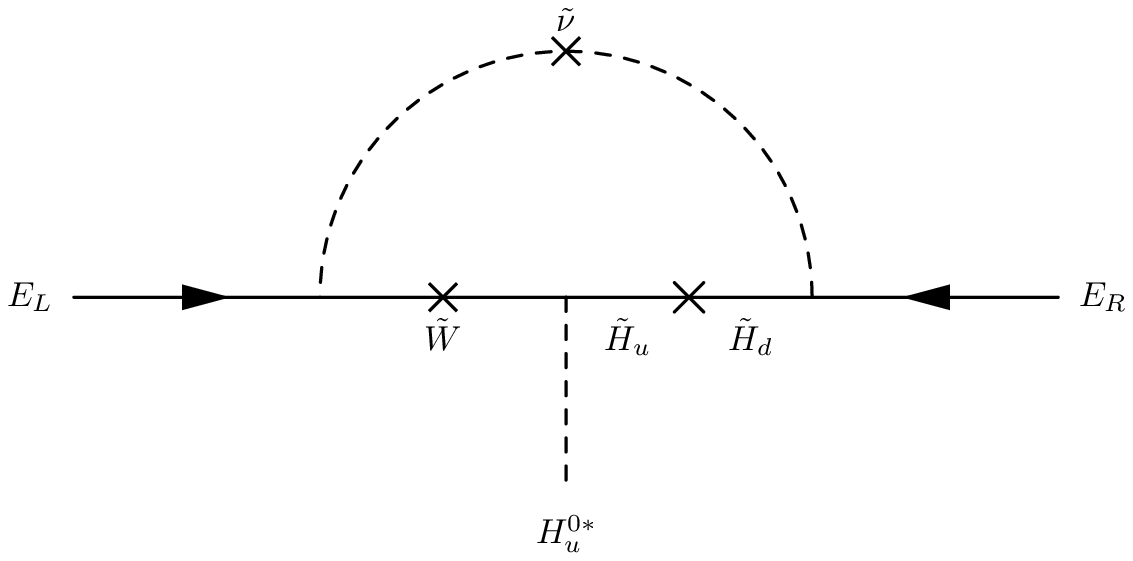}}
\hfill\subbottom{\includegraphics[width=0.49\textwidth]{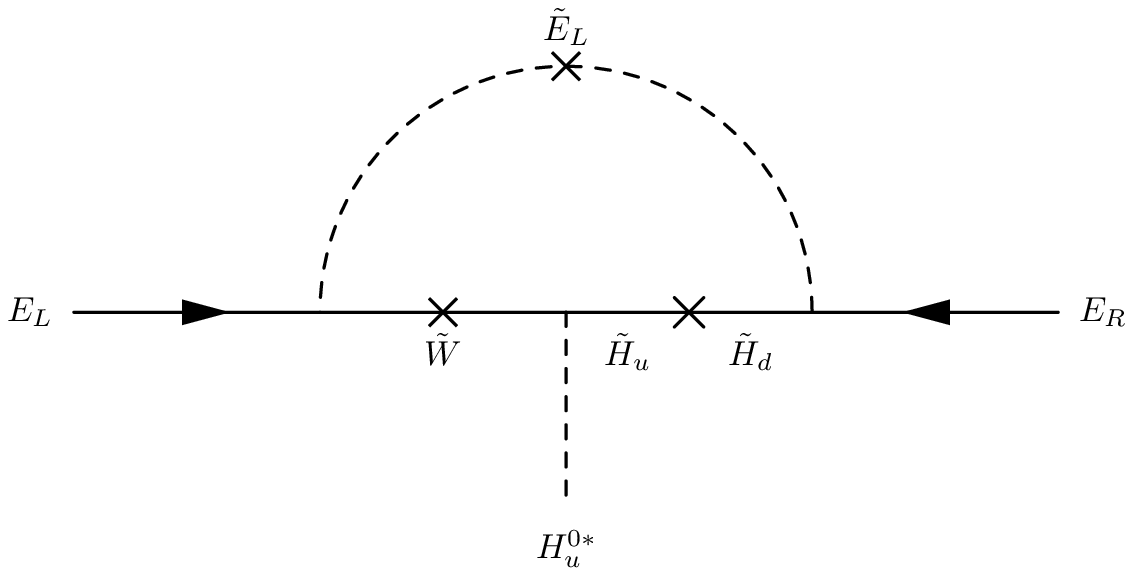}}
\caption[Diagrams contributing to $\epsilon_2$]{Diagrams contributing to $\epsilon_2$. Arrows correspond to chiralities; crosses 
on  scalar lines represent  LFV mass insertions $(\Delta m_{\widetilde{L}}^2)_{ij}$, while those on fermion lines denote chirality flips.}\label{1}
\end{figure}

\begin{figure}
\begin{center}
\includegraphics[width=0.50\textwidth]{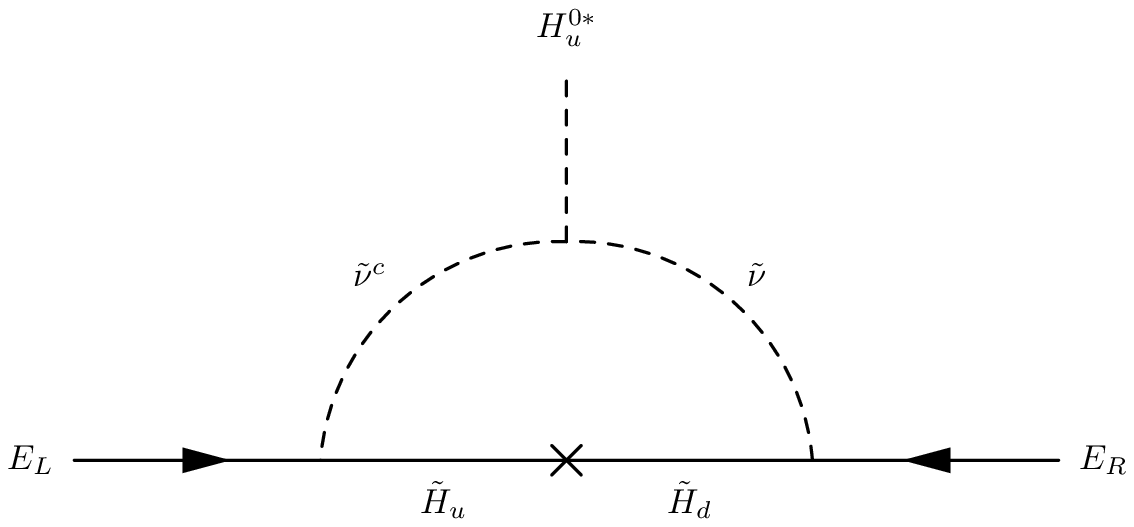}
\caption[Right-handed sneutrino contribution  to $\epsilon'_2$]{Right-handed sneutrino contribution  to $\epsilon'_2$. 
This contribution is particularly relevant when 
$\widetilde \nu^C$ is light.}\label{2}
\end{center}
\end{figure}

The effective Lagrangian describing the couplings of the neutral Higgs fields
to the charged leptons is given by
\begin{equation}
-\mathcal{L}^\text{eff}=\bar E^i_R Y_{e}^{ii} \left[ 
\delta_{ij} H_d^0 + \left(\epsilon_1 \delta_{ij} + 
\epsilon_{2ij} (Y_\nu^\dagger Y_\nu)_{ij} \right) H_u^{0*}
\right] E^j_L + \text{h.c.}  \,. 
\label{Leff}
\end{equation}
In the above,  the first term corresponds to the usual Yukawa interaction, 
while the coefficient $\epsilon_1$ encodes
the corrections to the charged lepton Yukawa couplings. In the basis 
where  the charged lepton Yukawa couplings are diagonal, 
the last term in eq.~(\ref{Leff}), i.e. $\epsilon_{2ij} (Y_\nu^\dagger Y_\nu)_{ij}$, is in general  
non-diagonal, thus
providing  a new source of charged
lepton flavour violation through Higgs mediation. Its origin can be diagrammatically understood from fig.\ref{1},
where flavour violation is parametrized  via a mass insertion
$(\Delta m_{\widetilde{L}}^2)_{ij}$ (see eq. (\ref{slepmixing})).
The coefficient $\epsilon_2$ can be estimated as
\begin{align}
\epsilon_{2ij} =& 
\frac{\alpha^\prime}{8\pi} \xi  \mu M_1 
\left[
2F_2\left(M_1^2,m_{\widetilde E_{Lj}}^2,
m_{\widetilde E_{Li}}^2,m_{\widetilde E_{Ri}}^2\right)
-F_2 \left(\mu^2, m^2_{\widetilde E_{Lj}}, m^2_{\widetilde E_{Li}},M_1^2\right)
\right] +\nonumber\\
&
\frac{\alpha_2}{8\pi}\xi  \mu M_2 
\left[F_2\left(\mu^2,m^2_{\widetilde E_{Lj}}, m^2_{\widetilde E_{Li}}, M_2^2\right) +
2F_2\left(\mu^2,m_{\widetilde\nu_{Lj}}^2,m_{\widetilde \nu_{Li}}^2,M_2^2\right)
\right] \,,
\end{align}
where
\begin{equation}
F_2\left(x,y,z,w\right) =
-\frac{x \ln x}{(x-y)(x-z)(x-w)} -\frac{y\ln y}{(y-x)(y-z)(y-w)}+ 
(x\leftrightarrow z,
y\leftrightarrow w)\, . 
\end{equation}
Here, $M_1$ and $M_2$ are the masses of the electroweak gauginos at low energies while $\alpha^\prime=g^{\prime 2}/4\pi$ and $\alpha_2=g^2/4\pi$ are the reduced $\mathrm{U}(1)_Y$ and $\mathrm{SU}(2)_L$ couplings, respectively. On the other hand, the flavour conserving loop-induced form factor 
$\epsilon_{1}$ 
(notice that the diagrams of fig.\ref{1} contribute to this form factor, 
but without the slepton flavour mixings in the internal lines) 
can be expressed as~\cite{Babu:2002et, Dedes:2002rh}
\begin{align}
\epsilon_{1}=&\frac{\alpha'}{8\pi}\mu M_1 \left[2 F_1\left(M_1^2,
m_{\widetilde E_L}^2,
m_{\widetilde E_R}^2\right) -
F_1\left(M_1^2,\mu^2,m^2_{\widetilde E_L}\right) + 2F_1\left(M_1^2,\mu^2,
m^2_{\widetilde E_R}\right)\right] \nonumber \\ 
&+\frac{\alpha_2}{8\pi}\mu M_2\left[F_1\left(\mu^2,
m^2_{\widetilde E_L}, M_2^2\right) 
+ 2F_1\left(\mu^2,m_{\widetilde \nu_L}^2,M_2^2\right)\right],
\label{eq:eps1} 
\end{align}
with
\begin{equation}
F_1\left(x,y,z\right)=-\frac{xy\ln (x/y)+yz\ln (y/z)+zx\ln (z/x)}{(x-y)(y-z)(z-x)} \,.
\end{equation}
In the standard (type I) seesaw mechanism, the diagrams of 
fig.~\ref{1} provide the only source of Higgs-mediated
lepton flavour violation. However, in the framework of the inverse SUSY seesaw,  
there is an additional diagram that 
may even account for the dominant Higgs-mediated
lepton flavour violation contribution: the sneutrino-chargino mediated loop, depicted in fig.~\ref{2}.
Due to the large masses of $\widetilde \nu^C$ in the type I seesaw, this process provides negligible contributions, and is hence not taken into account.

The new contribution from the sneutrino-chargino mediated loop gives an effective Lagrangian term which reads
\begin{equation}
 -\mathcal{L}^\prime= \imath Y_{e}^{ii} (Y_\nu^\dagger Y_\nu)_{ij} A_\nu^\dagger \mu^* \epsilon^\text{tot}_{2ij} \bar E^i_R I H_u^{0\ast }
 E^j_L + \text{h.c.}\,,
\end{equation}
with
\begin{equation}
 I=\int \frac{\mathrm{d}^4 k}{(2\pi)^4} \frac{\slashed p + \slashed k + \mu}{(p+k)^2-\mu^2}\frac{\slashed p + \slashed k - \mu}{(p+k)^2-\mu^2}\frac{1}{k^2-M_{\widetilde \nu^C}^2}\frac{1}{(k+q)^2-m_{\widetilde \nu_L}^2}\,.
\end{equation}
Using the loop integrals defined in~\cite{Brignole:2004ah},
\begin{equation}
 I=\frac{1}{16\pi^2}\left[-J_4\left(\mu^2,\mu^2,M_{\widetilde \nu^C}^2,m_{\widetilde \nu_L}^2\right)+\mu^2 I_4\left(\mu^2,\mu^2,M_{\widetilde \nu^C}^2,m_{\widetilde \nu_L}^2\right)\right]
\end{equation}
which gives
\begin{equation}
 I=\frac{-\imath}{16\pi^2} I_3\left(\mu^2,M_{\widetilde \nu^C}^2,m_{\widetilde \nu_L}^2\right)\,.
\end{equation}
Consequently, the effective Lagrangian terms encoding 
lepton flavour violation are accordingly modified as  
\begin{equation}
-\mathcal{L}^\text{LFV} = \bar E^i_R Y_{e}^{ii}\epsilon^\text{tot}_{2ij} (Y_\nu^\dagger Y_\nu)_{ij} H_u^{0*}
 E^j_L + \text{h.c.}  \,, 
\label{Leff1}
\end{equation}
where $\epsilon^\text{tot}_2=\epsilon_2+\epsilon_2'$, $\epsilon_2'$ being  the 
contribution from the new diagram.
This contribution can be expressed as
\begin{equation}
\epsilon'_{2ij}= \frac{1}{16\pi^2} \mu A_\nu 
F_1(\mu^2,m^2_{\widetilde \nu_i},M^2_{\widetilde \nu^c_j}),
\label{newdiag}
\end{equation}
since the soft trilinear  term for the neutral sleptons is parametrized by $A_\nu Y_\nu$, where
$A_\nu$ is a flavour independent real mass term, and $\mu$ is real because we consider scenarios with no extra source of CP violation, namely the CMSSM and the NUHM.

A quick estimate reveals that in the inverse SUSY seesaw, $\epsilon^\text{tot}_{2}$ 
is enhanced by a factor of order $\sim 10$ compared to the standard seesaw. The large enhancement of $\epsilon^\text{tot}_{2}$  
will have an impact regarding all 
Higgs-mediated lepton flavour violating observables. 
The computation of the cLFV observables requires specifying the 
couplings of the physical Higgs bosons to the leptons, in particular
$\bar E^i_{R}E^j_{L}H_k$ (where $H_k = h,H,A$). The effective 
Lagrangian describing this interaction can be derived from 
eq.~(\ref{Leff}), and reads~\cite{Babu:2002et, Dedes:2002rh}
\begin{equation}
-\mathcal{L}^\text{eff}_{i\neq j} =
(2G_F^2)^{1/4} \,
\frac{m_{E_i} \kappa^E_{ij}}{\cos^2\beta}
\left(\bar E^i_{R}\,E^j_{L}\right)
\left[\cos(\alpha-\beta) h + \sin(\alpha-\beta) H - i A\right]+\text{h.c.}
\,,
\label{Leffl}
\end{equation}
where $\alpha $ is the CP-even Higgs mixing angle and 
\begin{equation}
\kappa^E_{ij} = \frac{\epsilon^\text{tot}_{2ij} (Y^\dagger_\nu Y_\nu)_{ij}
}{
\left[1+\left(\epsilon_1+\epsilon^\text{tot}_{2ii}
(Y^\dagger_\nu Y_\nu)_{ii}\right)\tan\beta\right]^2 }\ \label{kappa}.
\end{equation}

As clear from the above equation, large values of $\epsilon^\text{tot}_2$ lead to an increase 
of $\kappa^E_{ij}$. Given that the cLFV branching ratios are proportional to $({\kappa^E_{ij}})^2$,
a sizeable enhancement, as large as two orders of magnitude, is expected for all Higgs-mediated LFV observables in this framework.

\subsubsection{Higgs-mediated lepton flavour violating observables}\label{hmlfv}
Here we focus our attention on the cLFV observables where the dominant contribution to flavour 
violation arises from the 
Higgs penguin diagrams, in particular those involving $\tau$-leptons (due to the comparatively large value 
of  $Y_\tau$).

In what follows, we discuss some of these LFV decays in detail.
\begin{itemize}
\item
$\tau \rightarrow 3\mu$

\noindent
In the large $\tan\beta$ regime, 
Higgs-mediated flavour violating diagrams
would be particularly important in this decay mode. The branching ratio can be expressed as~\cite{Babu:2002et, Dedes:2002rh} 
\begin{align}
\mathcal{B}(\tau\to3\mu)&=
\frac{G_F^2 \,m_\mu^2 \,m_\tau^7 \,\tau_\tau}{1536\,\pi^3 \cos^6\beta}\,
 |\kappa_{\tau\mu}^E |^2 
\left[\left(\frac{\sin(\alpha-\beta)\cos\alpha}{m_{H}^2} -
\frac{\cos(\alpha-\beta)\sin\alpha}{m_{h}^2}\right)^2 
+\frac{\sin^2\beta}{m_A^4}\right]\nonumber\\
& 
\simeq \frac{G_F^2 \,m_\mu^2\, m_\tau^7\, \tau_\tau}{768\,\pi^3\, m_A^4}
 |\kappa_{\tau\mu}^E |^2 \tan^6\beta \,.
\end{align}
In the above,  $\tau_\tau$ is the $\tau$ life time and the approximate result has been obtained in the large $\tan\beta$ regime.  For other Higgs-mediated lepton flavour violating  3-body decays, 
$\tau \rightarrow e\mu\mu$,
$\tau \rightarrow 3e$ or 
$\mu \rightarrow 3e$, their corresponding branching ratios  can easily be obtained  with the appropriate
kinematic factors and the proper flavour changing factor $\kappa$. 
While $\mathcal{B}(\tau \rightarrow e\mu\mu)$ can be as large as  
$\mathcal{B}(\tau \rightarrow 3\mu)$ when $(Y^\dagger_\nu Y_\nu)_{13}\sim O(1)$ 
(which is possible in the case of  an inverted hierarchical light neutrino spectrum), 
other flavour 
violating decays  with final state electrons  such as $\mu \rightarrow 3e$ 
are 
considerably suppressed due to the smallness of the electron Yukawa couplings. 
\item
$B_s\to \ell_i \ell_j$

\noindent
$B$ mesons can also have  Higgs-mediated LFV decays, which are  significantly enhanced in the large $\tan\beta$ regime. The corresponding branching fraction is given by 
\begin{align}
\mathcal{B}(B_s\to \ell_i \ell_j) =& \frac{G_F^4 \,m^4_W}{8\,\pi^5}\,
|V_{tb}^*V_{ts}|^2\, m_{B_s}^5 \,f_{B_s}^2\, \tau_{B_s} \biggl (\frac{m_b}{m_b+
m_s}\biggr )^2 \nonumber \\
&\times \sqrt{ \biggl [1-\frac{(m_{\ell_i}+m_{\ell_j})^2}{m_{B_s}^2}\biggr ]
\biggl [1-\frac{(m_{\ell_i}-m_{\ell_j})^2}{m_{B_s}^2}\biggr ] }       
\nonumber \\
&\times  
\Biggl \{ \biggl (1-\frac{(m_{\ell_i}+m_{\ell_j})^2}{m_{B_s}^2}\biggr ) 
|c^{ij}_{S}|^2
+\biggl ( 1-\frac{(m_{\ell_i}-m_{\ell_j})^2}{m_{B_s}^2}\biggr ) |c_{P}^{ij}|^2
\Biggr \} \,,
\label{brll}
\end{align}
where $V_{ij}$ represents the Cabibbo-Kobayashi-Maskawa (CKM) matrix,  
$m_{B_s}$ and $\tau_{B_s}$ respectively denote the mass and lifetime of the 
$B_s$ meson, while $f_{B_s}=230\pm 30$ MeV~\cite{Bernard:2000ki} is the ${B_s}$ meson 
decay constant and $c_{P}^{ij}$, $c_{S}^{ij}$ are form factors.
As an example, the lepton flavour violating (double-penguin) $B_s\to \mu\tau$ decay can be computed with the following form factors~\cite{Dedes:2002rh}: 
\begin{align}
c_S^{\mu\tau}=c_P^{\mu\tau} =&
\frac{\sqrt{2}\,\pi^2}{G_F \,m^2_W}
\frac{m_\tau \, \kappa_{bs}^d \,\kappa_{\tau\mu}^{E}}
{\cos^4\beta \,\bar\lambda^t_{bs}}
\left[\frac{\sin^2(\alpha-\beta)}{m^2_{H}}+
\frac{\cos^2(\alpha-\beta)}{m^2_{h}} + \frac{1}{m^2_{A}} \right]
\nonumber \\
\approx&
\frac{8 \,\pi^2 \,m_\tau \, m_t^2}{ m^2_W}
\frac{ \epsilon_Y~ \kappa_{\tau\mu}^{E} ~\tan^4\beta }
{\left[1+(\epsilon_0+\epsilon_Y Y^2_{t} )\tan\beta\right]
\left[1+\epsilon_0 \tan\beta\right]}
\frac{1}{m^2_{A}} \,.
\label{ctm}
\end{align}
Here, $\kappa_{bs}^d$ represents the flavour mixing in the quark
sector while $\bar\lambda^t_{bs} = V^*_{tb}V_{ts}$. Similarly, 
$\epsilon_0$ and $\epsilon_Y$ are the down type quark form factors mediated
by gluino and squark exchange diagrams. 
The final result was, once again, derived in the large $\tan\beta$ regime.
The branching fractions of other flavour violating decays such as  
$\mathcal{B}(B_{d,s}\rightarrow \tau e)$, would receive identical contribution from the
Higgs penguins. 
Likewise, the branching ratio $\mathcal{B}(B_{d,s}\rightarrow \mu e)$  
can be calculated using the appropriate form factors and lepton masses;
as expected, these will be suppressed when compared to $\mathcal{B}(B_{d,s}\rightarrow \tau \mu)$.

\item
$\tau\to \mu P$

\noindent
Similar to what occurred in the previous processes, virtual Higgs exchange could also induce
decays such as $\tau \to \mu P$, where $P$ denotes a neutral pseudoscalar 
meson ($P=\pi, \eta, \eta')$. In the large $\tan\beta$ limit, where the  pseudoscalar Higgs couplings to down-type quarks are enhanced,
CP-odd Higgs boson exchanges provide the
dominant contribution to the $\tau \to \mu P$ decay. The corresponding coupling can be written as
\begin{equation}
-i (\sqrt{2} \,G_F)^{1/2}\tan\beta~ A(\xi_d \,m_d \,\bar d \,d +
\xi_s \, m_s \,\bar s \,s +\xi_b \,m_b \,\bar b \,b) + \mathrm{ h.c.} .
\end{equation}
Here, the parameters $\xi_d,\,\xi_s,\,\xi_b$ are  of order  $\mathcal{O}(1)$. 
Since we are mostly interested in the Higgs-mediated contributions, we estimate the amplitude of these processes in the limit when both $\tau \rightarrow 3 \mu$ and 
$\tau \to \mu P$ are indeed  dominated by the exchange of the scalar fields.
Accordingly, and following~\cite{Brignole:2004ah}, one can write 
\begin{align}
\frac{\mathcal{B}(\tau \to \mu \eta)}{\mathcal{B}(\tau \to 3 \mu )} 
\simeq	&
36 \,\pi^2 \left(\frac{f^8_\eta \,m^2_\eta}{m_\mu\, m^2_\tau}\right)^2  
(1 - x_\eta)^2 
\left[\xi_s +\frac{\xi_b}{3}\left(1 +\sqrt2 \,\frac{f^0_\eta}{f^8_\eta}
\right)\right]^2 , \\
\label{taumu_eta}  
\frac{\mathcal{B}(\tau \to \mu \eta^\prime)}{\mathcal{B}(\tau \to \mu \eta)} 
 \simeq &	  
\frac2 9  \left( \frac{f^0_{\eta^\prime}}{f^8_\eta}\right)^2 
\frac{m^4_{\eta^\prime}}{m^4_\eta} \left(\frac{1 - x_{\eta^\prime}}{1 - x_\eta} \right)^2
\left[\frac{
1 + \frac{3}{\sqrt2}\, \frac{f^8_{\eta^\prime}}{f^0_{\eta^\prime}} 
\left( \frac{\xi_s}{\xi_b}  +\frac13\right)}
{\frac{\xi_s}{\xi_b} + \frac13 + 
\frac{\sqrt2}{3} \,\frac{f^0_\eta}{f^8_\eta}}
\right]^2 , \\
\label{pieta}
\frac{\mathcal{B}(\tau\to \mu \pi)}{\mathcal{B}(\tau \to \mu \eta)} 
 \simeq & \frac4 3  \left(\frac{f_\pi}{f^8_\eta}\right)^2 \, 
 \frac{m^4_\pi}{m^4_\eta} ~   
(1 - x_\eta)^{-2}
\left[\frac{\frac{\xi_d}{\xi_b} \,\frac{1}{1+z} + \frac{1}{2}\,
(1 + \frac{\xi_s}{\xi_b}) 
\frac{1-z}{1+z}}
{\frac{\xi_s}{\xi_b} + \frac13 + 
\frac{\sqrt2}{3}\, \frac{f^0_\eta}{f^8_\eta}}
\right]^2 \,, 
\end{align}
where $z=m_u/m_d$, $m_\pi, \ f_\pi$ are the pion  mass and decay constant,  $m_{\eta,\eta^\prime}$ 	are  the  masses of $\eta,\ \eta^\prime$, $x_{\eta,\eta^\prime}= m_{\eta,\eta^\prime}^2/m_\pi^2$,  and 
$f^8_{\eta,\eta^\prime}$ and $f^0_{\eta,\eta^\prime}$ are evaluated from the corresponding matrix elements. 
As first discussed in~\cite{Sher:2002ew}, and taking $\xi_s,\xi_b \sim 1$ and fixing the other parameters as in~\cite{Brignole:2004ah}, one finds  $\frac{\mathcal{B}(\tau \to \mu \eta)}{\mathcal{B}(\tau \to 3 \mu )} \simeq
5$. The other branching fractions such as 
$\mathcal{B}(\tau \to \mu \eta^\prime,\mu \pi)$  are considerably  suppressed compared to 
$\mathcal{B}(\tau \to \mu \eta)$. 
While the ratio $\frac{\mathcal{B}(\tau \to \mu \eta^\prime)}{\mathcal{B}(\tau \to \mu \eta)}$ 
can be as large as $6\times 10^{-3}$, 
$\frac{\mathcal{B}(\tau\to \mu \pi)}{\mathcal{B}(\tau \to \mu \eta)}$ 
would approximately lie in the range  
$10^{-3}-4\times 10^{-3}$~\cite{Brignole:2004ah}. 
Since all these ratios 
are independent of $\kappa_{\tau\mu}^{E}$, the above quoted numbers can also be
applied to the present framework. 
However, an enhancement in the 
$\mathcal{B}(\tau \to 3 \mu)$, due to 
the large values of  $\kappa_{\tau\mu}^{E}$, 
would also imply sizeable values of $\mathcal{B}(\tau \to \mu \eta)$.

\item $H_k\to \mu\tau\;(H_k = h,H,A)$

\noindent
The branching ratios of flavour violating 
Higgs decays provide another interesting
probe of lepton flavour violation. 
Following~\cite{Brignole:2003iv},  the branching fraction $H_k\to \mu\tau$ 
(normalised
to the flavour conserving one  
$H_k\to \tau \tau$) can be cast as:
\begin{equation}
{\mathcal{B}(H_k\to \mu\tau)}
=  \tan^2\beta~ (|\kappa_{\tau\mu}^{E}|^2 ) 
 ~ C_\Phi~ { \mathcal{B}(H_k\to \tau\tau)}  \, ,
\end{equation}
where we approximated $1/\cos^2\beta \simeq \tan^2\beta$. 
The coefficients $C_\Phi$  are given by:
\begin{equation}
C_h = \left[\frac{\cos(\beta - \alpha)}{\sin\alpha}\right]^2, ~~~~
C_H = \left[\frac{\sin(\beta - \alpha)}{\cos\alpha}\right]^2, ~~~~
C_A = 1.
\end{equation}

\end{itemize}

\subsubsection{Numerical results for the Higgs-mediated contribution}

\begin{figure}
\begin{center}
\includegraphics[width=0.4\textwidth]{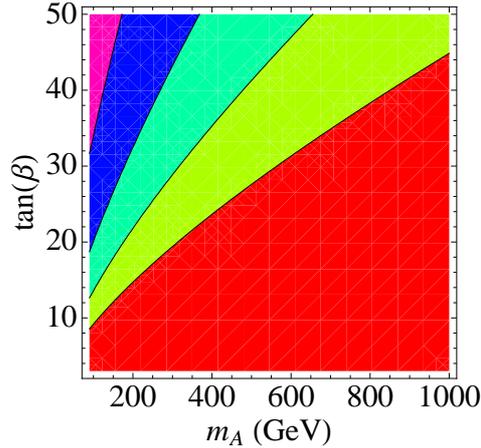}
\end{center}
\caption[$\mathcal{B}(\tau \rightarrow 3 \mu)$ in the SUSY inverse seesaw]{Branching ratio of the process $\tau \rightarrow 3 \mu$ as a function of $m_A$ (GeV) and $\tan \beta$.
From left to right, the contours correspond to $\mathcal{B}(\tau \rightarrow 3 \mu) =2.1 \times 10^{-8}$, $10^{-9}$, $10^{-10}$, $10^{-11}$. The purple region has already been experimentally excluded~\cite{Hayasaka:2010np}. Figure reprinted from~\cite{Abada:2011hm} by the author, with kind permission from Springer Science and Business Media.}\label{3}
\end{figure}

As expected from the analytical study above, 
$m_A$ and $\tan \beta$ are the most relevant parameters
in the Higgs-mediated flavour violating processes. To better illustrate this, 
in fig.~\ref{3} we study the dependence of $\mathcal{B}(\tau \rightarrow 3 \mu)$ on the aforementioned parameters. 
We have assumed a common value for the squark masses, $m_{\widetilde q}\sim \text{TeV}$, 
while for left- and right-handed sleptons we take $m_{\widetilde \ell}\sim 400~\text{GeV}$ and $M_{\widetilde \nu^c}\sim 3~\text{TeV}$ for the right handed sneutrinos. 
The contours correspond to different values of the branching ratios and the purple region has already been experimentally excluded. From this figure one can easily identify the regimes for $m_A$ and $\tan\beta$ which are associated with values of the LFV observables within reach of the present and
future experiments. 

In what follows, we numerically evaluate some LFV observables.  Concerning the 
CMSSM parameters 
(and instead of scanning over the parameter space), we have selected a 
few benchmark points~\cite{AbdusSalam:2011fc} that took into account 
the most recent LHC constraints~\cite{Chatrchyan:2011qs, *ATLAS:2011ad} at the time of this study. However, searches for supersymmetric particles and the Higgs mass measurements conducted at the LHC in 2012 now exclude these specific points~\cite{Strege:2012bt}. Nevertheless, they illustrate well the impact of the various parameters on the predicted branching ratios.
We have also considered the case in which the GUT scale universality conditions are 
relaxed for the 
Higgs sector, i.e.  scenarios of Non-Universal Higgs Masses (NUHM), as this allows to 
explore the 
impact of a light CP-odd Higgs boson.
In table~\ref{tab:sfp10.1}, we list the chosen points: CMSSM-A and CMSSM-B respectively 
correspond to 
the 10.2.2 and 40.1.1 benchmark points in~\cite{AbdusSalam:2011fc}, while NUHM-C is an example of a non-universal scenario.
\begin{table}[t]
\begin{center}
\begin{tabular}{|c||c||c|c||c|c|c|c|c|}
\hline
Point & $\tan\beta$ & $m_{1/2}$ & $m_0$ & $m^2_{H_U}$& $m^2_{H_D}$ &$A_0$  &$\mu$&$m_A$ \\
\hline
\hline
 CMSSM-A &10& 550 & 225 & $(225)^2$ &$(225)^2$ &0 &690 & 782 \\
\hline
\hline
CMSSM-B &40& 500 & 330 & $(330)^2$&$(330)^2$ &-500&698 &604 \\
\hline
NUHM-C &15& 550 & 225 &$(652)^2$ &$-(570)^2$ &0&478&150  \\
\hline
\hline
\end{tabular}
\caption[Benchmark points in our analysis]{Benchmark points used in the numerical analysis (dimensionful parameters in GeV).
CMSSM-A and CMSSM-B correspond to 10.2.2 and 40.1.1 benchmark 
points of~\cite{AbdusSalam:2011fc}. \label{tab:sfp10.1} }
\end{center}
\end{table}

For each point considered, the low-energy SUSY parameters were obtained using SuSpect~\cite{Djouadi:2002ze}.
 In what concerns the evolution of the soft-breaking right-handed sneutrino masses $M_{\widetilde \nu^C}^2$, we
   have assumed that the latter hardly run between the GUT scale and the low-energy one. 
  The flavour-violating charged 
slepton parameters (e.g. $(\Delta m_{\widetilde{L}}^2)_{ij}$ or $\xi$), were estimated at the 
leading order using eq.~(\ref{slepmixing}). Concerning NUHM, we use the same value of $\xi$ as for CMSSM-A. Here, we
are particularly interested to study the effect of light CP-odd Higgs boson and this naive
approximation will serve our purpose. Furthermore, we use the mass insertion approximation, 
assuming that mixing between left and right chiral 
slepton states are relatively small. In computing the branching fractions and the 
flavour violating factor $\kappa^E_{ij}$ 
we have assumed (physical) right-handed sneutrino masses
$M_{\widetilde \nu^C} \approx 3$ TeV and $\left(Y_\nu^{\dagger}Y_\nu \right) = 0.7$, in agreement with low-energy neutrino data as well as other low-energy  constraints, which are particularly relevant in the inverse seesaw case such as Non-Standard Neutrino Interactions bounds~\cite{Antusch:2006vwa, Antusch:2008tz}. Moreover, in our numerical analysis, we have 
fixed the trilinear soft breaking parameter $A_\nu = - 500$ GeV (at the SUSY scale).

We now proceed to present our results for the flavour violating observables
discussed previously. 
In table~\ref{4}, we collect the values of the different branching ratios, as obtained for the considered benchmark points of table~\ref{1}.  We have also presented the corresponding current experimental bounds and future sensitivity in table~\ref{cLFVexp}.

\begin{table}[htb]
\begin{center}
 \begin{tabular}{|c|c|c|c|}
  \hline
    LFV Process & CMSSM-A & CMSSM-B & NUHM-C\\
  \hline
    $\tau \rightarrow \mu \mu \mu$ & $1.4 \times 10^{-15}$ & $3.9 \times 10^{-11}$ & $8.0 \times 10^{-12}$ \\
    $\tau^- \rightarrow e^- \mu^+ \mu^-$ &  $1.4 \times 10^{-15}$ & $3.4 \times 10^{-11}$ & $8.0 \times 10^{-12}$ \\
    $\tau \rightarrow e e e$ &  $3.2 \times 10^{-20}$ & $9.2 \times 10^{-16}$ & $1.9 \times 10^{-16}$ \\
    $\mu \rightarrow e e e$ &   $6.3 \times 10^{-22}$ & $1.5 \times 10^{-17}$ & $3.7 \times 10^{-18}$ \\
    $\tau \rightarrow \mu \eta$ &  $8.0 \times 10^{-15}$ & $3.3 \times 10^{-10}$ & $4.6 \times 10^{-11}$ \\
    $\tau \rightarrow \mu \eta^\prime$ &  $4.3 \times 10^{-16}$ & $1.1\times 10^{-10}$ & $3.1 \times 10^{-12}$ \\
    $\tau \rightarrow \mu \pi^{0}$ &  $1.8 \times 10^{-17}$ & $8.5 \times 10^{-13}$ & $1.0 \times 10^{-13}$ \\
    $B^{0}_{d} \rightarrow \mu \tau$ & $2.7 \times 10^{-15}$ & $8.5 \times 10^{-10}$ & $2.7 \times 10^{-11}$ \\
    $B^{0}_{d} \rightarrow e \mu$ &  $1.2 \times 10^{-17}$ & $3.1 \times 10^{-12}$ & $1.2 \times 10^{-13}$ \\
    $B^{0}_{s} \rightarrow  \mu \tau$ & $7.7 \times 10^{-14}$ & $2.5 \times 10^{-8}$ & $7.8 \times 10^{-10}$ \\
    $B^{0}_{s} \rightarrow e \mu$ &  $3.4 \times 10^{-16}$ & $8.9 \times 10^{-11}$ & $3.4 \times 10^{-12}$ \\
    $h \rightarrow \mu \tau$ & $1.3 \times 10^{-8}$ & $2.6 \times 10^{-7}$ & $2.3 \times 10^{-6}$\\
    $A,H \rightarrow \mu \tau$ & $3.4 \times 10^{-6}$ & $1.3 \times 10^{-4}$ & $5.0 \times 10^{-6}$\\
  \hline
 \end{tabular}
\end{center}
  \caption[Higgs-mediated cLFV branching ratios]{Higgs-mediated contributions to the branching ratios of several lepton flavour violating 
processes, for the different benchmark points of table~\ref{1}.}\label{4}
\end{table}

Another interesting property of the Higgs-mediated processes is that the corresponding 
amplitude strongly depends on 
the chirality of the heaviest lepton (be it the decaying lepton, or the heaviest 
lepton produced in $B$ decays). 
Considering the decays of a left-handed lepton $\ell^i_{L} \rightarrow \ell^j_{R} X$, one finds that 
the corresponding branching ratios would be suppressed by a factor $m_{\ell^j}^2/m_{\ell^i}^2$ compared to those of the right-handed lepton $\ell^i_{R} \rightarrow \ell^j_{L} X$. 
This may induce an asymmetry that potentially allows to identify if 
 Higgs mediation is the dominant contribution to the LFV observables. Furthermore
this asymmetry would be 
more pronounced in the inverse seesaw framework.

It is important to stress that the numerical results summarised in  table~\ref{4} correspond to considering 
\textit{only} Higgs-mediated contributions. In the low $\tan \beta$ regime,
photon- and $Z^0$-penguin diagrams may induce comparable or even larger contributions to the observables, and potentially enhance the branching fractions. Thus, the results
for small $\tan \beta$ should be interpreted as
conservative estimates,  representing only partial
contributions. 
For large $\tan\beta$ values, Higgs penguins do indeed provide the leading contributions.
Comparing our results with those obtained for a
type I SUSY seesaw at high scales (or even with a TeV scale SUSY seesaw), 
we find a large enhancement
of the branching fractions in the inverse seesaw framework. 

\subsubsection{Z-mediated and other contributions}

Regarding other dipole contributions in the supersymmetric inverse seesaw, one would wonder if the $Z^0$-penguin could be enhanced or even dominate over the $\gamma$-penguin. For example, let us discuss the chargino-sneutrino 1-loop diagrams leading to $\ell_i \to 3
\ell_j$. The photon-penguin contribution can be written as
\begin{equation} \label{Achar}
A_a^{(c)L,R} = \frac{1}{16 \pi^2 m_{\tilde{\nu}}^2} \mathcal{O}_{A_a}^{L,R}s(x^2)\, ,  
\end{equation}
whereas the $Z^0$-contributions read
\begin{equation} \label{Fchar}
F_{X} = \frac{1}{16 \pi^2 g^2 \sin^2 \theta_W m^2_{Z}}\mathcal{O}_{F_X}^{L,R}t(x^2)\, ,
\end{equation}
with $X=\left\{LL,LR,RL,RR\right\}$.  In the above $\mathcal{O}_{y}^{L,R}$ denote combinations of rotation matrices and coupling
constants and $s(x^2)$ and $t(x^2)$ represent the Passarino-Veltman
loop functions which depend on $x^2 =
m_{\tilde{\chi}^-}^2/m_{\tilde{\nu}}^2$ (see
\cite{Arganda:2005ji}). Notice that the only mass scale involved in
the $A$ form factors is $m_\text{SUSY}$ (the photon being massless).
On the other hand, the mass scale in the $F_X$ form factors is set, in
this case, by the $Z^0$-boson mass. Since
$m_{Z}^{2} \ll m_\text{SUSY}^2$, the $Z^0$-penguin might, in principle,
dominate over the photon penguin.

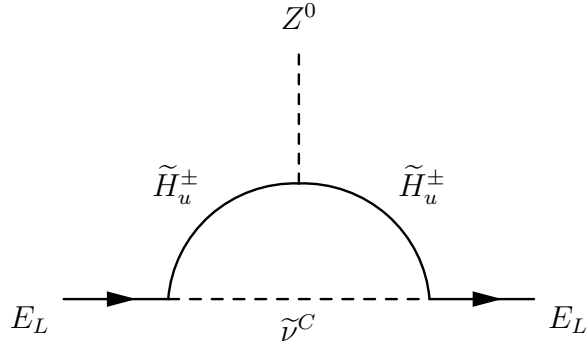
\begin{figure}[!t]
\centering
\begin{fmffile}{DiagZiss}
\begin{fmfgraph*}(220,120)
\fmflabel{$E_{L}$}{l1}
\fmflabel{$E_{L}$}{l2}
\fmflabel{$Z^{0}$}{phot}
\fmfforce{0.1w,0.1h}{l1}
\fmfforce{0.9w,0.1h}{l2}
\fmfforce{0.5w,0.9h}{phot}
\fmf{fermion}{l1,l1nu}
\fmf{dashes,tension=0.4,label=$\widetilde{\nu}^{C}$}{l1nu,nul2}
\fmf{fermion}{nul2,l2}
\fmffreeze
\fmf{plain,right=0.4,tension=0.6,label=$\widetilde{H}^{\pm}_u$}{Wphot,l1nu}
\fmf{plain,right=0.4,tension=0.6,label=$\widetilde{H}^{\pm}_u$}{nul2,Wphot}
\fmf{dashes}{Wphot,phot}
\end{fmfgraph*}
\end{fmffile}
\caption[Right-handed sneutrino contribution  to $Z^0$-penguins]{One of the right-handed sneutrino diagrams contributing  to $Z^0$-penguins in the supersymmetric inverse seesaw.} \label{DiagZiss}
\end{figure}

However, a specific cancellation appears between the different diagrams that contribute to the wino-sneutrino loop in the MSSM. This gives, at zeroth order in the chargino mixing angle, a combination of loop functions that does not depend on the masses of the particles running in the loops~\cite{Hirsch:2012ax}. One is then left with a Z-mediated contribution proportional to $( Z_V^\dagger Z_V)^{ij}$ where $Z_V$ is a $3 \times 3$ unitary
matrix that diagonalizes the mass matrix of the sneutrinos, which vanishes for $i \ne j$.  In conclusion, the first non-vanishing term in the expansion
appears at 2nd order in the chargino mixing angle, which naturally
leads to a suppression of $Z^0$-mediated contributions. This is the reason why
the photon contributions turn out to be dominant in the MSSM~\cite{Hisano:1995cp,Arganda:2005ji}. Nevertheless, in the supersymmetric inverse seesaw the right-handed sneutrino introduces new contributions like the one given by the diagram in fig.~\ref{DiagZiss}.
This new diagram can spoil the cancellation that occurs in the MSSM, leading to a large enhancement of the $Z^0$-mediated contributions~\cite{Abada:2012cq}. It was also realised in other low-scale (supersymmetric) seesaw models that, at the electroweak scale, the dominant contribution to $\mu-e$ conversion arises from box diagrams~\cite{Alonso:2012ji, Dinh:2012bp, *Ilakovac:2012sh}.

Since the singlet superfields couple to the up-type Higgs superfield, they might modify the phenomenology of the Higgs sector. We will illustrate this in the following sections, focusing on invisible decay channels of the NMSSM lightest pseudoscalar.

\section{Constraints on light CP-odd Higgs bosons}

Contrary to the MSSM, the NMSSM can admit a very light CP-odd Higgs boson $A_1$, with a mass $m_{A_1}\sim 1 - 10$~GeV~\cite{Maniatis:2009re, Ellwanger:2009dp}. This might open interesting phenomenological possibilities like new decays of the SM-like Higgs boson or new annihilation channels for the dark matter candidate. For example, if the lightest supersymmetric particle happens to be very light (a few GeV), then a light $A_1$ offers the possibility of $s$-channel LSP pair-annihilation into an on-shell $A_1$. 

In the NMSSM, the lightest CP-odd physical scalar $A_1$ can be decomposed as
\begin{equation}
\label{A1-def}
A_1 = \cos \theta_A A_\mathrm{MSSM} + \sin \theta_A A_S \, ,
\end{equation}
where $A_\mathrm{MSSM}$ is the MSSM part of the CP-odd scalar, which arises solely from the Higgs doublets, and $A_S$ is the part that arises from the new singlet superfield $\hat S$.  It is the singlet admixture, i.e.~the $\sin \theta_A $ projection, that allows the NMSSM pseudoscalar to be much lighter than what it could have been in the MSSM. On the other hand, if $A_1$ is very light, its detection crucially depends on its couplings to quarks and leptons, which depend on $\cos \theta_A$.  These couplings can be extracted from the following part of the Lagrangian~\cite{Dermisek:2010mg}:
\begin{equation}
  \label{cabbdef}
  \mathcal{L}_{{A_1}f\bar f} = X_{u(d)} \frac{g m_f}{2m_W}
  \bar f \gamma_5 f A_1 \, ,
\end{equation} 
where $g$ is the SU(2) gauge coupling,  $X_d (X_u) = \cos \theta_A \tan \beta \ (\cos \theta_A \cot \beta)$
for down-type (up-type) fermions and $\tan \beta$ is defined in eq.~(\ref{tanbeta}). However, experimental constraints put stringent bounds on the mass and couplings of the light pseudoscalar.

The constraints on $X_d$, defined in eq.~(\ref{cabbdef}), for $m_{A_1}$ approximately in the range of 1 to 10 GeV have been summarized in~\cite{Ellwanger:2009dp, Dermisek:2010mg}.  Measurements of $\Delta M_{d,s}$ (the $B_{d,s}-\bar B_{d,s}$ mass difference), $\mathcal{
  B}(\bar B \to X_s \gamma)$, $\mathcal{B}(B^+ \to \tau^+ \nu_\tau)$, and
particularly, $\mathcal{B}(\bar B_s \to \mu^+ \mu^-)$ severely constrain
$m_{A_1}$~\cite{Domingo:2007dx}.  The rates of these processes primarily
depend on the choice of $\tan\beta$ and $A_t$, the soft supersymmetry breaking
trilinear term. The constraints are in general weaker when these
parameters are small. Values of $m_{A_1}$ between 1 GeV and $m_b$ are
generally disfavoured from $B$-meson data~\cite{Ellwanger:2009dp}. Constraints on
$m_{A_1}$ also arise from radiative $\Upsilon$ decays
\cite{Love:2008aa, *Aubert:2009cp, *Aubert:2009cka}, namely, $\Upsilon (nS) \to \gamma A_1$,
$A_1 \to \mu^+\mu^- (\tau^+\tau^-)$ (further investigated and reviewed in
\cite{Dermisek:2010mg}). Severe constraints also arise as a consequence of
$\eta_b -A_1$ mixing~\cite{Drees:1989du, *Domingo:2009tb, *Domingo:2010am}.  The
different $m_{A_1}$ windows which are sensitive to different processes are listed
in table~\ref{cons}.
\begin{table}
\begin{center}
\begin{tabular}[ht]{|c|c|c|c|c|}
\hline
{\small Processes} &{\footnotesize$m_{A_1}<2m_{\tau}$} &\footnotesize
$[2m_{\tau}$,$ 9.2$ GeV$]$ &
\footnotesize $[9.2$ GeV,$M_{\Upsilon(1S)}]$
&\footnotesize $[M_{\Upsilon(1S)}$,$2m_B]$ \\
\hline
{ { \small $\Upsilon \to \gamma + (\mu^+ \mu^-, gg, s \bar s)$
}} 
& $\checkmark$
& $\times$ & $\times$ & $\times$  \\

{ { \small $\Upsilon \to \gamma \tau^+ \tau^-$ } }& $\times$ & 
$\checkmark$ &$\times$ & $\times$   \\

{{ \small $A_1$--$\eta_b$ mixing}} &$\times$  &$\times$  &$\checkmark$ &
$\times$     \\

{{ \small $e^+ e^- \rightarrow
 Z + 4\tau$}} &$\times$ & $\checkmark$  & $\checkmark$
& $\checkmark$    \\

{{ \small $e^+ e^- \rightarrow b \bar b \tau^+ \tau^-$}}  &$\times$  
&$\times$  &$\checkmark$ &$\checkmark$   \\
\hline
\end {tabular}
\caption[Constraints on the $A_1$ mass]{\small Various visible processes constraining different $m_{A_1}$ windows.
  The ``$\checkmark$'' symbol in a given entry attests the existence of 
  important or meaningful constraints from a given process, while the ``$\times$'' symbol implies otherwise. Reprinted from~\cite{Abada:2010ym} by the author, copyright 2011, with permission from Elsevier.}
\label{cons}
\end{center}
\end{table} 
The table also shows the ranges where the LEP (ALEPH~\cite{Schael:2010aw} for $e^+ e^- \rightarrow
 Z + 4\tau$  and OPAL~\cite{Abbiendi:2001kp, *Abbiendi:2002in} for $e^+ e^- \rightarrow b \bar b \tau^+ \tau^-$ and $e^+ e^- \rightarrow
 Z + 4\tau$) constraints are applicable. The
origin of all these constraints can be traced to the visible decay modes of
$A_1$. We should add that, since our study was completed, a new search based on dimuon decays of $A_1$ was published by the CMS collaboration adding constraints in the $0.25-3.55\;\mathrm{GeV}$ mass range~\cite{Chatrchyan:2012cg}.

However, the situation may dramatically change if $A_1$ has dominant invisible
decay modes.  Its decay into a pair of stable neutralinos (if kinematically
possible) is one example.  However, the BaBar Collaboration has searched for radiative $\Upsilon$-decays
where a large missing mass is accompanied by a monochromatic photon, and from
its non-observation has set a (preliminary) 90\% C.L. upper limit on $\mathcal{B}(\Upsilon(3S) \to \gamma A_1) \times \mathcal{B}(A_1 \to~\mathrm{invisible})$
at $(0.7-31)\times10^{-6}$ for $m_{A_1}$ in the range of 3 to 7.8 GeV~\cite{Aubert:2008as} and a 90\% C.L. upper limit on $\mathcal{B}(\Upsilon(3S) \to \gamma A_1) \times \mathcal{B}(A_1 \to~\mathrm{invisible})$ at $(1.9-4.5)\times10^{-6}$ for $m_{A_1}$ in the range of 0 to 8.0 GeV and $(2.7-37)\times10^{-6}$ for $m_{A_1}$ in the range of 8 to 9.2 GeV~\cite{delAmoSanchez:2010ac}.

This led us to consider the possibility of invisible decay channels
that would allow a light $A_1$ to escape detection even outside the range of 3 to
7.8 GeV.  We show that if we extend the NMSSM by two additional gauge singlets
with non-vanishing lepton numbers, they would not only provide a substantial
invisible decay channel for $A_1$ but, in addition, would also generate small
neutrino masses through lepton number violating ($\Delta L =2$) interactions.
Consequently, the visible decay branching ratios of $A_1$ would thus be reduced. As a
result, the constraints on $X_d$ would be weakened and a light $A_1$ could then be
comfortably accommodated.

\section{Embedding the inverse seesaw in the NMSSM}

The inverse seesaw mechanism can be embedded in the NMSSM by adding a pair of gauge singlet superfields $\widehat \nu^C$ and $\widehat X$, for each generation, which carry lepton numbers $L=-1$ and $L=+1$, respectively. This leads to the superpotential
\begin{equation}
 \widehat f = \widehat{f}_\mathrm{NMSSM}+\widehat{f}^\prime\,,
\end{equation}
where $\widehat{f}_\mathrm{NMSSM}$ is the NMSSM superpotential defined in eq.~(\ref{eq:SuperPotNMSSM}) and the additional terms involving the new singlet superfields are
\begin{equation}
 \widehat{f}^\prime= \varepsilon_{ab} Y_\nu^{ij}  \widehat \nu^C_i \widehat{L}_j^a \widehat{H}_u^b + (\lambda_{\nu})^i \widehat{S} \widehat \nu^C_i \widehat X_i + \frac{1}{2} \mu_X^i \widehat X_i \widehat X_i\,,\label{model-exp}
\end{equation}
with the generation indices ($i,j = 1,2,3$) and $Y_\nu$ the neutrino Yukawa coupling. The terms $\widehat \nu^C \widehat X$ and $\widehat X \widehat X$ can be written in a generation diagonal basis without
any loss of generality for our study. Once the scalar component of
$\widehat S$ acquires a vev $s$, not only the conventional $\mu$-term
is generated with $\mu = \lambda s$ (see discussion in Chapter~\ref{chapSUSY}), but a lepton number
  conserving mass term $M_{R}\overline{\nu_R}X$ is generated as well,
with $M_{R} = \lambda_{\nu} s$. Another lepton number
  conserving mass term $m_D \overline{\nu_L} \nu_R$ emerges, with $m_D =
Y_\nu v_u/\sqrt{2}$.

The crucial term relevant for the inverse seesaw is the $\Delta L = 2$
term involving $\mu_X$, which is the only mass dimensional term in the
superpotential. We assume that the $Z_3$ symmetry of the
superpotential is absent only in this term and treat $\mu_X$ as an extremely tiny effective mass parameter
generated by some unknown dynamics. From eqs.~(\ref{Mlight}, \ref{Mheavy}), one can see that if $\mu_X$ is sufficiently small then the heavy neutrinos can have masses around $10\;\mathrm{GeV}$ or lighter, potentially influencing the decay pattern of $A_1$.

We proceed to compute the branching ratios of $A_1$ into invisible modes
comprising the $\nu_L$, $\nu_R$ and the $X$ states.  Rigorously, one should first diagonalize the mass matrix of
eq.~(\ref{ISSmatrix}) to determine the physical neutrino states. However,
for our purpose it suffices to consider only one generation of fermions and estimate the branching fractions of $A_1$ into
the $\nu_L \nu_R$ and $\nu_R X$ interaction states.
Recall from eqs.~(\ref{A1-def}) and (\ref{model-exp}) 
that the decay of $A_1$ into $\nu_L \nu_R$
depends on how large the doublet component of $A_1$ is, i.e. on how large
$\cos \theta_A$ is, whereas the decay into $\nu_R X$ depends on the size of the $A_S$ component of $A_1$, i.e. on the magnitude of $\sin \theta_A$. From this, we can deduce the couplings between the lightest pseudoscalar and fermions
\begin{align}
 A_1\tau_L\tau_R:&\frac{\imath m_\tau}{v} \tan \beta \cos \theta_A\,,\label{couplingTau}\\
 A_1 t_L t_R:&-\frac{\imath m_t}{v \tan \beta} \cos \theta_A\,,\label{couplingTop}\\
 A_1 \nu_L \nu_R:&-\frac{\imath m_D}{v \tan \beta} \cos \theta_A\,,\label{couplingNuL}\\
 A_1 \nu_R X:&\frac{\imath M_R}{\sqrt{2} s}\sin\theta_A\,,\label{CouplingNuR}\\
 A_1 X X:&\frac{i\mu_X}{\sqrt{2} s} \sin\theta_A\,,\label{CouplingX}
\end{align}
where the $\sqrt{2}$ difference between couplings~(\ref{CouplingNuR}, \ref{CouplingX}) and the others comes from a $\sqrt{2}$ difference in the definition of $v$ and $s$ ($\langle S \rangle = s$ while $\langle H_u \rangle=v_u/\sqrt{2}$). Neglecting the phase space effects,  the branching ratios of the $A_1$ decays into invisible modes normalised to the visible ones are given by
\begin{eqnarray}
 \frac{\mathcal{B}\left(A_1 \rightarrow \nu_L \nu_R \right)}
{\mathcal{B}\left( A_1 \rightarrow 
f \bar f \right)+\mathcal{B}\left( A_1 \rightarrow c \bar{c} \right)} &\simeq& 
\frac{m_{D}^2}{m_{f}^2 \tan^4 \beta + m_{c}^2} \, , \label{a1nuN}\\
 \frac{\mathcal{B}\left( A_1 \rightarrow \nu_R X \right)}
{\mathcal{B}\left( A_1 \rightarrow 
f \bar f \right)+\mathcal{B}\left( A_1 \rightarrow c \bar{c} \right)} &\simeq& 
\tan^2\theta_A \frac{M_{R}^2}{m_{f}^2 \tan^2 \beta + m_{c}^2 \cot^{2} \beta} 
\frac{v^2}{2 s^2} \, . \label{a1sN}
\end{eqnarray}
Notice that the dominant visible decay modes of $A_1$ are $f\bar f\;(f=\mu, \tau, b)$
and $c \bar c$.  The $c \bar c$ mode is only numerically
relevant if $m_{A_1} < 2 m_b$ and $\tan \beta$ is small. Note that the
branching ratio into $\nu_R X$ dominates over that into
$\nu_L \nu_R$ for two reasons. First, there is a
$\tan^2\theta_A$ prefactor for the former which can be rather large if
$A_1$ has a dominant singlet component. Second, if the $m_f^2$ term
in the denominator of the branching ratio expressions is numerically
relevant, then the $\nu_L \nu_R$ channel suffers a suppression by
an additional $\tan^2\beta$ factor.

For a numerical illustration, we make two choices for $\tan \beta = (3,
20)$, and fix $\cos \theta_A = 0.1$, which yield $X_d = \cos \theta_A \tan \beta = (0.3, 2)$.
We recall that the upper limit on $X_d$ for $m_{A_1}<8$~GeV in the
minimal NMSSM has been obtained primarily from radiative
$\Upsilon$-decays, and the limit is between 0.7 to 3.0 for $\tan \beta =
50$, while it is 30 or above for $\tan \beta = 1.5$~\cite{Domingo:2008rr}.
A value of $X_d = 2$ is in fact slightly above the upper limit for
$m_{A_1}$ in the range of 4 to 8 GeV.  In the present scenario, $A_1$ has
a significant branching ratio into invisible modes which, in turn,
considerably relaxes the upper bound on $X_d$. Here, we do not choose a
very large value for $\tan \beta$ since that would increase the branching ratio
of $A_1$ into visible modes.  The value of $m_{A_1}$ is chosen to be
somewhat larger than $M_R$, so that the phase space suppression, given
by the factor 
\begin{equation}
\left(\frac{1-(\frac{2m_f}{m_{A_1}})^2}{1-(\frac{2M_R}{m_{A_1}})^2}\right)^{1/2}\,, 
\end{equation}
is not
numerically significant. We consider two values for $M_R = (5, 30)$
GeV. The rationale behind choosing $M_R = 5$ GeV is that it allows
to explore $m_{A_1} <10~$ GeV, a regime where constraints from $\Upsilon$-
and $B$-decays are particularly restrictive, see table
\ref{cons}. On the other hand, the choice $M_R = 30$ GeV implies that
${A_1}$ is moderately heavy ($m_{A_1} >30~$ GeV) which corresponds to the
range where LEP and $B$-decay constraints are relevant.  We display
our results in table~\ref{table2}.
\begin{table}[t]
\begin{center}\
\begin{tabular}{|c|c|c|c|c|c|}
\hline
\hline
&\multicolumn{2}{c|}{$\tan \beta = 20$, $\cos \theta_A = 0.1$} 
& \multicolumn{2}{c|}{$\tan \beta = 3$, $\cos \theta_A = 0.1$}\\
\hline
$M_R$ (GeV) & $5$ &$30$& $5$ &$30$ \\
\hline
\hline
$\mathcal{B}\left(A_1\rightarrow \nu_L \nu_R\right)$ 
& $ 7\times 10^{-5}$   &  $3\times 10^{-6}$   & $4\times 10^{-3}$  
&$1\times 10^{-4}$  \\
\hline 
$\mathcal{B}\left(A_1\rightarrow \nu_R X\right)$ 
& 0.7 & 0.9 & $\sim$ 1 & $\sim$ 1  \\
\hline
\hline
\end{tabular}
\end{center}
\caption[Invisible $A_1$ branching ratios]{\small Invisible branching ratios of the lightest NMSSM 
  pseudoscalar for $m_D=10$ GeV, $M_R = (5, 30)$ GeV and $\mu_X = 1$ eV. Reprinted from~\cite{Abada:2010ym} by the author, copyright 2011, with permission from Elsevier.}
\label{table2}
\end{table}
For numerical illustration, we have
assumed $s \sim \mathcal O(v)$.  

The main conclusion of this study is that if
$\cos \theta_A$ is small, $A_1$ has a dominant singlet component (which is
generally the case when $A_1$ is light~\cite{Ellwanger:2009dp}), then for a
significant part of the parameter space, $A_1$ can have a sizeable
invisible branching ratio which would weaken many of the constraints
discussed in the beginning of this section. However, it is important to stress that
$\cos \theta_A$ should not be excessively small, since in that case the purely
singlet $A_1$ would be completely decoupled from the visible sector.

The possibility of having a very light (of order 10
GeV) pseudoscalar in the NMSSM has many interesting phenomenological consequences. On the one hand, the additional fermionic singlets associated with the inverse seesaw provide a substantial invisible decay channel to
the lightest pseudoscalar which helps relaxing or even evading some of the
tight constraints from $\Upsilon$- and $B$-decays. On the other hand, they
naturally set up the stage for implementing the inverse seesaw mechanism in
order to generate light neutrino masses. Besides, the mixing between the MSSM
part of the CP-even Higgs and the singlet CP-even component and the
sizeable branching ratio of $A_1 \to ~\mathrm{invisible}$ opens the possibility of large branching ratios for invisible decays of the SM-like Higgs, which is constrained by the LHC and Tevatron measurements~\cite{Belanger:2013kya}.\\ \\

To summarise this chapter, should supersymmetry turn out to be realised in Nature, it would be necessary to extend minimal supersymmetric models in order to account for neutrino masses. Among other possibilities, this can be done by embedding the inverse seesaw mechanism in these frameworks, which is a very appealing perspective since all the new Physics can be at the TeV scale. In turn, this might lead to interesting phenomenological consequences like dominant invisible decays for the lightest pseudoscalar of the NMSSM or enhanced charged lepton flavour violating signals, within reach of the future generation of experiments. However, only some contributions have been considered in the supersymmetric inverse seesaw, and a complete study is required due to the contribution of the TeV-scale right-handed neutrinos to many processes. Finally, it would also be interesting to consider the interplay between the low-energy observables presented here and high-energy observables like  cLFV neutralino decays.

\chapter*{Conclusion}
\epigraph{The story so far:\\
In the beginning the Universe was created.\\
This has made a lot of people very angry and has been widely regarded as a bad move.}{\textit{The Restaurant at the End of the Universe}\\ \textsc{Douglas Adams}}

Neutrino oscillations call for the introduction of a neutrino mass generation mechanism. As we discussed in this thesis, an attractive possibility, among many others, is the inverse seesaw. Many reasons make the inverse seesaw mechanism very appealing.  First, it has a naturally low-scale which can make it testable at the LHC via the production of the heavy neutrinos. Second, the possibility of a comparatively low seesaw scale in association with sizeable Yukawa couplings can lead to a number of phenomenological effects such as lepton flavour violating processes, non-standard interactions and lepton universality violation. In turn, this can be used to constrain the size of the neutrino Yukawa couplings and the seesaw scale.

In~\cite{Abada:2012mc}, we have shown that in low-scale seesaw extensions of the Standard Model including fermionic singlets, the new heavy neutrinos can induce a violation of lepton flavour universality within reach of current experiments. A modified $W\ell\nu$ vertex leads to tree-level deviations from the SM prediction in observables mediated by a $W^\pm$ boson. This effect has two possible sources that are not mutually exclusive and depend on the heavy neutrino hierarchy. If the sterile neutrinos are lighter than the decaying particle but still much heavier than the three active neutrinos, then a difference in the kinematic factor of the processes appears leading to a phase space effect. On the contrary, if sterile neutrinos are heavier than the decaying particle, they cannot be produced in the final state and the non-unitarity of the $3\times 3$ submatrix that relates charged leptons and active neutrinos induces a deviation from lepton universality.

In~\cite{Abada:2011hm, Abada:2012cq}, we have considered a model where the inverse seesaw is embedded in the MSSM and addressed its impact on charged lepton flavour violating observables. We found that the presence of a TeV-scale right-handed neutrino opens the possibility of new chargino-sneutrino loops that contribute to cLFV observables. Moreover, the large neutrino Yukawa couplings, which are natural in the inverse seesaw mechanism, may enhance lepton flavour violation via a larger slepton mixing, which enhances cLFV branching ratios. In this thesis manuscript, we focused on Higgs-mediated contributions and showed that they can lead to branching ratios within reach of future experiments.

In~\cite{Abada:2010ym}, we have studied the phenomenological consequences of embedding the inverse seesaw in the NMSSM. More precisely, we focused on the possible new invisible decay channels of the lightest pseudoscalar $A_1$ that are opened when right-handed neutrinos have a mass around the GeV. Since most of the constraints on the mass and couplings of the lightest CP-odd Higgs boson are based on the assumption of a $100\%$ decay branching ratio into visible channels, the interesting possibility of dominant invisible decays weakens these constraints.

However, the complementarity of direct and indirect searches for new Physics behind neutrino mass generation is a vast topic. We are currently considering the effect of a modified $W\ell\nu$ vertex on observables other than $R_K$ and $R_\pi$. Besides, the $Z\nu\nu$ vertex can be modified in a similar fashion, leading to potentially interesting effects on observables like the invisible $Z^0$ decay width or the branching ratio of $K^+ \rightarrow \pi^+ \bar \nu \nu$ whose experimental uncertainty is expected to be reduced in the near future (NA62 and ORKA experiments). When in comes to cLFV in the supersymmetric inverse seesaw, we have considered new contributions (Higgs- and $Z^0$-mediated). Other contributions should be systematically taken into account into a future project.

As of today,  many problems of the SM remain unsolved. The discovery of the Higgs boson at the LHC appears to confirm the mechanism of EW symmetry breaking at work in the SM. However, the absence of signals of Physics beyond the Standard Model (other than neutrino oscillations) puts strong constraints on many new Physics scenarios. No dark matter candidate has been observed and no definitive explanation of the baryonic asymmetry of the Universe nor the family structure has been found. Particle Physics has entered an age where high precision tests of the SM and many experimental upcoming measurements will probe the SM to an unprecedented accuracy. In addition the search for new Physics much above the TeV scale via higher-order effects will also be carried on numerous fronts. Moreover, the measurement of a large $\theta_{13}$ mixing angle in the neutrino sector opens the possibility to determine the CP violating phase and resolve the neutrino mass hierarchy at current and future experiments. Due to  its specificity and its position at the intersection of the three frontiers, high-energy, high-intensity and cosmology, the neutrino sector offers exciting prospects for the years to come.

\bibliographystyle{myrsc}
\bibliography{/home/cedric/Dropbox/Articles/References}

\appendix
\chapter{Synopsis}

De nos jours, les neutrinos sont les fermions les plus abondants de l'Univers. Cela ne les empêche pas de demeurer des particules mystérieuses dont les propriétés sont encore mal connues. Par exemple, l'existence d'une masse non nulle a été confirmée indirectement par l'observation d'oscillations entre différentes saveurs plus de trente ans après la découverte du neutrino. Cependant, leur masse n'a pu être directement mesurée pour l'instant malgré une forte activité expérimentale dans ce domaine. 

De nombreux modèles ont été proposés pour expliquer le mécanisme par lequel les neutrinos acquièrent une masse. Durant cette thèse, nous nous sommes concentrés sur les extensions du Modèle Standard mettant en jeu des fermions singulets de jauge. En particulier, nous nous sommes intéressés au mécanisme de seesaw inverse dont nous avons étudié les conséquences sur des observables de basse et de haute énergie.

Ce synopsis suivra la structure du manuscrit principal qui a été rédigé en anglais. Ainsi, dans une première section, nous décrirons le Modèle Standard de la physique des particules qui forme le socle de notre compréhension du monde subatomique. La section suivante sera consacrée au secteur des neutrinos avec une résumé de la théorie des oscillations et des mécanismes générant les masses des neutrinos. La troisième section détaillera la paramétrisation de la matrice de mélange leptonique tandis que la quatrième section listera les contraintes qui pèsent sur l'existence de neutrinos stériles. La cinquième présentera l'effet du seesaw inverse sur différentes tests de l'universalité leptonique. La sixième partie est constituée d'une courte introduction à la supersymétrie alors que la septième introduira les modèles de seesaw inverse supersymétriques. Enfin, des conséquences du seesaw inverse supersymétriques sur les observables violant la saveur leptonique et les contraintes sur l’existence d'un boson de Higgs CP-impaire léger.

\section{Le Modèle Standard}

Le Modèle Standard (MS) de la physique des particules est l'une des théories les mieux testées en physique. Fondé sur la mécanique quantique et la relativité restreinte, il décrit le monde subatomique dans un cadre cohérent utilisant les formalismes issus de la théorie quantique des champs et des théories de jauge. Une brève introduction historique se trouve dans la section~\ref{SMhistory}.

En tant que théorie de jauge, toutes les interactions du Modèle Standard sont déterminées par les symétries sous-jacentes. L'invariance vis-à-vis d'une transformation locale est préservée par l'introduction de champs de jauge vectoriels réels et d'une dérivée covariante qui remplace la dérivée ordinaire. En notant $t^a\,(a=1,...,N)$ les matrices hermitiennes formant une représentation unitaire des $N$ générateurs de l'algèbre de Lie associée à la transformation et $A_\mu^a\,(a=1,...,N)$ les bosons de jauge, la dérivée covariante se note
\begin{equation}
 D_\mu=\partial_\mu+\imath g t^a A_\mu^a\,,
\end{equation}
avec $g$ est une constante de couplage réelle. Le champs de jauge est alors rendu dynamique par l'introduction d'un tenseur
\begin{equation}
 F_{\mu\nu}^a=\partial_\mu A^a_\nu - \partial_\nu A^a_\mu - g f^{abc} A_\mu^b A_\nu^c\,.
\end{equation}
Il est important de noter que l'invariance du Lagrangien sous des transformations de jauge locales interdit également la présence de termes de masse explicites pour les bosons vecteurs.

Le groupe de jauge du Modèle Standard est $\mathrm{SU}(3)_c\times\mathrm{SU}(2)_L\times\mathrm{U}(1)_Y$ et correspond à deux secteurs indépendants. Le premier décrit l'interaction forte via le groupe $\mathrm{SU}(3)_c$, où $c$ signifie couleur. Il s'agit de la charge élémentaire en Chromodynamique Quantique. Les interactions faibles et électromagnétiques sont, quant à elles, unifiées dans le second secteur dont le groupe de jauge est $\mathrm{SU}(2)_L\times\mathrm{U}(1)_Y$. La théorie électrofaible viole maximalement la parité puisqu'elle n'agit que sur les particules de chiralité gauche, ce que l'indice $L$ dénote. Enfin, l'hypercharge $Y$ d'un champ est reliée à sa charge électrique $Q$ par la relation de Gell-Mann--Nishijima
\begin{equation}
\label{GMNf}
 Q=I_3+\frac{Y}{2}\,,
\end{equation}
avec $I_3$ la troisième composante de l'isospin faible. Comme nous le verrons ci-dessous, la symétrie électrofaible est brisée à basse énergie en $\mathrm{U}(1)_{em}$, ce qui permet aux bosons vecteurs de l'interaction faible d'acquérir une masse tandis que le photon reste non-massif.

Le contenu en particule du Modèle Standard est tout d'abord dicté par son groupe de jauge. Les bosons vecteurs correspondant aux différents générateurs sont regroupés dans la table~\ref{SMbosonsF}
\begin{table}[tb]
  \begin{center}
    \begin{tabular}{|c|c|c|c|c|}
     \hline
	Champ & Masse (GeV) & Spin & Charge élec. & Rep. de $\mathrm{SU}(3)_c$\\
     \hline
     \hline
	$G$ & $0$ & $1$ & $0$ & $\mathbf{8}$\\
     \hline
	$W^\pm$ & $80.385 \pm 0.015$ & $1$ & $\pm1$ & $\mathbf{1}$\\
	$Z^0$ & $91.1876 \pm 0.0021$ & $1$ & $0$ & $\mathbf{1}$\\
     \hline
	$\gamma$ & $0$ & $1$ & $0$ & $\mathbf{1}$\\
     \hline
	$\phi^0$ & $\begin{array}{lr} ATLAS: & 125.5\pm0.7\\ CMS: & 125.8\pm0.6\end{array}$ & $0$ & $0$ & $\mathbf{1}$\\
     \hline
    \end{tabular}
    \caption[Contenu bosonique du MS]{\label{SMbosonsF} Contenu bosonique du Modèle Standard après brisure de la symétrie électrofaible. Les masses sont extraites du PDG~\cite{Beringer:1900zz}, à l'exception de la masse du boson de ``Higgs''~\cite{ATLAS-CONF-2013-014, *CMS-PAS-HIG-13-002}.}
 \end{center}
\end{table}
avec le boson de Higgs, responsable de la brisure de symétrie électrofaible qui sera décrite ci-dessous. Les fermions appartiennent, quant à eux, aux représentations unitaires irréductibles des différents groupes de Lie et peuvent donc être classés selon la représentation à laquelle ils appartiennent, comme dans la table~\ref{SMfermionsF}.
\begin{table}[tb]
  \begin{center}
    \begin{tabular}{|c|c|c|c|c|}
     \hline
	Champ & Masse (GeV) & Charge élec. & $I_3$ & Rep. de $\mathrm{SU}(3)_c$\\
     \hline
     \hline
	$\nu_e$ & $<2\times 10^{-9}$ & $\phantom{-}0$ & $\phantom{-}1/2$ & $\mathbf{1}$\\
	$e$ & $(5.10998928 \pm 0.00000011)\times 10^{-4}$ & $-1$ & $-1/2$ & $\mathbf{1}$\\
     \hline
	$\nu_\mu$ & $<1.9\times 10^{-4}$ & $\phantom{-}0$ & $\phantom{-}1/2$ & $\mathbf{1}$\\
	$\mu$ & $(1.056583715 \pm 0.000000035)\times 10^{-1}$ & $-1$ & $-1/2$ & $\mathbf{1}$\\
     \hline
	$\nu_\tau$ & $<1.82\times10^{-2}$ & $\phantom{-}0$ & $\phantom{-}1/2$ & $\mathbf{1}$\\
	$\tau$ & $1.77682 \pm 0.00016$ & $-1$ & $-1/2$ & $\mathbf{1}$\\
     \hline
     \hline
	$u$ & $(2.27 \pm 0.14)\times 10^{-3}\,(\overline{\mathrm{MS}})$ & $\phantom{-}2/3$ & $\phantom{-}1/2$ & $\mathbf{3}$\\
	$d$ & $(4.78 \pm 0.09)\times 10^{-3}\,(\overline{\mathrm{MS}})$ & $-1/3$ & $-1/2$ & $\mathbf{3}$\\
     \hline
	$c$ & $1.275 \pm 0.004\,(\overline{\mathrm{MS}})$ & $\phantom{-}2/3$ & $\phantom{-}1/2$ & $\mathbf{3}$\\
	$s$ & $(9.43 \pm 0.12)\times 10^{-2}\,(\overline{\mathrm{MS}})$ & $-1/3$ & $-1/2$ & $\mathbf{3}$\\
     \hline
	$t$ & $173.5 \pm 1.0$ & $\phantom{-}2/3$ & $\phantom{-}1/2$ & $\mathbf{3}$\\
	$b$ & $4.18 \pm 0.03\,(\overline{\mathrm{MS}})$ & $-1/3$ & $-1/2$ & $\mathbf{3}$\\
     \hline
    \end{tabular}
    \caption[Contenu fermionique du MS]{\label{SMfermionsF} Le contenu fermionique du Modèle Standard. Toutes les masses sont extraites du PDG~\cite{Beringer:1900zz}.}
 \end{center}
\end{table}
Il est néanmoins important de noter que le nombre de générations n'est pas contraint dans le Modèle Standard mais que le nombre de fermions par génération ainsi que la représentation à laquelle ils appartiennent assurent l'absence d'anomalie. De plus, tous les fermions ont deux chiralités, gauche et droite, à l'exception du neutrino qui ne possède pas de composant droit, le rendant non-massif\footnote{Cela est néanmoins en contradiction avec l'observation expérimentale du phénomène d'oscillation entre différentes saveurs qui sera discuté dans la partie suivante.} par construction.

Au début des années 1960, Yoichiro Nambu~\cite{PhysRevLett.4.380} et Jeffrey Goldstone~\cite{Goldstone1961} réalisèrent qu'une théorie peut ne pas respecter une symétrie conservée par son Lagrangien. En effet, l'état fondamental peut ne pas être invariant sous cette symétrie. Cette idée appliquée aux théories de jauge non-abéliennes est le fondement du mécanisme de Higgs. Un doublet scalaire $\phi$ évoluant dans un potentiel
\begin{equation}
 V(\phi)=\mu^2 \phi^\dagger \phi + \lambda (\phi^\dagger \phi)^2\,,
\end{equation}
admet un état fondamental qui brise l'invariance sous $\mathrm{SU}(2)_L$ si $\mu^2 < 0$. Il prend alors une valeur moyenne dans le vide 
\begin{equation}
 \langle \phi \rangle = \frac{1}{\sqrt{2}} \binom{0}{v}\,,
\end{equation}
avec
\begin{equation}
 v=\sqrt{\frac{-\mu^2}{\lambda}}\,.
\end{equation}
Par le biais de son couplage aux bosons de jauge électrofaibles et aux fermions, il va alors générer des masses non-nulles pour ceux-ci.

Ayant décrit les principaux mécanismes et composants du Modèle Standard, il devient possible d'écrire son Lagrangien avant brisure de la symétrie électrofaible. Les tenseurs associées aux huit gluons $G^a_\mu\,(a=1,...,8)$, aux trois bosons de jauge $W^a_\mu\,(a=1,2,3)$ provenant de $\mathrm{SU}(2)_L$ et au boson vecteur de $\mathrm{U}(1)_Y$, noté $B_\mu$, sont décrits par
\begin{eqnarray}
 G^a_{\mu\nu}&=&\partial_\mu G^a_\nu - \partial_\nu G^a_\mu - g_s f^{abc} G_\mu^b G_\nu^c\,,\\
 W_{\mu\nu}^a&=&\partial_\mu W^a_\nu - \partial_\nu W^a_\mu - g \varepsilon^{abc} W_\mu^b W_\nu^c\,,\\
 B_{\mu\nu}&=&\partial_\mu B_\nu - \partial_\nu B_\mu\,,
\end{eqnarray}
avec $g_s$ et $g$ les constantes de couplage de $\mathrm{SU}(3)_c$ et $\mathrm{SU}(2)_L$, respectivement, $f^{abc}=-\imath \mathrm{Tr}([\lambda_a,\lambda_b]\lambda_c)/4$ et $\varepsilon^{abc}$ le tenseur de Levi-Civita en trois dimensions. Dans un souci d'exhaustivité, le constante de couplage de $\mathrm{U}(1)_Y$ sera notée $g\prime$. Il est alors possible de définir la dérivée covariante comme
\begin{equation}
 D_\mu=\partial_\mu+\imath g_s \frac{\lambda^a}{2} G_\mu^a+\imath g \frac{\sigma^a}{2} W_\mu^a+\imath g\prime \frac{Y}{2} B_\mu\,.
\end{equation}
Le Lagrangien du Modèle Standard peut être décomposé selon
\begin{equation}
 \mathcal{L}_{SM} = \mathcal{L}_{gauge} + \mathcal{L}_{matter} + \mathcal{L}_{Higgs} + \mathcal{L}_{Yukawa}\,,
\end{equation}
où, en notant $L=\binom{\nu_L}{e_L}$, $Q=\binom{u_L}{d_L}$ et $\phi=\binom{\phi^+}{\phi^0}$ les doublets de $\mathrm{SU}(2)_L$, les différentes contributions sont données par
\begin{eqnarray}
 \mathcal{L}_{gauge} &=& -\frac{1}{4} G^a_{\mu\nu} G^{a\mu\nu} -\frac{1}{4} W^a_{\mu\nu} W^{a\mu\nu} -\frac{1}{4} B_{\mu\nu} B^{\mu\nu}\,,\\
 \mathcal{L}_{matter} &=& \imath \sum_{i=e,\mu,\tau} \overline{L_i} \slashed{D} L_i + \imath \sum_{i=1,2,3} \overline{Q_i} \slashed{D} Q_i +  \imath \sum_{i=e,\mu,\tau} \overline{\ell_{Ri}} \slashed{D} \ell_{Ri}\\ \nonumber
    & & + \imath \sum_{i=u,c,t} \overline{q_{Ri}} \slashed{D} q_{Ri} + \imath \sum_{i=d,s,b} \overline{q_{Ri}} \slashed{D} q_{Ri}\,,\\
 \mathcal{L}_{Higgs} &=& (D^\mu \phi)^\dagger (D_\mu \phi) - \mu^2 \phi^\dagger \phi - \lambda (\phi^\dagger \phi)^2\,,\label{HiggsSMf}\\
 \mathcal{L}_{Yukawa} &=& -\sum_{i,j=e,\mu,\tau} \left( Y^{ij}_\ell \overline{L_i} \phi e_{Rj} + h.c. \right) -
\sum_{i=1,2,3,j=u,c,t} \left( Y^{ij}_u \overline{Q_i} \widetilde{\phi} q_{Rj} + h.c. \right)\\ \nonumber
    & & - \sum_{i=1,2,3,j=d,s,b} \left( Y^{ij}_d \overline{Q_i} \phi q_{Rj} + h.c. \right)\,,
\end{eqnarray}
avec $\widetilde{\phi}=\imath \sigma_2 \phi^*$.

Le neutrino possède un statut particulier parmi les fermions du Modèle Standard. En effet, il est le seul à n'interagir que par interactions faibles, n'admet qu'une chiralité gauche et possède une masse au moins six ordres de grandeur plus faible que celles des autres fermions. De plus, il est électriquement neutre, ce qui lui ouvre la possibilité d'être un fermion de Majorana. Les différents phénomènes expérimentaux impliquant des neutrinos sont décrits dans le Chapitre~\ref{ChapNuExp}. Nous nous concentrerons ici sur les oscillations de neutrinos ainsi que sur la génération de leurs masses.

\section{Des oscillations de saveurs aux neutrinos massifs}

Durant les années 1970, la possibilité d'oscillations entre différents saveurs de neutrino apparut dans l'expérience d'Homestake~\cite{Cleveland:1998nv}. Cette hypothèse fut confirmée en 1998 par la collaboration Super-Kamiokande~\cite{Fukuda:1998mi} qui mit en évidence un déficit de neutrinos muoniques dépendant de la distance zénithale compatible avec une oscillation entre deux saveurs $\nu_\mu \rightarrow \nu_\tau$. Plus tard, les expériences SNO~\cite{Ahmad:2001an, *Ahmad:2002jz} et KamLAND~\cite{Araki:2004mb} confirmeront l'existence d'oscillations entre d'autres saveurs.

L'explication la plus simple à ce phénomène est de considérer que les neutrinos sont des particules massives dont les états propres de masse ne sont pas alignés avec ceux des leptons chargés. Cela se traduit par l'apparition d'une matrice de mélange dans les courants chargés
\begin{equation}
 J^\mu_W=\frac{1}{\sqrt{2}}\overline{\nu_i}U^*_{ji}\gamma^\mu P_L \ell_j\,,
\end{equation}
avec $U$ la matrice de Pontecorvo-Maki-Nakagawa-Sakata~(PMNS) et $\nu_i$, $\ell_j$ les états propres de masse. Il est possible de définir les états propres de saveur comme les neutrinos associés à la transition $\ell_\alpha^- \rightarrow \nu_\alpha$, dans la base où les leptons chargés sont diagonaux. Les états propres de saveur sont alors reliés aux états propres de masse par la relation
\begin{equation}
 |\nu_\alpha \rangle = \sum_{i=1}^3 U^*_{\alpha i} |\nu_i \rangle\,,
\end{equation}
quand les neutrinos sont décrits par des ondes planes. Si les états propres de masses ne sont pas dégénérés, les ondes planes associées ont une évolution temporelle différente via l'équation de Schrödinger. Par conséquent, leur superposition va différer de l'état propre de saveur initial, générant les oscillations entre différentes saveurs. Par exemple, la probabilité de transition pour des neutrinos ultra-relativistes est donnée par~\cite{giunti2007fundamentals}
\begin{equation}
  P_{\nu_\alpha \rightarrow \nu_\beta}(L,E)=\sum_{k,j=1}^{3} U^*_{\alpha k} U_{\beta k} U_{\alpha j} U^*_{\beta j} \mathrm{exp}\left(-\imath \frac{\Delta m^2_{kj}L}{2E}\right)\,,\label{NuOscillationF}
\end{equation}
avec $L$ la distance entre la source et le détecteur, $E\simeq|\vec{p}|$ l'énergie du neutrino et la différence de masse au carré
\begin{equation}
 \Delta m^2_{kj} = m^2_k - m^2_j\,.
\end{equation}
L'équation~(\ref{NuOscillationF}) indique clairement que le probabilité de transition est non-nulle uniquement si un mélange existe et que la différence de masse au carré est non-nulle.

\begin{figure}[!t]
\centering
 \includegraphics[width=0.8\textwidth]{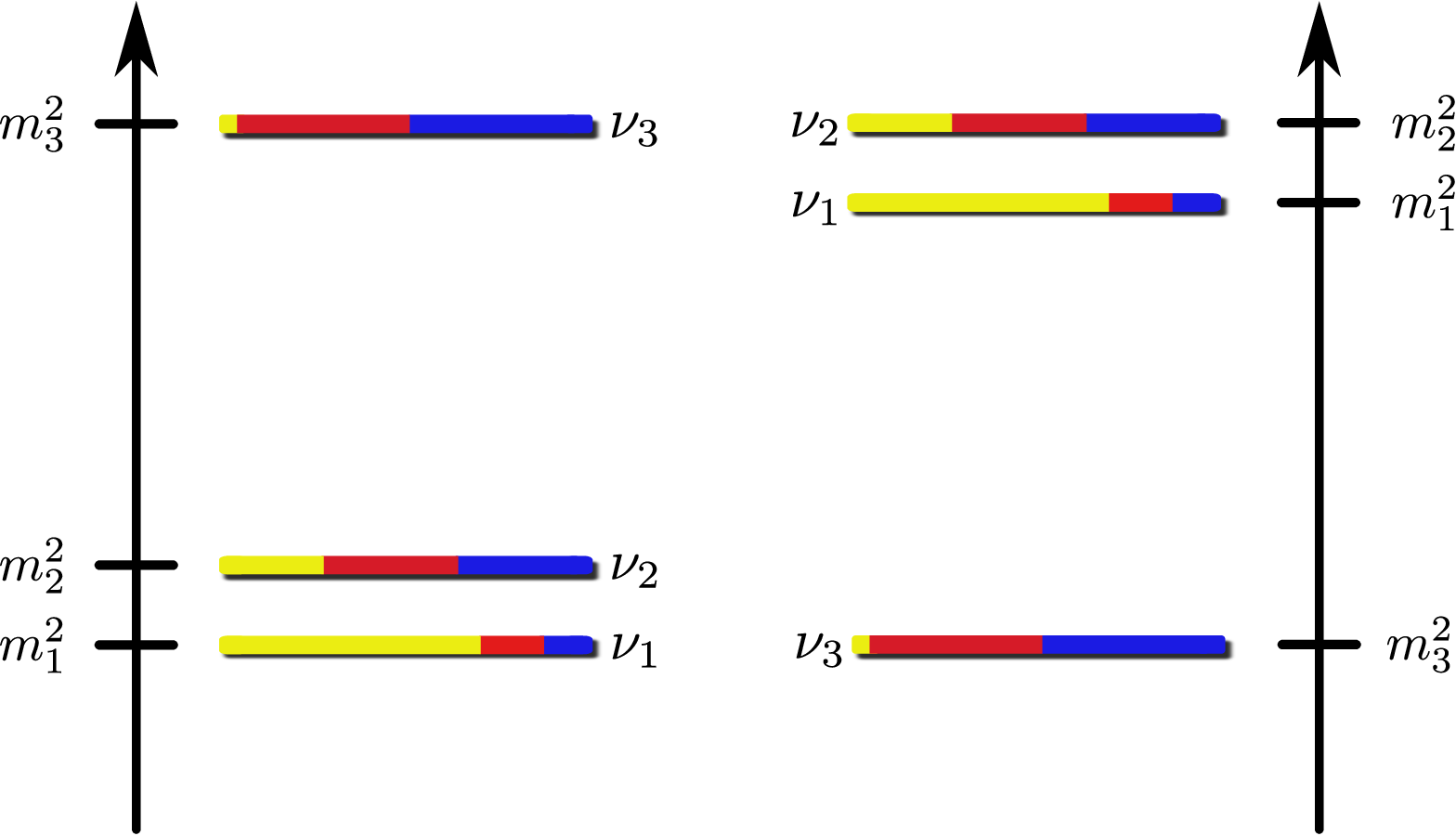}
\caption[Hiérarchies des neutrinos]{Composition des états propres de masse et spectre des neutrinos dans les hiérarchies normale (gauche, NH) et inverse (droite, IH). Pour chaque état propre, la contribution du neutrino électronique est jaune, celle du neutrino muonique est rouge et celle du neutrino tau est bleue.  Figure par kismalac sur Wikimedia Commons.}
\label{NeutrinoHierarchyF}
\end{figure}

La matrice PMNS peut être paramétrée selon
\begin{equation}
 U_{PMNS}= U_D \times \mathrm{diag}(1\,, e^{\imath \alpha_{21}/2}\,, e^{\imath \alpha_{31}/2})\,,
\end{equation}
où  $\alpha_{21}$ et $\alpha_{31}$ sont deux phases de Majorana qui sont absentes si les neutrinos sont des fermions de Dirac. La partie de la matrice PMNS qui est similaire à la matrice CKM s'écrit
\begin{equation}
U_D=\left(\begin{array}{c c c} c_{13}c_{12} & c_{13}s_{12} & s_{13}e^{-i\delta}\\ -c_{23}s_{12}-s_{23}s_{13}c_{12}e^{i\delta} & c_{23}c_{12}-s_{23}s_{13}s_{12}e^{i\delta} & s_{23}c_{13}\\ s_{23}s_{12}-c_{23}s_{13}c_{12}e^{i\delta} & -s_{23}c_{12}-c_{23}s_{13}s_{12}e^{i\delta} &
c_{23}c_{13}\end{array}\right)\,.
\label{PMNSf}
\end{equation}
avec $c_{ij}=\cos \theta_{ij}$, $s_{ij}=\sin \theta_{ij}$ et $\delta$ une phase violant CP. Il est intéressant de noter que la probabilité de transition est indépendant des phases $\alpha_{21}$ et $\alpha_{31}$  et ne permet donc pas de tester le caractère Dirac ou Majorana du neutrino. De nos jours, tous les paramètres déterminant les oscillations de neutrinos ont été mesurés expérimentalement, à l'exception de la phase $\delta$. La collaboration NuFIT a publié un ajustement global des paramètres aux différentes expériences~\cite{GonzalezGarcia:2012sz} 
\begin{align}
 \sin^2\theta_{12}&=0.306^{0.012}_{-0.012}\,, 					 & \Delta m^2_{12}&=7.45^{+0.19}_{-0.16}\times10^{-5}\;\mathrm{eV}^2\,,\nonumber\\
 \sin^2\theta_{23}&=0.437^{+0.061}_{-0.031}\,, 					& \Delta m^2_{32}&=-2.410^{+0.062}_{-0.063}\times10^{-3}\;\mathrm{eV}^2\;(IH)\,,\nonumber\\
\sin^2\theta_{13}&=0.0231^{+0.0023}_{-0.0022}\,,				& \Delta m^2_{31}&=+2.421^{+0.022}_{-0.023}\times10^{-3}\;\mathrm{eV}^2\;(NH)\,,
\end{align}
où NH et IH correspondent respectivement aux hiérarchies normale et inverse du spectre des neutrinos. Ces deux hiérarchies sont représentées dans la figure~\ref{NeutrinoHierarchyF}.

L'observation d'oscillations de neutrinos implique qu'au moins deux neutrinos sont massifs. Néanmoins, le Modèle Standard ayant été construit comme un modèle minimal, il ne contient donc pas de neutrino droit. Cela interdit l'écriture d'une masse de Dirac
\begin{equation}
 m(\overline{\nu_L}\nu_R+\overline{\nu_R}\nu_L)\,.
\end{equation}
Le neutrino étant un fermion électriquement neutre, il peut vérifier la condition de Majorana
\begin{equation}
 \psi=\psi^C\,,
\end{equation}
avec \begin{equation}
 \psi^C=\xi \mathcal{C} \overline{\psi}^T\,,\label{ChargeConjugationF}
\end{equation}
où $\mathcal{C}$ est la matrice de conjugaison de charge et $\xi$ est une phase arbitraire qui peut être réabsorbée par une redéfinition du champ $\psi$. Une masse de Majorana s'écrit alors \begin{equation}
 \frac{1}{2} m (\overline{\psi_L^C}\psi_L + \overline{\psi_L} \psi_L^C)\,.\label{MajoranaMassF}
\end{equation}
Cependant, dans le Modèle Standard, le neutrino appartient à un doublet de $\mathrm{SU}(2)_L$ d'hypercharge $Y=-1$. Par conséquent, un terme $\bar L L^C$ n'est pas invariant de jauge et le Lagrangien du Modèle Standard ne peut pas contenir de terme de masse pour le neutrino.

À l'instar des autres fermions, il est possible d'étendre le Modèle Standard par l'ajout de neutrinos droits et du terme de Yukawa correspondant. Cependant, le neutrino droit est un singulet de jauge et aucune symétrie n'interdit la présence d'une masse de Majorana pour celui-ci. En considérant trois neutrinos droits, les termes qui vont générer la masse des neutrinos s'écrivent avant brisure de le symétrie électrofaible
\begin{equation}
 \mathcal{L}_{\mathrm{type\;I}}= -\sum_{i,j} \left( Y^{ij}_\nu \overline{L_i} \widetilde{\phi} \nu_{Rj} + \frac{1}{2} M_{R}^{ij} \overline{\nu_{Ri}^C} \nu_{Rj} + h.c. \right)\,,
\end{equation}
avec $i$ et $j$ les indices de saveur qui vont de 1 à 3, $Y_\nu$ la matrice complexe de Yukawa des neutrinos et $M_R$ une matrice complexe symétrique. Après brisure de symétrie électrofaible, ces termes se réduisent dans la base $N_L=(\nu_{L1},...,\nu_{L3}\,,\;\nu_{R1}^C,...,\nu_{R3}^C)^T$ à
\begin{equation}
\mathcal{L}_{\mathrm{type\;I}}= -\frac{1}{2} \overline{N_L^C} M_{\mathrm{type\;I}} N_L + h.c. = -\frac{1}{2} \overline{N_L^C} \left(\begin{array}{c c} 0 & m_D\\ m_D^T & M_R\end{array}\right) N_L + h.c.\,,
\end{equation}
où $m_D=Y_\nu v / \sqrt{2}$ ce qui correspond dans la limite $M_R \gg m_D$ au seesaw de type I~\cite{Minkowski:1977sc, *GellMann:1980vs, *yanagida1979horizontal, *Mohapatra:1979ia}. La matrice de masse des neutrinos $M_{\mathrm{type\;I}}$ peut alors être diagonalisée par bloc ce qui donne
\begin{equation}
 M_{\mathrm{light}}\simeq -m_D M_R^{-1} m_D^T\,,\quad\quad M_{\mathrm{heavy}}\simeq M_R\,.
\end{equation}
Une masse des neutrinos légers, majoritairement composés des champs gauches, de l'ordre de l'électron-volt est alors générée en prenant $M_R\sim10^{15}\;\mathrm{GeV}$ et $Y_\nu\sim\mathcal{O}(1)$. Une autre possibilité est de considérer une échelle de seesaw de l'ordre du TeV avec un couplage de Yukawa $Y_\nu\sim10^{-6}$. Cependant, dans le seesaw de type I, les opérateurs générant la masse des neutrinos et les signatures de basse énergie comme la violation de saveur leptonique sont corrélés. La nécessaire suppression de la masse des neutrinos légers va alors réduire l'amplitude des observables de basse énergie. De plus, cette suppression va également réduire la section efficace des neutrinos lourds, rendant leur production et détection au LHC difficile.

\begin{figure}
\begin{center}
\begin{fmffile}{ISS}
\begin{fmfgraph*}(200,120)
\fmflabel{$H$}{h1}
\fmflabel{$L$}{l1}
\fmflabel{$H$}{h2}
\fmflabel{$L$}{l2}
\fmfbottom{l1,l2}
\fmftop{h1,t1,t2,h2}
\fmfv{decor.shape=cross, decor.size=6thick, label=$\mu_X$, label.angle=90}{middle}
\fmfv{decor.shape=cross, decor.size=6thick, label=$M_R$, label.angle=90}{h1h3}
\fmfv{decor.shape=cross, decor.size=6thick, label=$M_R$, label.angle=90}{h2h4}
\fmf{fermion}{l1,h1l1}
\fmf{fermion}{l2,h2l2}
\fmf{fermion,label=$\nu_R$, label.side=left}{h1l1,h1h3}
\fmf{fermion,label=$\nu_R$}{h2l2,h2h4}
\fmf{fermion,tension=1.35,label=$X$}{h1h3,middle}
\fmf{fermion,tension=1.35,label=$X$, label.side=left}{h2h4,middle}
\fmf{dashes}{h1,h1l1}
\fmf{dashes}{h2,h2l2}
\fmffreeze
\end{fmfgraph*}
\end{fmffile}\\
\caption[Mécanisme de seesaw inverse]{Diagramme générant la masse des neutrinos légers dans le seesaw inverse.}
\label{ISSf}
\end{center}
\end{figure}
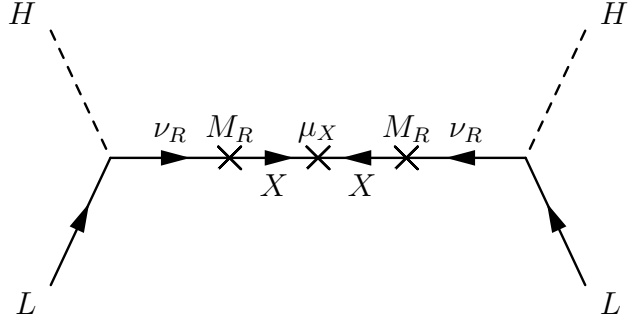

Il est néanmoins possible de découpler ces opérateurs dans le mécanisme de seesaw inverse~\cite{Mohapatra:1986aw, *Mohapatra:1986bd, *Bernabeu1987303}. Le Modèle Standard est alors étendu par l'ajout de deux types de singulets fermioniques de même nombre leptonique. Ce mécanisme et l'étude de sa phénoménologie dans différents modèles ont été au c{\oe}ur de ma thèse. Décrit par le diagramme~\ref{ISSf},
il ajoute au Lagrangien du MS les termes suivants
\begin{equation}
 \mathcal{L}_\mathrm{ISS} = - \sum_{i,j} \left( Y^{ij}_\nu \overline{L_i} \widetilde{\phi} \nu_{Rj} + M_R^{ij} \overline{\nu_{Ri}} X_j + \frac{1}{2} \mu_{R}^{ij} \overline{\nu_{Ri}^C} \nu_{Rj} + \frac{1}{2} \mu_{X}^{ij} \overline{X_{i}^C} X_{j} + h.c. \right)\,,
\label{ISSlagrangianF}
\end{equation}
avec $Y_\nu$ le couplage de Yukawa des neutrinos, $M_R$ une matrice de masse complexe qui conserve le nombre leptonique et $\mu_X$, $\mu_R$ deux matrices de masse de Majorana complexes qui violent la conservation du nombre leptonique par deux unités. Après brisure de la symétrie électrofaible, la matrice de masse des neutrinos est donnée dans la base $(\nu_L\,,\;\nu_R^C\,,\;X)$ par
\begin{equation}
 M_{\mathrm{ISS}}=\left(\begin{array}{c c c} 0 & m_D & 0 \\ m_D^T & \mu_R & M_R \\ 0 & M_R^T & \mu_X \end{array}\right)\,.\label{ISSmatrixF}
\end{equation}
Pour simplifier la discussion, cette matrice de masse peut être diagonalisée en ne considérant qu'une seule génération. Dans la limite $\mu_R\,,\mu_X \ll m_D, M_R$, les valeurs propres sont
\begin{align}
 m_1 &= \frac{m_{D}^2}{m_{D}^2+M_{R}^2} \mu_X\,,\label{MlightF}\\
 m_{2,3} &= \pm \sqrt{M_{R}^2+m_{D}^2} + \frac{M_{R}^2 \mu_X}{2 (m_{D}^2+M_{R}^2)} + \frac{\mu_R}{2}\,.\label{MheavyF}
\end{align}
Dans le cas du seesaw inverse, la petitesse de la masse des neutrinos est reliée à la petitesse de $\mu_X$ et cette petitesse est naturelle au sens de 't Hooft~\cite{'tHooft:1979bh} puisque, quand $\mu_X$ et $\mu_R$ valent zéro, la conservation du nombre leptonique est restaurée. Dans la suite, nous négligerons $\mu_R$ car il ne contribue à la masse des neutrinos et aux observables étudiées que par des contributions sous-dominantes. La petitesse de la masse des neutrinos étant proportionnelle à $\mu_X$, l'inverse seesaw peut présenter en même temps des couplages de Yukawa $Y_\nu\sim \mathcal{O}(1)$ et des neutrinos stériles avec des masses de l'ordre du TeV. Cela va avoir des conséquences phénoménologiques importantes comme un mélange actif-stérile accru ou la possibilité de produire directement les neutrinos lourds au LHC.

\section{Paramétrisation du mélange leptonique}

Puisque la matrice de masse des neutrinos $M_\nu$ est une matrice $n \times n$ complexe et symétrique du fait du caractère Majorana du neutrino, elle peut être décomposée en utilisant la factorisation de Takagi
\begin{equation}
 V_\nu^T M_\nu V_\nu = \mathrm{diag}(m_1\,,...\,,m_n)\,,
\end{equation}
avec $V_\nu$ une matrice unitaire et $D$ la matrice diagonale dont les éléments sont les racines carrées des valeurs propres de $M_\nu^\dagger M_\nu$. Dans un mécanisme d'inverse seesaw comportant 3 paires de neutrinos singulets, $n=9$ et $V_\nu$ est une matrice unitaire $9\times9$ qui relie les états propres faibles eux états propres de masse par
\begin{equation}
  \left(\begin{array}{c} \nu_L \\ \nu_R^C \\ X \end{array} \right) = V_\nu \left(\begin{array}{c} \nu_1 \\ \vdots \\ \nu_9 \end{array} \right)\,.
\end{equation}
La matrice de masse des leptons chargés peut être factorisée en utilisant une décomposition en valeurs singulières $V_L^\dagger M_\ell V_R = \mathrm{diag}(m_e\,,m_\mu\,,m_\tau)$
\begin{equation}
 (e_L)_{\mathrm{weak}}= V_L \left(\begin{array}{c} e_L \\ \mu_L \\ \tau_L \end{array} \right)_{\mathrm{mass}}\,,\quad\quad\quad (e_R)_{\mathrm{weak}}= V_R \left(\begin{array}{c} e_R \\ \mu_R \\ \tau_R \end{array} \right)_{\mathrm{mass}}\,,
\end{equation}
avec $V_L$ et $V_R$ des matrices unitaires $3\times3$.

En général, les neutrinos et les leptons chargés ne sont pas diagonaux dans la même base, ce qui induit un mélange entre différentes saveurs à la base des oscillations de neutrinos et d'autres processus violant la saveur leptonique. Aux énergies très inférieures à la masse du boson $W^\pm$, le courant faible chargé s'écrit
\begin{align}
 J_{W_{ij}}^{\mu+} &= \frac{1}{\sqrt{2}} \overline{\nu_{i}} (V_\nu^\dagger V_L)_{ij} \gamma^\mu P_L \ell_j \nonumber \\
   &= \frac{1}{\sqrt{2}} \overline{\nu_{i}} (U^\dagger)_{ij} \gamma^\mu P_L \ell_j\,,
\label{UdefF}
\end{align}
tandis que la matrice de mélange $U$ est explicitement donnée par
\begin{equation}
 U_{ji}=\sum_{k=1}^3 V_{L_{kj}}^* V_{\nu_{ki}}\,,
\end{equation}
avec $j$ compris entre 1 et 3 pour les leptons chargés et $i$ compris entre 1 et 9 pour les neutrinos. La matrice $U$ est alors rectangulaire, ce qui l'empêche d'être unitaire. En effet, on a
\begin{align}
 U U^\dagger = \mathbf{1}\,,\\
 U^\dagger U \neq \mathbf{1}\,,
\end{align}
parce que
\begin{align}
 (U^\dagger U)_{\alpha \alpha} &= \sum_{i=1}^{3} (U^\dagger)_{\alpha i} U_{i \alpha} \nonumber \\
      &= \sum_{i,j,k=1}^{3} V_{\nu_{j\alpha}}^* V_{L_{ji}} V_{L_{ki}}^* V_{\nu_{k\alpha}} \nonumber \\
      &= \sum_{j=1}^{3} V_{\nu_{j\alpha}}^* V_{\nu_{j\alpha}} \neq \mathbf{1}\,.
\end{align}
Cependant, dans la limite où les neutrinos lourds découplent de la théorie à basse énergie, la matrice de mélange leptonique est $3\times3$, unitaire et correspond à la matrice PMNS.

La paramétrisation de Casas-Ibarra~\cite{Casas:2001sr} est une méthode qui reconstruit une texture générique compatible avec les données expérimentales pour les couplages de Yukawa des neutrinos. Elle a été initialement introduite dans le seesaw de type I~\cite{Casas:2001sr}; mais elle peut directement être étendue au seesaw inverse. Dans ce dernier, la matrice de masse des neutrinos légers est donnée par
\begin{equation}
 M_{\mathrm{light}} \simeq Y_\nu {M_R^T}^{-1} \mu_X M_R^{-1} Y_\nu^T \frac{v^2}{2}\,,
\end{equation}
qui est factorisée par la matrice PMNS
\begin{equation}
 U_{PMNS}^T M_{\mathrm{light}} U_{PMNS} = \mathrm{diag}(m_1\,, m_2\,, m_3)\,.
\end{equation}
En définissant la matrice $M=M_R \mu_X^{-1} M_R^T$, la matrice de masse des neutrinos légers prend une forme similaire à celle du seesaw de type I
\begin{equation}
 M_{\mathrm{light}}\simeq Y_\nu M^{-1} Y_\nu^T \frac{v^2}{2}\,.
\end{equation}
Si la matrice $M$ est décomposée par une matrice unitaire $V$ selon
\begin{equation}
 M=V^\dagger \mathrm{diag}(M_1\,, M_2\,, M_3) V^*\,,
\end{equation}
alors la paramétrisation de Casas-Ibarra peut directement être appliquée au seesaw inverse:
\begin{equation}
 Y_\nu^T=\frac{\sqrt{2}}{v} V^\dagger \mathrm{diag}(\sqrt{M_1}\,,\sqrt{M_2}\,,\sqrt{M_3})\; R\; \mathrm{diag}(\sqrt{m_1}\,,\sqrt{m_2}\,,\sqrt{m_3}) U^\dagger_{PMNS}\,,
\label{CasasIbarraISSf}
\end{equation}
avec $R$ une matrice orthogonale complexe. Supposer $R=\mathbf{1}$ est équivalent à l'hypothèse selon laquelle $Y_\nu$ et $M$ sont simultanément diagonales, tandis que supposer que la matrice de masse des neutrinos est réelles correspond à $R \in \mathrm{O}_3(\mathbb{R})$. La paramétrisation de Casas-Ibarra sera très utile dans l'étude de la phénoménologie du seesaw inverse car elle permet de scanner l'espace de paramètre tout en s'assurant que chaque point est en accord avec les données expérimentales, économisant par la même du temps de calcul.

\section{Contraintes sur l'existence de neutrinos stériles}

Dans le mécanisme d'inverse seesaw décrit précédemment, le Modèle Standard est étendu par l'ajout de trois paires de singulets de jauge. Cela ajoute donc six neutrinos stériles dont les couplages et les masses sont contraints par diverses expériences et observations. Tout d'abord, et bien que nous ne considérons pas ce régime de masse, les couplages des neutrinos stériles aux neutrinos actifs sont contraints par les oscillations de neutrinos~\cite{Smirnov:2006bu} pour des masses inférieures à $100\;\mathrm{eV}$. D'autres limites, plus générales, proviennent de la recherche directe de neutrinos stériles~\cite{Atre:2009rg, Kusenko:2009up}. Ces recherches peuvent se faire par l'étude du spectre de l'électron émis lors de désintégrations $\beta$~\cite{Shrock:1980vy} ou de celui du lepton produit dans la désintégration à deux corps d'un méson~\cite{Shrock:1980vy, Shrock:1980ct}. Une autre possibilité est d'essayer de détecter les produits visibles de la désintégration d'un neutrino stérile~\cite{Kusenko:2004qc,Liventsev:2013zz} ou de mesurer précisément la largeur de désintégration invisible du boson $Z^0$~\cite{Dittmar:1989yg}.

La non-unitarité de la matrice de mélange leptonique est aussi sujette à de nombreuses contraintes. En notant $\widetilde U_\text{PMNS}$ le bloc $3\times 3$ qui correspond au mélange entre leptons chargés et neutrinos actifs, la déviation à l'unitarité de $\widetilde U_\text{PMNS}$ peut être paramétrée par
\begin{equation}\label{eq:U:eta:PMNSf}
\widetilde U_\text{PMNS} = (\mathbf{1} - \eta) U_\text{PMNS}\,,
\end{equation}
avec $U_\text{PMNS}$ la matrice de mélange quand seulement trois neutrinos massifs sont présents. Des limites sur $\eta$ ont été dérivées en étudiant des interactions non-standards pour les neutrinos~\cite{Antusch:2008tz}.

La présence d'un mélange actif-stérile peut aussi générer une violation conséquente de la saveur leptonique chargée, ce qui est contraint par l'expérience MEG~\cite{Adam:2013mnn}. D'autres limites peuvent être dérivées des recherches du boson de Higgs~\cite{BhupalDev:2012zg}, de sa largeur de désintégration invisible et des données de précision électrofaibles~\cite{delAguila:2008pw, Atre:2009rg}. De plus, si les neutrinos stériles sont des fermions de Majorana, alors d'autres pistes peuvent être explorées comme la désintégration double $\beta$ sans neutrino~\cite{Smirnov:2006bu} ou des signaux spécifiques comportant des particules de même charge au LHC~\cite{Atre:2009rg}.

Enfin, en supposant la validité du modèle cosmologique standard, des limites très contraignantes peuvent être extraites de nombreuses observations~\cite{Smirnov:2006bu, Kusenko:2009up}. Cependant, ces contraintes cosmologiques disparaissent en considérant des scénarios non-standards comme ceux comportant une température de réchauffage basse~\cite{Gelmini:2008fq}. Dans notre analyse numérique, nous négligerons parfois ces contraintes cosmologiques et le mentionnerons alors explicitement.

\section{Impact des neutrinos stériles sur les tests de l'universalité leptonique}

Une conséquence fondamentale de la structure de jauge du Modèle Standard est l'u\-ni\-ver\-sa\-li\-té des constantes de couplage. Puisque les interactions de jauge ne distinguent pas les diverses générations, les leptons de différentes saveurs ont des couplage identiques. La moindre déviation par rapport aux estimations faites dans le MS indique la présence de nouvelle Physique. Comme les leptons chargés se couplent au photon et aux bosons vecteurs de l'interaction faible, les différents tests de l'universalité leptonique peuvent être catégorisés selon le boson de jauge en jeu: photon, $W^\pm$ ou $Z^0$. Puisque notre intérêt porte sur l'effet des neutrinos singulets, nous nous concentrerons sur les processus comportant un boson $W^\pm$.

La première observable qui vient à l'esprit est simplement la désintégration leptonique du boson $W^\pm$. Les différentes largeurs de désintégration ont été mesurées au LEP-II. L'universalité leptonique est testée via les rapports~\cite{Alcaraz:2006mx}
\begin{align}
 \frac{\mathcal{B}(W \rightarrow \mu \bar{\nu}_{\mu})}{\mathcal{B}(W \rightarrow e \bar{\nu}_e)}&=0.994\pm0.020\,,\\
 \frac{\mathcal{B}(W \rightarrow \tau \bar{\nu}_{\tau})}{\mathcal{B}(W \rightarrow e \bar{\nu}_e)}&=1.074\pm0.029\,,\\
 \frac{\mathcal{B}(W \rightarrow \tau \bar{\nu}_{\tau})}{\mathcal{B}(W \rightarrow \mu \bar{\nu}_{\mu})}&=1.080\pm0.028\,,
\end{align}
qui présentent un léger écart à l'universalité pour la troisième génération. En prenant en compte la masse des neutrinos et l'existence de neutrinos stériles, la largeur de désintégration dans une saveur donnée s'écrit
\begin{equation}
 \Gamma(W \to \ell_i \nu) = \sum_j \Gamma_{VFF}(m_W,m_{\ell_i},m_{\nu_j},a_L^{ij},0)
\end{equation}
où la fonction $\Gamma_{VFF} = \Gamma_{VFF}(m_V,m_{F_1},m_{F_2},c_L,c_R)$ est
\begin{align}
 &\Gamma_{VFF} = \frac{\lambda^{1/2}(m_V, m_{F_1}, m_{F_2})}{48 \pi m_V^3} \\
 &\times \left[\left(|c_L|^2+|c_R|^2\right) \left(2 m_V^2-\frac{\left(m_{F_1}^2-m_{F_2}^2\right)^2}{m_V^2}-m_{F_1}^2-m_{F_2}^2\right)+12 m_{F_1}
   m_{F_2} \Re\left(c_L c_R^*\right)\right] \, . \nonumber
\end{align}
Le couplage $a_L$ est donné par $a_L^{ij} = 2^{3/4} m_W \sqrt{G_F} U_{ij}$.

Les désintégrations leptonique des mésons pseudo-scalaires sont également intéressantes car elles sont supprimés dans le MS par la chiralité de l'état final. Cependant, cette suppression disparaît si un neutrino droit est accessible dans l'état final. Les rapports
\begin{equation}
 R_P =\frac{\Gamma (P^+ \to e^+ \nu)}{\Gamma (P^+ \to \mu^+\nu)}\,,
\end{equation}
sont quasi-indépendants des incertitudes hadroniques et ont été mesurés pour différents mésons~\cite{Lazzeroni:2012cx, Beringer:1900zz}
\begin{align}
 R_K^\text{exp}&=(2.488 \pm 0.010)\times 10^{-5}\,,\label{RKexpF}\\
 R_\pi^\text{exp} &=(1.230 \pm 0.004)\times 10^{-4}\,.\label{RpiexpF}
\end{align}
L'universalité de la troisième génération peut être testée par l'intermédiaire du rapport
\begin{equation}
 R_{D_s}=\frac{\Gamma (D_s^+ \to \tau^+ \nu)}{\Gamma (D_s^+ \to \mu^+\nu)}\,.
\label{RDsF}
\end{equation}
dont la valeur expérimentale est~\cite{Beringer:1900zz}
\begin{equation}
 R_{D_s}=9.20\pm0.46\,.
\end{equation}
L'expression analytique de $R_P$ à l'arbre est alors donnée par
\begin{equation}\label{eq:RPresultf}
R_P = \frac{\sum_i F^{i1} G^{i1}}{\sum_k F^{k2} G^{k2}}\,, \quad \text{avec}
\end{equation}
\begin{align}
 F^{ij} &= |U_{ji}|^2 \quad \text{et}\nonumber \\  
 G^{ij} &= \left[m_P^2 (m_{\nu_i}^2+m_{l_j}^2) - (m_{\nu_i}^2-m_{l_j}^2)^2 \right] \left[ (m_P^2 - m_{l_j}^2 -
  m_{\nu_i}^2)^2 - 4 m_{l_j}^2 m_{\nu_i}^2 \right]^{1/2}\,,
  \label{eq:FGf}
\end{align}
où il est sous-entendu que la somme se fait sur tous les neutrinos cinématiquement accessibles. Il est alors possible d'obtenir un large écart à l'universalité via la non-unitarité de la matrice $\widetilde U_{\mathrm{PMNS}}$ mais aussi par un effet d'espace de phase si certains neutrinos sont cinématiquement accessibles. Une évaluation de la déviation par rapport à la prédiction du MS est donnée dans la figure~\ref{RkResults} pour le cas du seesaw inverse.

L'universalité du couplage du $\tau$ est également testable en considérant la désintégration d'un lepton $\tau$ dans un méson pseudo-scalaire. Des rapports similaires à $R_P$ peuvent être construits
 \begin{equation}
 R_{P,\mu}^\tau=\frac{\mathcal{B}(\tau^-\rightarrow P^- \nu)}{\mathcal{B}(P^+ \to \mu^+\nu)}\,,\quad\mathrm{et}\quad  R_{P,e}^\tau=\frac{\mathcal{B}(\tau^-\rightarrow P^- \nu)}{\mathcal{B}(P^+ \to e^+\nu)}\,,
\label{RPltauF}
\end{equation}
où les incertitudes hadroniques s'annulent à une bonne approximation, comme dans $R_P$. La formule analytique correspondante est alors donnée par
\begin{equation}
R_{P,\ell_j}^\tau = \frac{m_P^3}{2 m_\tau^3} \frac{\operatornamewithlimits{\sum}_{i=1}^{N_\text{max}^{(P)}} F^{i\tau} \tilde G^{i\tau}}{\operatornamewithlimits{\sum}_{k=1}^{N_\text{max}^{(\ell_j)}} F^{kj} G^{kj}}\,, \quad \text{avec}
\end{equation}
\begin{equation}
 \tilde G^{i\tau} = \left[(m_\tau^2-m_{\nu_i}^2)^2-m_P^2 (m_\tau^2+m_{\nu_i}^2)\right] \left[(m_\tau^2 - m_P^2 - m_{\nu_i}^2)^2-4 m_P^2 m_{\nu_i}^2\right]^{1/2}\,,
\end{equation}
et $F^{kj}$ et $G^{kj}$ donnés par l'équation~(\ref{eq:FGf}).

L'universalité leptonique peut aussi être testée par le rapport des largeurs de  désintégration leptonique du lepton $\tau$. La moyenne du PDG des différentes mesures donne~\cite{Beringer:1900zz}
\begin{equation}
 R_\tau=\frac{\mathcal{B}(\tau^-\rightarrow \mu^- \bar \nu_\mu \nu_\tau)}{\mathcal{B}(\tau^- \rightarrow e^- \bar \nu_e \nu_\tau)}=0.979 \pm 0.004\,.
\end{equation}
Dans le cas de neutrinos de Majorana, en sommant sur tout les neutrinos cinématiquement accessibles, la largeur de désintégration de  $\ell_i \rightarrow \ell_j \nu \nu$ est donnée par
\begin{align}
 \Gamma &= \sum_{\alpha=1}^{\mathrm{N_{max}}(\ell_j)} \sum_{\beta=1}^{\alpha} \Gamma_{\alpha\beta}\,,\\
 \mathrm{avec} & \nonumber\\
 \Gamma_{\alpha\beta} &= \frac{G_F^2 (2-\delta_{\alpha\beta})}{m_{\ell_i}^3 (2\pi)^3} \int_{(m_{\ell_j}+m_{\nu_\alpha})^2}^{(m_{\ell_i}-m_{\nu_\beta})^2} \mathrm{d}s_{j\alpha} \left[\frac{1}{4} |U_{i\alpha}|^2 |U_{j\beta}|^2 (s_{j\alpha} -m_{\ell_j}^2-m_{\nu_\alpha}^2) (m_{\ell_i}^2+m_{\nu_\beta}^2-s_{j\alpha}) \right.\nonumber\\
& \left. \quad \quad \quad \quad \quad + \frac{1}{2} \Re(U_{i\alpha}^* U_{j\beta} U_{i\beta} U_{j\alpha}^*) m_{\nu_\alpha} m_{\nu_\beta} \left(s_{j\alpha}-\frac{m_{\nu_\alpha}^2+m_{\nu_\beta}^2}{2}\right) \right]\nonumber\\
& \quad \quad \quad \quad \quad \times \frac{1}{s_{j\alpha}} \lambda^{1/2}(s^{1/2}_{j\alpha}, m_{\ell_j}, m_{\nu_\alpha}) \lambda^{1/2}(s^{1/2}_{j\alpha}, m_{\ell_i}, m_{\nu_\beta})\nonumber\\
& + \alpha \leftrightarrow \beta\,.\label{tau3lF}
\end{align}
Étant purement leptonique, cette observables a également l'avantage de comporter une erreur théorique très faible.

Notre analyse a montré que d'importants écarts à l'universalité leptonique peuvent être observés dans $R_K$~\cite{Abada:2012mc}, parfaitement observables à l'expérience NA62. Ceci contraste fortement avec d'autres modèles de nouvelle physique qui peinent à générer une déviation observable. Il est important de noter que, du fait des larges déviations possibles, les observables testant l'universalité leptonique peuvent être utilisées pour contraindre la déviation de $\widetilde U_\mathrm{PMNS}$ à l'unitarité. Une étude détaillée des observables autres que $R_K$ et $R_\pi$ est actuellement en cours.

Le seesaw inverse est une extension du MS très attractive car elle génère des neutrinos massifs tout en gardant des couplages de Yukawa naturels et une échelle de seesaw proche de l'échelle électrofaible. Cependant, le MS souffre de problèmes autres que l'absence de masse de neutrinos, par exemple le problème de hiérarchie ou l'absence de candidat pour la matière noire. Dans la partie suivante, nous présenterons le concept de supersymétrie qui permet de résoudre certains de ces problèmes.

\section{Une introduction à la supersymétrie}

Les modèles supersymétriques sont parmi les extensions du Modèle Standard les plus étudiées. Cela est lié aux nombreux aspects attractifs de la supersymétrie. Ainsi, elle donne une description unifiée des bosons et des fermions, les décrivant comme les différentes composantes d'un supermultiplet. La supersymétrie est aussi l'extension la plus générale de l'algèbre de Poincaré en supposant un seul générateur spinoriel. Les théories supersymétriques ont également un meilleur comportement à haute énergie que le MS. Du fait du théorème de non-renormalisation, les divergences quadratiques disparaissent, ce qui apporte une solution au problème de hiérarchie. Un autre aspect attractif réside dans l'unification des constantes de couplage à haute énergie, qui est bien meilleure que dans le MS. Si la supersymétrie est jaugée et considérée comme une transformation locale, alors le graviton émerge naturellement dans le spectre et connecte naturellement la physique des particules avec la gravité. La supersymétrie offre une explication à la brisure de symétrie électrofaible par le biais des corrections radiatives qui génèrent $\mu < 0$ à basse énergie. Enfin, dans les modèles où la R-parité est conservée, la particule supersymétrique la plus légère est un candidat pour la matière noire.

L'extension supersymétrique minimale et viable du MS est le MSSM (Minimal Supersymmetric Standard Model). Il est basé sur le même groupe de jauge $\mathrm{SU}(3)_c\times\mathrm{SU}(2)_L\times\mathrm{U}(1)_Y$, conserve la R-parité et son contenu en particules est donné dans les tables~\ref{MSSMgaugeF}
\begin{table}[t]
  \begin{center}
    \begin{tabular}{|c|c|c|c|}
     \hline
	Groupe de jauge & Superchamp vectoriel & Spineur & Vecteur \\
     \hline
     \hline
	$U(1)_Y$ & $\widehat{B}$ & $\vphantom{\widetilde{b^{B^B}}}\widetilde{b}$ & $B_\mu$ \\
	$SU(2)_L$ & $\widehat{W}^i$ & $\widetilde{w}^i$ & $W^i_\mu$ \\
	$SU(3)_c$ & $\widehat{G}^\alpha$ & $\widetilde{g}^\alpha$ & $G^\alpha_\mu$ \\
     \hline
    \end{tabular}
    \caption[Superchamps vectoriels du MSSM]{\label{MSSMgaugeF} Superchamps vectoriels du MSSM. Les composantes tildées sont impaires sous la R-parité et connues collectivement sous le nom de jauginos.}
 \end{center}
\end{table}
et~\ref{MSSMmatterF}.
\begin{table}[t]
  \begin{center}
    \begin{tabular}{|c|c|c|c|c|c|}
     \hline
	Superchamp chiral & Scalaire & Spineur & $SU(3)_c$ & $SU(2)_L$ & $U(1)_Y$ \\
     \hline
     \hline
	$\widehat{Q}_i=\binom{\vphantom{\widetilde{b^{B}}}\widehat{u}_{L_i}}{\widehat{d}_{L_i}}$ & $\widetilde{Q}_i$ & $Q_i$ & $\mathbf{3}$ & $\mathbf{2}$ & $\phantom{-}\frac{1}{3}$ \\
	$\widehat{U}_i$ & $\widetilde{U}_i$ & $U_i$ & $\mathbf{3}^*$ & $\mathbf{1}$ & $-\frac{4}{3}$ \\
	$\widehat{D}_i$ & $\widetilde{D}_i$ & $D_i$ & $\mathbf{3}^*$ & $\mathbf{1}$ & $\phantom{-}\frac{2}{3}$ \\
     \hline
	$\widehat{L}=\binom{\vphantom{\widetilde{b^{B}}}\widehat{\nu}_{L_i}}{\widehat{\ell}_{L_i}}$ & $\widetilde{L}_i$ & $L_i$ & $\mathbf{1}$ & $\mathbf{2}$ & $-1$ \\
	$\widehat{E}_i$ & $\widetilde{E}_i$ & $E_i$ & $\mathbf{1}$ & $\mathbf{1}$ & $\phantom{-}2$ \\
     \hline
	$\widehat{H}_u=\binom{\vphantom{\widetilde{b^{B}}}\widehat{h}_{u}^+}{\widehat{h}_u^0}$ & $H_u$ & $\widetilde{H}_u$ & $\mathbf{1}$ & $\mathbf{2}$ & $\phantom{-}1$ \\
	$\widehat{H}_d=\binom{\widehat{h}_{d}^0}{\widehat{h}_d^-}$ & $H_d$ & $\widetilde{H}_d$ & $\mathbf{1}$ & $\mathbf{2}$ & $-1$ \\
     \hline
    \end{tabular}
    \caption[Superchamps chiraux du MSSM]{\label{MSSMmatterF} Superchamps chiraux du MSSM et leurs transformations de jauge, avec $i$ un indice comprit entre 1 et 3 qui correspond aux différentes générations. Les composantes tildées sont impaires sous la R-parité et se nomme squarks, sleptons et higgsinos, étant respectivement les superpartenaires des quarks, leptons et bosons de Higgs.}
 \end{center}
\end{table}
La seule véritable différence se trouve dans le secteur du Higgs qui contient deux doublets de $\mathrm{SU}(2)$. Cela s'explique par l'impossibilité d'incorporer un supermultiplet conjugué au superpotentiel, ce dernier étant une fonction holomorphique, interdisant de fait l'utilisation de $\widetilde{\phi}=\imath \sigma_2 \phi^*$ comme dans le MS. De plus, le higgsino est un fermion supplémentaire d'hypercharge $Y=1$ ce qui génère une anomalie d'Adler-Bell-Jackiw~\cite{Adler:1969gk, *Bell:1969ts}. La solution est alors d'introduire un second doublet de Higgs avec une hypercharge opposée.  Le superpotentiel étant une fonction de superchamps gauches uniquement, les fermions droits sont introduits via le conjugué de charge de la composante gauche. Par exemple, si le spineur de Dirac de l'électron est donné par $e=e_L+e_R$, alors la composante droite de l'électron est introduite via le spineur de Majorana $E_1=(e_R)^C + e_R$. Par conséquent, le sélectron est donné par $\widetilde{E}_1=\widetilde{e}_R^*$.

Puisque le MSSM a été construit comme un modèle conservant la R-parité~\cite{Barbier:2004ez}, il est défini par le superpotentiel suivant:
\begin{align}
\widehat{f}_\mathrm{MSSM}= \varepsilon_{ab} \left[
Y^{ij}_d \widehat{D}_i \widehat{Q}_j^b  \widehat{H}_d^a
              +Y^{ij}_{u}  \widehat{U}_i \widehat{Q}_j^a \widehat{H}_u^b 
              + Y^{ij}_e \widehat{E}_i \widehat{L}_j^b  \widehat{H}_d^a- \mu \widehat{H}_d^a \widehat{H}_u^b \right]\,,
\label{eq:SuperPotMSSMf}
\end{align}
avec $a$ et $b$ des indices de $\mathrm{SU}(2)$, $i\,,j$ correspondant aux différentes générations et les indices de couleur ayant été omis. Gardant des couplages identiques pour les particules appartenant au même supermultiplet, la supersymétrie est brisée par les termes de brisure douce du MSSM qui sont 
\begin{align}
 -\mathcal{L}_\mathrm{MSSM}^\mathrm{soft} =& m_{H_u}^2 H_u^\dagger H_u+m_{H_d}^2 H_d^\dagger H_d+(m_L^2)_{ij}\widetilde{L}_i^\dagger\widetilde{L}_j +(m_{E}^2)_{ij}\widetilde{E}_{i}^{\dagger}\widetilde{E}_{j} \nonumber \\
&+(m_Q^2)_{ij}\widetilde{Q}_i^\dagger\widetilde{Q}_j+(m_{U}^2)_{ij}\widetilde{U}_{i}^{\dagger}\widetilde{U}_{j}+(m_{D}^2)_{ij}\widetilde{D}_{i}^{\dagger}\widetilde{D}_{j} \nonumber \\
&+B \mu \varepsilon_{ab} H_d^a H_u^b + h.c.\nonumber \\
&-\varepsilon_{ab}\left[(A_uY^u)_{ij}\widetilde{U}_i \widetilde{Q}_j^a H_u^b + (A_dY^d)_{ij}\widetilde{D}_i \widetilde{Q}_j^b  H_d^a+ (A_eY^e)_{ij}\widetilde{E}_i \widetilde{L}_j^b  H_d^a+h.c.\right]\nonumber \\
&+M_1\overline{\tilde{b}}\tilde{b}+M_2\overline{\widetilde{w}}_i\widetilde{w}_i+M_3\overline{\widetilde{g}}_\alpha\widetilde{g}_\alpha + h.c.\,,\label{softMSSMf}
\end{align}
avec des matrices de masse scalaire au carré $3\times 3$ hermitiennes, $M_i$ les masses complexes des jauginos, $A$ des matrices $3 \times 3$ complexes qui correspondent aux couplages trilinéaires  ainsi que  $m_{H_u}^2$, $m_{H_u}^2$ et $B$ des paramètres réels.

Après brisure de la symétrie électrofaible, les composantes neutres des doublets de Higgs développent une valeur moyenne dans le vide (vev) non-nulle. Notant $v_u$ et $v_d$ respectivement les vev des doublets $H_u$ et $H_d$, elles vérifient $v=\sqrt{v_u^2 + v_d^2}$, avec $v$ la vev du doublet de Higgs du MS. Alors les vev $v_u$ et $v_d$ sont reliées par
\begin{equation}
 \tan \beta = \frac{v_u}{v_d}\,.\label{tanbetaF}
\end{equation}
À basse énergie, le spectre du MSSM contient alors deux bosons pairs sous CP et un boson impair, en supposant l'absence de violation de CP dans le secteur du Higgs. Les higgsinos neutres vont se mélanger avec le bino et le wino neutre, générant quatre neutralinos, tandis que les higgsinos chargés se mélangent aux winos chargés pour générer quatre charginos. Ceci est repris dans la table~\ref{mixingMSSMf}.
\begin{table}[t]
  \begin{center}
    \begin{tabular}{|c|c|c|c|}
     \hline
	Secteur & Spin & État propre de jauge & État propre de masse \\
     \hline
     \hline
	Higgs & 0 & $\vphantom{\widetilde{b^{B}}}{h}_{u}^+\,,{h}_{u}^0\,,{h}_{d}^-\,,{h}_{d}^0$ & $h\,,H\,,A\,,H^\pm$ \\
     \hline
	Neutralinos & 1/2 & $\vphantom{\widetilde{b^{B^B}}}\widetilde{W}^3\,,\widetilde{B}\,,\widetilde{h}_{u}^0\,,\widetilde{h}_{d}^0$ & $\widetilde{\chi}^0_1\,,\widetilde{\chi}^0_2\,,\widetilde{\chi}^0_3\,,\widetilde{\chi}^0_4$ \\
     \hline
	Charginos & 1/2 & $\vphantom{\widetilde{b^{B^B}}}\widetilde{W}^1\,,\widetilde{W}^2\,,\widetilde{h}_{u}^+\,,\widetilde{h}_{d}^-$ & $\widetilde{\chi}^\pm_1\,,\widetilde{\chi}^\pm_2$ \\
     \hline
    \end{tabular}
    \caption[Bosons de Higgs, charginos et neutralinos du MSSM]{\label{mixingMSSMf}Bosons de Higgs, charginos et neutralinos dans le MSSM après brisure la de symétrie électrofaible. $h$ et $H$ sont CP-pairs, alors que $A$ est CP-impair.}
 \end{center}
\end{table}

Puisque les matrices de masse des squarks et des sleptons sont en général complexes avec de larges éléments non-diagonaux, des courants non-chargés peuvent générer des transitions de saveur, ce qui n'a pas été observé expérimentalement. Ce premier problème est connu sous le nom de problème de saveur. Il existe un second problème, nommé problème CP, qui est lié à la présence de nouvelle source de violation de CP dans ces matrices complexes. Une solution commune à ces deux problèmes est de considérer que les masses des squarks et sleptons sont universelles et réelles tandis que les couplages trilinéaires sont proportionnels aux couplages de Yukawa. Ceci est naturellement réalisé par les mécanismes brisant la supersymétrie  par la gravité. Un premier modèle considéré durant cette thèse est le CMSSM qui possède seulement cinq paramètres à l'échelle de grande unification:
\begin{equation}
 m_0\,,m_{1/2}\,,A_0\,,\tan\beta\,,\mathrm{sign}(\mu)\,,
\end{equation}
avec
\begin{align}
 &m_0^2=m_{H_u}^2=m_{H_d}^2\,,\label{CMSSM1F}\\
 &m_0^2 \mathbf{1}= m_Q^2 = m_U^2= m_D^2= m_L^2 = m_E^2\,,\label{CMSSM2F}\\
 &m_{1/2}=M_1=M_2=M_3\label{CMSSM3F}\\
 &A_0 \mathbf{1}=A_u=A_d=A_e\,.\label{CMSSM4F}
\end{align}
Un second modèle, le NUHM, relâche les contraintes sur le secteur du Higgs où
\begin{equation}
 m_0^2\neq m_{H_u}^2\neq m_{H_d}^2\,,
\end{equation}
ce qui introduit un sixième paramètre, par exemple $m_A$.

Cependant, le MSSM souffre aussi de certains problèmes, à l'instar du problème de $\mu$. Ce paramètre du MSSM doit être compris dans l'intervalle $100\;\mathrm{GeV}\leq\mu\leq M_\mathrm{SUSY}$~\cite{Ellwanger:2009dp} afin d'avoir un modèle phénoménologiquement viable. Mais ce paramètre n'est protégé par aucune symétrie et devrait naturellement être proche de zéro, de l'échelle de grande unification ou de l'échelle de Planck. Une solution simple et élégante est d'introduire un superchamps chiral singulet de jauge dont la composante scalaire prend une vev non-nulle après brisure de la supersymétrie, ce qui est amène au NMSSM (Next-to-Minimal Supersymmetric Standard Model). Ce modèle est défini par le  superpotentiel suivant
\begin{align}
\widehat{f}_\mathrm{NMSSM}= \varepsilon_{ab} \left[
Y^{ij}_d \widehat{D}_i \widehat{Q}_j^b  \widehat{H}_d^a
              +Y^{ij}_{u}  \widehat{U}_i \widehat{Q}_j^a \widehat{H}_u^b 
              + Y^{ij}_e \widehat{E}_i \widehat{L}_j^b  \widehat{H}_d^a- \lambda \widehat{S} \widehat{H}_d^a \widehat{H}_u^b \right] -\frac{\kappa}{3}\widehat{S}^3\,,
\label{eq:SuperPotNMSSMf}
\end{align}
avec $\lambda$ et $\kappa$ des couplages adimensionnés. Le Lagrangien de brisure douce de la supersymétrie est alors
\begin{align}
 -\mathcal{L}_\mathrm{NMSSM}^\mathrm{soft} =& m_{H_u}^2 H_u^\dagger H_u+m_{H_d}^2 H_d^\dagger H_d+ m_S^2 S^\dagger S +(m_L^2)_{ij}\widetilde{L}_i^\dagger\widetilde{L}_j +(m_{E}^2)_{ij}\widetilde{E}_{i}^{\dagger}\widetilde{E}_{j} \nonumber \\
&+(m_Q^2)_{ij}\widetilde{Q}_i^\dagger\widetilde{Q}_j+(m_{U}^2)_{ij}\widetilde{U}_{i}^{\dagger}\widetilde{U}_{j}+(m_{D}^2)_{ij}\widetilde{D}_{i}^{\dagger}\widetilde{D}_{j} \nonumber \\
&+\left(\lambda A_\lambda \varepsilon_{ab} H_d^a H_u^b S +\frac{\kappa}{3} A_\kappa S^3 + h.c.\right)\nonumber \\
&-\varepsilon_{ab}\left[(A_uY^u)_{ij}\widetilde{U}_i \widetilde{Q}_j^a H_u^b + (A_dY^d)_{ij}\widetilde{D}_i \widetilde{Q}_j^b  H_d^a+ (A_eY^e)_{ij}\widetilde{E}_i \widetilde{L}_j^b  H_d^a+h.c.\right]\nonumber \\
&+M_1\overline{\tilde{b}}\tilde{b}+M_2\overline{\widetilde{w}}_i\widetilde{w}_i+M_3\overline{\widetilde{g}}_\alpha\widetilde{g}_\alpha + h.c.\,,\label{softNMSSMf}
\end{align}
avec $A_\lambda$ et $A_\kappa$ les nouveaux couplages trilinéaires associés à l'introduction du superchamp singulet $\widehat{S}$. Quand la composante scalaire de ce dernier prend une vev non-nulle, elle génère un paramètre $\mu$ effectif
\begin{equation}
 \mu_{\mathrm{eff}}=\lambda s\,,
\end{equation}
qui est naturellement du même ordre que l'échelle de brisure de la supersymétrie et apporte, par conséquent, une solution au problème de $\mu$.

Puisque le NMSSM contient un champs chiral supplémentaire, son spectre à basse énergie est plus étendu que celui du MSSM. Le superchamp étant un singulet, il va seulement introduire de nouveaux champs neutres ce qui conduit à trois bosons de Higgs CP-pairs $H_{1\,,2\,,3}$, deux bosons de Higgs CP-impaires $A_{1\,,2}$ et cinq neutralinos $\widetilde{\chi}^0_{1\,,...\,,5}$. Cela peut conduire à une phénoménologie très différente~\cite{Ellwanger:2011aa, *Belanger:2012tt, Cao:2012fz, Belanger:2005kh, *Gunion:2005rw, *Cerdeno:2007sn, *Kraml:2008zr, *Vasquez:2012hn} et réduire l'ajustement excessif des paramètres du modèle.

Malheureusement, les deux modèles présentés ci-dessus ne contiennent pas de mécanisme générant une masse pour les neutrinos. Il convient donc d'étendre ces modèles et d'étudier la phénoménologie associée, ce que nous avons fait dans le cas du seesaw inverse supersymétrique.

\section{Modèles de seesaw inverse supersymétrique}

Conjuguer l'inverse seesaw avec des modèles supersymétriques est particulièrement attractif car toute le nouvelle Physique se trouve alors à l'échelle du TeV. Cela conduit à d'intéressantes conséquences phénoménologiques pour les désintégrations double $\beta$ sans neutrino~\cite{Awasthi:2013we}, le secteur du Higgs~\cite{Abada:2010ym, Gogoladze:2012jp, *Bandyopadhyay:2012px, BhupalDev:2012zg, Banerjee:2013fga}, les processus violant la saveur leptonique chargée~\cite{Deppisch:2004fa, Hirsch:2009ra, Abada:2011hm, Abada:2012cq}, les observables au LHC et à un futur collisionneur linéaire~\cite{Mondal:2012jv, *Das:2012ze}, le neutrino droit comme candidat pour la matière noire~\cite{Cerdeno:2011qv, *An:2011uq, *BhupalDev:2012ru, *DeRomeri:2012qd, Banerjee:2013fga} ou un modèle de leptogénèse~\cite{Blanchet:2010kw}.

L'inverse seesaw peut être inclus dans le MSSM par l'addition de deux superchamps chiraux singulets de jauge portant un nombre leptonique opposé. Si trois paires de superchamps singulets sont considérées, $\widehat{\nu}^C_i$ et $\widehat{X}_i$ ($i=1,2,3$)\footnote{Nous utilisons la notation:
  $\widetilde{\nu}^C=\widetilde{\nu}_R^* $.}, alors le modèle de seesaw inverse supersymétrique est défini par le superpotentiel suivant:
\begin{equation}
\widehat f= \widehat f_\mathrm{MSSM} + \varepsilon_{ab} Y^{ij}_\nu \widehat{\nu}^C_i \widehat{L}^a_j \widehat{H}_u^b+M_{R_{ij}}\widehat{\nu}^C_i\widehat{X}_j+
\frac{1}{2}\mu_{X_{ij}}\widehat{X}_i\widehat{X}_j\,,
\label{eq:SuperPotMSSMISSf}
\end{equation}
avec $\widehat f_\mathrm{MSSM}$ le superpotentiel du MSSM défini dans l'équation~(\ref{eq:SuperPotMSSMf}) et $i,j = 1,2,3$ les indices de saveur. Le terme de masse $M_{R_{ij}}$ conserve le nombre leptonique tandis que $\mu_{X_{ij}}$ le viole de deux unités. Le Lagrangien de brisure douce de la supersymétrie est alors
\begin{align}
-\mathcal{L}^\mathrm{soft}&=-\mathcal{L}_\mathrm{MSSM}^\mathrm{soft} 
         +   \widetilde\nu^{c}_i m^2_{\widetilde \nu^C_{ij}}\widetilde\nu^{c*}_j
         + \widetilde X^{\dagger}_i m^2_{X_{ij}} \widetilde X_j
     + (A_{\nu}^{ij} Y_\nu^{ij} \varepsilon_{ab}
                 \widetilde\nu^C_i \widetilde L^a_j H_u^b +
                B_{M_R}^{ij} M_{R_{ij}}\widetilde\nu^C_i \widetilde X_j \nonumber\\
      &+\frac{1}{2}B_{\mu_X}^{ij} \mu_{X_{ij}} \widetilde X_i \widetilde X_j
      +\mathrm{h.c.}),
\label{eq:softSUSYf}
\end{align}
avec  ${\mathcal L}_\mathrm{MSSM}^\mathrm{ soft}$ le Lagrangien de brisure douce du MSSM. $B_{M_R}^{ij}$ et $B_{\mu_X}^{ij}$ sont de nouveaux paramètres dus à la présence des nouveaux partenaires scalaires des neutrinos stériles. Il est important de noter que, tandis que le premier de ces termes conserve le nombre leptonique, le second le viole de deux unités. Supposant un mécanisme de brisure de la supersymétrie indépendant de la saveur, nous avons considéré des conditions aux limites universelles à haute énergie (par exemple, à l'échelle de grande unification)
\begin{equation} m_\phi = m_0\,, M_\text{gaugino}= M_{1/2}\,,
A_{i}= A_0 \, \mathbb{I}\,,B_{\mu_X}=B_{M_R}= B_0 \, \mathbb{I}\,.  
\end{equation}
De manière tout à fait générale, nous travaillons dans une base où $M_R$ est diagonale à l'échelle de brisure de la supersymétrie
\begin{equation}
    M_{R} = \mathrm{diag}\;M_{R_{ii}}\,.
\end{equation}
Nous avons également supposé lors de nos évaluations numériques que $\mu_X$ est diagonale, une hypothèse simplificatrice motivée par le fait que les observables violant la saveur leptonique ne dépendent qu'indirectement de $\mu_X$. Enfin, il est important de remarquer que le terme effectif de masse du sneutrino droit s'écrit
\begin{equation}
M^2_{\widetilde \nu^C} \simeq m^2_{\widetilde \nu^C} + M_{R}^2 + { |Y_\nu|^2 v_u^2}\,.
\end{equation}
Si $M_R \sim {\mathcal{O}}$(TeV), alors le terme de masse sera proche du TeV, à l'opposé du seesaw supersymétrique de type I~\cite{Arganda:2005ji}. Nous nous sommes particulièrement intéressés au rôle d'un tel sneutrino léger dans les contributions médiées par les bosons de Higgs et $Z^0$ pour les observables violant la saveur leptonique.

L'inverse seesaw peut aussi être inclus dans le NMSSM via l'ajout de paires de superchamps chiraux singulets de jauge comme dans le MSSM. Le superpotentiel de ce modèle est alors
\begin{equation}
 \widehat f = \widehat{f}_\mathrm{NMSSM}+\varepsilon_{ab} Y_\nu^{ij}  \widehat \nu^C_i \widehat{L}_j^a \widehat{H}_u^b + (\lambda_{\nu})^i \widehat{S} \widehat \nu^C_i \widehat X_i + \frac{1}{2} \mu_X^i \widehat X_i \widehat X_i\,,\label{model-expF}
\end{equation}
où $\widehat{f}_\mathrm{NMSSM}$ est le superpotentiel du NMSSM, défini à l'équation~(\ref{eq:SuperPotNMSSMf}) et $i,j = 1,2,3$ sont les indices de saveur. Pour notre étude, les termes $\widehat \nu^C \widehat X$ et $\widehat X \widehat X$ peuvent être écrits dans une base où ils sont diagonaux sans perte de généralité. Lorsque la composante scalaire de $\widehat S$ acquière une vev, non seulement un terme $\mu$ effectif est généré mais aussi un terme de masse conservant le nombre leptonique $M_{R}\overline{\nu_R}X$ avec $M_{R} = \lambda_{\nu} s$. Nous nous sommes intéressés dans ce cas à l'impact de neutrinos stériles très légers, ce qui est permis dans l'inverse seesaw, sur les modes de désintégration de $A_1$ quand celui-ci a une masse inférieure à $10\;\mathrm{GeV}$.

\section{Violation de la saveur leptonique dans le seesaw inverse supersymétrique}

L'observation des oscillations de neutrinos a établi de manière indiscutable l'absence de conservation de la saveur leptonique neutre. En l'absence d'un principal fondamental interdisant la violation de la saveur leptonique chargée (cLFV), il est entendu que les extensions du Modèle Standard qui génèrent les masses et le mélange des neutrinos induisent aussi une violation de la conservation de la saveur pour les leptons chargés. Puisque la prédiction du Modèle Standard est très fortement supprimée car la violation de saveur est générée par des corrections quantiques, l'observation de signaux de cLFV serait une preuve indiscutable de l'existence d'une nouvelle physique touchant le secteur leptonique. De nombreuses expériences recherchent ces signaux et les résultats utiles à notre étude sont repris dans la table~\ref{cLFVexpF}.

\begin{table}[t]
\begin{center}
 \begin{tabular}{|c|c|c|}
  \hline
    Processus LFV  & Limite actuelle & Sensibilité future \\
  \hline
    $\tau \rightarrow \mu \mu \mu$ & $2.1\times10^{-8}$~\cite{Hayasaka:2010np} & $8.2 \times 10^{-10}$~\cite{O'Leary:2010af} \\
    $\tau^- \rightarrow e^- \mu^+ \mu^-$ &  $2.7\times10^{-8}$~\cite{Hayasaka:2010np} & $\sim 10^{-10}$~\cite{O'Leary:2010af} \\
    $\tau \rightarrow e e e$ & $2.7\times10^{-8}$~\cite{Hayasaka:2010np} &  $2.3 \times 10^{-10}$~\cite{O'Leary:2010af} \\
    $\mu \rightarrow e e e$ &  $1.0 \times 10^{-12}$~\cite{Bellgardt:1987du} &  $\sim10^{-16}$~\cite{Blondel:2013ia} \\
    $\tau \rightarrow \mu \eta$ & $2.3\times 10^{-8}$~\cite{collaboration:2010ipa} & $\sim 10^{-10}$~\cite{O'Leary:2010af}  \\
    $\tau \rightarrow \mu \eta^\prime$ & $3.8\times 10^{-8}$~\cite{collaboration:2010ipa} & $\sim 10^{-10}$~\cite{O'Leary:2010af} \\
    $\tau \rightarrow \mu \pi^{0}$ & $2.2\times 10^{-8}$~\cite{collaboration:2010ipa} & $\sim 10^{-10}$~\cite{O'Leary:2010af}  \\
    $B^{0}_{d} \rightarrow \mu \tau$ & $2.2\times 10^{-5}$~\cite{Aubert:2008cu} &  \\
    $B^{0}_{d} \rightarrow e \mu$ & $6.4\times 10^{-8}$~\cite{Aaltonen:2009vr} & $1.6\times 10^{-8}$~\cite{Bonivento:1028132}  \\
    $B^{0}_{s} \rightarrow e \mu$ & $2.0\times 10^{-7}$~\cite{Aaltonen:2009vr} & $6.5\times 10^{-8}$~\cite{Bonivento:1028132}  \\
    $\mu^-, \mathrm{Ti} \rightarrow e^-, \mathrm{Ti}$ &  $4.3\times 10^{-12}$~\cite{Dohmen:1993mp} & $\sim10^{-18}$~\cite{mori1996experimental} \\
    $\mu^-, \mathrm{Au} \rightarrow e^-, \mathrm{Au}$ & $7\times 10^{-13}$~\cite{Bertl:2006up} & \\
  \hline
 \end{tabular}
\end{center}
  \caption[Limites expérimentales sur les rapports d'embranchement cLFV]{Limites expérimentales actuelles et sensibilités futures pour certaines observables cLFV médiées  par les bosons de Higgs et $Z^0$.}\label{cLFVexpF}
\end{table}

Pour tout seesaw supersymétrique, le couplage de Yukawa des neutrinos va induire des termes non-diagonaux dans la matrice de masse des sleptons via les corrections dues aux équations du groupe de renormalisation~\cite{Borzumati:1986qx, Hisano:1995cp, Hisano:1998fj}. Par exemple, avec les conditions aux limites du CMSSM, ces corrections à la matrice de masse des sleptons gauches sont données par
\begin{eqnarray}
(\Delta m_{\widetilde{L}}^2)_{ij}&\simeq&
-\frac{1}{8\pi^2}(3m_0^2+A_0^2) 
(Y_\nu^\dagger L Y_\nu)_{ij} \,, ~~ L=\ln\frac{M_{GUT}}{M_{R}} \,
\nonumber \\ 
&=&\xi (Y^\dagger_\nu Y_\nu)_{ij}\,,
\label{slepmixingF}
\end{eqnarray} 
quand un spectre dégénéré est considéré pour les neutrinos droits $M_{R_i}=M_{R}$. Par rapport au seesaw de type I, l'inverse seesaw permet d'avoir simultanément $Y_\nu \sim 1$ et $M_R \sim \mathcal{O}(\mathrm{TeV})$, ce qui augmente le facteur $\xi$ et au final va induire d'importants effets dans les observables violant la saveur leptonique à basse énergie.

\begin{figure}
\begin{center}
\includegraphics[width=0.50\textwidth]{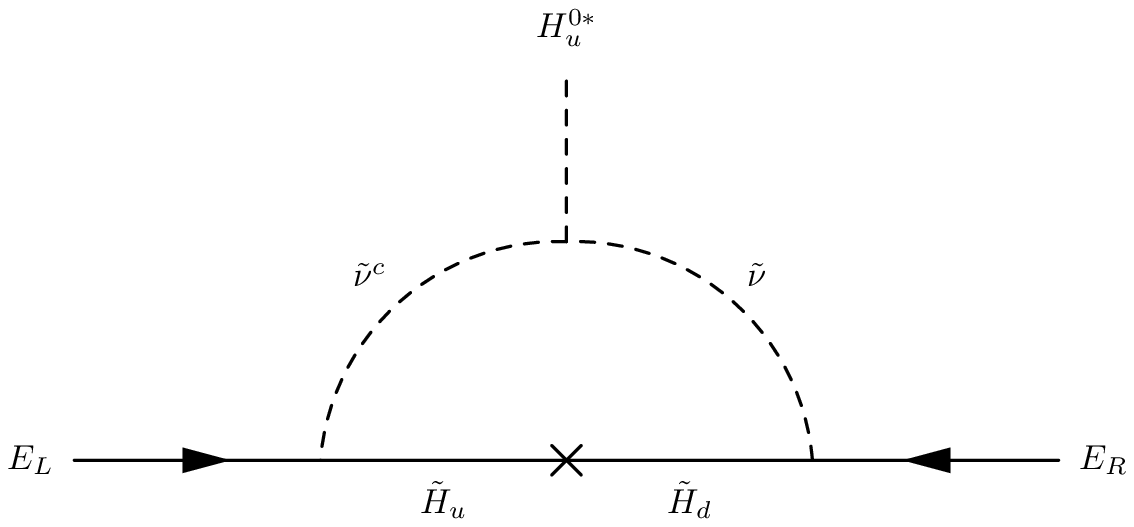}
\caption[Contribution du sneutrino doit à $\epsilon'_2$]{Contribution du sneutrino droit à $\epsilon'_2$. Cette contribution est particulièrement importante quand $\widetilde \nu^C$ est léger.}\label{2F}
\end{center}
\end{figure}

En présence de sneutrinos droits ayant une masse comparable à celle des autres sfermions, les processus contenant un sneutrino droit ne sont plus supprimés comme ils le sont dans le seesaw supersymétrique de type I. Cela va avoir un effet marqué sur les observables médiées par les bosons de Higgs et $Z^0$ en générant de nouvelles contributions. Dans le cas des diagrammes pingouins comportant un boson de Higgs, le Lagrangien effectif qui décrit les couplages des bosons neutres aux leptons chargés est
\begin{equation}
-\mathcal{L}^\text{eff}=\bar E^i_R Y_{e}^{ii} \left[ 
\delta_{ij} H_d^0 + \left(\epsilon_1 \delta_{ij} + 
\epsilon_{2ij} (Y_\nu^\dagger Y_\nu)_{ij} \right) H_u^{0*}
\right] E^j_L + \text{h.c.}  \,. 
\label{LeffF}
\end{equation}
Le dernier terme de ce Lagrangien, $\epsilon_{2ij} (Y_\nu^\dagger Y_\nu)_{ij}$, est en général non-diagonal, ce qui génère une violation de la saveur leptonique. Une nouvelle contribution $\epsilon'_2$ est issue de la boucle sneutrino-chargino représentée dans la figure~\ref{2F}
et elle peut s'écrire
\begin{equation}
\epsilon'_{2ij}= \frac{1}{16\pi^2} \mu A_\nu 
F_1(\mu^2,m^2_{\widetilde \nu_i},M^2_{\widetilde \nu^c_j}),
\label{newdiagF}
\end{equation}
avec
\begin{equation}
F_1\left(x,y,z\right)=-\frac{xy\ln (x/y)+yz\ln (y/z)+zx\ln (z/x)}{(x-y)(y-z)(z-x)} \,.
\end{equation}
Nous avons observé qu'elle peut dominer les autres contributions médiées par un boson de Higgs, étant jusqu'à dix fois plus importante. Nous avons mené une analyse numérique plus poussée de diverses observables dont les résultats sont regroupés dans la table~\ref{4}. Il est intéressant de remarquer que ces valeurs ne correspondent qu'aux contributions médiées par un boson de Higgs et dépendent fortement des valeurs de $\tan \beta$ et $m_A$. De plus, elles sont très sensibles à la chiralité du lepton le plus lourd. Ainsi, le rapport d'embranchement de la désintégration $\ell^i_{L} \rightarrow \ell^j_{R} X$ est supprimé par un facteur $m_{\ell^j}^2/m_{\ell^i}^2$ par comparaison avec $\ell^i_{R} \rightarrow \ell^j_{L} X$. Cela peut induire une asymétrie qui permettrait d'identifier si la contribution générée par les diagrammes pingouins contenant un boson de Higgs domine.

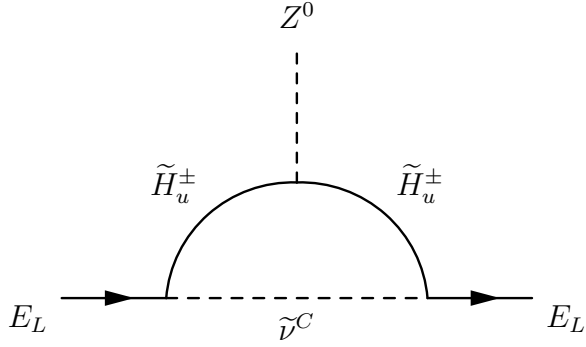
\begin{figure}[!t]
\centering
\begin{fmffile}{DiagZiss}
\begin{fmfgraph*}(220,120)
\fmflabel{$E_{L}$}{l1}
\fmflabel{$E_{L}$}{l2}
\fmflabel{$Z^{0}$}{phot}
\fmfforce{0.1w,0.1h}{l1}
\fmfforce{0.9w,0.1h}{l2}
\fmfforce{0.5w,0.9h}{phot}
\fmf{fermion}{l1,l1nu}
\fmf{dashes,tension=0.4,label=$\widetilde{\nu}^{C}$}{l1nu,nul2}
\fmf{fermion}{nul2,l2}
\fmffreeze
\fmf{plain,right=0.4,tension=0.6,label=$\widetilde{H}^{\pm}_u$}{Wphot,l1nu}
\fmf{plain,right=0.4,tension=0.6,label=$\widetilde{H}^{\pm}_u$}{nul2,Wphot}
\fmf{dashes}{Wphot,phot}
\end{fmfgraph*}
\end{fmffile}
\caption[Contribution du sneutrino droit au pingouins avec un $Z^0$]{Un des diagrammes contenant un sneutrino droit contribuant aux diagrammes pingouins avec un boson $Z^0$ dans le seesaw inverse supersymétrique.} \label{DiagZissF}
\end{figure}

Cependant, la présence de sneutrinos droits avec une masse proche du TeV va aussi augmenter les contributions médiées par le boson $Z^0$. En effet, dans le MSSM étendu par un seesaw de type I à haute énergie, une compensation spécifique a lieu entre les diagrammes qui contribuent à la boucle wino-sneutrino. À l'ordre zéro dans l'angle de mélange du chargino, la combinaison des fonctions de boucle ne dépend pas des masses des particles dans la boucle~\cite{Hirsch:2012ax}. La contribution médiée par le boson $Z^0$ est alors proportionnelle à $( Z_V^\dagger Z_V)^{ij}$ avec $Z_V$ la matrice $3 \times 3$ unitaire qui diagonalise la matrice de masse des sneutrinos. Or ce terme s'annule pour $i \ne j$. C'est cette suppression qui assure la domination habituelle de la contribution médiée par le photon. Cependant la présence de diagrammes supplémentaires, comme celui de la figure~\ref{DiagZissF},
dans le seesaw inverse supersymétrique, fait disparaître cette compensation. Les contributions médiées par le boson $Z^0$ augmentent alors fortement ~\cite{Abada:2012cq}, pouvant même fournir la contribution dominante. 

\section{Désintégrations invisibles d'un bosons de Higgs CP-impair}

Dans ce cas, nous nous intéressons à une conséquence de l'incorporation de l'inverse seesaw dans le NMSSM et calculons les rapports d'embranchement de $A_1$, le boson de Higgs CP-impair le plus léger, dans un état final contenant deux neutrinos. Pour illustrer notre propos, il n'est pas nécessaire de diagonaliser la matrice de masse des neutrinos et il est donc possible de calculer les rapports d'embranchement dans les états propres d'interaction $\nu_L \nu_R$ et $\nu_R X$. 
Dans le NMSSM, $A_1$ peut être décomposé selon
\begin{equation}
\label{A1-defF}
A_1 = \cos \theta_A A_\mathrm{MSSM} + \sin \theta_A A_S \,.
\end{equation}
Les équation~(\ref{A1-defF}) et (\ref{model-expF}) montrent que la largeur de désintégration de $A_1$ en $\nu_L \nu_R$ dépend de $\cos \theta_A$, tandis que la largeur de désintégration en $\nu_R X$ dépend de $\sin \theta_A$. Il est alors possible de déduire les couplages entre le pseudo-scalaire le plus léger et différents fermions
\begin{align}
 A_1\tau_L\tau_R:&\frac{\imath m_\tau}{v} \tan \beta \cos \theta_A\,,\label{couplingTauF}\\
 A_1 t_L t_R:&-\frac{\imath m_t}{v \tan \beta} \cos \theta_A\,,\label{couplingTopF}\\
 A_1 \nu_L \nu_R:&-\frac{\imath m_D}{v \tan \beta} \cos \theta_A\,,\label{couplingNuLf}\\
 A_1 \nu_R X:&\frac{\imath M_R}{\sqrt{2} s}\sin\theta_A\,,\label{CouplingNuRf}\\
 A_1 X X:&\frac{i\mu_X}{\sqrt{2} s} \sin\theta_A\,,\label{CouplingXf}
\end{align}
où la différence de $\sqrt{2}$ entre les couplages~(\ref{CouplingNuRf}, \ref{CouplingXf}) et les autres provient d'une différence de $\sqrt{2}$ dans les définitions de $v$ et $s$ ($\langle S \rangle = s$ alors que $\langle H_u \rangle=v_u/\sqrt{2}$). Si les effets d'espace de phase sont négligés, alors les rapports d'embranchement de $A_1$ dans les modes de désintégrations invisibles normalisés aux modes visibles sont donnés par
\begin{eqnarray}
 \frac{\mathcal{B}\left(A_1 \rightarrow \nu_L \nu_R \right)}
{\mathcal{B}\left( A_1 \rightarrow 
f \bar f \right)+\mathcal{B}\left( A_1 \rightarrow c \bar{c} \right)} &\simeq& 
\frac{m_{D}^2}{m_{f}^2 \tan^4 \beta + m_{c}^2} \, , \label{a1nuNf}\\
 \frac{\mathcal{B}\left( A_1 \rightarrow \nu_R X \right)}
{\mathcal{B}\left( A_1 \rightarrow 
f \bar f \right)+\mathcal{B}\left( A_1 \rightarrow c \bar{c} \right)} &\simeq& 
\tan^2\theta_A \frac{M_{R}^2}{m_{f}^2 \tan^2 \beta + m_{c}^2 \cot^{2} \beta} 
\frac{v^2}{2 s^2} \, . \label{a1sNf}
\end{eqnarray}
Les modes de désintégrations visibles dominant sont $f\bar f\;(f=\mu, \tau, b)$ et $c \bar c$, ce dernier n'étant numériquement significatif que pour $m_{A_1} < 2 m_b$ et $\tan \beta$ petit.

L'analyse numérique donnée dans la table~\ref{table2} nous apprend que, si $\cos \theta_A$ est petit, le pseudo-scalaire le plus léger est dominé par sa composante singulet et se désintègre majoritairement dans des canaux invisibles. Cela a pour conséquence de relaxer les contraintes sur la masse et les couplages de $A_1$. Une autre conséquence importante est d'augmenter la largeur de désintégration invisible du scalaire le plus léger, via la chaîne $H_1\rightarrow A_1 A_1$, $A_1\rightarrow \nu_R X$.\\

Durant cette thèse, nous nous sommes intéressés à un mécanisme spécifique qui génère des masses et mélanges non-nuls pour les neutrinos. Ce mécanisme, très attractif car il peut simultanément avoir des couplages de Yukawa naturels et une échelle de nouvelle physique de l'ordre du TeV, est le seesaw inverse. Nous nous sommes concentrés tout particulièrement sur sa phénoménologie lorsqu'il est inclus dans le Modèle Standard, le MSSM et le NMSSM. Dans le MS, nous avons étudié son impact sur les tests de l'universalité leptonique et avons montré que des expériences comme NA62 et sa mesure du rapport $R_K$ peuvent déjà contraindre ce modèle. Dans le MSSM, nous nous sommes intéressés aux observables violant la saveur leptonique et à l'impact d'un sneutrino droit à l'échelle du TeV. Nous avons alors remarqué que les rapports d’embranchement cLFV peuvent atteindre la sensibilité expérimentale des expériences actuelles et futures. Enfin, dans le NMSSM, nous avons mis en évidence la possibilité d'obtenir une désintégration majoritairement invisible de $A_1$, ce qui relaxe les contraintes expérimentales sur la masse et les couplages de ce boson.

\end{document}